\documentclass[12pt]{article}

\usepackage{color}
\usepackage{amsmath,amssymb}
\usepackage{amscd}
\usepackage{graphicx} 
\usepackage{epstopdf}
\usepackage{latexsym}
\usepackage{cite}
\usepackage{ascmac}
\usepackage{cases}
\usepackage{mathrsfs}
\usepackage{multirow}
\usepackage[vcentermath]{youngtab}

\allowdisplaybreaks[1]

\newcommand{\eq}[1]{(\ref{#1})}
\newcommand{\nn}{\nonumber}
\newcommand{\ds}{\displaystyle}

\newcommand{\vev}[1]{\left\langle #1 \right\rangle}
\newcommand{\ket}[1]{\bigl|#1\bigr>}

\newcommand{\del}{\partial}

\newcommand{\asymeq}{\underset{asym}{\simeq}}

\newcommand{\bP}{\boldsymbol{P}}
\newcommand{\bQ}{\boldsymbol{Q}}

\DeclareMathOperator{\tr}{tr}

\DeclareMathOperator{\sgn}{sgn}

\DeclareMathOperator{\GCD}{gcd}

\newcommand{\nol}
  {
    \begin{array}{r}
    \raisebox{-2.1mm}{\mbox{\scriptsize{$\star$}}} \\ 
    \raisebox{3.1mm}{\mbox{\scriptsize{$\star$}}}
    \end{array}
   \!\,\!\!}
\newcommand{\nor}
  {\!\,\!\!
    \begin{array}{l}
    \raisebox{-2.1mm}{\mbox{\scriptsize{$\star$}}} \\ 
    \raisebox{3.1mm}{\mbox{\scriptsize{$\star$}}}
    \end{array}
   }

\renewcommand{\thefootnote}{\fnsymbol{footnote}}

\makeatletter
\@addtoreset{equation}{section}

\begin{document}


\begin{titlepage}
\thispagestyle{empty} 
\begin{flushright}
January 2016 \\
\vspace{.1cm}
arXiv:1601.nnnnn \\
\vspace{.1cm}
YITP-15-28 
\end{flushright}

\vspace{1.8cm}

\begin{center}
\noindent{\Large \textbf{
Wronskians, dualities and FZZT-Cardy branes
}}
\end{center}

\vspace{1cm}

\begin{center}
\noindent{Chuan-Tsung Chan\footnote{ctchan@thu.edu.tw}$^{,a}$, 
Hirotaka Irie\footnote{irie@yukawa.kyoto-u.ac.jp; hirotaka.irie@gmail.com}$^{,a,b}$, Benjamin Niedner\footnote{benjamin.niedner@physics.ox.ac.uk; benjamin.niedner@gmail.com}$^{,c}$
and Chi-Hsien Yeh\footnote{d95222008@ntu.edu.tw}$^{,d,e}$ }
\end{center}

\vspace{0.5cm}

\begin{center}
{\em
$^{a}$Department of Applied Physics, 
Tunghai University, Taichung 40704, Taiwan \\
\vspace{.3cm}
$^{b}$Yukawa Institute for Theoretical Physics, 
Kyoto University, Kyoto 606-8502, Japan \\
\vspace{.3cm}
$^{c}$Rudolf Peierls Centre for Theoretical Physics, \\
University of Oxford, Oxford OX1 3NP, UK \\
\vspace{.3cm}
$^{d}$National Center for Theoretical Sciences, \\
National Tsing-Hua University, Hsinchu 30013, Taiwan \\
\vspace{.3cm}
$^{e}$Department of Physics and Center for Advanced Study in Theoretical Sciences, \\
National Taiwan University, Taipei 10617, Taiwan, R.O.C.

}
\end{center}

\vspace{0.5cm}

\begin{abstract}
The resolvent operator plays a central role in matrix models. For instance, with utilizing the loop equation, all of the perturbative amplitudes including correlators, the free-energy and those of instanton corrections can be obtained from the spectral curve of the resolvent operator. However, at the level of non-perturbative completion, the resolvent operator is generally {\em not} sufficient to recover all the information from the loop equations. Therefore it is necessary to find {\em a sufficient set of operators} which provide the missing non-perturbative information. 

In this paper, we study {\em generalized Wronskians} of the Baker-Akhiezer systems as a manifestation of these new degrees of freedom. In particular, we derive their isomonodromy systems and then extend several spectral dualities to these systems. In addition, we discuss how these Wronskian operators are naturally aligned on the Kac table. Since they are consistent with the Seiberg-Shih relation, we propose that these new degrees of freedom can be identified as FZZT-Cardy branes in Liouville theory. This means that FZZT-Cardy branes are the bound states of {\em elemental FZZT branes} (i.e.~the twisted fermions) rather than the bound states of principal FZZT-brane (i.e.~the resolvent operator). 
\end{abstract}

\end{titlepage}

\newpage

\renewcommand{\thefootnote}{\arabic{footnote}}
\setcounter{footnote}{0}


\tableofcontents


\section{Introduction and summary} 
Matrix models are tractable toy models for studying non-perturbative aspects of string theory \cite{DSL,BIPZ,Mehta,KazakovSeries,Kostov1,Kostov2,Kostov3,
BDSS,BMPNonP,TwoMatString,GrossMigdal2,
DouglasGeneralizedKdV,TadaYamaguchiDouglas,
Moore,FIK,GinspargZinnJustin,Shenker,fkn1,fkn2,fkn3,
DVV,MultiCut,David0,MSS,David,EynardZinnJustin,DKK,
fy1,fy2,fy3,MultiCutUniversality,McGreevyVerlinde,
Martinec,KMS,AKK,KazakovKostov,MMSS,SeSh2,
HHIKKMT,SatoTsuchiya,IshibashiKurokiYamaguchi,
fis,fim,fi1,EynardOrantin,EynardMarino,irie2,
CISY1,CIY1,CIY2,CIY3,CIY4,CIY5}. Not only their perturbation theory 
is controllable to all orders \cite{EynardOrantin,EynardMarino} 
(including correlators, free-energy and those of instanton corrections) 
but also non-perturbative completions (such as their associated Stokes phenomena) 
have revealed intriguing and solvable features in these models \cite{CIY2,CIY3,CIY4,CIY5}. 
Such solvable aspects of matrix models have pushed forward 
our fundamental understanding of string theory beyond the perturbative perspectives. 
Although its beautiful framework is already considered to be ``well-established'', 
it is our belief that solvability of matrix models is now opening 
the next stage of development toward a non-perturbative definition of string theory. 

In this paper, we will explore a new horizon toward the fundamental understanding of such solvable aspects. 
The key question which we would like to address is {\em what are the non-perturbative degrees of freedom in string theory/matrix models?} We now have quantitative control over non-perturbative completions of matrix models (especially by the theory of  isomonodromy deformations \cite{ItsBook}). So we would like to investigate how the non-perturbative completions help us in identifying the fundamental degrees of freedom of matrix models and in extracting the physical consequence of their existence. 

Among various aspects of the solvability, we would like to consider the formulation of {\em the loop equations} \cite{BIPZ,KazakovSeries}. In this formulation, {\em the resolvent operator} has been the central player in the past studies \cite{BIPZ}. In particular, what we try to tackle is the folklore (or a commonsense) about the resolvent operator: the resolvent operator is believed to contain all the information of matrix models and is considered as the only fundamental degree of freedom. However, as it has been noticed in several non-perturbative analyses \cite{CIY2,CIY3,CIY4,CIY5}, the resolvent operator is not sufficient to {\em fully} describe the system non-perturbatively: {\em There are missing degrees of freedom} which are necessary in order to complete all the information of the matrix models. This new type of degrees of freedom is the main theme of this paper. In fact, as far as the authors know, the study on this aspect is almost missing in the literature but it should push forward our understanding of matrix models toward new regimes of non-perturbative study. 

The rest of this introduction is organized as follows: We first discuss {\em why the resolvent operator is not sufficient} in Section \ref{Introduction:WhyNotSufficient}. We then discuss {\em what are the missing degrees of freedom and how they are described} 
in Section \ref{Introduction:WhatIsTheOtherDOF}. 
In fact, the developments in non-critical string theory \cite{Polyakov}, especially of Liouville theory \cite{Polyakov,KPZ,DDK,DOZZ,Teschner,FZZT,ZZ,SeSh,KOPSS}, come out to provide an interesting clue. Our short answer to the question is that they are {\em Wronskians of the Baker-Akhiezer systems}. This Wronskian construction is motivated from {\em the FZZT-Cardy branes} \cite{SeSh} in Liouville theory. Accordingly, we will encounter three variants of Wronskians. In Section \ref{Introduction:WhatIsTheOtherDOF}, therefore, we briefly present {\em the whole picture of our proposal and the roles of these variants of Wronskians} as a summary of this paper. The organization of this paper is presented in Section \ref{Introduction:Organization}. 

Although this paper focuses only on the (double-scaled) two-matrix models which describe $(p,q)$ minimal string theory, the fundamental ideas can also be applied to any other kinds of matrix models. There is no big difference at least if they are described by the various types of topological recursions \cite{EynardOrantin,EynardMarino,QuantumSpectralCurves,KnotTheory,GeneralizedRecursions,GlobalRecursions,EynardKimura}.

\subsection{Why is the resolvent not sufficient? \label{Introduction:WhyNotSufficient}}

\subsubsection{In perturbation theory}
We first recall the folklore about the resolvent in perturbation theory, starting from 
the case of one-matrix models (say, of $X$) for simplicity (See e.g.~\cite{MatrixReview}). The loop equations are given by the Schwinger-Dyson equations about the correlators of the resolvent operator $\del_{x} \hat \phi(x)$, 
\begin{align}
\vev{ \prod_{j=1}^M \del_{x}\hat{\phi} (x_j)} = \dfrac{\ds \int dX e^{-N \tr V(X)} \prod_{j=1}^M \del_{x}\hat{\phi} (x_j)}{\ds \int dX e^{-N \tr V(X)} }, \qquad \del_x\hat{\phi}(x) \equiv  \tr \frac{1}{x-X}. 
\end{align}
The explicit form of the loop equations can be found in \cite{EynardOrantin}, for example. 
The loop equations are usually studied in the large $N$ (or topological genus) expansion, therefore one obtains recursive relations among the perturbative correlators, 
\begin{align}
\vev{ \prod_{j=1}^M \del_{x}\hat{\phi} (x_j)}_{\rm c} \asymeq \sum_{h=0}^\infty N^{2-2h-M}\vev{ \prod_{j=1}^M \del_{x}\hat{\phi} (x_j)}_{\rm c}^{(h)} \quad \bigl( N\to \infty\bigr).   \label{Eq:ColleratorsOfResolvents}
\end{align}
One of the best ways to understand these loop equations is the Eynard-Orantin topological recursion \cite{EynardOrantin}. In particular, all the perturbative amplitudes (Eq.~\eq{Eq:ColleratorsOfResolvents}) are obtained by solving the topological recursion (i.e.~loop equations), and the only input data is the spectral curve of the resolvent operator: 
\begin{align}
\mathscr S:\quad 
F(x,Q)=0,\qquad Q(x) \equiv \vev{\del_{x} \hat \phi(x)}^{(0)}_{\rm c}. 
\end{align}
From these amplitudes, it is also possible to obtain the perturbative free-energy to all orders \cite{EynardOrantin}. In view of this, the perturbative free-energy can be viewed as a function of spectral curve $\mathscr S$: 
\begin{align}
\mathcal F_{\rm pert}(\mathscr S) = \sum_{h=0}^{\infty} N^{2-2h} \mathcal F_{h}(\mathscr S)\qquad \Bigl(\mathcal F_{h}(\mathscr S) \equiv \vev{1}_{\rm c}^{(h)},\quad h\geq 0\Bigr). \label{Introduction:Eqn:PerturbativeFreeEnergy}
\end{align}
This means that the set of loop equations (i.e.~topological recursions) and the leading behavior of the resolvent operator (i.e.~spectral curves) are the only necessary information for reconstructing all the perturbative amplitudes.  Therefore, the folklore is completely proved in perturbation theory. 

\subsubsection{In non-perturbative completion \label{SubSubSection:IntroductionNonpertComp}}

From a non-perturbative perspective, on the other hand, such a consideration is inadequate. Shortly speaking, the resolvent operator is not sufficient at the level of {\em non-perturbative completions}. In order to see this, we first clarify what is ``non-perturbative loop equations'' and specify how matrix-model amplitudes are reconstructed from the non-perturbative loop equations and the resolvent operator. In contrast to the perturbative loop equations, the ``non-perturbative'' loop equations may be represented as ``string equations'' \cite{DSL} or ``Virasoro constraints'' \cite{fkn1,DVV}, and so on. Among them, we focus on ``Baker-Akhiezer system'' \cite{DouglasGeneralizedKdV,TadaYamaguchiDouglas} (or the related ``isomonodromy system'' \cite{Moore,FIK}) as an alternative description of the non-perturbative loop equations. 

For simplicity, we now focus on {\em two-matrix models} \cite{Mehta}, especially the models after the double scaling limit \cite{DSL,TwoMatString}. In fact, as the topological recursion shows us, the double scaling limit is not such an essential operation \cite{EynardOrantin}. Essentially, it means that the discussion within the two-matrix models mostly can be  applied to other kinds of matrix models. This fact can be also observed from the viewpoint of Baker-Akhiezer systems. 

In $(p,q)$ critical points of the two-matrix models, the corresponding Baker-Akhiezer system is the following systems of linear partial differential equations \cite{DouglasGeneralizedKdV,TadaYamaguchiDouglas}: 
\begin{align}
\zeta \psi(t;\zeta) = \bP(t;\del) \psi(t;\zeta),\qquad g\frac{\del}{\del \zeta}\psi(t;\zeta) = \bQ(t;\del) \psi(t;\zeta), \label{Eq:BAsystemFor11}
\end{align}
where $\del \equiv g \del_t$, and $\bP(t;\del)$ and $\bQ(t;\del)$ are the $p$-th and $q$-th order differential operators: 
\begin{align}
\bP(t;\del) = 2^{p-1} \del^p + \sum_{n=2}^p u_n(t) \, \del^{p-n},\qquad 
\bQ(t;\del) = \beta_{p,q}\Bigl[2^{q-1} \del^q + \sum_{n=2}^q v_n(t) \, \del^{q-n} \Bigr]. \label{Eq:Introduction:PQoperators}
\end{align}
The coefficients $\{u_n\}_{n=2}^p$ and $\{v_n\}_{n=2}^q$ are given by various derivatives of free-energy with respect to KP flow parameters (See \cite{fkn1,fkn2}). 
The integrability condition of these two differential equations gives rise to the Douglas equation \cite{DouglasGeneralizedKdV}, 
\begin{align}
\bigl[ \bP(t;\del),\bQ(t;\del)\bigr] = g 1, 
\end{align}
which results in the string equation of the matrix models. The string equations are differential equations about $\{u_n(t)\}_{n=2}^p$ and $\{v_n(t)\}_{n=2}^q$, and therefore the equations about free-energy. 
In this formulation, the degree of freedom of the resolvent operator is inherited from the determinant operator \cite{GrossMigdal2}: 
\begin{align}
\psi(t;\zeta) = \vev{e^{\hat \phi(t;\zeta)}} \quad \overset{\rm DSL}{\Longleftarrow} \quad \vev{\det \bigl(x-X\bigr)}_{n\times n} = \vev{e^{\tr \ln (x-X)}}_{n\times n}, \label{Eq:ResolentIntroduction1}
\end{align}
which is a solution to the linear differential equation system Eq.~\eq{Eq:BAsystemFor11}. 
Given this set-up, we would like to discuss {\em 1) how to understand ``non-perturbative completions'' from the Baker-Akhiezer systems and the determinant operator}, and {\em 2) how much information about non-perturbative completions can be extracted from the determinant operator. } One of the best ways to investigate these questions is to employ the theory of isomonodromic deformations \cite{JimboMiwaUeno} and the inverse monodromy approach \cite{DZmethod,ItsKapaev} (See e.g.~\cite{ItsBook} for details). The investigations of the questions with utilizing these formalism are given in \cite{CIY4,CIY5}. 

\paragraph{Non-perturbative completions and the Stokes data}

If one talks about non-perturbative ``corrections'', it means the non-perturbative corrections generated by instanton configurations. For instance, the non-perturbative corrections to the perturbative free-energy Eq.~\eq{Introduction:Eqn:PerturbativeFreeEnergy} are generally given by D-instantons \cite{David0,David,EynardZinnJustin}\cite{fy1,fy2,fy3}, and can be expanded to all-order \cite{EynardMarino} as follows: 
\begin{align} 
&\mathcal Z\bigl(g, \{\theta_a\}_{a=1}^{\mathfrak g}\bigr) \asymeq \sum_{n_1,n_2,\cdots,n_{\mathfrak g} \geq 0} \theta_1^{n_1} \times \cdots \times \theta_{\mathfrak g}^{n_{\mathfrak g}} \, \exp\Bigl[ \sum_{h=0}^\infty g^{2h-2}\mathcal F_{h}(\mathscr S_{n_1,\cdots,n_{\mathfrak g}})\Bigr] \qquad \bigl(g\to +0 \bigr) \nn\\
&\Leftrightarrow \qquad \mathcal F \bigl(g,\{ \theta_a \}_{a=1}^{\mathfrak g}\bigr) \asymeq \sum_{n=0}^\infty g^{2n-2} \mathcal F_n(\mathscr S) + \sum_{a=1}^{\mathfrak g} \theta_a\, g^{\gamma_a} \exp\Bigl[\sum_{n=0}^\infty g^{n-1} \mathcal F_n^{(a)}(\mathscr S)\Bigr] + O(\theta^2), \label{EqIntroExpansionOfFreeEnergy}
\end{align}
where $\{\mathscr S_{n_1,n_2,\cdots,n_{\mathfrak g}}\}$ are deformed spectral curves from the original spectral curve $\mathscr S$ with filling fractions $\{n_a/N\}_{a=1}^{\mathfrak g}$, and $\bigl\{\theta_a\bigr\}_{a=1}^{\mathfrak g}$ are the fugacities of the D-instantons (D-instanton fugacities). These perturbative/non-perturbative corrections are obtained, for example, by evaluating asymptotic expansion of the string equation about $g$ (or $t$) \cite{DSL,Shenker}. This form of fugacity-dependence is also shown in \cite{fy1,fy2,fy3} by the free-fermion formulation. The above all-order expansion of partition function (i.e.~of free-energy) is called non-perturbative partition function \cite{EynardMarino}. From the viewpoint of non-perturbative corrections, the D-instanton fugacities are arbitrary parameters, although the evaluation of matrix models only provides some particular values of fugacities \cite{HHIKKMT}. In this sense, in terms of the asymptotic expansion Eq.~\eq{EqIntroExpansionOfFreeEnergy}, we are totally blind to the physical distinction of the value of these fugacities. {\em An important theme of non-perturbative completion is to understand non-perturbative significance of the D-instanton fugacities with making a quantitative connection between the non-perturbative asymptotic expansions Eq.~\eq{EqIntroExpansionOfFreeEnergy} and the exact functions (i.e.~non-perturbative completions) such as matrix-model integrals.} This consideration takes us to the regime beyond the non-perturbative corrections. 

One way to complete the non-perturbative information associated with the non-perturbative corrections   Eq.~\eq{EqIntroExpansionOfFreeEnergy} is to make the connection between all the asymptotic expansion of the partition function in any direction of $g$, 
\begin{align}
g \to 0 \times e^{i\xi} \in \mathbb C\qquad \bigl( \xi \in \mathbb R \bigr) \qquad \Rightarrow \qquad \bigl\{\theta_a^{(\xi)} \bigr\}_{a=1}^{\mathfrak g}\qquad \bigl( \xi \in \mathbb R \bigr), 
\end{align}
as well as the expansion Eq.~\eq{EqIntroExpansionOfFreeEnergy} around the positive zero, $g\to + 0 \in \mathbb R$. That is, we specify all the connection rules of the asymptotic expansions. This is one of the themes to understand the behaviors of transcendental functions such as Painlev\'e equations (see e.g.~\cite{ItsBook}). In this consideration, the D-instanton fugacities $\bigl\{\theta_a\bigr\}_{a=1}^{\mathfrak g}$ themselves are not a good parametrization since the values of fugacities depend on the scheme of asymptotic expansion and also depend on the spectral curve that we start with. 

According to the theory of isomonodromy deformations, as its advantage is emphasized in \cite{ItsBook}, the integration constants of the string equation are parametrized by the Stokes data of the $p$ linearly independent solutions $\bigl\{\psi^{(j)}(t;\zeta)\bigr\}_{j=1}^p$ of the Baker-Akhiezer system Eq.~\eq{Eq:BAsystemFor11}. Given the integration constants (or the Stokes data), one can evaluate the connection formula of the partition function expanded along any direction of $g \to 0 \times e^{i\xi} \in \mathbb C$ ($\xi \in \mathbb R$). 
What is more, the set of all the possible Stokes data forms an algebraic variety of Stoke multipliers, each point of which possesses a clear geometric meaning (by the Riemann-Hilbert graph/spectral networks) on the spectral curve \cite{ItsBook}. In this sense, the Stokes data of the Baker-Akhiezer system Eq.~\eq{Eq:BAsystemFor11} defines the algebraic variety of Stokes multipliers which parametrizes non-perturbative completions of the asymptotic expansion Eq.~\eq{EqIntroExpansionOfFreeEnergy}. This algebraic variety is referred to as {\em the total solution space of the string equation} or {\em the space of general non-perturbative completions in the string equation}. Therefore, the study of Stokes phenomena associated with the Baker-Akhiezer systems is one of the nice ways to understand the significance of the D-instanton fugacities Eq.~\eq{EqIntroExpansionOfFreeEnergy} and non-perturbative completions of matrix models/string theory \cite{CIY2,CIY3,CIY4,CIY5}. 

\paragraph{Non-perturbative completions and the determinant (or resolvent) operator}

The question is then how much information can be extracted from the determinant (i.e.~the resolvent) operator. 
A naive consideration tells us that the determinant operator carries only a part of the information, because the determinant operator $\psi(t;\zeta)$ is only {\em one of the solutions}, 
\begin{align}
\psi(t;\zeta) = \psi^{(1)}(t;\zeta), \label{Eq:ResolentIntroduction2}
\end{align}
among the $p$ linearly independent solutions $\bigl\{\psi^{(j)}(t;\zeta)\bigr\}_{j=1}^p$ of the Baker-Akhiezer system Eq.~\eq{Eq:BAsystemFor11}. In other words, the Stokes data of other independent solutions, say $\psi^{(j)} (t;\zeta)$ ($j\neq 1$) can be freely adjusted in general, even if the Stokes data of the determinant operator $\psi(t;\zeta)$ is determined by the matrix models \cite{CIY2}.%
\footnote{In fact, there are also exceptional cases, where the determinant/resolvent operator completely constraints the Stokes data of other solutions. One example is given by the topological models such as $(p,1)$ minimal topological string theory, as shown in \cite{CIY5}. The topological $(p,1)$ series is identified with the higher-derivative extension of Airy system. It is well-known that, in the Airy system, the Biry function can be expressed by the Airy function,
\begin{align}
{\rm Bi}(\zeta) = e^{\frac{\pi}{6}i} {\rm Ai}(e^{\frac{2\pi}{3}i} \zeta) + 
e^{-\frac{\pi}{6}i} {\rm Ai}(e^{-\frac{2\pi}{3}i} \zeta), \qquad \frac{d^2 f(\zeta)}{d\zeta^2} -\zeta \, f(\zeta)=0 \qquad \Bigl(f(\zeta) = {\rm Ai}(\zeta),\, {\rm Bi}(\zeta) \Bigr), 
\end{align} 
even though these two functions are linearly independent solutions of the Airy equation. This is due to an accidental $\mathbb Z_{3}$-symmetry of the Airy equation. This accidental symmetry completely reduces the total solution space of the string equations to the single unique solution. This kind of conditions on the solution space is referred to as {\em environmental conditions}. Of course, this is a specialty of the topological series. In $(p,q)$ minimal string theory, such an accidental symmetry appears iff the theory is the topological $(p,1)$ series \cite{CIY5}. In this sense, topological models (such as Gaussian matrix-integrals) are too simple and not suitable as the playground of non-perturbative completions. }  
Consequently, there exist ambiguities in determining non-perturbative completions based only on the non-perturbative loop equation (i.e.~the Baker-Akhiezer system) and the determinant (or resolvent) operator. In order to judge the folklore, however, we should further discuss {\em what is the non-perturbative completions describing the matrix models} \cite{CIY2,CIY3,CIY4,CIY5}.

\subsubsection{In non-perturbative completions of matrix models}

The matrix models are defined by an integral over the set of $N\times N$ normal matrices $\mathscr C_X^{(N)}$ associated with a complex contour $\mathscr C_x$: 
\begin{align}
\mathcal Z(N;\mathscr C_x) = \int_{\mathscr C_X^{(N)}} dX \, e^{-N\tr V(X)} \qquad \bigl( N \sim g^{-1} \bigr).  \label{Introduction:Eqn:OneMatrixIntegrals}
\end{align}
Since $\mathscr C_X^{(N)}$ is the set of normal matrices, any elements $X \in \mathscr C_X^{(N)}$ can be diagonalized by a unitary matrix $U \in U(N)$, such as 
\begin{align}
{}^\forall X = U 
\begin{pmatrix}
 x_1 \cr 
 & x_2 \cr 
 & & \ddots \cr
 & & & x_N
\end{pmatrix} 
U^\dagger \in \mathscr C_X^{(N)}, \qquad 
{}^\exists U\in U(N). 
\end{align}
If we only consider gauge invariant (i.e.~$U(N)$-independent) observables in the matrix models, the integration Eq.~\eq{Introduction:Eqn:OneMatrixIntegrals} over matrix ensembles $\mathscr C_X^{(N)}$ then reduces to the integrals over the $N$ eigenvalues $\{x_j\}_{j=1}^N$. All of them lie in the same contour $x_j\in \mathscr C_x \subset \mathbb C$ ($j=1,2,\cdots,N$). On the other hand, the asymptotic behavior ($|x|\to \infty$) of the matrix potential $V(x)$ specifies particular sectors (angular regions) within which the matrix integral over any eigenvalues $\{x_j\}_{j=1}^N$ is finite. In this way, we can define a set of convergent contours $\{\gamma_a\}_{a=1}^{\mathfrak h}$ which connect various sectors, 
\begin{align}
  \qquad \int_{\gamma_a} dx e^{-NV(x)} < \infty \qquad \bigl( a=1,2,\cdots,\mathfrak h\bigr).  
\end{align}
The most general contour $\mathscr C_x$ can be expressed as a formal sum of these convergent contours $\{ \gamma_a\}_{a=1}^{\mathfrak h}$ with arbitrary weights $\{c_a\}_{a=1}^{\mathfrak h}$ of complex numbers, 
\begin{align}
\mathscr C_x = \sum_{a=1}^{\mathfrak h} c_a \gamma_a.   \label{Eqn:Introduction:SpaceOfContoursINOMM}
\end{align}
That is, $x_j \in \mathscr C_x$ means that 
\begin{align}
\int_{\mathscr C_x} dx_j \Bigl( \cdots \Bigr) = \sum_{a=1}^{\mathfrak h} c_a \int_{\gamma_a} dx_j \Bigl( \cdots \Bigr), \qquad j=1,2,\cdots,N. 
\end{align}
This is known as the general definition of matrix models, which shares the same string equation. 
The point is that the coefficients $\{c_a\}_{a=1}^{\mathfrak h}$ (i.e.~the choice of the contour $\mathscr C_x$) are related to the integration constants of the string equation \cite{David,HHIKKMT}. Therefore, the contour (i.e.~the weight coefficients $\{c_a\}_{a=1}^{\mathfrak h}$) provides a parametrization of ``the non-perturbative completions describing the matrix models''.

From this point of view, it can be seen that the space of contours in matrix models is generally much smaller than the total solution space of the string equation \cite{CIY4}. The dimensions are even smaller than that of the total solution space of the string equation: 
\begin{align}
\mathfrak g > \mathfrak h. 
\end{align}
Therefore, {\em not all the non-perturbative completions are realized within the matrix models}. By taking into account this fact, we continue the discussion about the folklore given in Section \ref{SubSubSection:IntroductionNonpertComp}. 

\paragraph{The completion space of one-matrix models v.s.~of two-matrix models}

{\em The space of non-perturbative completions describing matrix models} is investigated in \cite{CIY2,CIY3,CIY4,CIY5} within the isomonodromy formulation. The strategy there is to capture the matrix-model space from the asymptotic behavior of the determinant/resolvent operator in matrix models: it is observed that $k$ non-perturbative branch cuts of the resolvent operator Eq.~\eq{Eq:ResolentIntroduction1} should be formulated radially and symmetrically in the double-scaled $k$-cut matrix models. This consideration results in the multi-cut boundary condition on the Stokes data \cite{CIY2}. 

\begin{figure}[htbp]
\centering
\includegraphics[scale=1.045]{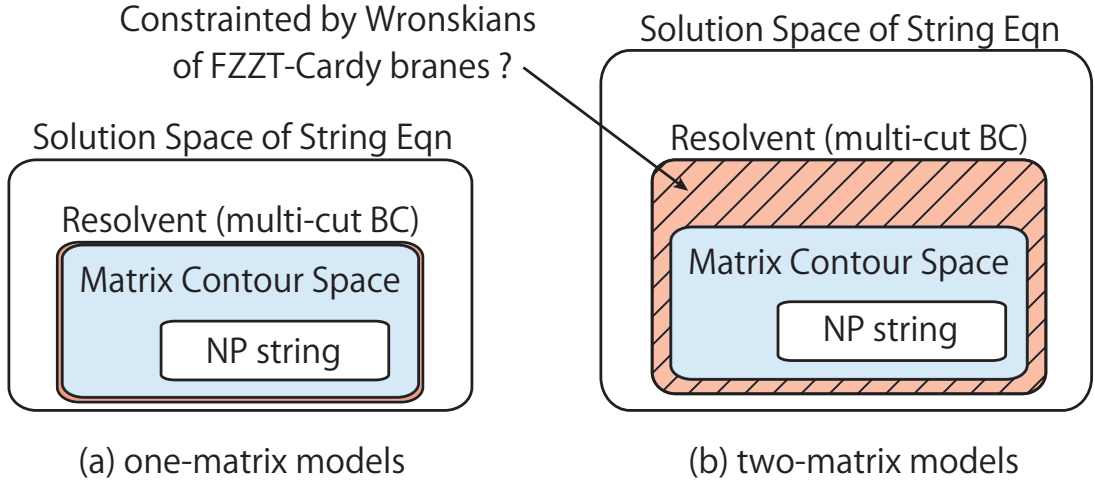}
\caption{\footnotesize Solution space of the string equations and of matrix models: (a) is of one-matrix models; (b) is of two-matrix models. The total solution space of the string equation is shown by the largest box. The next largest box is given by the multi-cut boundary condition (and with some environmental constraints) of the determinant/resolvent operator. The contour space of matrix models is shown as a box labeled by ``Matrix Contour Space''. The shaded region is the solution space which cannot be captured by the resolvent operator (and also by environmental constraints) itself. Such a region is governed by the other independent degrees of freedom, such as Wronskian operators discussed in this paper. The smallest box labeled by ``NP string'' is the completion space of non-perturbative string theory given by duality constraints. }
\label{Figure:SolutionSpaceOfStringEquations}
\end{figure}

In particular, in the one-matrix models, this multi-cut boundary condition of the determinant/resolvent operator correctly reduces the total solution space of the string equation to the space of the contours in the one-matrix models \cite{CIY4} (described in Figure \ref{Figure:SolutionSpaceOfStringEquations}-a). Therefore (in fact, as a speciality of one-matrix models) the resolvent operator represents the sufficient degree of freedom in the loop equations to specify the non-perturbative completions describing the one-matrix models. In this sense, the folklore about the resolvent operator is valid in the one-matrix models. In addition, it is also found in \cite{CIY3} that this multi-cut boundary condition turns out to be an analogy of the so-called ODE/IM correspondence of \cite{ODEIMCorrespondence}. Hence, the space of non-perturbative completions describing matrix models is generally a sub-manifold (of the solution space of the string equation) which is accompanied with quantum integrability associated with T-systems \cite{CIY3}. 

{\em The situation is however different in the case of two-matrix models}. The resolvent operator does not capture the full view of the contour space of two-matrix models \cite{CIY5}. This is already apparent in the $p-q$ dual counterparts of the one-matrix models (which are realized in two-matrix models). We refer this model as {\em the dual one-matrix models}, since this model is simple and also important. In \cite{CIY5}, a trial to specify the space of non-perturbative completions for the two-matrix models is investigated. In particular, not only the multi-cut boundary condition of the resolvent operator, but also {\em the existence of the perturbative $(p,q)$ minimal string theory as a meta-stable vacuum of the non-perturbative completions} is imposed. In the perturbative analysis, one might be tempted to consider that meta-stable vacua are just solutions of the saddle point equation. However, in non-perturbative completions, the contribution of the meta-stable vacuum to the path-integral is another issue. By the existence of such a contribution, we claim that such a meta-stable vacuum exists as a relevant vacuum in the non-perturbative completions. This should be imposed because we focus on the non-perturbative completions of perturbative $(p,q)$ minimal string theory, and also this situation is already realized in our double-scaled two-matrix models. This means that such a perturbative $(p,q)$ minimal string theory is included inside the string theory landscape of the non-perturbative completion \cite{CIY5}. With these conditions, one can reduce the total solution space to some tractable subspace which is close to the contour space of the two-matrix models. However, the resulting subspace still possesses apparently irrelevant solutions which cannot be associated with the contours of two-matrix-model integrals (described in Figure \ref{Figure:SolutionSpaceOfStringEquations}-b). Even though it can be qualitatively excluded in the simplest models such as the dual one-matrix models, the origin of associated quantitative conditions is still unclear and therefore cannot provide the sufficient non-perturbative information \cite{CIY5}. 

\paragraph{Two possible scenarios: environmental and/or new degrees of freedom}
There are basically two possibilities to solve this problem. One possibility is that we still miss some other environmental constraints associated with matrix models. In \cite{CIY5}, the existence of the perturbative $(p,q)$ minimal string theory as a meta-stable vacuum of the non-perturbative completions is imposed. Therefore, it might be still possible to find such ``a new environmental condition'' of matrix models to specify the contour space of the two-matrix models. If this is the case, the folklore is true itself, and the resolvent operator then may come back to retain its unique role in the non-perturbative completions of matrix models. 

The other possibility is that, even if such environmental constraints exist, it would not be enough to specify the system. In this case, the missing information should be supplied by {\em other degrees of freedom} which are independent from the resolvent operator. As discussed before, the resolvent operator only constitutes a part of the Stokes data in these systems. Therefore, if there exist other physical operators in matrix models, the remaining non-perturbative information is naturally provided by such physical operators.  Interestingly, it is also observed in \cite{CIY2} that there naturally exists an analog of the multi-cut boundary condition even for the other solutions of the Baker-Akhiezer systems, $\bigl\{\psi^{(j)}(t;\zeta)\bigr\}_{j=1}^p$, called {\em complementary boundary conditions} \cite{CIY2,CIY3}. Therefore, it is natural to suspect that these solutions $\bigl\{\psi^{(j)}(t;\zeta)\bigr\}_{j=1}^p$ of the Baker-Akhiezer systems also play an equally significant role as an independent object. This paper focuses on this second possibility. 

In either or both scenarios, such considerations should lead us to the complete and quantitative identification of the contour space of the two-matrix models from the total solution space of the string equation. The physical significance of the contour space in two-matrix models is suggested in \cite{CIY5} which is {\em duality constraints on string theory}. It was found in \cite{CIY5} that most of non-perturbative completions associated with the contour space of the one-matrix models do not have their counterpart in the contour space of the dual one-matrix models. That is, {\em the dual counterpart of a matrix contour does not exist in the dual one-matrix models}. The dimension of the completion space of the dual one-matrix models is generally much smaller than that of one-matrix models. This means that string duality is generally broken at the level of non-perturbative completion of string theory. Therefore, this discovered phenomenon can be used to further restrict the space of non-perturbative completion of string theory, which results in {\em non-perturbative string theory} \cite{CIY5}. In the discussion given in \cite{CIY5}, a qualitative argument is used to specify the contour space of the dual one-matrix models. Although it is apparent in the dual one-matrix models, it is quite non-trivial and far from our intuition for the general two-matrix models. Therefore, it is important to find out the quantitative form of constraint equations (either or both by environmental conditions and/or by the new physical degrees of freedom) which reduces the total solution space of the string equation to the contour space of the two-matrix models. This point would be elucidated by future investigations based on the results of this paper. This is one of the major motivations of the investigation presented in this paper.

\subsection{What are the other degrees of freedom? \label{Introduction:WhatIsTheOtherDOF}}

{\em What is then the physical meaning of the new degrees of freedom?} Our starting point for this study is along the proposal of \cite{CIY3}: such degrees of freedom would be {\em FZZT-Cardy branes} known in Liouville theory. In particular, we consider that the FZZT-Cardy branes are multi-body states of these $p$ independent solutions $\bigl\{\psi^{(j)}(t;\zeta)\bigr\}_{j=1}^p$. These multi-body states are given by {\em Wronskians} of the Baker-Akhiezer systems. Since the missing non-perturbative information is originally shared by the $p$ independent solutions $\bigl\{\psi^{(j)}(t;\zeta)\bigr\}_{j=1}^p$, such information is now inherited by the Wronskian functions, which correspond to the FZZT-Cardy branes. Accordingly, our basic proposal on the non-perturbative understanding of matrix models is that we should {\em replace the resolvent operator by the set of Wronskians living in the Kac table}
\begin{align}
\begin{array}{lcl}
\underline{\text{Determinant/Resolvent operators}} & & \underline{\text{Wronskian functions}} \cr
\ds \quad \vev{\det \bigl(x-X\bigr)} = \vev{e^{\tr \ln(x-X)}} & \qquad \longrightarrow \qquad  & \ds 
 \quad \Bigl\{ W^{(r,s)}_\varnothing(t;\zeta) \Bigr\}_{
\begin{subarray}{c}
1\leq r\leq p-1 \cr
1 \leq s \leq q-1
\end{subarray}
}
\end{array} 
\end{align}
as the independent degrees of freedom for non-perturbative description of matrix models. 

There are also several other discussions on the FZZT-Cardy branes in the literature \cite{BasuMartinec,Gesser,BourgineIshikiRim,AtkinWheater,AtkinZohren}. 
In particular, we comment on the differences from these constructions/proposals of the FZZT-Cardy branes \cite{BourgineIshikiRim,AtkinWheater,AtkinZohren}: 
\begin{itemize}
\item In \cite{BourgineIshikiRim}, the FZZT-Cardy brane amplitudes are constructed by the multi-point functions of resolvent operators. They adjusted the amplitudes based on the Seiberg-Shih relation \cite{SeSh} (reviewed in Section \ref{Subsubsection:SeibergShihRelations}) as a guide for the FZZT-Cardy brane. Their construction is essentially different from ours. One of the critical differences is about the non-perturbative degrees of freedom: while their computations are based on principal FZZT-branes, our Wronskian functions are constituted of elemental FZZT-branes. 
\item In \cite{AtkinWheater,AtkinZohren}, the FZZT-Cardy brane amplitudes are constructed using the spin-model interpretation of random surface. In particular, they proposed a specific form of new resolvent operators which describe the FZZT-Cardy branes. Since our construction is based on the $p$ independent solutions $\bigl\{\psi^{(j)}(t;\zeta)\bigr\}_{j=1}^p$ of the Baker-Akhiezer systems, such a concrete realization of new (corresponding) resolvent operators is missing. Therefore, it is interesting to further study a possible relation between these two approaches, and to find a systematic derivation and a precise dictionary. 
\end{itemize}

\subsubsection{Three variants of Wronskian functions and the Kac table}

As motivated from the FZZT-Cardy branes, three variants of Wronskian functions are considered in this paper. Since it comes from the Liouville theory, these Wronskian functions are naturally aligned in the Kac table: 
\begin{align}
\begin{array}{c}
 W^{(r,s)}_\varnothing(t;\zeta) \qquad \leftrightarrow \qquad \text{$(r,s)$ FZZT-Cardy brane}  \cr
\bigl( 1\leq r\leq p-1,\qquad 1\leq s \leq q-1\bigr), 
\end{array}
\end{align}
and are shown in Figure \ref{Figure:WronskiansOnKactable}. 
The three variants of Wronskian functions are summarized as follows: 

\begin{figure}[htbp]
\centering
\includegraphics[scale=1.045]{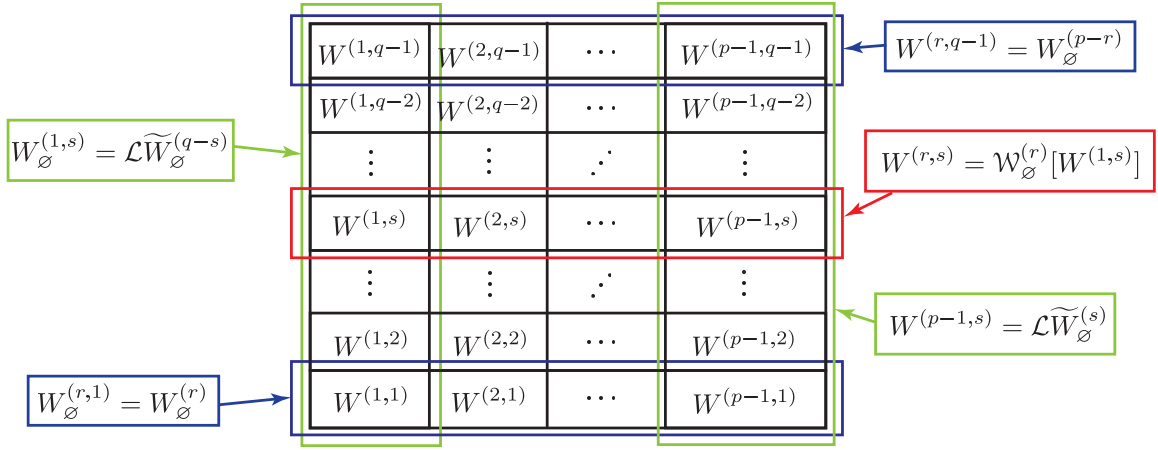}
\caption{\footnotesize Three variants of Wronskian functions $W^{(r,s)}(t;\zeta)$ and the Kac table}
\label{Figure:WronskiansOnKactable}
\end{figure}

\begin{itemize}
\item [\bf 1) ] \underline{\bf The rank-$r$ Wronskians of $\psi$: $W^{(r,1)}_{\varnothing}(t;\zeta) = W_\varnothing^{(r)}[\psi](t;\zeta)$} 
(in Section \ref{Section:FromSStoWronskians} and \ref{Section:WronskiansForR1FZZTCardyBranes})

 $W_\varnothing^{(r)}(t;\zeta)$ is the representative notation of rank-$r$ Wronskians of $\psi$:  
\begin{align}
W_\varnothing^{(r)}[\psi](t;\zeta) \qquad \leftrightarrow \qquad \bigl\{W^{[j_1,j_2,\cdots,j_r]}_\varnothing[\psi](t;\zeta)\bigr\}_{1\leq j_1<\cdots<j_r\leq p}, 
\end{align}
where 
\begin{align}
W^{[j_1,j_2,\cdots,j_r]}_\varnothing[\psi](t;\zeta) \equiv &  
\begin{vmatrix}
\del^{r-1}\psi^{(j_1)}(t;\zeta) & \del^{r-1}\psi^{(j_2)}(t;\zeta) & \cdots & \del^{r-1}\psi^{(j_r)}(t;\zeta) \cr
\vdots & \vdots & \ddots & \vdots \cr
\del \psi^{(j_1)}(t;\zeta) & \del \psi^{(j_2)}(t;\zeta) & \cdots & \del \psi^{(j_r)}(t;\zeta) \cr 
\psi^{(j_1)}(t;\zeta) & \psi^{(j_2)}(t;\zeta) & \cdots & \psi^{(j_r)}(t;\zeta) 
\end{vmatrix} \nn\\
=& \sum_{\sigma\in \mathfrak S_r} \sgn(\sigma)\, \del^{\sigma(1)-1} \psi^{(j_1)}(t;\zeta) \times \cdots \times 
\del^{\sigma(r)-1} \psi^{(j_r)}(t;\zeta). 
\end{align}
By construction, these Wronskian functions are consistent with the Liouville theory amplitudes (i.e.~the Seiberg-Shih relation \cite{SeSh}, reviewed in Section \ref{Subsubsection:SeibergShihRelations}) to all order in perturbative expansion. 

\item [\bf 2) ] \underline{\bf The Laplace-Fourier transformation of the rank-$s$ Wronskians of $\chi$:}\\  \underline{$W^{(1,s)}_{\varnothing}(t;\zeta) = \mathcal L \widetilde W_\varnothing^{(q-s)}(t;\zeta)$} (in Section \ref{Section:1SFZZTCardyBranesAndKacTable})

 $\mathcal L \widetilde W_\varnothing^{(r)}(t;\zeta)$ is the representative notation of the Laplace-Fourier transformation of rank-$s$ Wronskians of $\chi$: 
\begin{align}
\mathcal L \widetilde W_\varnothing^{(s)}(t;\zeta) \qquad \leftrightarrow \qquad \bigl\{\mathcal L W^{[l_1,l_2,\cdots,l_s]}_\varnothing[\chi](t;\zeta)\bigr\}_{1\leq l_1<\cdots<l_s\leq q}, 
\end{align}
where 
\begin{align}
\mathcal L W^{[l_1,l_2,\cdots,l_s]}_\varnothing[\chi](t;\zeta) \equiv \int_{\gamma} d\eta \, e^{\frac{\beta_{p,q}}{g} \zeta \eta} \, W^{[l_1,l_2,\cdots,l_s]}_\varnothing[\chi](t;\eta). \label{Eqn:Introduction:LaplaceTransformedW}
\end{align}
The wave-functions $\chi(t;\eta)\leftrightarrow \{\chi^{(l)}(t;\eta)\}_{l=1}^q$ are the $p-q$ dual counterpart of $\psi(t;\zeta)\leftrightarrow \{\psi^{(j)}(t;\zeta)\}_{j=1}^p$. These Laplace-Fourier transformed Wronskian functions come out, because of our proposal on generalization of spectral $p-q$ duality for all the Wronskian functions, which is given in Section \ref{SubSubSection:PQdualityLaplace}

\item [\bf 3) ] \underline{\bf The rank-$r$ Schur-differential Wronskians of $W^{(1,s)}$:}\\  \underline{$W^{(r,s)}_{\varnothing}(t;\zeta) = \mathscr W^{(r)}[W^{(1,s)}](t;\zeta)$} (in Section \ref{Section:FromSStoWronskians} and \ref{Section:TowardInteriorOfKacTable})

 $\mathscr W^{(r)}_\varnothing [W^{(1,s)}](t;\zeta)$ is the representative notation of the Schur-differential Wronskians (defined in Section \ref{Section:FromSStoWronskians} and \ref{Section:TowardInteriorOfKacTable}) of the Wronskian functions $W^{(1,s)}$: 
\begin{align}
&\mathscr W^{(r)}_{\varnothing} \bigl[W^{(1,s)}\bigr](t;\zeta) \equiv \sum_{
\pi \in \binom{rs}{s|s|\cdots|s} 
}
\sgn(\pi) \,  \mathcal S_{\pi^{(1)}_\varnothing}(\del)\overbrace{W^{(1,s)}_{\varnothing}(t;\zeta)}^{[j_1^{(1)},\cdots,j_s^{(1)}]}\times  \cdots \times  \mathcal S_{\pi^{(r)}_\varnothing}(\del)\overbrace{W^{(1,s)}_{\varnothing}(t;\zeta)}^{[j_1^{(r)},\cdots,j_s^{(r)}]}, \label{Introduction:Eqn:SDWronskians}
\end{align}
where $\binom{rs}{s|s|\cdots|s}$ and $\bigl\{ \pi_\varnothing^{(a)}\bigr\}_{a=1}^r$ are defined in Section \ref{Subsubsection:SchurdifferentialWronskians}, and $\mathcal S_\lambda(\del)$ is defined in Section \ref{Subsubsection:Schur-Derivatives} (more generally the normal-ordered Schur-differential Wronskains are required, see Section \ref{Subsection:HorizontalConnectionsAndMultipointCorrelators}). This new type of Wronskians is proposed again by respecting the Seiberg-Shih relation, discussed in Section \ref{Subsection:HorizontalConnectionsAndMultipointCorrelators}. 
\end{itemize}
In order to see how our proposal works, we analyze the differential equation systems (including isomonodromy systems) of these Wronskian functions in Section \ref{Section:WronskiansForR1FZZTCardyBranes} (Wronskian), in Section \ref{Section:1SFZZTCardyBranesAndKacTable} (Laplace-Fourier transformed) and in Section \ref{Section:TowardInteriorOfKacTable} (Schur-differential Wronskians). 

Importantly, several consistency checks are carried out in Section \ref{Section:ConsistencyOfDualities}. By passing the consistency checks, we obtain the Kac table labeling of these Wronskian functions.

\subsection{Organization of this paper \label{Introduction:Organization}}

The organization of this paper is as follows: 

\begin{itemize}
\item In Section \ref{Section:FromSStoWronskians}, we discuss how the Wronskian comes out in the construction: 
\begin{itemize}
\item Since our construction is based on the Seiberg-Shih relation, we first review it with the spectrum of D-branes in Liouville theory in Section \ref{Subsection:ReviewOfSeShrelation}. 
\item In Section \ref{Subsection:FromSStoWronskian}, we start to construct the first Wronskian functions from the Seiberg-Shih relation. Since we need elemental FZZT brane operators which represent different degrees of freedom from the resolvent operator, we employ the free-fermion formulation of \cite{fkn1,fkn2,fkn3,fy1,fy2,fy3}, which is directly related to the Baker-Akhiezer systems in Section \ref{Subsubsection:TwistedFermionAsElementalFZZTbranes}. 
\item Technical details on the Wronskians are discussed in Section \ref{Subsection:FurthermoreWronskians}. In addition, correlators of Wronskian functions are also studied in Section \ref{Subsection:MultipointCorrelatorsOfWronskians}, which will be a motivation of Schur-differential Wronskian Eq.~\eq{Introduction:Eqn:SDWronskians}. 
\end{itemize}
\end{itemize}
\begin{itemize}
\item In Section \ref{Section:SpectralCurvesOfFZZTCardyBranes}, we discuss the spectral curves of the $(r,s)$ FZZT-Cardy branes: 
\begin{itemize}
\item In Section \ref{Section:PQdualityLegendre}, we discuss the interpretation of the $p-q$ duality as the Legendre transformation on spectral curve. This can be used not only to obtain the spectral curve of general $(r,s)$ FZZT-Cardy branes in Section \ref{SubSection:SpectralCurveCardy}, but also to generalize the spectral $p-q$ duality to the general $(r,s)$ FZZT-Cardy branes in Section \ref{SubSubSection:PQdualityLaplace}. 
\item In order to carry out a consistency check of our Wronskians associated with the spectral curves, the spectral curves of $(r,s)$ FZZT-Cardy branes are evaluated from the Liouville-theory viewpoint in Section \ref{SubSection:SpectralCurveCardy}. 
\item Associated with the spectral curves of FZZT-Cardy branes, several issues on ZZ-Cardy branes are also discussed in Section \ref{Subsection:ZZCardyBranes}. 
\end{itemize}
\end{itemize}
\begin{itemize}
\item In Section \ref{Section:WronskiansForR1FZZTCardyBranes}, we discuss differential equations of the $(r,1)$ FZZT-Cardy branes: 
\begin{itemize}
\item As an extension of Baker-Akhiezer systems, Schur-differential equations of the $(r,1)$ FZZT-Cardy branes are discussed in Section \ref{Subsection:SchurDifferentialEqnR1} and the associated isomonodromy systems are derived in Section \ref{Subsection:IMSforR1FZZTCardyBranes}. The results of this analysis on isomonodromy systems are summarized in Appendix \ref{Section:ExampleOfIMS}. 
\item Section \ref{Subsection:ChargeConjugationR1Complex} is devoted to a new duality relation among the isomonodromy systems of $(r,1)$ FZZT-Cardy branes, called charge conjugation. The charge conjugation matrices are discussed in Section \ref{SubSubSection:ChargeConjugation}; the charge conjugation transformation of Wronskian functions is discussed in Section \ref{Subsubsection:ChargeConjuTransOfWronskian/Fermions}; the charge conjugation transformation is also interpreted as B\"acklund transformation of the string equation in Section \ref{Subsubsection:Backlundtransformation}. We also come back to the charge conjugation matrices in Section \ref{Subsection:GeneralChargeConjugation} with a generalization. 
\end{itemize}
\end{itemize}
\begin{itemize}
\item In Section \ref{Section:1SFZZTCardyBranesAndKacTable}, we discuss differential equations of the $(1,s)$ FZZT-Cardy branes: 
\begin{itemize}
\item For the Wronskian description of the $(1,s)$ FZZT-Cardy branes, the Legendre transformation of Section \ref{Section:PQdualityLegendre} is extended to spectral dualities of FZZT-Cardy branes in Section \ref{SubSubSection:PQdualityLaplace}, which is called generalized FZZT-Cardy branes. This gives rise to the Laplace-Fourier transformed Wronskians, Eq.~\eq{Eqn:Introduction:LaplaceTransformedW}. 
\item  With utilizing the generalized spectral $p-q$ duality, Schur-differential equations and isomonodromy systems of the $(1,s)$ FZZT-Cardy branes are analyzed in Section \ref{Subsection:SchurDifferentialIMSin1SFZZTCardyBranes}. The results of this analysis are summarized in Appendix \ref{Section:ExampleOfIMS}. 
\item The concept of general charge conjugation is given by the dual-space pairing of isomonodromy systems in Section \ref{Subsection:GeneralChargeConjugation}. Accordingly, the concept of dual charge conjugation is introduced. The dual charge conjugation matrices are summarized in Appendix \ref{Section:ExampleOfIMS} (under the subsection of ``Duality matrices''). 
\end{itemize}
\end{itemize}
\begin{itemize}
\item In Section \ref{Section:ConsistencyOfDualities}, we discuss consistency of dualities and construction of the Kac table: 
\begin{itemize}
\item In Section \ref{Subsection:ConsistencyOfChargeConjugations}, the consistency of charge conjugation and dual charge conjugation is discussed. The equivalence is shown by the existence of a gauge transformation among the isomonodromy systems. The resulting gauge transformation matrices are summarized in Appendix \ref{Section:ExampleOfIMS} (under the subsection of ``Duality matrices''). 
\item In Section \ref{Subsection:ConsistencyOfDualityWithReflectionRelations}, the role of the reflection relation on the Wronskian functions is discussed. We discuss  consistency relations associated with the reflection relation, and how it guarantees the consistent construction of the Kac table. 
\end{itemize}
\end{itemize}
\begin{itemize}
\item In Section \ref{Section:TowardInteriorOfKacTable}, we discuss construction of the general $(r,s)$ FZZT-Cardy branes: 
The motivation of Schur-differential Wronskians Eq.~\eq{Introduction:Eqn:SDWronskians} is discussed in Section \ref{Subsection:HorizontalConnectionsAndMultipointCorrelators} and the proposed form is given in Section \ref{SDWronskianFZZTCardyProposal}. The consistency check is shown for the example of the $(p,q)=(3,4)$-system in Section \ref{ConsistencyOfSDWronskiansProposal}. 
\item Section \ref{Section:ConclusionAndDiscussion} is devoted to conclusion and discussion. 
\item We have four appendices: 
\begin{itemize} 
\item Appendix A gives a brief summary of basic notations in two-matrix models. 
\item Appendix B lists the results of our analysis on isomonodromy systems of various FZZT-Cardy branes. 
\item Appendix C explains the calculations of the constraint equations over partitions/Young diagrams. 
\item Appendix D is a note on boundary entropy of FZZT-Cardy branes. 
\end{itemize}
\end{itemize}

\section{From the Seiberg-Shih relation to Wronskians \label{Section:FromSStoWronskians}}

In this section, we present some basic background of our construction, starting with Liouville field theory. We first review the spectrum of D-branes and the Seiberg-Shih relation in $(p,q)$ minimal string theory. We then discuss an interpretation of the Seiberg-Shih relation as multi-brane configurations, which naturally leads us to the Wronskian construction of the FZZT-Cardy branes \cite{CIY3}. The rest of this section is devoted to development of the basic techniques about Wronskian functions, which will be used throughout this paper.

\subsection{A review of the Seiberg-Shih relation \label{Subsection:ReviewOfSeShrelation}}
\subsubsection{The spectrum of D-branes}

D-branes in minimal string theory are classified into two categories: FZZT-Cardy branes and ZZ-Cardy branes; and, in addition, there exist the dual counterparts of these two types of branes \cite{SeSh}. These are summarized in Table~\ref{tab:brane}. 

\begin{table}[htb]
  \begin{center}
    \begin{tabular}{|c||c|c|c|} \hline
      Branes & Liouville field theory & $(p,q)$ minimal CFT & $bc$-ghost \\ \hline 
\parbox[c][0.9cm][c]{0cm}{} FZZT-Cardy 
&  $\left|\zeta\strut\right>_{\rm FZZT}\,\, (\zeta\in \mathbb C )$ \cite{FZZT} 
& \multirow{2}{*}{   \parbox[c][1.4cm][c]{0cm}{}
       $\bigl\{ \left|(r,s)\strut\right>_{\rm Cardy} \bigr\}^{qr-ps>0}_{
                \begin{subarray}{c}
                   1\leq r\leq p-1 \cr 
                   1\leq s\leq q-1
                \end{subarray}
            }$ \cite{CardyStates}}  
& \multirow{2}{*}{   \parbox[c][1.4cm][c]{0cm}{}
            $\left|{\rm gh}\strut\right>_{\rm I}$
       } \\ \cline{1-2}
     \parbox[c][0.9cm][c]{0cm}{} ZZ-Cardy 
     &  $\bigl\{\left|(m,n)\strut\right>_{\rm ZZ} \bigr\}_{m,n \in \mathbb N}$ \cite{ZZ}
     &  
     & \\ \hline \hline
\parbox[c][0.9cm][c]{0cm}{} dual FZZT-Cardy 
&  $\widetilde{\left|\eta\right>}_{\rm FZZT}\,\, (\eta\in \mathbb C )$ \cite{FZZT} 
& \multirow{2}{*}{ \parbox[c][1.4cm][c]{0cm}{}
      $\bigl\{ \widetilde{\left|(s,r)\right>}_{\rm Cardy} \bigr\}^{ps-qr>0}_{
               \begin{subarray}{c}
                 1\leq s\leq q-1 \cr 
                 1\leq r\leq p-1
               \end{subarray}
           }$ \cite{CardyStates}
    }  
& \multirow{2}{*}{   \parbox[c][1.4cm][c]{0cm}{}
            $\left|{\rm gh}\strut\right>_{\rm I}$
       } \\ \cline{1-2}
     \parbox[c][0.9cm][c]{0cm}{} dual ZZ-Cardy 
&  $\widetilde{\bigl\{\left|(n,m)\right>}_{\rm ZZ} \bigr\}_{n,m \in \mathbb N}$ \cite{ZZ}
&  
& \\ \hline
    \end{tabular}
        \caption{Branes in Liouville theory}
    \label{tab:brane}
  \end{center}
\end{table}

\begin{itemize}
\item $(p,q)$ minimal string theory is constructed by the tensor-product of Liouville field theory, $(p,q)$ minimal CFT and the conformal $bc$ ghost. Therefore, the D-brane boundary states are also given by their tensor-product \cite{SeSh}:
\begin{align}
\text{\underline{FZZT-Cardy brane:}}&&
\left|(r,s)\strut\right>_{\zeta} \equiv \left|\zeta\strut\right>_{\rm FZZT}\otimes \left|(r,s)\strut\right>_{\rm Cardy}\otimes \left|{\rm gh}\strut\right>_{\rm I},
\end{align}
and ZZ-Cardy branes are \cite{SeSh}:
\begin{align}
\text{\underline{ZZ-Cardy brane:}}&&
\left|(r,s)\strut\right>_{(m,n)} \equiv \left|(m,n)\strut\right>_{\rm ZZ}\otimes \left|(r,s)\strut\right>_{\rm Cardy}\otimes \left|{\rm gh}\strut\right>_{\rm I}. 
\end{align}
Their dual counterparts will be introduced later. 
\item These branes are aligned along {\em the Kac table}, which is uniquely determined by the following pair of indices $(r,s)$: 
\begin{align}
1\leq r\leq p-1,\quad 1\leq s\leq q-1, \qquad qr-ps>0. 
\end{align}
These indices originate from the Cardy states of $(p,q)$ minimal CFT which are associated with the representations of the Virasoro algebra \cite{CardyStates}. It is also useful to consider a periodic extension: 
\begin{align}
\left|(r,s) \strut\right>_{\rm Cardy} = \left|(r+p,s) \strut\right>_{\rm Cardy}  = \left|(r,s+q) \strut\right>_{\rm Cardy}. 
\end{align}
The restriction $qr-ps>0$ is then understood as a choice of the representative under {\em the reflection relation}: 
\begin{align}
\left|(r,s) \strut\right>_{\rm Cardy} = \left|(p-r,q-s) \strut\right>_{\rm Cardy}.  \label{Eqn:LiouvilleReflectionRelation}
\end{align}
Therefore, the $(r,s)$ labeling in FZZT-Cardy branes $\left|(r,s)\strut\right>_{\zeta}$ and ZZ-Cardy branes $\left|(r,s)\strut\right>_{(m,n)}$ also satisfy the same reflection relation. 
\item In addition, $(p,q)$ minimal string theory possesses $p-q$ duality which is generated by exchanging $p$ and $q$ of the $(p,q)$ index \cite{fkn3}: 
\begin{align}
(p,q) \quad \leftrightarrow \quad (q,p).
\end{align}
$(p,q)$ minimal CFT is self-dual under this duality transformation, but it is convenient to transpose the Kac table: 
\begin{align}
\left|(r,s)\strut\right>_{\rm Cardy} \quad \leftrightarrow \quad \widetilde{\left|(s,r)\right>}_{\rm Cardy} \quad \Bigl( = \left|(r,s)\strut\right>_{\rm Cardy} \Bigr). 
\end{align}
Therefore, nothing will change except for the Kac table labeling. 
\item On the other hand, Liouville field theory is invariant under a strong/weak self-duality about the Liouville coupling $b$ \cite{DOZZ,Teschner}: 
\begin{align}
b \quad \leftrightarrow \quad \frac{1}{b} \qquad \Bigl(b=\sqrt{\frac{p}{q}}\Bigr). 
\end{align}
In particular, FZZT-Cardy (and ZZ-Cardy) branes are mapped to their dual branes, which are called dual FZZT-Cardy (and dual ZZ-Cardy) branes \cite{SeSh}: 
\begin{align}
\text{\underline{dual FZZT-Cardy brane:}}&&
&\widetilde{\left|(s,r)\right>}_{\eta} \equiv \widetilde{\left|\eta\right>}_{\rm FZZT}\otimes \widetilde{\left|(s,r)\right>}_{\rm Cardy}\otimes \left|{\rm gh}\strut\right>_{\rm I}, \\
\text{\underline{dual ZZ-Cardy brane:}}&&
&\widetilde{\left|(s,r)\strut\right>}_{(n,m)} \equiv \widetilde{\left|(n,m)\right>}_{\rm ZZ}\otimes \widetilde{\left|(s,r)\right>}_{\rm Cardy}\otimes \left|{\rm gh}\strut\right>_{\rm I}.
\end{align}
In this sense, the full D-brane spectrum of $(p,q)$ minimal string theory includes both FZZT-Cardy (and ZZ-Cardy) branes and their dual FZZT-Cardy (and dual ZZ-Cardy) branes. 
\item In particular, within two-matrix models, the $p-q$ duality is introduced in \cite{fkn3} and understood as an exchange of the matrices $X$ and $Y$ (See \cite{CIY5}). It is one of the spectral dualities of \cite{BEH}. Therefore, this duality is also called spectral $p-q$ duality. Originally, the $(1,1)$-FZZT brane is introduced to describe the resolvent operator of matrix models \cite{FZZT}, and spectral $p-q$ duality suggests the following identification: 
\begin{align}
 \left|(1,1)\strut\right>_{\zeta} \quad \leftrightarrow \quad  \tr \ln (x-X),\qquad 
\widetilde{\left|(1,1)\right>}_{\eta} \quad \leftrightarrow \quad  \tr \ln(y-Y)
\end{align}
In this sense, one can understand that the $(1,1)$-FZZT brane (and also the $(1,1)$-dual FZZT brane) plays a central role in matrix models. 

\end{itemize}

Since there are several names available for these branes, these are also summarized in Table \ref{tab:branename}. FZZT-Cardy branes are originally called $(r,s)$-FZZT branes \cite{SeSh}. In particular, the $(1,1)$-FZZT brane is given a distinct name, called principal FZZT brane (or just FZZT brane). General $(r,s)$-FZZT branes are also called $(r,s)$-Cardy branes in order to emphasize the $(r,s)$ labeling of Cardy states in minimal CFT. The names of ZZ-Cardy branes are also the same. These are listed in Table \ref{tab:branename}. 

\begin{table}[htb]
  \begin{center}
    \begin{tabular}{|c||c|c|c|} \hline
      Branes & other names \\ \hline \hline
     \parbox[c][0.7cm][c]{0cm}{} $(1,1)$-FZZT-Cardy: $\left|(1,1)\strut\right>_{\zeta}$& principal FZZT-brane / FZZT-brane \\ \hline
     \parbox[c][0.7cm][c]{0cm}{} $(r,s)$-FZZT-Cardy: $\left|(r,s)\strut\right>_{\zeta}$& $(r,s)$-FZZT brane / $(r,s)$-Cardy brane  \\ \hline \hline
          \parbox[c][0.7cm][c]{0cm}{} $(1,1)$-$(m,n)$ ZZ-Cardy: $\left|(1,1)\strut\right>_{(m,n)}$& principal $(m,n)$ ZZ-brane / $(m,n)$ ZZ-brane \\ \hline
     \parbox[c][0.7cm][c]{0cm}{} $(r,s)$-$(m,n)$ ZZ-Cardy: $\left|(r,s)\strut\right>_{(m,n)}$& $(r,s)$-$(m,n)$ ZZ brane   \\ \hline
    \end{tabular}
        \caption{Other names of the branes}
    \label{tab:branename}
  \end{center}
\end{table}

In addition, in this paper, the $(r,s)$ labeling of FZZT-Cardy branes (and of ZZ-Cardy branes) is often specified by ``$(r,s)$-type''. Therefore, we also use the following names: 
\begin{align}
\text{$(r,s)$-FZZT-Cardy branes} \quad &= \quad \text{$(r,s)$-type FZZT-Cardy branes}, \nn\\
\text{$(r,s)$-$(m,n)$-ZZ-Cardy branes} \quad &= \quad \text{$(r,s)$-type $(m,n)$ ZZ-Cardy branes}. 
\end{align}

\subsubsection{Seiberg-Shih relation \label{Subsubsection:SeibergShihRelations}}

We first focus on the FZZT-Cardy brane. Among these types of FZZT-Cardy branes, Seiberg and Shih pointed out that the following relation holds up to BRST exact contributions \cite{SeSh}: 
\begin{align}
\left|(r,s)\strut\right>_\zeta &= \sum_{
\begin{subarray}{c}
k=-(r-1) \cr \text{step 2}
\end{subarray}
}^{(r-1)} \,\sum_{
\begin{subarray}{c}
l=-(s-1) \cr \text{step 2}
\end{subarray}
}^{(s-1)} \left|(1,1)\strut\right>_{\zeta_{k,l}(\zeta)}, \label{Eq:SeibergShihRelation}
\end{align}
where $\zeta_{k,l}(\zeta)$ are the functions of $\zeta$ which are obtained through the coordinate $\tau$, 
\begin{align}
\zeta = \sqrt{\mu} \cosh(p\tau),\qquad \zeta_{k,l} = \sqrt{\mu} \cosh\bigl(p \bigl[\tau + \pi i \bigl(\frac{k}{p} + \frac{l}{q} \bigr)\bigr]\bigr). 
\label{Eq:SeibergShihRelZeta}
\end{align}
Therefore, $\zeta_{k,l}$ indicates analytic continuations in the $\zeta$-coordinate. From this relation, the boundary states of $(r,s)$-FZZT-Cardy branes are given by a superposition of the boundary states $\left|(1,1)\strut\right>_\zeta$ (of the $(1,1)$-FZZT-Cardy brane, i.e.~FZZT-brane) with simultaneously making analytical continuations of $\zeta$. 

Similarly, any types of dual FZZT-Cardy branes can be expressed (up to BRST exact contributions) \cite{SeSh} as 
\begin{align}
\widetilde{\left|(s,r)\strut\right>}_\eta &= \sum_{
\begin{subarray}{c}
l=-(s-1) \cr \text{step 2}
\end{subarray}
}^{(s-1)} \sum_{
\begin{subarray}{c}
k=-(r-1) \cr \text{step 2}
\end{subarray}
}^{(r-1)} \widetilde{\left|(1,1)\strut\right>}_{\eta_{l,k}(\eta)}, \label{Eq:dualSeibergShihRelation}
\end{align}
where $\eta_{l,k}(\eta)$ are the functions of $\eta$ which are obtained by solving through the coordinate $\tau$, 
\begin{align}
\eta = \sqrt{\tilde \mu} \cosh(q\tau),\qquad \eta_{l,k} = \sqrt{\tilde \mu} \cosh\bigl(q \bigl[\tau + \pi i \bigl(\frac{l}{q} + \frac{k}{p} \bigr)\bigr]\bigr). 
\label{Eq:SeibergShihRelEta}
\end{align}
The dual cosmological constant $\tilde \mu$ and the usual cosmological constant $\mu$ are related \cite{DOZZ,FZZT,SeSh} as 
\begin{align}
\mu = {\tilde \mu}^{b^2}. 
\end{align}

As is mentioned in the introduction, there was a folklore about the resolvent operator, and it was believed that the resolvent is a necessary and sufficient operator to analyze matrix models. Because of this folklore as well as the Seiberg-Shih relation, it naturally suggests that the principal FZZT brane is the only independent degree of freedom; and the other FZZT-Cardy branes are dependent and are described by the principal one \cite{SeSh}. Nevertheless, there are many arguments both for and against this statement \cite{BasuMartinec, Gesser, BourgineIshikiRim, AtkinWheater, AtkinZohren}. 

On the other hand, our consideration about this issue is as follows \cite{CIY3}: As mentioned in the introduction, the boundary states obtained by analytic continuation of the FZZT-brane Boundary state $\left|(1,1)\strut\right>_\zeta$ are no longer attributed to the original FZZT-brane non-perturbatively. Therefore, the FZZT-Cardy branes constructed by the Seiberg-Shih relation should be  considered as independent degrees of freedom from the principal FZZT-brane. 

Of course, Liouville theory can describe minimal string theory only perturbatively, and therefore any non-perturbative definition of FZZT-Cardy branes is not well-established. However,the Kac table of the FZZT-Cardy branes (observed in perturbation theory) should survive even in non-perturbative completions. This is our main criterion for identification of the FZZT-Cardy branes and is discussed in Section \ref{Section:ConsistencyOfDualities} and Section \ref{Section:TowardInteriorOfKacTable}.

\subsection{From the Seiberg-Shih relation to Wronskians \label{Subsection:FromSStoWronskian}}
\subsubsection{Seiberg-Shih relation as multi-brane states \label{SubSection:SSrelAsMultiBranes}}

In order to understand the physical meaning of the Seiberg-Shih relation, we first recall the D-brane combinatorics \cite{DCombinatorics,fy3}. If a boundary state is given by a sum of two boundary states, 
\begin{align}
\left|B\right> = \left|B_1\right> + \left|B_2\right>, \label{Eq:BSsummation}
\end{align}
then the corresponding D-brane operators (or determinant operators in matrix models) $\Theta_B, \Theta_{B_1}$ and $\Theta_{B_2}$ satisfy the following product formula: 
\begin{align}
\Theta_B = \Theta_{B_1} \Theta_{B_2}.  \label{Eq:ThetaBB1B2}
\end{align}

Although this fact has been already seen in the literature, it is worth recalling the proof with utilizing single trace operators in matrix models. We first note that, given a determinant operator $\Theta$, the large $N$ expansion of the correlators (of $\Theta$) in terms of its corresponding single trace operator $\hat \phi$ (i.e.~$\Theta = e^{\hat \phi}$) follows the same pattern of the D-brane combinatorics: 
\begin{align}
\vev{e^{\hat \phi}} &= \exp\Bigl[\vev{e^{\hat \phi}-1}_{\rm c} \Bigr] = \exp\Bigl[\sum_{n=1}^\infty \frac{1}{n!} \vev{\hat \phi^n}_{\rm c} \Bigr] 
= \exp\Bigl[\sum_{n=1}^\infty \sum_{h=0}^\infty N^{2-2h-n}\frac{1}{n!} \vev{\hat \phi^n}_{\rm c}^{(h)} \Bigr] \nn\\
&=\exp\Bigl[ N \vev{\hat \phi\,}_{\rm c}^{(0)} + \frac{1}{2!} \vev{\hat \phi^2}_{\rm c}^{(0)} + N^{-1}\Bigl( \vev{\hat \phi\,}_{\rm c}^{(1)} +  \frac{1}{3!}\vev{\hat \phi^3}_{\rm c}^{(0)}\Bigr) + \cdots\Bigr], 
\end{align}
where $\vev{\cdots}_{\rm c}$ is the connected amplitude as usual. From this expansion formula, the relation of boundary states, Eq.~\eq{Eq:BSsummation}, can be translated into the relation of the corresponding single trace operators $\hat \phi_B,\hat \phi_{B_1}$ and $\hat \phi_{B_2}$ as 
\begin{align}
\hat \phi_B = \hat \phi_{B_1} + \hat \phi_{B_2} \qquad \Leftrightarrow \qquad \left|B\right> = \left|B_1\right> + \left|B_2\right>. 
\end{align}
Therefore, the relation among the corresponding determinant operators $\Theta_B, \Theta_{B_1}$ and $\Theta_{B_2}$) is shown as 
\begin{align}
\Theta_B = e^{\hat \phi_B} = e^{\hat \phi_{B_1} + \hat \phi_{B_2}} = e^{\hat \phi_{B_1}} e^{\hat \phi_{B_2}} = \Theta_{B_1} \Theta_{B_2}. 
\end{align}
This is the relation, Eq.~\eq{Eq:ThetaBB1B2}. 

Then, we come back to our case: From this consideration, we can interpret that the Seiberg-Shih relation implies that the FZZT-Cardy branes can be expressed by a multi-body state of more primitive degrees of freedom. In order to show this, we write the determinant operators corresponding to the boundary states $\left|(1,1)\strut\right>_{\zeta_{k,l}(\zeta)}$ and $\widetilde{\left|(1,1)\strut\right>}_{\eta_{l,k}(\eta)}$ in the following way: 
\begin{align}
\hat \psi_{\rm asym}\bigl(\zeta_{k,l}(\zeta)\bigr)  = e^{\hat \phi_{\rm asym}(\zeta_{k,l}(\zeta))} \qquad &\text{i.e.}\quad
\hat \phi_{\rm asym}\bigl(\zeta_{k,l}(\zeta)\bigr)\quad \Leftrightarrow \quad \left|(1,1)\strut\right>_{\zeta_{k,l}(\zeta)},  \\
\hat \chi_{\rm asym}\bigl(\eta_{l,k}(\eta)\bigr)  = e^{\hat {\tilde \phi}_{\rm asym}(\eta_{l,k}(\eta))} \qquad &\text{i.e.}\quad
\hat {\tilde \phi}_{\rm asym}\bigl(\eta_{l,k}(\eta)\bigr)\quad \Leftrightarrow \quad \widetilde{\left|(1,1)\strut\right>}_{\eta_{l,k}(\eta)}. 
\end{align}
The determinant operator of $(r,s)$-FZZT-Cardy brane and of its dual are then given by the multi-point operators of these D-brane operators: 
\begin{align}
\text{\underline{$(r,s)$-FZZT-Cardy brane:}}&&  \Theta^{(r,s)}_{\rm Cardy}(\zeta) = \prod_{
\begin{subarray}{c}
k=-(r-1) \cr \text{step 2}
\end{subarray}
}^{(r-1)} \,\prod_{
\begin{subarray}{c}
l=-(s-1) \cr \text{step 2}
\end{subarray}
}^{(s-1)} 
\hat \psi_{\rm asym}\bigl( \zeta_{k,l}(\zeta)\bigr)
\\
\text{\underline{$(s,r)$-dual FZZT-Cardy brane:}}&& \widetilde \Theta^{(s,r)}_{\rm Cardy}(\eta)
=\prod_{
\begin{subarray}{c}
l=-(s-1) \cr \text{step 2}
\end{subarray}
}^{(s-1)} \prod_{
\begin{subarray}{c}
k=-(r-1) \cr \text{step 2}
\end{subarray}
}^{(r-1)}
\hat \chi_{\rm asym}\bigl(\eta_{l,k}(\eta)\bigr)
\end{align}
Note that we put ``asym'' in $\hat \psi_{\rm asym}(\zeta)$ and $\hat \phi_{\rm asym}(\zeta)$, in order to emphasize that these operators are always defined by asymptotic expansions around $\zeta\to\infty$, which means ``within perturbation theory.'' What we are going to investigate now is how to provide their non-perturbative realization which does not depend on the specific choice of spectral curves. 

\subsubsection{Twisted fermions as elemental FZZT branes \label{Subsubsection:TwistedFermionAsElementalFZZTbranes}}
In Section \ref{SubSection:SSrelAsMultiBranes}, we have performed the analytic continuations of the asymptotic expansion $\hat \psi_{\rm asym}(\zeta)$ around $\zeta\to\infty$. This operation reminds us of the introduction of twisted free fermions in minimal string theory \cite{fkn1,fkn2,fkn3,fy1,fy2,fy3}. The $p$-th twisted fermions $\bigl\{ \hat \psi^{(j)}(\zeta)\bigr\}_{j=1}^p$ are introduced as 
\begin{align}
\hat \psi^{(j)}(\zeta) = \hat \psi_{\rm asym}(e^{-2\pi i (j-1)}\zeta)\qquad \bigl(j\in \mathbb Z/p\mathbb Z\bigr). \label{Eq:pTwistedFermion}
\end{align}
In the same way the dual $q$-th twisted fermions are introduced as 
\begin{align}
\hat \chi^{(l)}(\eta) = \hat \chi_{\rm asym}(e^{-2\pi i (l-1)}\eta)\qquad \bigl(l\in \mathbb Z/q\mathbb Z\bigr). \label{Eq:qTwistedFermion}
\end{align}
These fermion operators are the determinant (i.e.~D-brane) operators associated with the $p$ independent solutions (or $q$ independent solutions for the dual side) of the Baker-Akhiezer system: 
\begin{align}
\zeta \psi^{(j)}(t;\zeta) = \bP(t;\del) \psi^{(j)}(t;\zeta),\qquad g\frac{\del}{\del\zeta} \psi^{(j)}(t;\zeta) = \bQ(t;\del) \psi^{(j)}(t;\zeta), \label{Eqn:Baker-AkhiezerSystemsSection222psi}\\
\eta \chi^{(l)}(t;\eta) = \widetilde{\bP}(t;\del) \chi^{(l)}(t;\eta),\qquad g\frac{\del}{\del\eta} \chi^{(l)}(t;\eta) = \widetilde \bQ(t;\del) \chi^{(l)}(t;\eta),\label{Eqn:Baker-AkhiezerSystemsSection222chi}
\end{align}
where 
\begin{align}
\psi^{(j)}(t;\zeta) = \vev{\hat \psi^{(j)}(\zeta)}_t\quad \bigl(j\in \mathbb Z/p\mathbb Z\bigr),\qquad 
\chi^{(l)}(t;\eta) = \vev{\hat \chi^{(l)}(\eta)\rule{0pt}{2.5ex}}_t\quad \bigl(l\in \mathbb Z/q\mathbb Z\bigr), \label{Eq:PsiChiExactSolutions}
\end{align}
with the associated Lax operators: 
\begin{align}
\widetilde \bP(t;\del) = \frac{(-1)^{q}}{\beta_{p,q}} \bQ^{\rm T}(t;\del),\qquad \widetilde \bQ(t;\del) = (-1)^{q}\beta_{p,q} \bP^{\rm T}(t;\del). 
\label{Eqn:PQdualityInLaxOperators}
\end{align}
Basics of these Baker-Akhiezer systems and of these Lax operators are summarized in Appendix \ref{Appendix:Two-matrixModels} (See also \cite{CIY5} for reference therein). 
Here ``$t$'' of $\vev{\cdots\strut}_t$ denotes the background (i.e.~KP flows) of minimal string theory (See \cite{fy1,fy2,fy3}), and corresponds to ``$n$'' of $\vev{\cdots\strut}_{n\times n}$ in finite $N$ matrix models (See Appendix \ref{Appendix:Two-matrixModels}). 

In terms of the twisted free fermions, the statement about mutual independence is stated as follows: The twisted fermions are analytically continued under asymptotic expansion around $\zeta\to \infty$ as shown in Eq.~\eq{Eq:pTwistedFermion} and Eq.~\eq{Eq:qTwistedFermion}. On the other hand, analytic continuations of the exact solutions Eq.~\eq{Eq:PsiChiExactSolutions} (without using any asymptotic expansions) cannot be connected to each other, due to the Stokes phenomena, 
\begin{align}
\psi^{(j)}(t; e^{-2\pi i}\zeta) \neq \psi^{(j+1)}(t; \zeta). 
\end{align}
In this sense, these twisted fermions carry their independent degree of freedom. In the following, these twisted fermions are called {\em elemental FZZT-branes}, since these are a sort of basic elements for the FZZT-Cardy branes and each elemental brane possesses each individual degree of freedom. 

We should note that, according to the Seiberg-Shih relation, not all the FZZT-Cardy branes can be simply represented by the twisted fermions. It is because some of the analytic continuations (Eq.~\eq{Eq:SeibergShihRelZeta} and Eq.~\eq{Eq:SeibergShihRelEta}) depend on the spectral curves. Among these branes, only $(r,1)$-type FZZT-Cardy branes $\Theta_{\rm Cardy}^{(r,1)}(\zeta)$ (and $(s,1)$-type dual FZZT-Cardy branes $\widetilde \Theta_{\rm Cardy}^{(s,1)}(\eta)$) are defined without knowing the details of spectral curves. It is because the relevant analytic continuations ($\bigl\{\zeta_{k,0} \bigr\}_{k}$ and $\bigl\{ \eta_{l,0} \bigr\}_l$) are just rotating the coordinate $\zeta$ (and $\eta$) around the asymptotic infinity: 
\begin{align}
\left\{
\begin{array}{l}
\zeta_{k,0} = \sqrt{\mu} \cosh \bigl(p\bigl[\tau + \frac{\pi i k}{p}\bigl]\bigr) = e^{\pi i k} \zeta, \cr
\qquad \quad k= -(r-1), -(r-1)+2, \cdots, (r-1)-2, (r-1) \cr
\eta_{l,0} = \sqrt{\tilde \mu} \cosh \bigl(q\bigl[\tau + \frac{\pi i l}{q}\bigl]\bigr) = e^{\pi i l} \eta \cr
 \qquad \quad l = -(s-1), -(s-1)+2,\cdots, (s-1)-2, (s-1)
\end{array}
\right.. 
\end{align}
If $k$ (or $l$) is an even integer (i.e.~$r$ or $s\in 2 \mathbb Z+1$), we consider the asymptotic expansion around the positive real axis ($\zeta\to +\infty\in \mathbb R$); if $k$ (or $l$) is an odd integer (i.e.~$r$ or $s\in 2 \mathbb Z$), we consider the expansion around the negative real axis ($\zeta\to -\infty\in \mathbb R$).

In order to adjust the twisted free-fermion formalism to this suggested form, we introduce the twisted fermions with half-integer index, 
\begin{align}
\psi^{(j+\frac{1}{2})}(\zeta) &\equiv \psi^{(j)}(e^{-\pi i}\zeta) \qquad \Bigl(j\in  \mathbb Z/p\mathbb Z\Bigr), \\
\chi^{(l+\frac{1}{2})}(\eta) &\equiv \chi^{(l)}(e^{-\pi i}\eta) \qquad \Bigl(l\in  \mathbb Z/q\mathbb Z\Bigr). 
\end{align}
Note that this does not introduce any new fermion degrees of freedom. Therefore, the following two sets of wave functions equivalently form the complete set of solutions to the linear differential equations: 
\begin{align}
\bigl\{ \hat  \psi^{(j)}(\zeta) \bigr\}_{j=1}^{p}\, \leftrightarrow \, \bigl\{ \hat  \psi^{(j+\frac{1}{2})}(\zeta) \bigr\}_{a=1}^{p};\qquad 
\bigl\{ \hat  \chi^{(l)}(\eta) \bigr\}_{j=1}^{q}\, \leftrightarrow \, \bigl\{ \hat  \chi^{(l+\frac{1}{2})}(\eta) \bigr\}_{a=1}^{q}, 
\end{align}
and therefore we obtain the free-fermion realization of the $(r,1)$-type FZZT-Cardy branes and of the $(s,1)$-type dual FZZT-Cardy branes: 
\begin{align}
\Theta_{\rm Cardy}^{(r,1)} (\zeta)
=& \prod_{
\begin{subarray}{c}
k=-(r-1) \cr \text{step 2}
\end{subarray}
}^{(r-1)}
\hat \psi_{\rm asym}(e^{\pi i k}\zeta) 
= \prod_{
\begin{subarray}{c}
k=-(r-1) \cr \text{step 2}
\end{subarray}
}^{(r-1)}
\hat \psi^{(1+\frac{k}{2})}(\zeta), \label{Eqn:R1FZZTCardyFFrealization}\\
\widetilde \Theta_{\rm Cardy}^{(s,1)} (\eta)
=& \prod_{
\begin{subarray}{c}
l=-(s-1) \cr \text{step 2}
\end{subarray}
}^{(s-1)}
\hat \chi_{\rm asym}(e^{\pi i l}\eta) 
= \prod_{
\begin{subarray}{c}
l=-(s-1) \cr \text{step 2}
\end{subarray}
}^{(s-1)}
\hat \chi^{(1+\frac{l}{2})}(\eta). \label{Eqn:S1dualFZZTCardyFFrealization}
\end{align}

\subsubsection{Multi-point correlators as Wronskians \label{Subsubsection:MultipointCorrelatorsAsWronskians}}

Next, we rephrase the free-fermion realization Eq.~\eq{Eqn:R1FZZTCardyFFrealization} and Eq.~\eq{Eqn:S1dualFZZTCardyFFrealization} by using ``matrix-model'' amplitudes or wave functions, that is, by using {\em Wronskians}. 
In practice, we show the following relations: 
\begin{align}
\vev{\prod_{a=1}^n \hat \psi^{(j_a)}(\zeta)}_t =&  
\begin{vmatrix}
\del^{n-1}\psi^{(j_1)}(t;\zeta) & \del^{n-1}\psi^{(j_2)}(t;\zeta) & \cdots & \del^{n-1}\psi^{(j_n)}(t;\zeta) \cr
\vdots & \vdots & \ddots & \vdots \cr
\del \psi^{(j_1)}(t;\zeta) & \del \psi^{(j_2)}(t;\zeta) & \cdots & \del \psi^{(j_n)}(t;\zeta) \cr 
\psi^{(j_1)}(t;\zeta) & \psi^{(j_2)}(t;\zeta) & \cdots & \psi^{(j_n)}(t;\zeta) 
\end{vmatrix}
\nn\\
=& \underset{1\leq a,b\leq n}{\det}\Bigl[ \del^{n-a} \psi^{(j_b)}(t;\zeta) \Bigr] \equiv  W^{[j_1,j_2,\cdots,j_n]}_\varnothing[\psi](t;\zeta),
\end{align}
where $j_a \neq j_b \, (a\neq b)$. The dual side also satisfies the same relation: 
\begin{align}
\vev{\prod_{a=1}^n \hat \chi^{(l_a)}(\eta)}_t =&  
\begin{vmatrix}
\del^{n-1}\chi^{(l_1)}(t;\eta) & \del^{n-1}\chi^{(l_2)}(t;\eta) & \cdots & \del^{n-1}\chi^{(l_n)}(t;\eta) \cr
\vdots & \vdots & \ddots & \vdots \cr
\del \chi^{(l_1)}(t;\eta) & \del \chi^{(l_2)}(t;\eta) & \cdots & \del \chi^{(l_n)}(t;\eta) \cr 
\chi^{(l_1)}(t;\eta) & \chi^{(l_2)}(t;\eta) & \cdots & \chi^{(l_n)}(t;\eta) 
\end{vmatrix}
\nn\\
=& \underset{1\leq a,b\leq n}{\det}\Bigl[ \del^{n-a} \chi^{(l_b)}(t;\eta) \Bigr] = W^{[l_1,l_2,\cdots,l_n]}_\varnothing[\chi](t;\eta),
\end{align}
where $l_a \neq l_b \, (a\neq b)$. A more general definition will be mentioned later.%
\footnote{Note that the signature associated with the ordering of the product symbol, $\prod$, is fixed by defining
\begin{align}
\prod_{a=1}^M \hat f_a = \hat f_1 \times \hat f_2 \times \cdots \times  \hat f_M
\end{align}
}

We here consider the following more general $M$-point formula ($M\equiv\sum_{j=1}^p n_j$): 
\begin{align}
\vev{\prod_{j=1}^p\Biggl[\prod_{a=1}^{n_j} \hat \psi^{(j)}(\zeta_{a}^{(j)})\Biggr] }_t = \frac{ W^{[1^{n_1},2^{n_2},\cdots,p^{n_p}]}_\varnothing [\psi](t;\cup_{j=1}^p\{\zeta_a^{(j)}\}_{a=1}^{n_j})}{\ds \prod_{j=1}^p  \Delta^{(n_j)}\bigl(\bigl\{\zeta_a^{(j)}\bigr\}_{a=1}^{n_j}\bigr)},
\label{WronskianFormulaGeneral}
\end{align}
where $\Delta^{(n)}(\zeta)$ is the Van der Monde determinant, 
\begin{align}
\Delta^{(n_j)}\bigl(\bigl\{\zeta_a^{(j)}\bigr\}_{a=1}^{n_j}\bigr)\equiv \prod_{1\leq a <b\leq n_j} \bigl(\zeta_a^{(j)}-\zeta_b^{(j)}\bigr).
\end{align}
We have also used the following abbreviation:  
\begin{align}
[1^{n_1},2^{n_2},\cdots,p^{n_p}] &= [\underbrace{1,\cdots,1}_{n_1},\underbrace{2,\cdots,2}_{n_2},\cdots,\underbrace{p,\cdots,p}_{n_p}], \\
(t;\cup_{j=1}^p\{\zeta_a^{(j)}\}_{a=1}^{n_j}) &= (t;\underbrace{\zeta_1^{(1)},\cdots,\zeta_{n_1}^{(1)}}_{n_1}, \underbrace{\zeta_1^{(2)},\cdots,\zeta_{n_2}^{(2)}}_{n_2},\cdots,\underbrace{\zeta_1^{(p)},\cdots,\zeta_{n_p}^{(p)}}_{n_p}). 
\end{align}
Note that the coordinates $\{\zeta_a^{(j)}\}_{a=1}^{n_j}$ ($j=1,2,\cdots,p$) of wave-functions $\bigl\{\psi^{(j)}(\zeta_a^{(j)})\bigr\}_{a=1}^{n_j}$ ($j=1,2,\cdots,p$) are generally different from each other. 

The Van der Monde determinants in the denominator of  Eq.~\eq{WronskianFormulaGeneral} come from the definition of correlators in the free-fermion formalism \cite{fy1,fy2,fy3}, that is, the normal ordering of the fermions. More concretely, by regarding the determinant operators $\bigl\{\hat \psi^{(j)}(\zeta)\bigr\}_{j=1}^p$ as fermion operators, and by temporarily presenting the normal orderings of the free-fermion correlators, the matrix model correlators and free-fermion correlators are related as follows \cite{fy1,fy2,fy3}: 
\begin{align}
\vev{\prod_{a=1}^{n} \hat \psi^{(j_a)}(\zeta_{a}) }_t  = \vev{:\! \prod_{a=1}^{n} \hat \psi^{(j_a)}(\zeta_{a})\!: }_t^{\rm (FF)}. 
\end{align}
Accordingly, the Van der Monde determinants appear in removing the normal ordering \cite{fy1,fy2,fy3}: 
\begin{align}
\vev{:\!\! \prod_{j=1}^p\Biggl[\prod_{a=1}^{n_j} \hat \psi^{(j)}(\zeta_{a}^{(j)})\Biggr] \!\!: }_t^{\rm (FF)} = \frac{ \vev{\prod_{j=1}^p\Biggl[\prod_{a=1}^{n_j} :\!\hat \psi^{(j)}(\zeta_{a}^{(j)}) \!: \Biggr] }_t^{\rm (FF)} }{\ds \prod_{j=1}^p  \Delta^{(n_j)}\bigl(\bigl\{\zeta_a^{(j)}\bigr\}_{a=1}^{n_j}\bigr) }. 
\end{align}
Here, in order to emphasize the normal ordering, we intentionally express $\hat \psi^{(j)}(\zeta_{a}^{(j)})$ as $:\! \hat \psi^{(j)}(\zeta_{a}^{(j)}) \!:$. 
Therefore, the formula Eq.~\eq{WronskianFormulaGeneral}, which we are going to show, becomes the following: 
\begin{align}
\vev{\prod_{j=1}^p\Biggl[\prod_{a=1}^{n_j} :\!\hat \psi^{(j)}(\zeta_{a}^{(j)}) \!: \Biggr] }_t^{\rm (FF)} =  W^{[1^{n_1},2^{n_2},\cdots,p^{n_p}]}_\varnothing [\psi](t;\cup_{j=1}^p\{\zeta_a^{(j)}\}_{a=1}^{n_j}).  \label{Eq:WronskianFormulaPreGeneralAsym}
\end{align}
Up to this formula, we have not used any asymptotic expansion. 

By taking the asymptotic expansions of these fermions (i.e.~Eq.~\eq{Eq:pTwistedFermion}), the expression Eq.~\eq{Eq:WronskianFormulaPreGeneralAsym} is (perturbatively) equivalent to the following correlators: 
\begin{align}
\vev{\prod_{a=1}^{M} :\! \hat \psi^{(1)}(\zeta_{a}) \!:  }_t^{\rm (FF)} = W^{[1^{M}]}_\varnothing [\psi](t;\{\zeta_a\}_{a=1}^{M}). \label{Eq:WronskianFormulaPre}
\end{align}
Therefore, we now focus on this formula Eq.~\eq{Eq:WronskianFormulaPre}. Since this correlator is given by the standard determinant operators $\det(x-X)$, the exact formula is already known \cite{Morozov}, which is expressed by difference Wronskians of the orthogonal polynomials of the matrix models: 
\begin{align}
\vev{\prod_{a=1}^M \det (x_a-X)}_{n\times n} &= \frac{
\begin{vmatrix}
\alpha_n(x_1) & \alpha_n(x_2) & \cdots & \alpha_n(x_M) \cr
\alpha_{n+1}(x_1) & \alpha_{n+1}(x_2) & \cdots & \alpha_{n+1}(x_M) \cr
\vdots & \vdots & \ddots & \vdots \cr
\alpha_{n+M-1}(x_1) & \alpha_{n+M-1}(x_2) & \cdots & \alpha_{n+M-1}(x_M)
\end{vmatrix}
}{\Delta^{(M)}(x)} \nn\\
&=\frac{\ds 
\det_{1\leq a,b \leq M} \bigl[ \alpha_{n+a-1}(x_b)\bigr]
}{\Delta^{(M)}(x)},  \label{Eq:CorrelatorOfalphaMorozovFormula}
\end{align}
where $\alpha_n(x) = \vev{\det (x-X) \strut}_{n\times n}$. By taking the double scaling limit,%
\footnote{Here note the scaling law of the shift operator $Z$: $Z=e^{-\del_n} = e^{{\rm a}_L^{1/2} \del}\, ({\rm a}_L\to 0)$. }
one obtains the following formula: 
\begin{align}
\vev{\prod_{a=1}^{M}  \hat \psi^{(1)}(\zeta_{a}) }_t = \frac{ W^{[1^{M}]}_\varnothing [\psi](t;\{\zeta_a\}_{a=1}^{M})}{\Delta^{(M)}(\zeta)}.
\end{align}
Therefore, we have shown the validity of Eq.~\eq{Eq:WronskianFormulaPre} and that  Eq.~\eq{Eq:WronskianFormulaPreGeneralAsym} holds under the asymptotic expansion. 

The final task is to argue that the formula, Eq.~\eq{Eq:WronskianFormulaPreGeneralAsym}, is even valid at the level of non-perturbative completion. This can be seen by focusing on the global connection rules (i.e.~Stokes phenomenon) of the correlator around $\zeta_a\to \infty$ (for each $a=1,2,\cdots,M$). The formula is then shown by assuming the cluster property of the correlator: If one focuses on each fermion operator $\hat \psi^{(j_a)}(\zeta_a)$ and considers the behavior around $\zeta_a\to \infty$, the strength of correlation (between the singled-out fermion and the other fermions) diminishes (i.e.~the cluster property). Therefore, the Stokes phenomenon of the correlator around $\zeta_a\to \infty$ should be governed by that of the singled-out wave-function, $\psi^{(j_a)}(t;\zeta_a) = \vev{\hat \psi^{(j_a)}(\zeta_a)}_t$.%
\footnote{Of course, correlators in two-dimensional field theory are given by a logarithm, $\sim \ln(\zeta-\zeta_0)$, and these fermions will correlate at an infinite distance. However, the existence of the other fermions cannot affect the global behavior of the singled-out fermion. } 
This can happen only when Eq.~\eq{Eq:WronskianFormulaPreGeneralAsym} holds non-perturbatively. 

As a result, we conclude that the one-point functions of (dual) FZZT-Cardy branes ($\Theta_{\rm Cardy}^{(r,1)}(\zeta)$ and $\widetilde \Theta_{\rm Cardy}^{(s,1)}(\eta)$) are given by the generalized Wronskians of the Baker-Akhiezer system of the matrix models: 
\begin{align}
\vev{\Theta_{\rm Cardy}^{(r,1)}(\zeta)}_t &=  W^{[j_1,j_2,\cdots,j_{r}]}_\varnothing[\psi](t;\zeta) &\left(j_a = 1 + \frac{(r-1)-2(a-1)}{2} \right),
 \\
\vev{\widetilde \Theta_{\rm Cardy}^{(s,1)}(\eta)}_t &=  W^{[l_1,l_2,\cdots,l_s]}_\varnothing[\chi](t;\eta) &\left( l_b = 1+ \frac{(s-1)-2(b-1)}{2}\right).  \label{Eqn:CardyBranesAsWronskianIdentification}
\end{align}
Since elemental FZZT branes carry independent degrees of freedom, the independence of elemental FZZT branes is now inherited by these FZZT-Cardy branes.

\subsection{More about the Wronskians \label{Subsection:FurthermoreWronskians}} 

In Section \ref{Section:WronskiansForR1FZZTCardyBranes}, we discuss differential-equation systems of the FZZT-Cardy branes. This section is thus devoted to the preparation of some technical material about the Wronskians. Note that the notion of generalized Wronskians itself dates back to Schmidt \cite{Schmidt39} and some recent works are found in \cite{Towse, Anderson, GattoSchrbak}. 

\subsubsection{Abbreviation and notation \label{Subsubsection:AbbreviationAndNotations}} 

We make a comment on some abbreviations of notations for the Wronskians. In many cases, the context of indices $[j_1,j_2,\cdots,j_r]$ in the Wronskians is not so important, since Wronskians with different indices still satisfy the same differential equation. The number of branes which constitute the Wronskian (i.e.~``$r$'') is rather important information. Therefore, we employ the following abbreviation: 
\begin{align}
W^{[j_1,j_2,\cdots,j_{r}]}_\varnothing[\psi](t;\zeta) \quad \to \quad W^{(r)}_\varnothing[\psi](t;\zeta). 
\end{align}
In particular, these Wronskians represented by $W^{(r)}_\varnothing[\psi](t;\zeta)$ are referred to as {\em rank-$r$ Wronskians}. 
Depending on the situations, the following abbreviations are also used: 
\begin{align}
\left\{
\begin{array}{l}
W^{(r)}_\varnothing[\psi](t;\zeta) \quad \to \quad W^{(r)}_\varnothing(t;\zeta)
\quad \to \quad W^{(r)}_\varnothing \cr
W^{(s)}_\varnothing[\chi](t;\eta) \quad \to \quad \widetilde W^{(s)}_\varnothing(t;\eta)
\quad \to \quad \widetilde W^{(s)}_\varnothing
\end{array}
\right.
\end{align}
In Section \ref{Subsubsection:GWronskiansYoungDiagrams}, we will introduce an index $\lambda$ of $W^{(r)}_\lambda$ (which represents a partition or Young diagram). This generalization of the Wronskians is called {\em generalized Wronskians} but we often simply call them ``Wronskians''. In some occasions, we also use the following abbreviation: 
\begin{align}
W^{(r)}_\lambda[\psi](t;\zeta) \quad \to \quad W^{(r)}_\lambda(t;\zeta)
\quad \to \quad W^{(r)}_\lambda \quad \to \quad \lambda. 
\end{align}
The last abbreviation is used in Appendix \ref{Appendix:DerivationOfIMS}. 


\subsubsection{Generalized Wronskians and Young diagrams \label{Subsubsection:GWronskiansYoungDiagrams}}

We also consider the following more general Wronskians (with general derivatives): 
\begin{align}
W_\lambda^{(r)}[\psi](t;\zeta) &= \begin{vmatrix}
\del^{r-1+\lambda_{r}} \psi^{(j_1)}(t;\zeta) & \del^{r-1+\lambda_{r}} \psi^{(j_2)}(t;\zeta) & \cdots & \del^{r-1+\lambda_{r}} \psi^{(j_r)}(t;\zeta) \cr
\vdots & \vdots & \ddots & \vdots \cr
\del^{1+\lambda_{2}} \psi^{(j_1)}(t;\zeta) & \del^{1+\lambda_{2}} \psi^{(j_2)}(t;\zeta) & \cdots & \del^{1+\lambda_{2}} \psi^{(j_r)}(t;\zeta) \cr 
\del^{\lambda_{1}}\psi^{(j_1)}(t;\zeta) & \del^{\lambda_{1}} \psi^{(j_2)}(t;\zeta) & \cdots & \del^{\lambda_{1}} \psi^{(j_r)}(t;\zeta) 
\end{vmatrix} \nn\\
&=\underset{1\leq a,b\leq r}{\det}\Bigl[ \del^{r-a+\lambda_{r-a+1}} \psi^{(j_b)}(t;\zeta) \Bigr]. \label{Eqn:GeneralizedWronskiansDefinitionInFurtherMoreWronskians}
\end{align}
Here $\lambda = (\lambda_r,\lambda_{r-1},\cdots,\lambda_2,\lambda_1)$ is a partition and satisfies 
\begin{align}
\lambda_r \geq \lambda_{r-1} \geq \cdots  \geq \lambda_{2} \geq \lambda_1 \geq 0.  \label{Eq:CanonicalPartitionRule}
\end{align}
If components of a partition $\lambda$ satisfy the above ordering, then $\lambda$ is said to be {\em standard partition}. Any partition can be represented by a Young diagram, as follows: 
\begin{align}
  (4,2,1,0,0) = {\tiny \yng(4,2,1)},\qquad (6,3,1,1) = {\tiny \yng(6,3,1,1)},\qquad (0,0,0)=\varnothing. 
\end{align}
The number $r$ of the partition $\lambda = (\lambda_r,\lambda_{r-1},\cdots,\lambda_2,\lambda_1)$ is called {\em the maximum length} and denoted by $\ell_{\rm max}(\lambda)$. The length of the Young diagram $\ell(\lambda)$ is defined as usual: 
\begin{align}
\lambda_r \geq \lambda_{r-1} \geq \cdots  
\geq \lambda_{r-\ell(\lambda)+1} \neq 0 = \lambda_{r-\ell(\lambda)} = \cdots 
= \lambda_1 = 0. 
\end{align}
Obviously, the length cannot be larger than the maximum length, $\ell_{\rm max}(\lambda)\geq \ell(\lambda)$. In particular, if the maximum length $\ell_{\rm max}(\lambda)$ is obvious (or not necessary to be specified), and if $\ell_{\rm max}(\lambda)> \ell(\lambda)$, then zeros in the partition are not explicitly shown: 
\begin{align}
(4,2,1,0,0)  \quad \to \quad (4,2,1). 
\end{align}

\subsubsection{Spaces of Young diagrams and linear extension of the index \label{Subsubsection:LinearExtensionYoungDiagramsW}}

It is then convenient to consider (infinite-dimensional) linear spaces of Young diagrams over $\mathbb C$. While the set of the partitions of the maximum length $r$ is denoted as 
\begin{align}
{\mathsf P \mathsf T}_{\ell\leq r}  \equiv &\, \bigl\{\lambda: \text{partitions } \big| \ell(\lambda)\leq r \bigr\} \nn\\
=&\, \bigl\{\lambda= (\lambda_r,\lambda_{r-1},\cdots,\lambda_2,\lambda_1) \big| \lambda_r \geq \lambda_{r-1} \geq \cdots  \geq \lambda_{2} \geq \lambda_1 \geq 0 \bigr\},  \label{Eqn:Definition:PartitionPTlleqr}
\end{align}
the linear space of Young diagrams of the maximum length $r$ is defined as 
\begin{align}
\mathcal Y_{\ell\leq r} \equiv \bigoplus_{\lambda \in {\mathsf P \mathsf T}_{\ell\leq r}} \mathbb C   \lambda. 
\end{align}
Accordingly, we also linearly extend the Young-diagram labeling $\lambda$ of $W^{(r)}_\lambda$ to elements of the linear space: 
\begin{align}
W_{\alpha \mu+\beta \nu}^{(r)} = \alpha \,  W_\mu^{(r)} + \beta\, W_\nu^{(r)}\qquad \bigl(\mu,\nu \in\mathcal Y_{\ell\leq r};\, \alpha,\beta \in \mathbb C\bigr). 
\end{align}
In particular, if $\lambda$ is given by a partition in the space (i.e.~$\lambda \in \mathsf P \mathsf T_{\ell \leq r} \subset \mathcal Y_{\ell \leq r}$), then $\lambda$ is referred to as {\em a pure basis}. 


\subsubsection{Standard v.s.~non-standard Young diagrams} 

One can also exchange rows of Wronskians. For example, if one exchanges the $a$-th row with the $(a+1)$-th row, there appears a minus sign, $(-1)\times$: 
\begin{align}
W_\lambda^{(r)}[\psi](t;\zeta) &= \begin{vmatrix}
\vdots &   \cr
\del^{a+\lambda_{a+1}} \psi^{(j_1)}(t;\zeta) &  \cdots \cr
\del^{a-1+\lambda_{a}} \psi^{(j_1)}(t;\zeta) & \cdots  \cr 
\vdots & 
\end{vmatrix} 
=(-1)\times \begin{vmatrix}
\vdots &   \cr
\del^{a+(\lambda_{a}-1)} \psi^{(j_1)}(t;\zeta) & \cdots  \cr 
\del^{a-1+(\lambda_{a+1}+1)} \psi^{(j_1)}(t;\zeta) &  \cdots \cr
\vdots & 
\end{vmatrix} 
\end{align}
This new Wronskian can be represented with a new (but not standard) partition/Young diagram $\lambda'$ as 
\begin{align}
W_{\lambda}^{(r)}=(-1)\times W_{\lambda'}^{(r)}. 
\end{align}
The new partition is given as follows: 
\begin{align}
\lambda = (\cdots,\lambda_{a+1},\lambda_a,\cdots) \quad \to \quad 
\lambda' = (\cdots,\overbrace{\lambda_a-1}^{\lambda'_{a+1}},\overbrace{\lambda_{a+1}+1}^{\lambda'_{a}},\cdots). 
\end{align}
For example, if one exchanges $\lambda_2$ with $\lambda_3$ in $\lambda=(5,4,1,1,0)$, one obtains 
\begin{align}
W^{(5)}_{\tiny \yng(5,4,1,1,0)} = (-1)\times W^{(5)}_{\tiny \yng(5,4,0,2,0)}. 
\end{align}
In this way, it is also convenient to consider a ``non-standard'' partition/Young diagram (i.e.~which does not satisfy Eq.~\eq{Eq:CanonicalPartitionRule}). In particular, if a (non-standard) partition $\lambda$ satisfies the following condition, then the Wronskian $W^{(r)}_\lambda$ vanishes: 
\begin{align}
\lambda_{a+1} + 1 = \lambda_a \qquad \bigl(1\leq {}^\exists a \leq r-1 \bigr) \qquad \Rightarrow \qquad W^{(r)}_\lambda =-W^{(r)}_\lambda= 0. 
\end{align}
For example: 
\begin{align}
W^{(r)}_{\tiny \yng(3,1,2)} = 0,\qquad W^{(r)}_{\tiny \yng(0,0,1)}=0,\qquad W^{(r)}_{\tiny\reflectbox{\yng(0,1,0)}}=0,
\end{align}
where each partition is given as $\lambda=(3,1,2),\, (0,0,1), \, (0,-1,0)$. These Wronskians vanish since two adjacent rows have the same order of differential and the determinant vanishes. 

As a natural extension, we should also consider some non-standard partitions/Young diagrams which possess negative values. Such a partition/Young diagram generally exists when 
\begin{align}
\lambda_n \geq -(n-1) \qquad \bigl(n=1, 2, 3, \cdots, r\bigr).  \label{Eqn:ExistenceOfWronskiansLambda}
\end{align}
If $\lambda$ includes a negative component, the partition is said to be {\em a negative partition}. 

\subsubsection{Schur-derivatives on Wronskians \label{Subsubsection:Schur-Derivatives}}

We here temporarily introduce the derivative operator $\del_{(j)}$ which acts only on the corresponding wave function in $\{\psi^{(j)}(t;\zeta)\}_{j=1}^p$: 
\begin{align}
\del_{(j)} \psi^{(l)}(t;\zeta) = \delta_{j,l} \, \del \psi^{(l)}(t;\zeta) \qquad \bigl(1\leq j,l\leq p\bigr). 
\end{align}
With these derivatives, any general Wronskian $W^{(r)}_\lambda(t;\zeta)$ can be rewritten by the Schur polynomials $\mathcal S_{\lambda}(\del)$ as 
\begin{align}
W^{(r)}_\lambda(t;\zeta) =& \,(-1)^{\frac{r(r-1)}{2}} \det_{1\leq a,b \leq r}\Bigl[ \del_{(j_b)}^{a-1+\lambda_{a}}\Bigr] \prod_{b=1}^r \psi^{(j_b)}(t;\zeta)\nn\\
=& \, \mathcal S_{\lambda}(\del)\, W^{(r)}_{\varnothing}(t;\zeta),  \label{Eq:WlambdaAsSchurDerivativeW}
\end{align}
where the Schur polynomials $\mathcal S_{\lambda}(\del)$ are given as 
\begin{align}
\mathcal S_{\lambda}(\del) = \frac{\det_{1\leq a,b \leq r}\Bigl[ \del_{(j_b)}^{a-1+\lambda_{a}}\Bigr]}{\det_{1\leq a,b \leq r}\Bigl[ \del_{(j_b)}^{a-1}\Bigr]}. 
\end{align}
Usually, Schur polynomials are functions of Miwa variables, which are (again temporarily) defined as 
\begin{align}
\del_{[n]}\equiv  \frac{1}{n} \sum_{a=1}^r \bigl(\del_{(j_a)}\bigr)^n,\qquad \big(n=1,2,3,\cdots\bigr). 
\end{align}
The physical meaning of this expression (with Schur polynomials) is understood as follows: FZZT-Cardy branes are multi-body states of elemental FZZT branes. Therefore, differential operators acting on the states should be symmetric derivative operators on the multi-body Fock space, which are derivatives represented by Miwa variables or Schur polynomials. 

The derivative $\mathcal S_{\lambda}(\del)$ are referred to as {\em Schur-derivatives}. Following the similar consideration to the Wronskians in Section \ref{Subsubsection:LinearExtensionYoungDiagramsW}, the Young-diagram labeling of Schur-derivatives is also extended linearly: 
\begin{align}
\mathcal S_{\alpha \mu + \beta \nu}(\del) = \alpha \, \mathcal S_\mu (\del) + \beta \, \mathcal S_\nu (\del)\qquad \bigl(\mu,\nu \in\mathcal Y_{\ell\leq r};\, \alpha,\beta \in \mathbb C\bigr). 
\end{align}

\subsection{Multi-point correlators of Wronskians \label{Subsection:MultipointCorrelatorsOfWronskians}}
In Section \ref{Subsubsection:MultipointCorrelatorsAsWronskians}, we have shown the formula Eq.~\eq{WronskianFormulaGeneral}, which represents how the correlators of the elemental FZZT-branes, $\vev{\prod_b \hat \psi^{(j_b)}(\zeta)}_t$, can be expressed by the one-point wave functions $\bigl\{ \psi^{(j)}(t;\zeta) \bigr\}_{j=1}^p$. Here, we consider how the multi-point correlators of FZZT-Cardy branes, $\vev{\prod_b \Theta_{\rm Cardy}^{(r_b,1)}(\zeta_b)}_t$, can be expressed by the one-point wave functions $\bigl\{W^{(r)}_{\varnothing}(t;\zeta) \bigr\}_{r=1}^{p-1}$. 

\subsubsection{Wronskian correlators and Schur-differential Wronskians \label{Subsubsection:SchurdifferentialWronskians}}
As the most general situation, we consider a Wronskian $W^{[j_1,j_2,\cdots,j_r]}_{\varnothing}(t;\zeta)$ and write its corresponding ``D-brane operator'' $\hat W^{[j_1,j_2,\cdots,j_r]}(\zeta)$ by ``adding a hat on the head'' as 
\begin{align}
\hat W^{[j_1,j_2,\cdots,j_r]}(\zeta) \equiv \prod_{a=1}^r \hat \psi^{(j_a)}(\zeta),\qquad W^{[j_1,j_2,\cdots,j_r]}_\varnothing (t;\zeta) = \vev{\hat W^{[j_1,j_2,\cdots,j_r]}(\zeta)}_t. 
\end{align}
The correlators of these Wronskian operators are then expressed by the following ``Wronskian-like'' function: 
\begin{align}
\vev{\prod_{a=1}^M \overbrace{\hat W^{(r_a)}(\zeta_a)}^{[j_1^{(a)},\cdots,j_{r_a}^{(a)}]} }_t 
&= \frac{ \ds
\sum_{
\pi \in \binom{r_1+\cdots+r_M}{r_1|r_2|\cdots|r_M} 
}
\sgn(\pi) \, \overbrace{W^{(r_1)}_{\pi^{(1)}}(t;\zeta_1)}^{[j_1^{(1)},\cdots,j_{r_1}^{(1)}]}\times  \cdots \times  \overbrace{W^{(r_M)}_{\pi^{(M)}}(t;\zeta_M)}^{[j_1^{(M)},\cdots,j_{r_M}^{(M)}]}
}
{ \ds \prod_{1\leq a<b\leq M} (\zeta_a-\zeta_b)^{m_{ab}}} \nn\\
& \equiv \frac{\ds \overbrace{\mathscr W^{[r_1,r_2,\cdots,r_M]}_{\varnothing}[W](t;\zeta_1,\cdots,\zeta_M )}^{[j_1^{(1)},\cdots,j_{r_1}^{(1)}|\cdots|j_1^{(M)},\cdots,j_{r_M}^{(M)}]} }{\ds \prod_{1\leq a<b\leq M} (\zeta_a-\zeta_b)^{m_{ab}} }. \label{Eq:FurtherGeneralizedWronskians} 
\end{align}
The ``Wronskian-like'' function $\mathscr W$ is referred to as {\em Schur-differential Wronskian} for a reason mentioned later. 
Some notes are following: 
\begin{itemize}
\item [1) ] The integers $m_{ab}$ ($1\leq a<b\leq M$) are the overlap numbers which count the overlapping indices among $W^{(r_a)}$ and $W^{(r_b)}$. With giving the indices of each Wronskian as 
\begin{align}
\hat W^{(r_a)}(\zeta) = \hat W^{[j_1^{(a)},\cdots,j_{r_a}^{(a)}]}(\zeta), 
\end{align}
the overlap number is defined as 
\begin{align}
m_{ab} = \# \bigl[ \bigl\{j_n^{(a)}\bigr\}_{n=1}^{r_a} \cap \bigl\{j_n^{(b)}\bigr\}_{n=1}^{r_b} \bigr].  \label{Eq:OverlapNumber}
\end{align}
\item [2) ] The combinatorics $\pi\in \binom{r_1+\cdots+r_M}{r_1|r_2|\cdots|r_M}$ is a combination of dividing distinct $(r_1+\cdots+r_M)$ elements into $M$ groups, where each group constitutes $r_a$ elements ($a=1,2,\cdots,M$). The set of all combinatorics is given as 
\begin{align}
\binom{r_1+\cdots+r_M}{r_1|r_2|\cdots|r_M} =   \mathfrak S_{r_1} \times \cdots\times \mathfrak S_{r_M} \big\backslash \mathfrak S_{r_1+\cdots+r_M}.  \label{Eqn:PiEquivalentClassSigma}
\end{align} 
The total sum of $\{r_a\}_{a=1}^M$ is now denoted as $\ds r_{\rm tot} \equiv \sum_{a=1}^M r_a$. 
\begin{itemize}
\item [i. ]
For each element $\sigma \in \mathfrak S_{r_1+r_2+\cdots+r_M} = \mathfrak S_{r_{\rm tot}}$, one inserts $(M-1)$ divisions ``~$\big|$~'' as follows: 
\begin{align}
&\Bigl(\sigma(1),\sigma(2),\cdots,\sigma(r_{\rm tot})\Bigr) = \nn\\
& \qquad = \Bigl( \sigma(1),\cdots, \sigma(r_1) \Big| \sigma(r_1+1),\cdots,\sigma(r_1+r_2) \Big|\cdots \nn\\
&\qquad \qquad \qquad \cdots  \Big|\sigma(r_{\rm tot}-r_M+1), \cdots,\sigma(r_{\rm tot} -1), \sigma(r_{\rm tot})\Bigr). 
\end{align}
\item [ii. ]
By applying a proper element $\tau \equiv (\tau_1|\tau_2|\cdots|\tau_M)\in \mathfrak S_{r_1} \times \cdots\times \mathfrak S_{r_M} \subset \mathfrak S_{r_{\rm tot}}$, one rearranges the numbers inside each partition, 
\begin{align}
&\Bigl(\tau\sigma(1),\tau \sigma(2),\cdots,\tau \sigma(r_{\rm tot})\Bigr)\equiv  \nn\\
& \qquad \equiv \Bigl( \tau_1 \sigma(1),\cdots, \tau_1 \sigma(r_1) \Big| \tau_2 \sigma(r_1+1),\cdots, \tau_2 \sigma(r_1+r_2) \Big|\cdots \nn\\
&\qquad \qquad \qquad \cdots  \Big|\tau_M \sigma(r_{\rm tot}-r_M+1), \cdots,\tau_M \sigma(r_{\rm tot} -1), \tau_M \sigma(r_{\rm tot})\Bigr), \label{Eqn:RepresentativeOfPiEqTauSigma}
\end{align}
such that 
\begin{align}
&\tau_a \sigma\Bigl( \sum_{b=1}^{a-1} r_{b} + n\Bigr) \,\, < \,\, \tau_a \sigma\Bigl( \sum_{b=1}^{a-1} r_{b} +m\Bigr) \nn\\
&\qquad \bigl( 0<n<m\leq r_a; \, a=1,2,\cdots, M\bigr), 
\end{align}
where $\mathfrak S_{r_a}$ is the symmetric group which acts on the $r_a$ distinct integers $\Big\{\sigma(r_a+n-1)\Bigr\}_{n=1}^{r_a}$ inside the $a$-th partition: 
\begin{align}
\mathfrak S_{r_a} \ni \tau : \, \Big\{\sigma\Bigl(\sum_{b=1}^{a-1} r_{b} +n-1\Bigr)\Bigr\}_{n=1}^{r_a} \quad \overset{\ds \sim}{\longrightarrow} \quad \Big\{\sigma\Bigl(\sum_{b=1}^{a-1} r_{b}+n-1\Bigr)\Bigr\}_{n=1}^{r_a}. 
\end{align}
\item [iii. ] By this manipulation, we define the equivalence class, 
\begin{align}
\sigma_1 \sim \sigma_2 \qquad \bigl( \sigma_1,\sigma_2 \in \mathfrak S_{r_{\rm tot}}\bigr)
\qquad \Leftrightarrow \qquad 
\sigma_1 = \rho \sigma_2 \qquad \bigl({}^\exists \rho \in \mathfrak S_{r_1} \times \cdots\times \mathfrak S_{r_M}\bigr),
\end{align}
and obtain Eq.~\eq{Eqn:PiEquivalentClassSigma}. 
\item [iv. ]
The above $\tau \sigma \in \mathfrak S_{r_{\rm tot}}$ gives the standard representative of $\pi \in \binom{r_1+\cdots+r_M}{r_1|r_2|\cdots|r_M}$: 
\begin{align}
\pi\, (\equiv  \tau \sigma) \in \binom{r_1+\cdots+r_M}{r_1|r_2|\cdots|r_M}. 
\end{align}
In particular, from the $M$-division of the representative $\pi = \tau \sigma$ (i.e.~Eq.~\eq{Eqn:RepresentativeOfPiEqTauSigma}), we define the following $M$ different partitions $\bigl\{ \pi^{(a)}_\varnothing \bigr\}_{a=1}^M$: 
\begin{align}
\pi^{(a)}_\varnothing \equiv \bigl(\lambda^{(a)}_{r_{a}}, \cdots, \lambda^{(a)}_2, \lambda^{(a)}_1 \bigr) \qquad \bigl(a=1,2,\cdots, M\bigr), 
\end{align}
such that 
\begin{align}
\lambda^{(a)}_n \equiv \tau_a \sigma\Bigl(\sum_{b=1}^{a-1} r_{b} +n \Bigr) - n \qquad \bigl(n=1,2,\cdots, r_a\bigr). 
\end{align}
In particular, they satisfy the condition of standard partitions: 
\begin{align}
0 \leq \lambda^{(a)}_1 \leq \lambda^{(a)}_2 \leq \cdots \leq \lambda^{(a)}_{r_{a}}. 
\end{align}
\item [v. ] The signature $\sgn(\pi)$ of $\pi \in \binom{r_1+\cdots+r_M}{r_1|r_2|\cdots|r_M}$ is defined by the signature of the representative. That is, 
\begin{align}
\sgn(\pi) \equiv \sgn(\tau \sigma), 
\end{align}
of Eq.~\eq{Eqn:RepresentativeOfPiEqTauSigma}. 
\end{itemize} 
\item [3) ] This formula is equivalently expressed as 
\begin{align}
W_\varnothing^{[j_1^{(1)},\cdots,j_{r_1}^{(1)}|\cdots|j_1^{(M)},\cdots,j_{r_M}^{(M)}]}[\psi]\bigl(t;\bigl\{\zeta_a\bigr\}_{a=1}^M\bigr) = \frac{\ds \overbrace{\mathscr W^{[r_1,r_2,\cdots,r_M]}_{\varnothing}[W](t;\zeta_1,\cdots,\zeta_M )}^{[j_1^{(1)},\cdots,j_{r_1}^{(1)}|\cdots|j_1^{(M)},\cdots,j_{r_M}^{(M)}]} }{\ds \prod_{1\leq a<b\leq M} (\zeta_a-\zeta_b)^{m_{ab}} }. \label{Eqn:SDWronskianSection2Correlators}
\end{align}
Therefore, the formula Eq.~\eq{Eq:FurtherGeneralizedWronskians} is obtained by decomposing each Wronskian into the multi-body states of the elemental FZZT branes and by re-expressing it in the above Wronskian-like form. 
\end{itemize}

From the above note (3), a further generalization of this ``Wronskian-like'' function is also possible: We allow them to possess the labeling of partition/Young diagram $\lambda$: 
\begin{align}
\overbrace{\mathscr W^{[r_1,r_2,\cdots,r_M]}_{\lambda}[W](t;\zeta_1,\cdots,\zeta_M )}^{[j_1^{(1)},\cdots,j_{r_1}^{(1)}|\cdots|j_1^{(M)},\cdots,j_{r_M}^{(M)}]} \equiv \sum_{
\pi \in \binom{r_1+\cdots+r_M}{r_1|r_2|\cdots|r_M} 
}
\sgn(\pi) \, \overbrace{W^{(r_1)}_{\pi^{(1)}_\lambda}(t;\zeta_1)}^{[j_1^{(1)},\cdots,j_{r_1}^{(1)}]}\times  \cdots \times  \overbrace{W^{(r_M)}_{\pi^{(M)}_\lambda}(t;\zeta_M)}^{[j_1^{(M)}, \cdots, j_{r_M}^{(M)}]},
\label{Eqn:SDWronskianSection2lambda}
\end{align}
where 
\begin{itemize}
\item [1) ] The partition $\lambda$ is that of the maximum length $r_{\rm tot}$: 
\begin{align}
\ell_{\rm max}( \lambda) = r_{\rm tot} \qquad \Bigl(r_{\rm tot} \equiv \sum_{a=1}^M r_a\Bigr). 
\end{align}
\item [2) ] The $M$ different partitions $\bigl\{ \pi^{(a)}_\lambda \bigr\}_{a=1}^M$ are defined by 
\begin{align}
\pi^{(a)}_\lambda \equiv \bigl(\lambda^{(a)}_{r_{a}}, \cdots, \lambda^{(a)}_2, \lambda^{(a)}_1 \bigr) \qquad \bigl(a=1,2,\cdots, M\bigr), \label{DefinitionOfPiLambdaEi1}
\end{align}
such that 
\begin{align}
\lambda^{(a)}_n \equiv \lambda_{\tau_a \sigma\bigl(\mbox{\scriptsize$\ds\sum_{b=1}^{a-1} r_{b} +n$}\bigr)}+ \tau_a \sigma\Bigl(\sum_{b=1}^{a-1} r_{b} +n \Bigr) - n  \qquad \bigl(n=1,2,\cdots, r_a\bigr). \label{DefinitionOfPiLambdaEi2}
\end{align}
In particular, they satisfy the condition of standard partitions: 
\begin{align}
0 \leq \lambda^{(a)}_1 \leq \lambda^{(a)}_2 \leq \cdots \leq \lambda^{(a)}_{r_{a}}. 
\end{align}
\item [3) ] This generalized function comes from the following rank-$r_{\rm tot}$ Wronskian: 
\begin{align}
W_\lambda^{[j_1^{(1)},\cdots,j_{r_1}^{(1)}|\cdots|j_1^{(M)},\cdots,j_{r_M}^{(M)}]}[\psi]\bigl(t;\bigl\{\zeta_a\bigr\}_{a=1}^M\bigr) = \frac{\ds \overbrace{\mathscr W^{[r_1,r_2,\cdots,r_M]}_{\lambda}[W](t;\zeta_1,\cdots,\zeta_M )}^{[j_1^{(1)},\cdots,j_{r_1}^{(1)}|\cdots|j_1^{(M)},\cdots,j_{r_M}^{(M)}]} }{\ds \prod_{1\leq a<b\leq M} (\zeta_a-\zeta_b)^{m_{ab}} }. 
\end{align}
That is, this is essentially from $r_{\rm tot}$-point functions of elemental FZZT branes. 
\end{itemize}

Note that, since each Wronskian $W^{(r_a)}_{\pi^{(a)}_\lambda}(t;\zeta_a)$ is expressed by the Schur-derivatives: 
\begin{align}
\cdots \times W^{(r_a)}_{\pi^{(a)}_\lambda}(t;\zeta_a) \times \cdots \quad = \quad \cdots  \times \mathcal S_{\pi^{(a)}_\lambda}(\del) \, W^{(r_a)}_{\varnothing}(t;\zeta_a)\times  \cdots, 
\end{align}
this ``Wronskian-like'' function $\mathscr W_\lambda^{[r_1,r_2,\cdots,r_M]}$ is understood as a generalization of the generalized Wronskians $W^{(r)}_\lambda$ by replacing the ordinary derivative ``$\del^n$'' with the Schur-derivative ``$\mathcal S_\lambda (\del)$''. In this sense, we refer to this new type of Wronskians as {\em Schur-differential Wronskians}.

\subsubsection{Abbreviation and notation \label{Subsubsection:AbbreviationSchurDifferentialWronskians}}

As in Section \ref{Subsubsection:AbbreviationAndNotations}, we abbreviate the index $[j_1,j_2,\cdots,j_s]$ of Eq.~\eq{Eq:FurtherGeneralizedWronskians} and Eq.~\eq{Eqn:SDWronskianSection2lambda}. To achieve this without generating any confusion, we express the Schur-differential Wronskians as follows: 
\begin{align}
\mathscr W^{[r_1,r_2,\cdots,r_M]}_{\lambda} \bigl[W\bigr](t;\bigl\{\zeta_a\bigr\}_{a=1}^M) = \sum_{
\pi \in \binom{r_1+\cdots+r_M}{r_1|r_2|\cdots|r_M} 
}
\sgn(\pi) \, W^{(r_1)}_{\pi^{(1)}_\lambda}(t;\zeta_1) \otimes  \cdots \otimes  W^{(r_M)}_{\pi^{(M)}_\lambda}(t;\zeta_M). 
\end{align}
We here use ``$\otimes$'' (i.e.~$W_\lambda^{(r_1)}(t;\zeta_1) \otimes W_{\mu}^{(r_2)}(t;\zeta_2)$) so that we can distinguish the following: 
\begin{align}
& W_\lambda^{(r)}(t;\zeta) \otimes W_{\mu}^{(r)}(t;\zeta) \quad \neq \quad W_{\mu}^{(r)}(t;\zeta) \otimes W_\lambda^{(r)}(t;\zeta) \nn\\
 &\Leftrightarrow \qquad \overbrace{W_\lambda^{(r)}(t;\zeta)}^{[j_1,j_2,\cdots,j_r]} \overbrace{W_{\mu}^{(r)}(t;\zeta)}^{[l_1,l_2,\cdots,l_r]} \quad \neq \quad  \overbrace{W_{\mu}^{(r)}(t;\zeta)}^{[j_1,j_2,\cdots,j_r]} \overbrace{W_\lambda^{(r)}(t;\zeta)}^{[l_1,l_2,\cdots,l_r]}. 
\end{align}
This notation will be used later in Section \ref{Section:TowardInteriorOfKacTable}. 

\subsubsection{Correlators of FZZT-Cardy branes \label{Subsubsection:CorrelatorsOfFZZTCardyBranes}}

Based on the discussions above, we can now write down arbitrary correlators of $(r,1)$ FZZT-Cardy branes. The important information is the overlap number Eq.~\eq{Eq:OverlapNumber}, which is essentially obtained by two-point functions. The two-point function of $(r,1)$ FZZT-Cardy branes is given as 
\begin{align}
\vev{\Theta_{\rm Cardy}^{(r_1,1)}(\zeta_1) \Theta_{\rm Cardy}^{(r_2,1)}(\zeta_2) }_t = \frac{\mathscr W^{[r_1,r_2]}_\varnothing [W] (t;\zeta_1,\zeta_2) }{\bigl[(-1)^{r_1+1} \zeta_1 - (-1)^{r_2+1} \zeta_2\bigr]^{{\rm min}[r_1,r_2]}}. 
\end{align}
That is, the overlap number is $m_{r_1,r_2}^{\rm Cardy} \equiv {\rm min}[r_1,r_2]$. The multi-point functions are then given as 
\begin{align}
\vev{\prod_{a=1}^M \Theta_{\rm Cardy}^{(r_a,1)} (\zeta_a) }_t 
&=\frac{\ds \mathscr W^{[r_1,r_2,\cdots,r_M]}_{\varnothing}[W](t;\zeta_1,\cdots,\zeta_M ) }{\ds \prod_{1\leq a<b\leq M} \bigl[(-1)^{r_a+1}\zeta_a-(-1)^{r_b+1}\zeta_b\bigr]^{{\rm min}[r_a,r_b]} }. 
\end{align}
This result can be compared with the calculations of Liouville theory. For example, \cite{KOPSS} gives the same pole structure, $\bigl\{ \zeta_a\bigr\}_{a=1}^M$, but the overlap number does not quite coincide. This discrepancy has not been solved yet. However, noting that the discrepancy comes from a convention of normal ordering for the free-fermion, it may be adjusted by changing it. For the criterion of the ``improved'' normal ordering, one may require whole non-perturbative-duality consistency of FZZT-Cardy branes, as is considered in \cite{CIY5}. Duality relations among different FZZT-Cardy branes are discussed in Section \ref{Section:ConsistencyOfDualities}.

\section{Spectral curves of FZZT-Cardy branes \label{Section:SpectralCurvesOfFZZTCardyBranes}}
In this section, we discuss several issues related to the spectral curves of $(r,s)$ FZZT-Cardy branes. For a given determinant operator $\Theta_B(\zeta)$, its single trace operator is denoted by $\hat \phi_B(\zeta)$ (i.e.~$\Theta_B (\zeta) = e^{\hat \phi_B(\zeta)}$). The spectral curve of the D-brane $\Theta_B(\zeta)$ is then given by the following algebraic equation: 
\begin{align}
\mathscr S_B:\quad 
F_B(\zeta,Q)=0,\qquad Q(\zeta)=\vev{\del_{\zeta} \hat \phi_B(\zeta)}^{(0)}.
\end{align}
In order to apply this consideration to our cases of the $(r,s)$ FZZT-Cardy brane $\Theta_{\rm Cardy}^{(r,s)}(\zeta)$ (and the $(s,r)$ dual FZZT-Cardy brane $\widetilde \Theta_{\rm Cardy}^{(s,r)}(\eta)$), we express their single trace operators as $\hat \phi_{\rm Cardy}^{(r,s)}(\zeta)$ (or $\hat {\tilde \phi}_{\rm Cardy}^{(s,r)}(\eta)$). Their spectral curves are then represented as 
\begin{align}
&\mathscr S_{\rm Cardy}^{(r,s)}:\quad 
F_{\rm Cardy}^{(r,s)}(\zeta,Q)=0,\qquad Q(\zeta)=\vev{\del_{\zeta} \hat \phi_{\rm Cardy}^{(r,s)}(\zeta)}^{(0)}, \\
&\widetilde {\mathscr S}_{\rm Cardy}^{(s,r)}:\quad 
\widetilde F_{\rm Cardy}^{(s,r)}(\eta,P)=0,\qquad P(\eta)=\vev{\del_{\eta} \hat {\tilde \phi}_{\rm Cardy}^{(s,r)}(\eta)}^{(0)}. 
\end{align}
We start our discussions with the relation between the $(r,s)$ FZZT-Cardy brane and its dual $(s,r)$ FZZT-Cardy brane, especially from the viewpoint of spectral curves. The concrete form of the spectral curves is then evaluated. In addition, we also comment on ZZ-Cardy branes associated with the spectral curves.

\subsection{Spectral $p-q$ duality as Legendre transformation \label{Section:PQdualityLegendre}}

We here consider spectral $p-q$ duality between the FZZT-Cardy branes and their duals as {\em Legendre transformation on the spectral curve}. What we focus on is the disk amplitudes of FZZT-Cardy brane, $\varphi_{\rm Cardy}^{(r,s)}(t;\zeta)$, and of its dual, $\widetilde \varphi_{\rm Cardy}^{(s,r)}(t;\eta)$: 
\begin{align}
\varphi_{\rm Cardy}^{(r,s)}(t;\zeta) \equiv& \vev{\hat \phi_{\rm Cardy}^{(r,s)}(\zeta)}_t^{(0)}\qquad \Bigl(\Theta_{\rm Cardy}^{(r,s)}(\zeta) = e^{\hat \phi_{\rm Cardy}^{(r,s)}(\zeta)}\Bigr),\\
\widetilde \varphi_{\rm Cardy}^{(s,r)}(t;\eta) \equiv & \vev{\hat {\tilde\phi}_{\rm Cardy}^{(s,r)}(\eta)}_t^{(0)}\qquad \Bigl(\widetilde \Theta_{\rm Cardy}^{(s,r)}(\eta) = e^{\hat {\tilde \phi}_{\rm Cardy}^{(s,r)}(\eta)}\Bigr). 
\end{align}
The Seiberg-Shih relation for the disk amplitudes is then given as 
\begin{align}
&\varphi_{\rm Cardy}^{(r,s)}(\zeta) = \sum_{
\begin{subarray}{c}
k=-(r-1) \cr \text{step 2}
\end{subarray}
}^{(r-1)} \,\sum_{
\begin{subarray}{c}
l=-(s-1) \cr \text{step 2}
\end{subarray}
}^{(s-1)} 
\varphi_{\rm Cardy}^{(1,1)}(\zeta_{k,l}(\zeta)), \\
&\widetilde \varphi_{\rm Cardy}^{(s,r)}(\eta) = 
\sum_{
\begin{subarray}{c}
l=-(s-1) \cr \text{step 2}
\end{subarray}
}^{(s-1)} \sum_{
\begin{subarray}{c}
k=-(r-1) \cr \text{step 2}
\end{subarray}
}^{(r-1)}
\widetilde \varphi_{\rm Cardy}^{(1,1)}(\eta_{l,k}(\eta)),
\end{align}
where $\zeta_{k,l}$ is given by Eq.~\eq{Eq:SeibergShihRelZeta} and $\eta_{k,l}$ is given by Eq.~\eq{Eq:SeibergShihRelEta}. We first consider the case of the principal FZZT-branes, and then extend it to the cases of the general $(r,s)$ FZZT-Cardy branes. 

\subsubsection{Legendre transformation for the $(1,1)$ FZZT-Cardy branes}
The essence of the $p-q$ duality between $\varphi_{\rm Cardy}^{(1,1)}(\zeta)$ and $\widetilde \varphi_{\rm Cardy}^{(1,1)}(\eta)$ is the following mutual relation among their integral representations on the spectral curve: 
\begin{align}
\varphi_{\rm Cardy}^{(1,1)}(\zeta) = \beta_{p,q}  \int^{\zeta} \mathcal Y \, d\mathcal X,\qquad 
\widetilde \varphi_{\rm Cardy}^{(1,1)}(\eta) = \beta_{q,p} \int^{\eta} \mathcal X\, d\mathcal Y\qquad \bigl(\beta_{p,q} = (-1)^{p+q} \beta_{q,p} \bigr), 
\end{align}
where both integrals are given by the common coordinate $(\mathcal X, \mathcal Y)$ of the spectral curve, 
\begin{align}
F_{\rm Cardy}^{(1,1)} (\mathcal X,\beta_{p,q}\mathcal Y) \,\, \propto \,\,
\widetilde F_{\rm Cardy}^{(1,1)} (\mathcal Y,\beta_{q,p}\mathcal X)= 0. \label{Eqn:PQdualFXYspectralcurves}
\end{align}
Since their difference is just exchanging $(\mathcal X, \mathcal Y)$ of the spectral curve, these are the pair of dual branes which belongs to the same spectral curve. What is more, this relation holds {\em on the whole algebraic curve} (i.e.~not only on a particular branch) Eq.~\eq{Eqn:PQdualFXYspectralcurves}. We consider this to be the essence of the $p-q$ duality from the viewpoint of spectral curves. 

Note that if one changes the normalization of $\zeta$ as $\zeta \to c \zeta$ then the spectral curve is replaced as $F_{\rm Cardy}^{(1,1)}(\mathcal X,\beta_{p,q}\mathcal Y) = 0 \to F_{\rm Cardy}^{(1,1)}(c \mathcal X, \beta_{p,q}\, c^{-1}  \mathcal Y)=0$. By this operation, the overall normalization of $\widetilde \varphi_{\rm Cardy}^{(1,1)}$ does not change. 

As a result, $\varphi_{\rm Cardy}^{(1,1)}(\zeta)$ and $\widetilde \varphi_{\rm Cardy}^{(1,1)}(\eta)$ are related to each other as follows: 
\begin{align}
\varphi_{\rm Cardy}^{(1,1)}(\zeta) &= \beta_{p,q}\int^{\zeta} \mathcal Y\, d\mathcal X = \beta_{p,q} \int^{\zeta\eta} d (\mathcal X \mathcal Y) - \beta_{p,q} \int^{\eta} \mathcal X \, d\mathcal Y \nn\\
&= \beta_{p,q} \zeta \eta - \widetilde \varphi_{\rm Cardy}^{(1,1)}(e^{\pi i q}\eta). \label{Eq:LegendreCardy11}
\end{align}
This is understood as the Legendre transformation,%
\footnote{This also gives the reverse relation: 
\begin{align}
\widetilde \varphi_{\rm Cardy}^{(1,1)}(\eta) &= \sup_{\zeta}
\Bigl( \beta_{q,p} \zeta \eta - \varphi_{\rm Cardy}^{(1,1)}(e^{\pi i p} \zeta) \Bigr). 
\end{align}
This is because the relation $\beta_{q,p}=(-1)^{p+q}\beta_{p,q}$ comes from the transpose of the KP Lax operators Eq.~\eq{Eqn:PQdualityInLaxOperators} in $p-q$ duality, which is given by a shift of $\tau$, $\tau \to \tau + \pi i$, so that 
\begin{align}
\zeta \to e^{\pi i p} \zeta = \sqrt{\mu} \cosh \bigl(p\bigl[\tau + \pi i\bigr]\bigr),\qquad \eta \to e^{\pi i q} \eta = \sqrt{\tilde \mu} \cosh \bigl(q\bigl[\tau + \pi i\bigr]\bigr). 
\end{align}
}
\begin{align}
\varphi_{\rm Cardy}^{(1,1)}(\zeta) &= \sup_{\eta}
\Bigl( \beta_{p,q} \zeta \eta - \widetilde \varphi_{\rm Cardy}^{(1,1)}(e^{\pi i q}\eta) \Bigr), \label{Eq:LegendrePhi11}
\end{align}
meaning that the new function $\varphi^{(1,1)}_{\rm Cardy}(\zeta)$ which is defined by Eq.~\eq{Eq:LegendrePhi11} from $\widetilde \varphi^{(1,1)}_{\rm Cardy}(\eta)$ of the spectral curve $\ds \bigl(\widetilde F^{(1,1)}_{\rm Cardy}(\eta, P)=0\,\,\Leftrightarrow\,\, P(\eta) = \frac{\del \widetilde \varphi^{(1,1)}_{\rm Cardy}(\eta)}{\del \eta}\bigr)$ is guaranteed to satisfy
\begin{align}
&\text{Eq.~\eq{Eq:LegendrePhi11}}\quad \Rightarrow \quad 
\left\{
\begin{array}{cl}
\ds \zeta = \frac{1}{\beta_{q,p}}\frac{\del \widetilde \varphi_{\rm Cardy}^{(1,1)}(\eta)}{\del \eta} &\quad  \text{: the equation for $\zeta$}\cr
\ds \frac{\del \varphi^{(1,1)}_{\rm Cardy}(\zeta)}{ \del \zeta} = \beta_{p,q}\, \eta & \quad \text{: the equation for $\del_\zeta \varphi^{(1,1)}_{\rm Cardy}(\zeta)$}
\end{array}
\right. 
\end{align}
That is, 
\begin{align}
 0 = \widetilde F^{(1,1)}_{\rm Cardy}\Bigl(\frac{Q}{\beta_{p,q}}, \beta_{q,p}\zeta\Bigr) \,\, \propto\,\,  F^{(1,1)}_{\rm Cardy}(\zeta,Q),
\qquad \Bigl(Q(\zeta) = \frac{\del \varphi^{(1,1)}_{\rm Cardy}(\zeta)}{\del \zeta}\Bigr). 
\end{align}
Note that in Eq.~\eq{Eq:LegendreCardy11} there appears ``$(-1)$'' in front of $\widetilde \varphi_{\rm Cardy}^{(1,1)}(\tau)$. This can be interpreted as charge conjugation of the free fermions (See also Section \ref{SubSubSection:ChargeConjugation} and \ref{SubSubSection:PQdualityLaplace}). 

\subsubsection{Legendre transformation for the $(r,s)$ FZZT-Cardy branes}

We next apply these relations to the cases of $(r,1)$ and $(1,s)$ FZZT-Cardy branes. The Legendre transformation is given by 
\begin{align}
\varphi^{(1,s)}_{\rm Cardy} (\zeta) & = \sup_{\eta}
\Bigl(\beta_{p,q} A_s\,  \zeta \,\eta  - \widetilde \varphi^{(s,1)}_{\rm Cardy} (e^{\pi i q}\eta)\Bigr), \nn\\
\varphi^{(r,1)}_{\rm Cardy} (\zeta) & = \sup_{\eta}
\Bigl(\beta_{p,q}\widetilde A_r\,  \zeta\,\eta  - \widetilde \varphi^{(1,r)}_{\rm Cardy} (e^{\pi i q}\eta)\Bigr),
\end{align}
where 
\begin{align}
A_s &\equiv (-1)^{s-1}
\sum_{
\begin{subarray}{c}
l=-(s-1) \cr \text{step 2}
\end{subarray}
}^{(s-1)}
\frac{\zeta_{0,l}}{\zeta} = (-1)^{s-1} \dfrac{\sin ( \pi {sp}/{q})}{\sin(\pi {p}/{q})},\\
\widetilde A_r &\equiv (-1)^{r-1}
\sum_{
\begin{subarray}{c}
k=-(r-1) \cr \text{step 2}
\end{subarray}
}^{(r-1)}
\frac{\eta_{0,k}}{\eta} =(-1)^{r-1} \dfrac{\sin ( \pi {rq}/{p})}{\sin(\pi {q}/{p})}
\end{align}
and one eventually obtains the following general formula: 
\begin{align}
\varphi^{(r,s)}_{\rm Cardy} (\zeta) =\sup_{\eta} \Bigl[ \beta_{p,q}\, A_s \widetilde A_r \times  \zeta \,\eta - \widetilde \varphi^{(s,r)}_{\rm Cardy} (e^{\pi i q}\eta) \Bigr]. \label{Eqn:GeneralLegendreTransformation}
\end{align}
Again as a result of the $p-q$ duality, this relation should hold {\em on the whole algebraic curves} of these FZZT-Cardy branes. For this reason, this relation guarantees the following relations along the spectral curves of $F_{\rm Cardy}^{(r,s)}(P,Q)=0$ and of $\widetilde F_{\rm Cardy}^{(s,r)}(\eta, P)=0$: 
\begin{align}
 0 = \widetilde F^{(s,r)}_{\rm Cardy}\Bigl(\frac{Q}{\beta_{p,q} A_s \widetilde A_r}, \beta_{q,p} A_s \widetilde A_r\zeta\Bigr) \,\, \propto\,\,  F^{(r,s)}_{\rm Cardy}(\zeta,Q),
\qquad \Bigl(Q(\zeta) = \frac{\del \varphi^{(r,s)}_{\rm Cardy}(\zeta)}{\del \zeta}\Bigr).  \label{EqnFandFtilde:RelationOfSpectralCurves}
\end{align}
That is, these spectral curves are mutually related as 
\begin{align}
\zeta = \frac{1}{\beta_{q,p} A_s \widetilde A_r} \frac{\del \widetilde \varphi^{(s,r)}_{\rm Cardy} (\eta)}{\del \eta}\qquad \Leftrightarrow \qquad  \eta = \frac{1}{\beta_{p,q} A_s \widetilde A_r } \frac{\del \varphi^{(r,s)}_{\rm Cardy} (\zeta)}{\del \zeta}. 
\end{align}
From this point of view, one can obtain the spectral curve of $F^{(1,s)}_{\rm Cardy}(\zeta,Q)=0$ by the Legendre transformation of $\widetilde F^{(s,1)}_{\rm Cardy}(\eta,P)=0$. Later this Legendre transformation is non-perturbatively extended to a spectral duality of their integrable systems in Section \ref{Section:1SFZZTCardyBranesAndKacTable}.

\subsection{Spectral curves of $(r,s)$ FZZT-Cardy branes \label{SubSection:SpectralCurveCardy}}

We here discuss details of the spectral curves of $(r,s)$ FZZT-Cardy branes. 

\subsubsection{Spectral curves of $(1,1)$ FZZT-Cardy branes: a review}

Let us first recall the discussion on the spectral curve of the principal FZZT brane. It can be obtained by the elemental FZZT-branes (i.e.~the $p$-th twisted fermions), 
\begin{align}
\psi^{(j)}(\zeta) = e^{\hat \phi^{(j)}(\zeta)},\qquad Q^{(j)}(\zeta) \equiv \vev{\del_{\zeta} \hat \phi^{(j)}(\zeta)}^{(0)} \qquad \bigl(j \in \mathbb Z/p\mathbb Z\bigr), 
\end{align}
as follows: 
\begin{align}
F_{\rm Cardy}^{(1,1)}(\zeta,Q)= \prod_{j=1}^{p} 
\Bigl(Q-Q^{(j)}(\zeta)\Bigr) = 0.
\end{align}
It is because $\bigl\{Q^{(j)}(\zeta)\bigr\}_{j=1}^p$ exhausts all the branches obtained by analytic continuation of the resolvent operator $Q^{(1)}(\zeta)$ of the $(1,1)$ FZZT-Cardy brane (See also \cite{fim}). In particular, if one chooses the conformal background \cite{MSS} and also chooses the string vacuum%
\footnote{See \cite{CIY4} and also Appendix \ref{Section:ExampleOfIMS} for the discussions of string vacua. } 
which is described by Liouville theory, the spectral curve becomes (\cite{Kostov1,SeSh}), 
\begin{align}
F_{\rm Cardy}^{(1,1)}(\zeta,Q)= \frac{1}{2^{p-1}}\Biggl[T_p\Bigl(\frac{Q}{\beta_{p,q}\,\mu^{q/2p}}\Bigr) - T_q\Bigl(\frac{\zeta}{\sqrt{\mu}}\Bigr)\Biggr]=0, \label{Eqn:KostovSolutionOfChebyshevPolynomialSpectralCurve}
\end{align}
where $T_p(x)$ is the $p$-th Chebyshev polynomial of the first kind. 

It is also convenient to introduce the resolvents of the half-integer twisted fermions: 
\begin{align}
\psi^{(j)}(\zeta) = e^{\hat \phi^{(j)}(\zeta)},\qquad Q^{(j)}(\zeta) \equiv \vev{\del_{\zeta} \hat \phi^{(j)}(\zeta)}^{(0)} \qquad \Bigl(j \in \frac{1}{2} +\mathbb Z/p\mathbb Z\Bigr), 
\end{align}
which are related to the spectral curve as follows:%
\footnote{Note that $Q^{(1+\frac{k}{2})}(\zeta) \equiv e^{-\pi i k }Q^{(1)}(e^{-\pi i k}\zeta)$ $\Leftrightarrow$ $\varphi^{(1+\frac{k}{2})}(\zeta) = \varphi^{(1)}(e^{-\pi i k}\zeta)$. }
\begin{align}
F_{\rm Cardy}^{(1,1)}(\zeta,Q)= \prod_{j=1}^{p} 
\Bigl(Q-Q^{(j)}(\zeta)\Bigr) = \prod_{j=1}^{p} 
\Bigl(Q+Q^{(j+\frac{1}{2})}(-\zeta)\Bigr) = 0.
\end{align}

\subsubsection{Spectral curves of $(r,1)$ FZZT-Cardy branes}

We next analyze the case of $(r,1)$ FZZT-Cardy branes. The spectral curves are defined as follows: 
\begin{align}
F_{\rm Cardy}^{(r,1)}(\zeta,Q) &= \prod_{1\leq j_1 <j_2<\cdots<j_r \leq p} \biggl\{ Q- (-1)^{r+1}Q^{(j_1,j_2,\cdots,j_r)}\Bigl((-1)^{r+1}\zeta\Bigr)\biggr\} = 0, \nn\\
&\text{with}\qquad Q^{(j_1,j_2,\cdots,j_r)}(\zeta) \equiv \sum_{a=1}^r Q^{(j_a)}(\zeta), 
\label{Eqn:R1SpectralCurveAlgebraicEquationDefinition}
\end{align}
where 
\begin{align}
\Theta_{\rm Cardy}^{(r,1)} = e^{\hat \phi_{\rm Cardy}^{(r,1)}(\zeta)},\qquad Q(\zeta) = \vev{\del_{\zeta} \hat \phi_{\rm Cardy}^{(r,1)}(\zeta)}^{(0)} = \sum_{
\begin{subarray}{c}
k=-(r-1) \cr \text{step 2}
\end{subarray}
}^{(r-1)} 
Q^{(1+\frac{k}{2})}(\zeta). 
\end{align}
The result of the spectral curves in the conformal background (and Chebyshev/Liouville vacuum as is before) is given as follows: 

\paragraph{Preparation} For a given index $(j_1,j_2,\cdots,j_r)$ satisfying 
\begin{align}
1 \leq  j_1 < j_2 < \cdots < j_r \leq p, 
\end{align}
we introduce the following radius/angle variables $(s_{j_1,j_2,\cdots,j_r}, \nu_{j_1,j_2, \cdots,j_r})$ as 
\begin{align}
s_{j_1,j_2,\cdots,j_r} \times \exp\Bigl[{ 2\pi i \frac{\nu_{j_1,j_2,\cdots,j_r}}{p}  } \Bigr] \equiv \sum_{a=1}^r e^{\frac{2\pi i }{p} q(j_a-1)}.  \label{Eqn:PhaseSummationSandNu}
\end{align}
These variables satisfy
\begin{align}
s_{j_1,j_2,\cdots,j_r} &= \sqrt{\sum_{a,b} \cosh\Bigl(2\pi \frac{q(j_b-j_a)}{p}\Bigr)} \in \mathbb R, 
\end{align}
and 
\begin{align}
\nu_{j_1, j_2, j_3, \cdots,j_r} = \nu_{1, (j_2-j_1+1), (j_3-j_1+1), \cdots, (j_r-j_1+1)} + q (j_1-1). 
\end{align}
According to this radius/angle variable, we categorize the index into three classes: 
\begin{itemize}
\item [I. ] The cases of $s_{j_1,j_2,\cdots,j_r} = 0$. The number of the solutions (i.e.~the roots of Eq.~\eq{Eqn:R1SpectralCurveAlgebraicEquationDefinition}) is denoted by $N_{\rm I}$ and is given by the following summation of binomial coefficients: 
\begin{align}
N_{\rm I} = \sum_{n=1}^{\mathfrak m_{p,r}} (-1)^{n-1} \sum_{1\leq a_1<\cdots<a_n\leq \mathfrak m_{p,r}}
\binom{\frac{p}
{k_{a_1}k_{a_2}\cdots k_{a_n}}}{\frac{r}{k_{a_1}k_{a_2}\cdots k_{a_n}}}, \label{Eqn:NIformula}
\end{align}
where $\{k_a\}_{a=1}^{\mathfrak m_{p,r}}$ are the prime factors of the greatest common divisor (gcd) of $(p,r)$: 
\begin{align}
\GCD(p,r) = k_1^{n_1} k_2^{n_2} \cdots k_{\mathfrak m_{p,r}}^{n_{\mathfrak m_{p,r}}} \qquad \bigl(k_a \text{: a prime number}\bigr). 
\end{align}
\item [II. ] The cases of $\nu_{j_1,j_2,\cdots,j_r} \in \mathbb Z/2$. The number of the solutions is given by $p N_{\rm II}$, where $N_{\rm II}$ is the number which counts the following ``marked'' index, 
\begin{align}
(1,j_2,j_3,\cdots,j_r)\qquad 1 (= j_1) <j_2<\cdots <j_r \leq p, 
\end{align}
satisfying the condition. 
These marked indices are symbolically denoted by $J^{\rm (II)}_a$ $(1\leq a \leq N_{\rm II})$, and we represent the corresponding variables as 
\begin{align}
s_a^{\rm (II)} \equiv s_{J_a^{\rm (II)}},\qquad \nu_a^{\rm (II)} \equiv \nu_{J_a^{\rm (II)}} \qquad \bigl(a=1,2, \cdots,N_{\rm II}\bigr). 
\end{align}
\item [III. ] The cases of $s_{j_1,j_2,\cdots,j_r} > 0$ and $\nu_{j_1,j_2,\cdots,j_r} \not\in \mathbb Z$. The number of the solutions is given by $2p N_{\rm III}$, where $2 N_{\rm III}$ is the number which counts the ``marked'' index, satisfying the condition. Note that these solutions always appear in complex conjugate pairs:
\begin{align}
(1,j_2,j_3,\cdots,j_r)  \quad \overset{\rm c.c.}{\longleftrightarrow} \quad (1,p+2-j_r,\cdots,p+2-j_3, p+2-j_2). 
\end{align}
This conjugation gives ``2'' of $2N_{\rm III}$. 
These marked indices are symbolically denoted by $J^{\rm (III)}_a$ $(1\leq a \leq N_{\rm III})$ and its conjugation $\bar J^{\rm (III)}_a$ $(1\leq a \leq N_{\rm III})$ so that they satisfy $\{J_a^{\rm (III)}\}_{a=1}^{N_{\rm III}} \cap \{\bar J_a^{\rm (III)}\}_{a=1}^{N_{\rm III}} = \emptyset$. We then use the following notation: 
\begin{align}
&s_a^{\rm (III)} \equiv s_{J_a^{\rm (III)}} = s_{\bar J_a^{\rm (III)}},\quad \nu_a^{\rm (III)} \equiv \nu_{J_a^{\rm (III)}} = -\nu_{\bar J_a^{\rm (III)}} \quad \bigl(a=1,2,\cdots,N_{\rm III}\bigr). 
\end{align}
\end{itemize}
By definition, these three integers satisfy 
\begin{align}
\binom{p}{r} = N_{\rm I} + p\Bigl( N_{\rm II} + 2 N_{\rm III}\Bigr). \label{Eqn:N1N2N3}
\end{align}

\paragraph{General formula} The spectral curve $F^{(r,1)}_{\rm Cardy}(\zeta,Q)=0$ of $(r,1)$ FZZT-Cardy brane is given as follows: 
\begin{align}
0= F_{\rm Cardy}^{(r,1)}(\zeta,Q) 
& \propto  Q^{N_{\rm I}} 
\Biggl[\prod_{a=1}^{N_{\rm II}} \biggl( T_p\Bigl(\frac{(-1)^{r+1}Q/s_{a}^{\rm (II)}}{\beta_{p,q} \mu^{q/2p}}\Bigr) - T_q\Bigl(\frac{(-1)^{r+1}\zeta}{\sqrt{\mu}}\Bigr) \biggr) \Biggr] \times \nn\\
\times&\,  \prod_{b=1}^{N_{\rm III}} \Bigg[\biggl( T_p^{(\nu_b^{\rm (III)})}\Bigl(\frac{(-1)^{r+1}Q/s_{b}^{\rm (III)}}{\beta_{p,q} \mu^{q/2p}}\Bigr) - T_q\Bigl(\frac{(-1)^{r+1}\zeta}{\sqrt{\mu}}\Bigr) \biggr) \times \nn\\
&\qquad \qquad \times  \biggl( T_p^{(-\nu_b^{\rm (III)})}\Bigl(\frac{(-1)^{r+1}Q/s_{b}^{\rm (III)}}{\beta_{p,q} \mu^{q/2p}}\Bigr) - T_q\Bigl(\frac{(-1)^{r+1}\zeta}{\sqrt{\mu}}\Bigr) \biggr) \Bigg], \label{Eqn:GeneralFormulaSpectralCurveR1}
\end{align}
where $T_n^{(\nu)}(x)$ is the deformed Chebyshev function \cite{CIY1} defined by 
\begin{align}
T_n^{(\nu)}(\cosh\theta) = \cosh\bigl(n \theta + 2\pi i \nu\bigr). 
\end{align}
This formula can be shown by noticing that the solutions of this algebraic equation are given by
\begin{align}
Q^{(j_1,j_2,\cdots,j_r)}(\zeta) &= s_{j_1,j_2,\cdots,j_r}  \times \beta_{p,q} \mu^{q/2p} \cosh \Bigl( q\tau +2\pi i \frac{\nu_{j_1,j_2,\cdots,j_r}}{p} \Bigr) \nn\\
&=s_{j_1,j_2,\cdots,j_r}  \times \beta_{p,q} \mu^{q/2p} \, T_{p/q}^{(\frac{\nu_{j_1,j_2,\cdots,j_r}}{p})}\Bigl(\frac{\zeta}{\sqrt{\mu}}\Bigr). 
\end{align}
where $\zeta = \sqrt{\mu} \cosh p \tau$. Note that similar algebraic equations are also found as spectral curves in $(p,q)$ minimal fractional superstring theory \cite{irie2} whose algebraic equations are derived in \cite{CIY1} with solving loop equations of the multi-cut two-matrix models.

\paragraph{Charge conjugation} From the general formula, one can observe that there is a duality between the spectral curves of $(r,1)$-type and $(p-r,1)$-type FZZT-Cardy branes: 
\begin{align}
0= F^{(p-r,1)}_{\rm Cardy }\bigl(\zeta,Q\bigr) = F^{(r,1)}_{\rm Cardy }\bigl(\zeta_{\mathcal C},- Q_{\mathcal C}\bigr),\qquad \bigl(\zeta_{\mathcal C}, Q_{\mathcal C}\bigr) =  \bigl( (-1)^p\zeta,(-1)^p Q\bigr). \label{Eqn:ChargeConjugationInSpectralCurves}
\end{align}
This essentially follows from the dual relation of the phase summation Eq.~\eq{Eqn:PhaseSummationSandNu}: 
\begin{align}
\Biggl[ \sum_{a=1}^r e^{\frac{2\pi i }{p} q(j_a-1)} \Biggr]
= (-1)\times  \Biggl[ \sum_{b=1}^{p-r} e^{\frac{2\pi i }{p} q(l_b-1)} \Biggr], 
\end{align}
where $(j_1,j_2,\cdots,j_r)$ and $(l_1,l_2,\cdots,l_{p-r})$ are any possible divisions of indices satisfying 
\begin{align}
\bigl\{ j_1,j_2,\cdots,j_r \bigr\} \cup \bigl\{ l_1,l_2,\cdots,l_{p-r} \bigr\} = \bigl\{1,2, \cdots, p\bigr\}. 
\end{align}
This duality of the spectral curve is referred to as {\em charge conjugation} which is again discussed also as a duality among Wronskians in Section \ref{SubSubSection:ChargeConjugation} and Section \ref{SubSubSection:PQdualityLaplace}. 

\paragraph{Example 1: The case of $r=2$} 
The three parameters of Eq.~\eq{Eqn:N1N2N3} are given by 
\begin{align}
\frac{p(p-1)}{2} &= \underbrace{0}_{N_{\rm I}} + p\Bigl( \underbrace{\frac{p-1}{2}}_{N_{\rm II}} +  \underbrace{2 \times0}_{2N_{\rm III}}\Bigr) \qquad \bigl(p \in 2\mathbb Z+1\bigr), \\
\frac{p(p-1)}{2} &= \underbrace{\frac{p}{2}}_{N_{\rm I}} + p\Bigl( \underbrace{\frac{p-2}{2}}_{N_{\rm II}} +  \underbrace{2 \times0}_{2N_{\rm III}}\Bigr) \qquad \bigl(p \in 2\mathbb Z\bigr). 
\end{align}
Therefore, one obtains 
\begin{align}
&\underline{\text{Odd $p$ case:}} \nn\\
&\qquad \qquad F_{\rm Cardy}^{(2,1)}(\zeta,Q) \propto \prod_{a=1}^{(p-1)/2}\Biggl[T_p\Bigl(\frac{-Q/s_a}{\beta_{p,q}\,\mu^{q/2p}}\Bigr) - T_q\Bigl(\frac{-\zeta}{\sqrt{\mu}}\Bigr)\Biggr] = 0, \label{Eq:SpectralCurveCardyPodd} \\
&\underline{\text{Even $p$ case:}} \nn\\
&\qquad \qquad F_{\rm Cardy}^{(2,1)}(\zeta,Q) \propto Q^{p/2}\prod_{a=1}^{(p-2)/2} \Biggl[T_p\Bigl(\frac{-Q/s_a}{\beta_{p,q}\,\mu^{q/2p}}\Bigr) - T_q\Bigl(\frac{-\zeta}{\sqrt{\mu}}\Bigr)\Biggr]=0, \label{Eq:SpectralCurveCardyPeven}
\end{align}
Here we define $s_a$ ($1\leq a \leq N_{\rm II}$) as 
\begin{align}
s_a  \equiv s_{1,1+a}^{\rm (II)} = 2 \cos\bigl(\frac{\pi qa}{p}\bigr) \qquad \bigl( 1\leq a \leq N_{\rm II} = \bigl\lfloor \frac{p-1}{2} \bigr\rfloor \bigr). 
\end{align}

\paragraph{Example 2: The case of $r=3$} The three parameters of Eq.~\eq{Eqn:N1N2N3} are given by 
\begin{align}
\frac{p(p-1)(p-2)}{3\cdot 2\cdot 1} &= \underbrace{0}_{N_{\rm I}} + p\Bigl( \underbrace{\frac{p-1}{2}}_{N_{\rm II}} +  \underbrace{\frac{(p-1)(p-5)}{6}}_{2N_{\rm III}}\Bigr) \qquad \bigl(p \not\in 3\mathbb Z, \quad p \in 2 \mathbb Z+1\bigr), \\
\frac{p(p-1)(p-2)}{3\cdot 2\cdot 1} &= \underbrace{0}_{N_{\rm I}} + p\Bigl( \underbrace{\frac{p-2}{2}}_{N_{\rm II}} +  \underbrace{\frac{(p-2)(p-4)}{6}}_{2N_{\rm III}}\Bigr) \qquad \bigl(p \not\in 3\mathbb Z, \quad p \in 2 \mathbb Z\bigr), \\
\frac{p(p-1)(p-2)}{3\cdot 2\cdot 1} &= \underbrace{\frac{p}{3}}_{N_{\rm I}} + p\Bigl( \underbrace{\frac{p-3}{2}}_{N_{\rm II}} +  \underbrace{\frac{(p-3)^2}{6}}_{2N_{\rm III}}\Bigr) \qquad \bigl(p \in 3\mathbb Z, \quad p \in 2 \mathbb Z+1\bigr), \\
\frac{p(p-1)(p-2)}{3\cdot 2\cdot 1} &= \underbrace{\frac{p}{3}}_{N_{\rm I}} + p\Bigl( \underbrace{\frac{p-4}{2}}_{N_{\rm II}} +  \underbrace{\frac{p^2-6p + 12}{6}}_{2N_{\rm III}}\Bigr) \qquad \bigl(p \in 3\mathbb Z, \quad p \in 2 \mathbb Z\bigr). 
\end{align}
Here we show two non-trivial examples: 
\begin{align}
&\underline{\text{$p =6$ cases:}} \quad (N_{\rm I} = 2, N_{\rm II} = 1, N_{\rm III} = 1) \nn\\
&\qquad F_{\rm Cardy}^{(3,1)}(\zeta,Q) \propto Q^2\Biggl[T_p\Bigl(\frac{Q/s_1}{\beta_{p,q}\,\mu^{q/2p}}\Bigr) - T_q\Bigl(\frac{\zeta}{\sqrt{\mu}}\Bigr)\Biggr] \Biggl[T_p\Bigl(\frac{Q}{\beta_{p,q}\,\mu^{q/2p}}\Bigr) - T_q\Bigl(\frac{\zeta}{\sqrt{\mu}}\Bigr)\Biggr]^2 = 0, \label{Eq:SpectralCurveCardy3P6} \\
&\underline{\text{$p =7$ cases:}} \quad (N_{\rm I} = 0, N_{\rm II} = 3, N_{\rm III} = 1) \nn\\
&\qquad F_{\rm Cardy}^{(3,1)}(\zeta,Q) \propto \Biggl[\prod_{a=1}^{3} \biggl(T_p\Bigl(\frac{Q/s_a}{\beta_{p,q}\,\mu^{q/2p}}\Bigr) - T_q\Bigl(\frac{\zeta}{\sqrt{\mu}}\Bigr) \biggr)\Biggr] \times f(\zeta,Q) = 0, \label{Eq:SpectralCurveCardy3P7}
\end{align}
with 
\begin{align}
f(\zeta,Q) &= \Biggl[T_p^{(\nu)}\Bigl(\frac{Q/\sqrt{2}}{\beta_{p,q} \mu^{q/2p}}\Bigr) - T_q\Bigl(\frac{\zeta}{\sqrt{\mu}}\Bigr) \Biggr]  \Biggl[T_p^{(-\nu)}\Bigl(\frac{Q/\sqrt{2}}{\beta_{p,q} \mu^{q/2p}}\Bigr) - T_q\Bigl(\frac{\zeta}{\sqrt{\mu}}\Bigr) \Biggr] \nn\\
=&\biggl(T_p\Bigl(\frac{Q/\sqrt{2}}{\beta_{p,q} \mu^{q/2p}}\Bigr) \biggr)^2 - \frac{13}{8\sqrt{2}} T_p\Bigl(\frac{Q/\sqrt{2}}{\beta_{p,q} \mu^{q/2p}}\Bigr) T_q\Bigl(\frac{\zeta}{\sqrt{\mu}}\Bigr) + \biggl(T_q\Bigl(\frac{\zeta}{\sqrt{\mu}}\Bigr)\biggr)^2 - \frac{343}{512}. 
\end{align}
Here $s_a$ and $\nu$ are given by 
\begin{align}
s_a \equiv s_{1,1+a,p-a+1}^{\rm (II)}= 1 + 2\cos\bigl(\frac{2\pi q a}{p}\bigr) \qquad \bigl(\text{$p=6$ or $p=7$}\bigr),
\end{align}
and $\nu \equiv \nu_{1,3,4}^{\rm (III)}$ with%
\footnote{Note that, since $q\not\in 7\mathbb Z$, 
\begin{align}
s_{1,3,4}^{\rm (III)} = \sqrt{3 + 2\cos \bigl(\frac{2\pi q}{7}\times 2\bigr) + 2\cos \bigl(\frac{2\pi q}{7}\times 3\bigr) + 2\cos \bigl(\frac{2\pi q}{7}\times 1\bigr)} = \sqrt{2}. 
\end{align}} 
\begin{align}
 e^{\frac{2\pi i \nu}{p}} = \frac{1+ e^{\frac{4\pi iq}{p}} + e^{\frac{6\pi iq}{p}} }{\sqrt{2} }\qquad \bigl(p=7\bigr). 
\end{align}

\subsubsection{Spectral curves of $(1,s)$ FZZT-Cardy branes}

The spectral curves of $(1,s)$ FZZT-Cardy branes are obtained by applying Eq.~\eq{EqnFandFtilde:RelationOfSpectralCurves}. Since the spectral curves of $(s,1)$ dual FZZT-Cardy branes are similarly given as 
\begin{align}
0= \widetilde F_{\rm Cardy}^{(s,1)}(\eta,P) 
& \propto  P^{\widetilde N_{\rm I}} 
\Biggl[\prod_{a=1}^{\widetilde N_{\rm II}} \biggl( T_q\Bigl(\frac{(-1)^{s+1}P/\widetilde s_{a}^{\rm (II)}}{\beta_{q,p} \widetilde \mu^{p/2q}}\Bigr) - T_p\Bigl(\frac{(-1)^{s+1}\eta}{\sqrt{\widetilde \mu}}\Bigr) \biggr) \Biggr] \times \nn\\
\times&\,  \prod_{b=1}^{\widetilde N_{\rm III}} \Bigg[\biggl( T_q^{(\widetilde \nu_b^{\rm (III)})}\Bigl(\frac{(-1)^{s+1}P/\widetilde s_{b}^{\rm (III)}}{\beta_{q,p} \widetilde \mu^{p/2q}}\Bigr) - T_p\Bigl(\frac{(-1)^{s+1}\eta}{\sqrt{\widetilde \mu}}\Bigr) \biggr) \times \nn\\
&\qquad \qquad \times  \biggl( T_q^{(-\widetilde \nu_b^{\rm (III)})}\Bigl(\frac{(-1)^{s+1}P/\widetilde s_{b}^{\rm (III)}}{\beta_{q,p} \widetilde \mu^{p/2q}}\Bigr) - T_p\Bigl(\frac{(-1)^{s+1}\eta}{\sqrt{\widetilde \mu}}\Bigr) \biggr) \Bigg], \label{Eqn:GeneralFormulaDualSpectralCurveS1}
\end{align}
the spectral curves of $(1,s)$ FZZT-Cardy branes are given as 
\begin{align}
0  &= F_{\rm Cardy}^{(1,s)}(\zeta,Q)   \propto \widetilde F^{(s,1)}_{\rm Cardy}\Bigl(\frac{Q}{\beta_{p,q} A_{s}}, \beta_{q,p} A_{s}\zeta\Bigr) \nn\\
&\quad \qquad  \propto  \zeta^{\widetilde N_{\rm I}} 
\Biggl[\prod_{a=1}^{\widetilde N_{\rm II}} \biggl( T_p\Bigl(\frac{(-1)^{s+1}Q/A_s}{\beta_{p,q}\, \mu^{q/2p} }\Bigr) - T_q\Bigl(\frac{(-1)^{s+1}\zeta A_s/\widetilde s_{a}^{\rm (II)}}{ \sqrt{ \mu} }\Bigr)  \biggr) \Biggr] \times \nn\\
&\quad \quad \qquad\times\,  \prod_{b=1}^{\widetilde N_{\rm III}} \Bigg[\biggl( T_p\Bigl(\frac{(-1)^{s+1}Q/A_s}{\beta_{p,q}\, \mu^{q/2p} }\Bigr) - T_q^{(\widetilde \nu_b^{\rm (III)})}\Bigl(\frac{(-1)^{s+1}\zeta A_s/\widetilde s_{b}^{\rm (III)}}{\sqrt{ \mu}}\Bigr)  \biggr) \times \nn\\
&\quad \quad \qquad \qquad \qquad \times  \biggl( T_p\Bigl(\frac{(-1)^{s+1}Q/A_s}{\beta_{p,q}\, \mu^{q/2p} }\Bigr) - T_q^{(-\widetilde \nu_b^{\rm (III)})}\Bigl(\frac{(-1)^{s+1}\zeta A_s/\widetilde s_{b}^{\rm (III)}}{\sqrt{ \mu}}\Bigr)  \biggr) \Bigg], \label{Eqn:GeneralFormulaSpectralCurve1S}
\end{align}
where 
\begin{align}
\binom{q}{s} = \widetilde N_{\rm I} + q \Bigl( \widetilde N_{\rm II} + 2 \widetilde N_{\rm III}\Bigr), 
\end{align}
and 
\begin{align}
\widetilde s_{j_1,j_2,\cdots,j_s} \times \exp\Bigl[{ 2\pi i \frac{\widetilde \nu_{j_1,j_2,\cdots,j_s}}{q}  } \Bigr]  \equiv  \sum_{a=1}^s e^{\frac{2\pi i }{q} p(j_a-1)}\qquad \bigl(1\leq j_1<\cdots<j_s \leq p\bigr). 
\end{align}
Note that the order of algebraic equation in $Q$ is given by 
\begin{align}
\text{(Order in $Q$)} =  p \Bigl( \widetilde N_{\rm II} + 2 \widetilde N_{\rm III}\Bigr). 
\end{align}

\subsubsection{Spectral curves of $(r,s)$ FZZT-Cardy branes \label{Subsubsection:GeneralFormulaOfSpectralCurveRS}}
We propose the spectral curves of $(r,s)$ FZZT-Cardy branes to be given as 
\begin{align}
0  &= F_{\rm Cardy}^{(r,s)}(\zeta,Q)   \propto 
\zeta^{\widetilde N_{\rm I} (N_{\rm II} + 2 N_{\rm III})} 
Q^{N_{\rm I}(\widetilde N_{\rm II} + 2 \widetilde N_{\rm III})} \times \nn\\
&\qquad \qquad \qquad \qquad \qquad  \times
f^{\rm (II,II)}_{r,s}(\zeta,Q) f^{\rm (II,III)}_{r,s}(\zeta,Q) f^{\rm (III,II)}_{r,s}(\zeta,Q) f^{\rm (III,III)}_{r,s}(\zeta,Q), 
\nn\\
&\qquad \text{with} \nn\\
&f^{\rm (II,II)}_{r,s}(\zeta,Q) \equiv  \prod_{a=1}^{N_{\rm II}}\prod_{b=1}^{\widetilde N_{\rm II}} 
\biggl( T_p\Bigl(\frac{(-1)^{r+s}Q/A_s s_a^{\rm (II)}}{\beta_{p,q}\, \mu^{q/2p} }\Bigr) - T_q\Bigl(\frac{(-1)^{r+s}\zeta A_s/\widetilde s_{a}^{\rm (II)}}{ \sqrt{ \mu} }\Bigr)  \biggr),  \nn\\
&f^{\rm (II,III)}_{r,s}(\zeta,Q) \equiv  \prod_{a=1}^{N_{\rm II}} \prod_{b=1}^{\widetilde N_{\rm III}} \Bigg[\biggl( T_p\Bigl(\frac{(-1)^{s+r}Q/A_s s_a^{\rm (II)}}{\beta_{p,q}\, \mu^{q/2p} }\Bigr) - T_q^{(\widetilde \nu_b^{\rm (III)})}\Bigl(\frac{(-1)^{r+s}\zeta A_s /\widetilde s_{b}^{\rm (III)}}{\sqrt{ \mu}}\Bigr)  \biggr) \times \nn\\
&\quad \quad \qquad \qquad \qquad \times  \biggl( T_p\Bigl(\frac{(-1)^{r+s}Q/A_s s_a^{\rm (II)}}{\beta_{p,q}\, \mu^{q/2p} }\Bigr) - T_q^{(-\widetilde \nu_b^{\rm (III)})}\Bigl(\frac{(-1)^{r+s}\zeta A_s/\widetilde s_{b}^{\rm (III)}}{\sqrt{ \mu}}\Bigr)  \biggr) \Bigg], \nn\\
&f^{\rm (III,II)}_{r,s}(\zeta,Q) \equiv  \prod_{a=1}^{N_{\rm III}} \prod_{b=1}^{\widetilde N_{\rm II}} \Bigg[\biggl( T_p^{(\nu_a^{\rm (III)})}\Bigl(\frac{(-1)^{s+r}Q/A_s s_a^{\rm (III)}}{\beta_{p,q}\, \mu^{q/2p} }\Bigr) - T_q\Bigl(\frac{(-1)^{r+s}\zeta A_s/\widetilde s_{b}^{\rm (II)}}{\sqrt{ \mu}}\Bigr)  \biggr) \times \nn\\
&\quad \quad \qquad \qquad \qquad \times  \biggl( T_p^{(-\nu_a^{\rm (III)})}\Bigl(\frac{(-1)^{r+s}Q/A_s s_a^{\rm (III)}}{\beta_{p,q}\, \mu^{q/2p} }\Bigr) - T_q\Bigl(\frac{(-1)^{r+s}\zeta A_s/\widetilde s_{b}^{\rm (II)}}{\sqrt{ \mu}}\Bigr)  \biggr) \Bigg], \nn\\ 
&f^{\rm (III,III)}_{r,s}(\zeta,Q) \equiv  \prod_{a=1}^{N_{\rm III}} \prod_{b=1}^{\widetilde N_{\rm III}} \Bigg[\biggl( T_p^{(\nu_a^{\rm (III)})}\Bigl(\frac{(-1)^{s+r}Q/A_s s_a^{\rm (III)}}{\beta_{p,q}\, \mu^{q/2p} }\Bigr) - T_q^{(\widetilde \nu_b^{\rm (III)})}\Bigl(\frac{(-1)^{r+s}\zeta A_s /\widetilde s_{b}^{\rm (II)}}{\sqrt{ \mu}}\Bigr)  \biggr) \times \nn\\
&\quad \quad \qquad \qquad \qquad \times  \biggl( T_p^{(-\nu_a^{\rm (III)})}\Bigl(\frac{(-1)^{r+s}Q/A_s s_a^{\rm (III)}}{\beta_{p,q}\, \mu^{q/2p} }\Bigr) - T_q^{(\widetilde \nu_b^{\rm (III)})}\Bigl(\frac{(-1)^{r+s}\zeta A_s/\widetilde s_{b}^{\rm (II)}}{\sqrt{ \mu}}\Bigr)  \biggr) \times \nn\\
&\quad \quad \qquad \qquad \qquad \times \biggl( T_p^{(\nu_a^{\rm (III)})}\Bigl(\frac{(-1)^{s+r}Q/A_s s_a^{\rm (III)}}{\beta_{p,q}\, \mu^{q/2p} }\Bigr) - T_q^{(-\widetilde \nu_b^{\rm (III)})}\Bigl(\frac{(-1)^{r+s}\zeta A_s /\widetilde s_{b}^{\rm (II)}}{\sqrt{ \mu}}\Bigr)  \biggr) \times \nn\\
&\quad \quad \qquad \qquad \qquad \times  \biggl( T_p^{(-\nu_a^{\rm (III)})}\Bigl(\frac{(-1)^{r+s}Q/A_s s_a^{\rm (III)}}{\beta_{p,q}\, \mu^{q/2p} }\Bigr) - T_q^{(-\widetilde \nu_b^{\rm (III)})}\Bigl(\frac{(-1)^{r+s}\zeta A_s/\widetilde s_{b}^{\rm (II)}}{\sqrt{ \mu}}\Bigr)  \biggr)
\Bigg]. \label{Eqn:GeneralFormulaSpectralCurveRS}
\end{align}
The point of this proposed formula is that this respects {\em the $p-q$ duality transformation} (i.e.~Eq.~\eq{EqnFandFtilde:RelationOfSpectralCurves}), and {\em charge conjugations} (i.e.~Eq.~\eq{Eqn:ChargeConjugationInSpectralCurves}): 
\begin{align}
1) & \quad 0 = \widetilde F^{(s,r)}_{\rm Cardy}\Bigl(\frac{Q}{\beta_{p,q} A_s \widetilde A_r}, \beta_{q,p} A_s \widetilde A_r\zeta\Bigr) \,\, \propto\,\,  F^{(r,s)}_{\rm Cardy}(\zeta,Q), \nn\\
2) & \quad 0= F^{(r,s)}_{\rm Cardy}\bigl(\zeta,Q\bigr) \propto F^{(p-r,s)}_{\rm Cardy}\bigl(\zeta_{\mathcal C}, -Q_{\mathcal C}\bigr)\qquad \Bigl( \bigl(\zeta_{\mathcal C}, Q_{\mathcal C}\bigr) =  \bigl( (-1)^p\zeta,(-1)^p Q\bigr) \Bigr), \nn\\
3) & \quad 0= \widetilde F^{(s,r)}_{\rm Cardy}\bigl(\eta,P\bigr) \propto \widetilde F^{(q-s,r)}_{\rm Cardy}\bigl(\eta_{\mathcal C},-P_{\mathcal C}\bigr)\qquad \Bigl( \bigl(\eta_{\mathcal C}, P_{\mathcal C}\bigr) =  \bigl( (-1)^q\eta,(-1)^q P\bigr) \Bigr). 
\end{align}
The last charge conjugation is equivalent to the following {\em dual charge conjugation}: 
\begin{align}
0= F^{(r,s)}_{\rm Cardy}\bigl(\zeta,Q\bigr) \propto F^{(r,q-s)}_{\rm Cardy}\bigl(-\zeta_{\mathcal D},Q_{\mathcal D}\bigr)\qquad \Bigl( \bigl(\zeta_{\mathcal D}, Q_{\mathcal D}\bigr) =  \bigl( (-1)^q\zeta,(-1)^q Q\bigr) \Bigr). 
\end{align}
Note that, in a naive consideration, one might think that it is natural to consider more higher-order algebraic equations rather than the above formula. However, such a construction results in some algebraic equation which does not respect the spectral $p-q$ duality and charge conjugation. 

In fact, in this formula, we have selected some branches combined in the formula. This indicates {\em non-perturbative decoupling of the spectral curves} (i.e.~at the level of isomonodromy systems). In Section \ref{Section:TowardInteriorOfKacTable}, we will see that the non-perturbative decoupling phenomenon is actually favored in our construction based on Wronskians. In particular, in the case of the $(3,4)$-system, we will see evidence that this decoupling phenomenon occurs and results in the general formula Eq.~\eq{Eqn:GeneralFormulaSpectralCurveRS} (in Section \ref{ConsistencyOfSDWronskiansProposal}). 

According to the formula Eq.~\eq{Eqn:GeneralFormulaSpectralCurveRS}, the orders of the algebraic equations are given by 
\begin{align}
\text{(Order in $Q$)} = \mathfrak d_{r,s}^{(p,q)}, \qquad \text{(Order in $\zeta$)} = \mathfrak d_{s,r}^{(q,p)}, 
\end{align}
where the integer $\mathfrak d_{r,s}^{(p,q)}$ is given by 
\begin{align}
\mathfrak d_{r,s}^{(p,q)} \equiv \binom{p}{r} \frac{1}{q} \left[ \binom{q}{s} - \sum_{n=1}^{\mathfrak m_{p,r}} (-1)^{n-1} \sum_{1\leq a_1<\cdots<a_n\leq \mathfrak m_{p,r}}
\binom{\frac{p}
{k_{a_1}k_{a_2}\cdots k_{a_n}}}{\frac{r}{k_{a_1}k_{a_2}\cdots k_{a_n}}} \right] \label{Eqn:FormulaForDimensionDeltaRS}
\end{align}
with the integer $\mathfrak m_{p,r}$ associated with the greatest common divisor of $(p,r)$ (See also Eq.~\eq{Eqn:NIformula}): 
\begin{align}
\GCD(p,r) = k_1^{n_1} k_2^{n_2} \cdots k_{\mathfrak m_{p,r}}^{n_{\mathfrak m_{p,r}}} \qquad \bigl(k_a \text{: prime number}\bigr). 
\label{Eqn:FormulaForDimensionDeltaRS_GCD}
\end{align}

\subsection{Notes on $(r,s)$-type ZZ-Cardy brane corrections \label{Subsection:ZZCardyBranes}}

In this subsection, we discuss ZZ-Cardy branes associated with the FZZT-Cardy branes. 

As is discussed by Seiberg and Shih \cite{SeSh} (also \cite{ZZ,Martinec}), ZZ-branes are associated with a non-trivial cycle integral of the resolvent (i.e.~the differential $Q(\zeta)d\zeta$) on the spectral curve. On the matrix-model side, such cycle integrals are identified with instanton actions of the single eigenvalue dynamics which is evaluated by the semi-classical saddle point approximation of the matrix models \cite{McGreevyVerlinde, Martinec, KMS, AKK}. 

This identification is nicely described by the free-fermion formulation \cite{fy1,fy2,fy3} and is analyzed in \cite{fis,fim} (See also for the two-matrix-model analysis \cite{KazakovKostov}). The following discussion is then based on the analysis given in \cite{fis,fim}. According to the general prescription \cite{fy1,fy2,fy3}, we identify $(r,s)$-type $(m,n)$ ZZ-branes as follows: ZZ-branes are stationary points of the D-instanton action in free-fermion formalism on the spectral curve, where the D-instanton action $\varphi_{\vev{a,b}}(\zeta)$ on the $(r,s)$ FZZT-Cardy branes is given by 
\begin{align}
\varphi_{\vev{a,b}}^{(r,s)}(\zeta) \equiv \int^\zeta d\zeta' \Bigl( Q_{a}(\zeta') - Q_{b}(\zeta') \Bigr)\qquad \Bigl(1\leq a, b \leq \mathfrak d_{r,s}^{(p,q)}\Bigr)\nn\\
\text{with}\quad F^{(r,s)}_{\rm Cardy}(\zeta,Q) = \prod_{a=1}^{\mathfrak d_{r,s}^{(p,q)} }\bigl(Q-Q_{a}(\zeta) \bigr) = 0,\qquad \qquad \qquad  
\end{align}
where $(a,b)$ represents a pair of branches of the spectral curve. 
The disk amplitude of the $(r,s)$-type $(m,n)$ ZZ-brane $\mathcal A_{{\rm ZZ},(m,n)}^{(r,s)}$ is given by 
\begin{align}
\mathcal A_{{\rm ZZ},(m,n)}^{(r,s)} = \varphi_{\vev{a,b}}^{(r,s)}(\zeta^*)  \qquad \Bigl( {}^\exists(a,b;\zeta^*) \quad \text{satisfying} \quad \frac{\del \varphi_{\vev{a,b}}^{(r,s)}(\zeta^*)  }{\del \zeta}= 0 \Bigr),  \label{Eqn:ZZbraneAsDinstantonOperatorsRS}
\end{align}
where we choose $(a,b;\zeta^*)$ properly to associate it with the indices $(m,n)$. For the principal ZZ-branes (i.e.~of $(1,1)$-type), all the $(m,n)$ labeling of Liouville theory is shown to be given by the indices $(a,b;\zeta^*)$ exhaustively \cite{fis,fim}. 

For the general cases, the precise identification has not yet been carried out; however this would be related to an interesting phenomenon at the level of non-perturbative completions. 

Naively, one can expect that all the ZZ-branes are given by Eq.~\eq{Eqn:ZZbraneAsDinstantonOperatorsRS} and no other types of instantons (which are not in the ZZ-brane spectrum of Liouville theory) would arise as the saddle points of the D-instanton action. This consideration is natural since these ZZ-Cardy branes are related to the instanton effects of different isomonodromy systems which are commonly describing the same string equation.%
\footnote{These isomonodromy systems are derived in Section \ref{Section:WronskiansForR1FZZTCardyBranes} and Section \ref{Section:1SFZZTCardyBranesAndKacTable}}
The instanton spectrum of the spectral curve should then be intrinsically determined by the string equation, which is essentially given by that of the principal FZZT-branes.%
\footnote{Note that instanton effects appear within the perturbation theory. Therefore, ZZ-branes are intrinsically perturbative objects. In this sense, there is no difference from the principal ZZ-branes. This is the most critical difference from the FZZT-Cardy branes. } 

However, from the inverse isomonodromy approach (i.e.~the Riemann-Hilbert problem, see \cite{ItsBook}), one can also expect that the above consideration may be too naive. 
There are a few comments in order: 
\begin{itemize}
\item Interestingly, the spectral curves of FZZT-Cardy branes are similar to those of fractional-supersymmetric minimal string theory \cite{CIY1}, where the spectral curve of the FZZT-(Cardy) branes are similarly factorized into various kinds of irreducible spectral curves. Therefore, there exist ``ZZ-branes'' which connect different irreducible branches of the spectral curves.%
\footnote{Note that the D-instanton/ZZ-brane spectrum of the isomonodromy systems is completely parallel to that of the free-fermion (i.e.~Eq.~\eq{Eqn:ZZbraneAsDinstantonOperatorsRS}). It can be easily seen by applying Riemann-Hilbert analysis to the isomonodromy systems \cite{ItsBook,CIY4,CIY5}. Therefore, all the D-instantons which are given by Eq.~\eq{Eqn:ZZbraneAsDinstantonOperatorsRS} shall appear as non-perturbative effects of minimal string theory.}

\item These ``connecting ZZ-branes'', at the first sight, seem not to be in the traditional ZZ-brane spectrum. Regarding this point, therefore, there are two possibilities:
\begin{itemize}
\item [1)] The connecting ZZ-branes are again expressed by a superposition of principal ZZ-brane amplitudes. This means that the ZZ-Cardy branes and instanton saddle points are completely prescribed by Eq.~\eq{Eqn:ZZbraneAsDinstantonOperatorsRS}. 
\item [2)] The connecting ZZ-branes are not expressed by a superposition of principal ZZ-brane amplitudes. This means that {\em these connecting ZZ-branes should be forbidden} by the fact that both systems are described by the same string equation. 
\end{itemize}
\end{itemize}
Let us examine the second possibility in detail. 
\begin{itemize}
\item If the second possibility is true, it means that there appear non-standard ZZ-branes in the correction of the bulk free-energy Eq.~\eq{EqIntroExpansionOfFreeEnergy}, 
\begin{align}
\mathcal F (g) \asymeq &\sum_{n=0}^\infty g^{2n-2} \mathcal F_n(\mathscr S_0) + \sum_{a=1}^{\mathfrak g} \theta_a\, g^{\gamma_a} \exp\Bigl[\sum_{n=0}^\infty g^{n-1} \mathcal F_n^{(a)}(\mathscr S_0)\Bigr] + \nn\\
&+ \underbrace{ \sum_{b=1}^{\mathfrak h} \xi_a\, g^{\gamma_a'} \exp\Bigl[\sum_{n=0}^\infty g^{n-1} \mathcal F_n^{(a)}(\mathscr S_0^{(r,s)})\Bigr]}_\text{exceptional ZZ branes} +  O(\theta^2),
\end{align}
for the general non-perturbative completion based on the spectral curve of FZZT-Cardy branes. 
These ``unexpected'' brane contributions are referred to as {\em exceptional ZZ branes}. Since this free-energy is purely describing bulk physics without any FZZT-kinds of branes in the bulk, such exceptional branes should not exist in the correction.

This may happen because this system is described by using FZZT-Cardy branes (not by the standard principal FZZT-branes) whose isomonodromy systems are of large size in general (shown later in Section \ref{Section:WronskiansForR1FZZTCardyBranes} and Section \ref{Section:1SFZZTCardyBranesAndKacTable}).

Although these isomonodromy systems describe the same string equation, such a fact already gives a constraint on the coefficient matrices of the isomonodromy systems. If one performs the Riemann-Hilbert analysis, the general spectrum of ZZ-branes follows from the spectral curves of FZZT-Cardy branes (which would be different from the spectrum of Liouville theory). Therefore, we mean that we should carefully choose the instanton corrections (i.e.~giving the physical constraint on the general isomonodromy systems) such that these two spectra coincide. 

Therefore, if one requires that these different $(r,s)$ FZZT-Cardy branes describe the same bulk physics, we should impose an exclusion principle to forbid such instanton corrections, which is understood as {\em another duality constraint} discussed in \cite{CIY5}. 

\item Another interesting fact is that the single minimal string theory is now described by different spectral curves: 
\begin{align}
F^{(1,1)}_{\rm Cardy}(\zeta,Q)=0 \qquad \Rightarrow \qquad \mathcal F(g) \qquad \Leftarrow \qquad F^{(r,s)}_{\rm Cardy}(\zeta,Q)=0, 
\end{align}
although these spectral curves are generally very different. However, all the spectral curves include the same irreducible curve which is essentially given by that of the principal FZZT-brane (i.e.~$F^{(1,1)}_{\rm Cardy}(\zeta,Q)=0$). 

Therefore, one of the easy solutions to this problem is that the irreducible curve of the principal FZZT-brane would be {\em non-perturbatively decoupled} from the other irreducible curves of the total spectral curve. It is interesting to evaluate this issue quantitatively, but is already far from our scope of this paper. Hence, we leave it for future investigation. 

\end{itemize}

\section{Wronskians for the $(r,1)$ FZZT-Cardy branes \label{Section:WronskiansForR1FZZTCardyBranes}}

In this section, based on the Wronskian description Eq.~\eq{Eqn:CardyBranesAsWronskianIdentification}, we start to discuss differential-equation systems which describe the FZZT-Cardy branes. There are two different schemes of differential-equation systems: One is given by linear differential equations with Schur-derivatives; and the other is given by linear differential equations with rational coefficients, i.e.~the isomonodromy systems. 

This section focuses only on the case of $(r,1)$-type FZZT-Cardy brane. The other types of FZZT-Cardy branes require additional consideration and are discussed in Section \ref{Section:1SFZZTCardyBranesAndKacTable} and Section \ref{Section:TowardInteriorOfKacTable}. 

Note that, in this section, we mostly adopt the abbreviations mentioned in Section \ref{Subsubsection:AbbreviationAndNotations}, 
because the following general rank-$r$ Wronskian functions 
\begin{align}
W^{[j_1,j_2,\cdots,j_r]}_{\varnothing}(t;\zeta)\qquad \bigl(1\leq j_1 < \cdots < j_r \leq p\bigr), \label{Eq:BinomPRwavefunctions}
\end{align} 
as well as $(r,1)$ FZZT-Cardy branes $\vev{\Theta_{\rm Cardy}^{(r,1)}(\zeta)}$ satisfy the same differential equation. Therefore, we just employ the following collective expression of the Wronskians: 
\begin{align}
\vev{\Theta_{\rm Cardy}^{(r,1)}(\zeta)} \qquad \to \qquad W^{(r)}_\varnothing(t;\zeta) \qquad \bigl(r=1,2,\cdots,p-1\bigr). 
\end{align}

\subsection{Schur-differential equations for $(r,1)$ FZZT-Cardy branes \label{Subsection:SchurDifferentialEqnR1}}

The first scheme for describing the FZZT-Cardy branes is understood as an extension of the Baker-Akhiezer system Eq.~\eq{Eqn:Baker-AkhiezerSystemsSection222psi} and Eq.~\eq{Eqn:Baker-AkhiezerSystemsSection222chi}. The explicit form of the Baker-Akhiezer systems is as follows: 
\begin{align}
&\underline{\text{Principal FZZT brane $\psi(t;\zeta)$:}} \nn\\
&\qquad \left\{
\begin{array}{c}
\ds \zeta \,\psi(t;\zeta) = \bP(t;\del) \,\psi(t;\zeta) \equiv \Bigl[2^{p-1} \del^p + \sum_{n=2}^{p} u_n(t) \del^{p-n} \Bigr] \psi(t;\zeta) \cr
\ds g \frac{\del}{\del \zeta} \psi(t;\zeta) = \bQ(t;\del) \,\psi(t;\zeta) \equiv \beta_{p,q}\Bigl[2^{q-1} \del^q + \sum_{n=2}^{q} v_n(t) \del^{q-n}\Bigr] \psi(t;\zeta)
\end{array}
\right. \\
&\underline{\text{Principal dual FZZT brane $\chi(t;\zeta)$:}} \nn\\
&\qquad \left\{
\begin{array}{c}
\ds \eta \,\chi(t;\eta) = \widetilde \bP(t;\del) \,\chi(t;\eta) \equiv \Bigl[2^{q-1} \del^q + \sum_{n=2}^{q} \widetilde v_n(t) \del^{q-n} \Bigr] \chi(t;\eta) \cr
\ds g \frac{\del}{\del \eta} \chi(t;\eta) = \widetilde \bQ(t;\del) \,\chi(t;\eta) \equiv \beta_{q,p}\Bigl[2^{p-1} \del^p + \sum_{n=2}^{p} \widetilde u_n(t) \del^{p-n}\Bigr] \chi(t;\eta)
\end{array}
\right.
\end{align}
where 
\begin{align}
\beta_{q,p} = (-1)^{p+q} \beta_{p,q}
\end{align}
and 
\begin{align}
&\left\{
\begin{array}{cc}
\ds \widetilde u_n(t) \equiv \sum_{j=0}^{n-2} (-1)^{n-j}\binom{p-n+j}{j}  u_{n-j}^{(j)}(t)
\qquad &\bigl(n=2,3,\cdots,p\bigr) \cr
\ds \widetilde v_n(t) \equiv \sum_{j=0}^{n-2} (-1)^{n-j}\binom{q-n+j}{j}  v_{n-j}^{(j)}(t)
\qquad  &\bigl(n=2,3,\cdots,q\bigr)  
\end{array}
\right.. \label{Eqn:Section4CoefficientsUVandTildeUV}
\end{align}
Compared to the Baker-Akhiezer systems, on the other hand, the new type of equations are constituted of the Schur-derivatives $\mathcal S_\lambda(\del)$ (not of the usual derivatives $\del^n$). In this sense, we refer to such a new type of differential equations as {\em Schur-differential equations}. 

\subsubsection{Schur-differential equations}
The Schur-differential equation system for the rank-$r$ Wronskians  $W^{(r)}_\varnothing(t;\zeta)$ is given by the following pair of partial differential equations: 
\begin{align}
\left\{
\begin{array}{rl}
\text{${\mathscr P}_\lambda^{(r)}$-eqn.}:&
\ds \zeta\,W_\lambda^{(r)}(t;\zeta) = {\mathscr P}_{\lambda,r}^{(r)} (t;\del) \, W_\varnothing^{(r)}(t;\zeta), \cr
\text{${\mathscr Q}_\lambda^{(r)}$-eqn.}:&
\ds g\frac{\del}{ \del \zeta} W_\lambda^{(r)}(t;\zeta) = \beta_{p,q} \sum_{a=1}^r {\mathscr Q}_{\lambda,a}^{(r)} (t;\del) \, W_\varnothing^{(r)}(t;\zeta),
\end{array}
\right.
\label{Eq:Pequation}
\end{align}
for a partition $\lambda=(\lambda_r,\cdots,\lambda_1)$ which can be generally negative (as Eq.~\eq{Eqn:ExistenceOfWronskiansLambda}). The first equation is referred to as {\em ${\mathscr P}_\lambda^{(r)}$-equation} and the second equation is as {\em ${\mathscr Q}_\lambda^{(r)}$-equation}. 
\begin{itemize}
\item [1) ] The coefficient operators $\bigl\{ {\mathscr P}_{\lambda,a}^{(r)} (t;\del)\bigr\}_{a=1}^r$ and $\bigl\{ {\mathscr Q}_{\lambda,a}^{(r)} (t;\del) \bigr\}_{a=1}^r$ are partial differential operators which are given by 
\begin{align}
\left\{
\begin{array}{l}
\ds
{\mathscr P}_{\lambda,a}^{(r)} (t;\del)  \equiv 2^{p-1} \mathcal S_{(\cdots,\underset{\text{$a$-th comp.}}{ \footnotesize \lambda_a + p},\cdots)}(\del) + \sum_{n=2}^{p+(\lambda_a + a-1)} U_n^{(\lambda_a + a-1)}(t) \, \mathcal S_{(\cdots,\underset{\text{$a$-th comp.}}{ \footnotesize \lambda_a + p-n},\cdots)}(\del),\cr
\ds
{\mathscr Q}_{\lambda,a}^{(r)} (t;\del) \equiv 2^{q-1} \mathcal S_{(\cdots,\underset{\text{$a$-th comp.}}{ \footnotesize \lambda_a + q},\cdots)}(\del) + \sum_{n=2}^{q+(\lambda_a + a-1)} V_n^{(\lambda_a + a-1)}(t) \, \mathcal S_{(\cdots,\underset{\text{$a$-th comp.}}{ \footnotesize \lambda_a + q-n},\cdots)}(\del).
\end{array}
\right.
 \label{Eqn:Schur-differentialOperatorPeq1}
\end{align}
\item [2) ] Because of Eq.~\eq{Eqn:ExistenceOfWronskiansLambda}, for any (generally negative) partition $\lambda = (\lambda_r,\cdots,\lambda_1)$, the following numbers are non-negative integers: 
\begin{align}
\lambda_a + a-1\geq 0\qquad (a=1,2,\cdots,r). 
\end{align}
\item [3) ] The coefficients $\bigl\{ U_n^{(\ell)}(t) \bigr\}_{n=2}^p$ and $\bigl\{ V_n^{(\ell)}(t) \bigr\}_{n=2}^q$ are given by the coefficients of the Baker-Akhiezer systems, $\bigl\{u_n(t)\bigr\}_{n=2}^p$ and $\bigl\{v_n(t)\bigr\}_{n=2}^q$, as
\begin{align}
U_n^{(\ell)}(t) \equiv \sum_{j=0}^{{\rm min}[\ell,n-2]} \binom{\ell}{j} \, \del^{j} u_{n-j}(t),\qquad 
V_n^{(\ell)}(t) \equiv \sum_{j=0}^{{\rm min}[\ell,n-2]} \binom{\ell}{j} \, \del^{j} v_{n-j}(t). 
\end{align}
\end{itemize}
Note that one can also replace the ${\mathscr P}_\lambda^{(r)}$-equation of Eq.~\eq{Eq:Pequation} by other types of ${\mathscr P}_\lambda^{(r)}$-equation:
\begin{align}
\zeta\,W_\lambda^{(r)}(t;\zeta) &= {\mathscr P}_{\lambda,a}^{(r)} (t;\del) \, W_\varnothing^{(r)}(t;\zeta) \qquad \bigl(a=1,2,\cdots,r-1\bigr).
\end{align}
However, one can see that such a alternative choice of ${\mathscr P}_\lambda^{(r)}$-equation is equivalent to the original ${\mathscr P}_\lambda^{(r)}$-equation of Eq.~\eq{Eq:Pequation}. Also note that the ${\mathscr P}_\lambda^{(r)}$-equation provides non-trivial equations even in the cases of partitions $\lambda$ with negative components. For example, $\lambda=(-1,0)$ gives 
\begin{align}
\text{${\mathscr P}_{(-1,0)}^{(r)}$-eqn.}:\qquad 
0=\zeta\,W_{\tiny\reflectbox{\yng(1,0)}}^{(r)}(t;\zeta) = {\mathscr P}_{{(-1,0)},r}^{(r)} (t;\del) \, W_\varnothing^{(r)}(t;\zeta),  \label{Eq:PeqNegativeLambda}
\end{align}
which does not depend on $\zeta$ explicitly. This kind of ${\mathscr P}_\lambda^{(r)}$-equation is important in the next section. On the other hand, the ${\mathscr Q}_\lambda^{(r)}$-equation does not give such a non-trivial equation when the partition possesses a negative component, because there happens cancellation of equations and the ${\mathscr Q}_\lambda^{(r)}$-equation trivially vanishes. 

Before moving to the next discussion, we also show the Schur-differential equations of the $(s,1)$-type dual FZZT-Cardy branes, which are denoted as 
\begin{align}
\left\{
\begin{array}{rl}
\text{$\widetilde {\mathscr P}_\lambda^{(r)}$-eqn.}:&
\ds \eta\,\widetilde W_\lambda^{(s)}(t;\eta) = \widetilde{\mathscr P}_{\lambda,s}^{(s)} (t;\del) \, \widetilde W_\varnothing^{(s)}(t;\eta) \cr
\text{$\widetilde {\mathscr Q}_\lambda^{(r)}$-eqn.}:&
\ds g\frac{\del}{ \del \eta} \widetilde W_\lambda^{(s)}(t;\eta) = \beta_{q,p} \sum_{a=1}^s \widetilde {\mathscr Q}_{\lambda,a}^{(s)} (t;\del) \, \widetilde W_\varnothing^{(s)}(t;\eta) 
\end{array}
\right., \label{Eqn:SchurDifferentialForDualBranes}
\end{align}
with 
\begin{align}
&\left\{
\begin{array}{l}
\ds \widetilde {\mathscr P}_{\lambda,a}^{(s)} (t;\del)  \equiv 2^{q-1} \mathcal S_{(\cdots,\underset{\text{$a$-th comp.}}{ \footnotesize \lambda_a + q},\cdots)}(\del) + \sum_{n=2}^{q+(\lambda_a + a-1)} \widetilde V_n^{(\lambda_a + a-1)}(t) \, \mathcal S_{(\cdots,\underset{\text{$a$-th comp.}}{ \footnotesize \lambda_a + q-n},\cdots)}(\del) \cr
\ds \widetilde {\mathscr Q}_{\lambda,a}^{(s)} (t;\del) \equiv 2^{p-1} \mathcal S_{(\cdots,\underset{\text{$a$-th comp.}}{ \footnotesize \lambda_a + q},\cdots)}(\del) + \sum_{n=2}^{q+(\lambda_a + a-1)} \widetilde U_n^{(\lambda_a + a-1)}(t) \, \mathcal S_{(\cdots,\underset{\text{$a$-th comp.}}{ \footnotesize \lambda_a + q-n},\cdots)}(\del) 
\end{array}
\right., \nn\\
&\text{and}\qquad  \widetilde U_n^{(\ell)}(t) \equiv \sum_{j=0}^{{\rm min}[\ell,n-2]} \binom{\ell}{j} \, \del^{j} \widetilde u_{n-j}(t),\qquad 
\widetilde V_n^{(\ell)}(t) \equiv \sum_{j=0}^{{\rm min}[\ell,n-2]} \binom{\ell}{j} \, \del^{j} \widetilde v_{n-j}(t), 
\end{align}
where the coefficients $\bigl\{\widetilde u_n(t)\bigr\}_{n=2}^p$ and $\bigl\{\widetilde v_n(t)\bigr\}_{n=2}^q$ are given in Eq.~\eq{Eqn:Section4CoefficientsUVandTildeUV}. 

\subsubsection{Derivatives on the Schur-differential equations}

In view of the similarity to the Baker-Akhiezer system, one might be tempted to consider an analogy of the Douglas equation in these Schur-differential equation systems. In order to define such a concept, one may need to consider an action of Schur-derivatives on the Schur-differential equation systems. However, it is not easy to find a standard definition of such a operation: 
\begin{align}
\mathcal S_\mu(\del) \times \Bigl[ f(t)\, \mathcal S_\lambda(\del) \Bigr] =\,\, \dots \qquad ?
\end{align}
The only exception happens for the action of $\mathcal S_{\tiny \yng(1)}(\del)$, which is just the simple derivative: 
\begin{align}
\square \times \Bigl[ f(t)\, \mathcal S_\lambda(\del) \Bigr]= \del f(t)\, \mathcal S_\lambda (\del) + f(t) \, \mathcal S_{\square \times \lambda}(\del). 
\end{align}
This operation gives the following equations: 
\begin{align}
\text{$\square^n\times {\mathscr P}_\lambda^{(r)}$-eqn.}:&&
\zeta\,W_{\square^n\times \lambda}^{(r)}(t;\zeta) &={\mathscr P}_{\square^n\times \lambda,r}^{(r)} (t;\del) \, W_\varnothing^{(r)}(t;\zeta), \\
\text{$\square^n\times{\mathscr Q}_\lambda^{(r)}$-eqn.}:&&
g\frac{\del}{ \del \zeta} W_{\square^n\times \lambda}^{(r)}(t;\zeta) &=  \beta_{p,q} \sum_{a=1}^r {\mathscr Q}_{\square^n\times \lambda,a}^{(r)} (t;\del) \, W_\varnothing^{(r)}(t;\zeta).
\end{align}
These can be followed by the relation, 
\begin{align}
U_{n}^{(\ell+1)}(t) = U_n^{(\ell)}+\del U_n^{(\ell)}(t), 
\end{align}
which can be obtained by $\binom{\ell+1}{j} = \binom{\ell}{j} + \binom{\ell}{j-1}$. 

Therefore, we still have not found out the Schur-differential counterpart of the Douglas equation of the Baker-Akhiezer systems, which shall be reserved for future investigation.

\subsection{Isomonodromy systems for $(r,1)$ FZZT-Cardy branes \label{Subsection:IMSforR1FZZTCardyBranes}}

\subsubsection{Schur-differential equations as constraint equations \label{Subsubsection:SchurAsConstraintsInIMS}}

One of the important aspects of the Schur-differential equations is to play a role of {\em constraint equations} for the Young-diagram space. This leads to the second scheme of differential-equation systems, i.e.~the isomonodromy systems. 

For a general (standard) partition $\lambda=(\lambda_r,\cdots,\lambda_1) \in {\mathsf P \mathsf T}_{\ell\leq r}\subset \mathcal Y_{\ell\leq r}$, the ${\mathscr P}_{\lambda'}^{(r)}$-equation ($\lambda' = (\lambda_r-p,\lambda_{r-1},\cdots,\lambda_1)$) can be generally expressed as%
\footnote{This is then easily manipulated in Mathematica. This is the way how we derive isomonodromy systems in the following. }
\begin{align}
&\zeta W_{\lambda'}^{(r)} =2^{p-1}  W^{(r)}_{\lambda} + \sum_{n=2}^{\lambda_r-\lambda_{r-1}} U_{n}^{(\lambda_r-p+r-1)}(t) \, W^{(r)}_{(\underset{\text{$r$-th comp.}}{\lambda_r-n},\lambda_{r-1},\cdots)} + \nn\\
&+\sum_{a=1}^{r-1} \Bigl[ \sum_{n_a=1}^{\lambda_{r-a}-\lambda_{r-a-1}} (-1)^{a} \times U_{\lambda_{r-a+1}-\lambda_{r-a}+n_a+a}^{(\lambda_r-p+r-1)} W^{(r)}_{(\lambda_{r-1}-1,\cdots,\lambda_{r-a}-1,\underset{\text{($r$-$a$)-th comp.}}{\lambda_{r-a}-n_a}, \lambda_{r-a-1},\cdots) }\Bigr],
\end{align}
where $\lambda_0=0$. Note that, due to Eq.~\eq{Eqn:ExistenceOfWronskiansLambda}, these equations exist when and only when 
\begin{align}
\lambda_r-p +r-1\geq 0 \qquad \Leftrightarrow \qquad \lambda_r \geq p-r+1. \label{Eqn:LambdaConditionExistenceOfPequation}
\end{align}
These $\mathscr P$-operators are then understood as recursive equations to solve $W_\lambda^{(r)}$ by other Wronskians $W_{\mu}^{(r)}$ with a smaller size of Young diagram ($|\lambda|-2\geq |\mu|$). 

\begin{itemize}
\item [1) ] In particular, Eq.~\eq{Eqn:LambdaConditionExistenceOfPequation} indicates that the Wronskians $W^{(r)}_\lambda(t;\zeta)$ associated with the following partitions/Young diagrams $\lambda \in {\mathsf P \mathsf T}_{r\times (p-r)}$:%
\footnote{For example, the set of ${\mathsf P \mathsf T}_{3\times2}$ is following: 
\begin{align}
{\mathsf P \mathsf T}_{3\times2} = \Bigl\{ \varnothing,\,\, {\tiny \yng(1)}\,, \,\, {\tiny \yng(2)}\,, \,\,  {\tiny \yng(1,1)}\,, \,\,  {\tiny \yng(2,1)} \,, \,\,  {\tiny \yng(1,1,1)} \,, \,\,  {\tiny \yng(2,2)}\,, \,\,  {\tiny \yng(2,1,1)} \,, \,\, {\tiny \yng(2,2,1)} \,, \,\,  {\tiny \yng(2,2,2)}\,\, \Bigr\}. 
\end{align}
Therefore, they are Young diagrams which can be included by a $3\times 2$-size rectangle, ${\tiny \, \yng(2,2,2)}\,$. 
}
\begin{align}
{\mathsf P \mathsf T}_{r\times (p-r)} \equiv \bigl\{\lambda: \text{partitions } \big| \ell(\lambda)\leq r,  \lambda_r\leq p-r \bigr\}  \subset {\mathsf P \mathsf T}_{\ell\leq r},
\end{align}
are the linearly independent Wronskians. 

The general Wronskians $W_\lambda^{(r)}$ ($\lambda \in {\mathsf P \mathsf T}_{\ell\leq r}$) are then expressed as a Wronskian associated with a vector of the following (finite-dimensional) vector space $\mathcal Y_{r\times (p-r)}$ over $\mathbb C$,
\begin{align}
\mathcal Y_{r\times (p-r)} \equiv \sum_{
\lambda \in {\mathsf P \mathsf T}_{r\times (p-r)}
}
\mathbb C \times \lambda \quad \subset \mathcal Y_{\ell\leq r},\qquad \dim_{\mathbb C} \mathcal Y_{r\times (p-r)} = \binom{p}{r}. 
\end{align}
The set ${\mathsf P \mathsf T}_{r\times (p-r)}$ then forms a basis system of the linear space $\mathcal Y_{r\times (p-r)}$, which is referred to as {\em the canonical basis} of the space. 
\item [2) ] The $t$-dependence of the system is then governed by the differential equation, 
\begin{align}
\del_t W_{\lambda}^{(r)}(t;\zeta) = W^{(r)}_{\square \times \lambda}(t;\zeta) \qquad \Bigl( \lambda \in \mathcal Y_{r\times (p-r)} \Bigr), 
\end{align}
which describes a motion of vectors in the linear space of Young diagrams $\mathcal Y_{r\times (p-r)}$. 
\item [3) ] The $\zeta$-dependence of the system is then governed by the ${\mathscr Q}_{\lambda}^{(r)}$-equations, which describes a motion of vectors in the linear space of Young diagrams $\mathcal Y_{r\times (p-r)}$. 
\end{itemize}
By this manipulation, we can derive the linear differential equations with rational coefficients for the $(r,1)$ FZZT-Cardy branes, which are $\binom{p}{r}\times \binom{p}{r}$ isomonodromy systems. 

\paragraph{An example: $(p,q)=(3,2)$ and the $(2,1)$ FZZT-Cardy brane}

As an example, we demonstrate the case of $\vev{\widetilde \Theta^{(2,1)}_{\rm Cardy}(-\eta)}_t = \widetilde W^{(2)}_{\varnothing}[\chi](t;\eta)$. This is the simplest and educational. Several equations are shown as follows: 
\begin{align}
\underline{\text{$ {\mathscr P}$-equation}}\qquad & \nn\\
0&=4 \widetilde W^{(2)}_{\tiny \yng(2)} + \widetilde v_2(t) \widetilde W^{(2)}_{\tiny \varnothing} \label{Eqn:Peq32Eg1} \\
0 &= 4 \widetilde W^{(2)}_{\tiny \yng(3)} + \widetilde v_2(t) \widetilde W^{(2)}_{\tiny \yng(1)} + \bigl(\widetilde v_3(t)+\widetilde v_2'(t) - \eta\bigr) \widetilde W^{(2)}_{\tiny \varnothing} \label{Eqn:Peq32Eg2} \\
0 &= 4 \widetilde W^{(2)}_{\tiny \yng(2,1)} - (\widetilde v_3(t) - \eta) \widetilde W^{(2)}_{\tiny \varnothing} \label{Eqn:Peq32Eg3} \\
&   \cdots \nn\\
\underline{\text{$ {\mathscr Q}$-equation}} \qquad &  \nn\\
g\beta_{3,2}^{-1}\, \frac{\del}{\del \eta} \widetilde W^{(2)}_{\tiny \varnothing} & = 2 \Bigl( \widetilde W^{(2)}_{\tiny \yng(2)} - \widetilde W^{(2)}_{\tiny \yng(1,1)}\Bigr) + 2\widetilde u_2(t)  \widetilde W^{(2)}_{\tiny \varnothing} \\
 g\beta_{3,2}^{-1}\, \frac{\del}{\del \eta} \widetilde W^{(2)}_{\tiny \yng(1)} & = 2  \widetilde W^{(2)}_{\tiny \yng(3)} + 2\widetilde u_2(t)  \widetilde W^{(2)}_{\tiny \yng(1)} + 2\widetilde u_2'(t) \widetilde W^{(2)}_{\tiny \varnothing} \\
 g\beta_{3,2}^{-1}\, \frac{\del}{\del \eta} \widetilde W^{(2)}_{\tiny \yng(1,1)} & = 2 \Bigl( \widetilde W^{(2)}_{\tiny \yng(3,1)} - \widetilde W^{(2)}_{\tiny \yng(2,2)}\Bigr) + 2\widetilde u_2(t)  \widetilde W^{(2)}_{\tiny \yng(1,1)} + \widetilde u_2'(t) \widetilde W^{(2)}_{\tiny \yng(1)} - \widetilde u_2''(t) \widetilde W^{(2)}_{\tiny \varnothing} \\
&   \cdots \nn
\end{align}
where 
\begin{align}
\widetilde u_2(t) = u_2(t),\qquad \widetilde v_2(t) = v_2(t),\qquad \widetilde v_3(t) = -v_3(t) + u_2'(t). 
\end{align}
Here, for example, one can see 
\begin{align}
\square \times \bigl(\text{Eq.~\eq{Eqn:Peq32Eg1}}\bigr) = \bigl(\text{Eq.~\eq{Eqn:Peq32Eg2}}\bigr) + \bigl(\text{Eq.~\eq{Eqn:Peq32Eg3}}\bigr). 
\end{align}
By using the ${\mathscr P}$-equations, the Wronskians with a large-size Young diagram can be recursively solved by those of smaller-size Young diagram. As a result, the following Young diagrams are independent degrees of freedom in this case: 
\begin{align}
{\mathsf P \mathsf T}_{2\times 1} = \Bigl\{ \varnothing, \, {\tiny \yng(1)}\, ,\, {\tiny \yng(1,1)}\, \Bigr\},
\end{align}
and it is enough to consider the following finite-dimensional linear space, 
\begin{align}
\mathcal Y_{2\times 1} = \mathbb C \varnothing + \mathbb C \, {\tiny \yng(1)}\, + \mathbb C \, {\tiny \yng(1,1)}\, , \qquad  \dim_{\mathbb C} \mathcal Y_{2\times 1} = \binom{3}{2} = 3, 
\end{align}
This system is then described by $3\times 3$ isomonodromy systems.

\subsubsection{Isomonodromy systems for $(r,1)$ FZZT-Cardy branes}

As we have seen above, the isomonodromy system for the $(r,1)$ FZZT-Cardy branes in $(p,q)$ minimal string theory can be obtained by introducing the following rank-$\binom{p}{r}$ vector function,
\begin{align}
\vec{W}^{(r)}(t;\zeta) \equiv 
\begin{pmatrix}
W^{(r)}_\varnothing(t;\zeta) \cr
W^{(r)}_{\tiny \yng(1)}(t;\zeta) \cr 
\vdots
\end{pmatrix}
\equiv 
\begin{pmatrix}
W^{(r)}_\lambda(t;\zeta)
\end{pmatrix}_{\lambda\in {\mathsf P \mathsf T}_{r\times (p-r)}}, 
\end{align}
and by considering the action of $\del_t$ and $\del_\zeta$. 

For later convenience, we also specify a ``canonical'' ordering for the bases of the vector space, $\lambda\in {\mathsf P \mathsf T}_{r\times (p-r)}$, as follows: 
\begin{align}
\lambda <\mu \qquad \Leftrightarrow \qquad 
\left\{
\begin{array}{c}
 |\lambda|< |\mu| \cr
\text{or}\quad \ds  \sum_{a=0}^{r-1} \lambda_{r-a} (p-r)^a  < \sum_{a=0}^{r-1} \mu_{r-a} (p-r)^a  \qquad \bigl( |\lambda|= |\mu|\bigr)
\end{array}
\right., \label{Eqn:CanonicalOrderingOfYoungDiagrams}
\end{align}
where $|\lambda|$ is the size of the Young diagram $\lambda$. 
For example, the elements of the basis system ${\mathsf P \mathsf T}_{3\times 2}$ are ordered as follows: 
\begin{align}
\varnothing < {\tiny \yng(1)} < {\tiny \yng(2)} < {\tiny \yng(1,1)} < {\tiny \yng(2,1)} < {\tiny \yng(1,1,1)} < {\tiny \yng(2,2)} < {\tiny \yng(2,1,1)} < {\tiny \yng(2,2,1)} < {\tiny \yng(2,2,2)}\,\,. 
\end{align}

The isomonodromy system is then given as the following linear differential equation for the vector function $\vec{W}^{(r)}(t;\zeta)$: 
\begin{align}
g\frac{\del \vec{W}^{(r)}(t;\zeta)}{\del t}  = \mathcal B^{(r,1)} (t;\zeta) \vec{W}^{(r)}(t;\zeta),\qquad 
g\frac{\del \vec{W}^{(r)}(t;\zeta)}{\del \zeta}  = \mathcal Q^{(r,1)} (t;\zeta) \vec{W}^{(r)}(t;\zeta), 
\end{align}
with $\binom{p}{r}\times \binom{p}{r}$ matrix-valued rational functions $\mathcal B^{(r,1)} (t;\zeta)$ and $\mathcal Q^{(r,1)} (t;\zeta)$ in $\zeta$.%
\footnote{
Note that the size of the isomonodromy system (i.e.~$\binom{p}{r}\times \binom{p}{r}$) is also consistent because the number of independent general rank-$r$ Wronskian functions (Eq.~\eq{Eq:BinomPRwavefunctions}) is also given by $\binom{p}{r}$. }
Examples of isomonodromy systems are shown in Appendix \ref{Section:ExampleOfIMS}. 

The isomonodromic Douglas equation is given by the integrability condition of this isomomonodromy system:  
\begin{align}
\bigl[g\del_t - \mathcal B^{(r,1)},g \del_\zeta-\mathcal Q^{(r,1)}\bigr] = 0. \label{Eq:IsomonodromicDouglasEquation}
\end{align}
It is important to note that the string equation obtained from this (isomonodromic) Douglas equation is the same string equation as the original Douglas equation of the Baker-Akhiezer system: 
\begin{align}
\bigl[ \bP(t;\del),\bQ(t;\del) \bigr] = g 1 \quad \Leftrightarrow \quad \bigl[g\del_t - \mathcal B^{(r,1)},g \del_\zeta-\mathcal Q^{(r,1)}\bigr] = 0
\qquad \bigl(r=1,2\cdots,p-1\bigr). 
\end{align}
From the viewpoint of physics, it is because the bulk physics itself does not change, whether it is from the principal FZZT-brane or from $(r,1)$ FZZT-Cardy branes. From the viewpoints of (generalized) Painlev\'e equations, we obtained different kinds of  isomonodromy systems which describe the same (generalized) Painlev\'e equation. 

If one just focuses on the string/Painlev\'e equations, it is not much different from the principal one. However, since the isomonodromy systems are different, the spectral curves are different. If one interprets the spectral curves as the spacetime of this minimal string theory \cite{fy3}, the spacetime generated by the general FZZT-Cardy brane is much broader than that of the principal FZZT-brane. It even describes a semi-classical multiverse behavior as connecting ZZ-branes (See Section \ref{Subsection:ZZCardyBranes}). 

The point is that these FZZT-Cardy branes are just branes which constitute the single minimal string theory. Therefore, these non-perturbative completions viewed from each FZZT-Cardy brane should coincide with each other. This is the anticipated new duality constraint of FZZT-Cardy branes discussed in \cite{CIY5}. Further analysis on this direction is reserved for future investigation.


\subsection{Charge conjugation of $(r,1)$ FZZT-Cardy branes \label{Subsection:ChargeConjugationR1Complex}}

As one may observe in Appendix \ref{Section:ExampleOfIMS}, the sizes of isomonodromy systems for both the $(r,1)$ FZZT-Cardy brane and the $(p-r,1)$ FZZT-Cardy brane are the same, and both are given by 
\begin{align}
\binom{p}{r}\times \binom{p}{r} = \binom{p}{p-r}\times \binom{p}{p-r}. 
\end{align}
Similarly, there is also a duality relation among spectral curves of these two FZZT-Cardy branes (i.e.~Eq.~\eq{Eqn:ChargeConjugationInSpectralCurves}). Therefore, one may expect that these two isomonodromy systems are also related by a duality. In this section, we discuss this duality from the viewpoint of the Wronskians. Later, more general discussions are given in Section \ref{Subsubsection:ChargeConjugationDualSpace}. This duality of isomonodromy systems is understood as {\em charge conjugation} by the following means: 

\subsubsection{Charge conjugation matrices of isomonodromy systems \label{SubSubSection:ChargeConjugation}}

We first discuss the charge conjugation among isomonodromy systems for the $(r,1)$ FZZT-Cardy branes, directly. In fact, as one may find in the examples of Appendix \ref{Section:ExampleOfIMS}, the following relation holds:%
\footnote{The form of the transformation itself is a conjecture in general, but it is reasonable since the transformation Eq.~\eq{Eq:ChargeConjugationBQ} naturally preserves the isomonodromic Douglas equation, Eq.~\eq{Eq:IsomonodromicDouglasEquation}. In this sense, this is a spectral duality of the isomonodromy systems.  }
\begin{align}
\mathcal B^{(p-r,1)}(t;\zeta)  = - \mathfrak C^{(r,1)} \,\,\Bigl(  \mathcal B^{(r,1)}(t;\zeta)  \Bigr)^{\rm T} {\mathfrak C^{(r,1)}}^{-1}, \nn\\
\mathcal Q^{(p-r,1)}(t;\zeta)  = - \mathfrak C^{(r,1)}\,\, \Bigl(  \mathcal Q^{(r,1)}(t;\zeta)  \Bigr)^{\rm T} {\mathfrak C^{(r,1)}}^{-1}, \label{Eq:ChargeConjugationBQ}
\end{align}
where $\mathfrak C^{(r,1)}$ is a $\binom{p}{r}\times \binom{p}{r}$ matrix which is defined by the following linear map acting on the space of Young diagrams, $\mathfrak C^{(r,1)}:\, \mathcal Y_{r\times (p-r)} \to \mathcal Y_{(p-r) \times r}$:%
\footnote{The linear map $\mathfrak C^{(r,1)}$ is linearly extended as $\mathfrak C^{(r,1)}(a \lambda + b \mu) = a \,\mathfrak C^{(r,1)} (\lambda) + b\, \mathfrak C^{(r,1)} ( \mu)$ for $a,b \in \mathbb C; \, \lambda,\mu \in \mathcal Y_{r\times (p-r)}$. Then, if the $\lambda$ is a pure basis (See Section \ref{Subsubsection:LinearExtensionYoungDiagramsW}), we apply the rule of Eq.~\eq{Eqn:DefinitionOfCmatrix}. }
\begin{align}
\mathfrak C^{(r,1)}(W^{(r)}_\lambda) = W^{(p-r)}_{\mathfrak C^{(r,1)}(\lambda)},\qquad \mathfrak C^{(r,1)}(\lambda) = (-1)^{|\lambda|}\times \lambda^{\rm CC}. \label{Eqn:DefinitionOfCmatrix}
\end{align} 
Here, $|\lambda|$ is the size of the Young diagram $\lambda$, and  $\lambda^{\rm CC}$ is referred to as {\em complementary conjugate Young diagram} which is given by combining the following two operations: 
\begin{align}
\lambda \leftrightarrow \lambda^{\bot}\text{ : complement}  &\quad \Leftrightarrow  \quad  \lambda^{\bot}_{a} \equiv (p-r) -\lambda_{r-a+1} \quad \bigl(a=1,2,\cdots,r\bigr). \\
\lambda \leftrightarrow \lambda^{\lor}\text{ : conjugate} & \quad \Leftrightarrow  \quad \lambda^{\lor}_a  \equiv {\text{max}}_b \, \bigl\{b\big| \lambda_{r-b+1}\geq a\bigr\} \quad \bigl(a=1,2,\cdots,p-r\bigr).
\end{align}
For example, the case of $p=9$ and $r=4$ is given as 
\begin{align}
\lambda = {\tiny \yng(4,2,1)} \quad \to \quad \lambda^{\bot} = {\tiny \yng(5,4,3,1)} \quad \to\quad  \lambda^{\rm CC} =\bigl(\lambda^{\bot}\bigr)^{\lor} ={\tiny \yng(4,3,3,2,1)}. 
\end{align}
One can also take $\lambda^{\lor}$ first, since the result is the same, $\lambda^{\rm CC} =\bigl(\lambda^{\bot}\bigr)^{\lor} = \bigl(\lambda^{\lor}\bigr)^{\bot}$. By using the canonical ordering of Young diagrams Eq.~\eq{Eqn:CanonicalOrderingOfYoungDiagrams}, one can see that 
\begin{align}
\mathfrak C^{(p-r,1)} = (-1)^{r(p-r)} {\mathfrak C^{(r,1)}}^{\rm T},\qquad \mathfrak C^{(p-r,1)} \mathfrak C^{(r,1)} = (-1)^{r(p-r)}. 
\end{align}

Note that the form of the matrix $\mathfrak C^{(r,1)}$ depends also on $(p,q)$ (even though it is not explicitly shown). This matrix $\mathfrak C^{(r,1)}$ is referred to as {\em charge conjugation matrix}. The meaning of charge conjugation matrices is discussed and is also extended to more general cases later in Section \ref{Subsubsection:ChargeConjugationDualSpace}. 

\subsubsection{Charge conjugation transformation of the Wronskians \label{Subsubsection:ChargeConjuTransOfWronskian/Fermions}}

We next discuss the charge conjugation transformation $\mathcal C$. Note that the charge conjugation transformation $\mathcal C$ and the charge conjugation matrix $\mathfrak C^{(r,1)}$ both stem from the same concept of ``charge conjugation''; but they represent different aspects of it. We define the charge conjugation transformation of the Wronskians $W_\varnothing^{[j_1,j_2,\cdots,j_r]}(t;\zeta)$ as 
\begin{align}
\mathcal C \bigl[ W^{[j_1,j_2,\cdots,j_r]}_\varnothing(t;\zeta)\bigr] \equiv
W^{[l_1, l_2,\cdots,l_{p-r}]}_\varnothing(t;\zeta), 
\end{align}
where $j_1<j_2<\cdots<j_r$ and $l_1<l_2<\cdots<l_{p-r}$ with 
\begin{align}
\bigl\{ j_1,j_2,\cdots,j_r \bigr\} \cup \bigl\{ l_1,l_2,\cdots,l_{p-r} \bigr\} = \bigl\{1,2, \cdots, p\bigr\},\label{Eqn:JLlabelingInSectionChargeConjugationTransformation}
\end{align}
and therefore, by definition, it satisfies $\mathcal C\circ \mathcal C = 1$.

One way to understand this transformation is by utilizing the twisted free fermions. From this point of view, this transformation can be interpreted as {\em particle/hole exchange}, {\em charge conjugation} and {\em brane/ghost-brane exchange} of $p$-th twisted fermions. 
\begin{itemize}
\item [1. ] Let us represent the bosonization of the fermions as $\big\{ \hat \phi^{(j)}(\zeta) \bigr\}_{j=1}^p$ (i.e.~$\hat \psi^{(j)}(\zeta) = \,:\!e^{\hat \phi^{(j)}(\zeta)}\!:$). If one chooses the conformal background (i.e.~\cite{MSS}), one of the loop equations (i.e.~$W_{1+\infty}$-constraints) is given as follows: 
\begin{align}
\sum_{j=1}^p\hat \phi^{(j)}(\zeta) = 0, 
\end{align}
without loss of generality (as a result of loop equations) \cite{fkn1,fkn2,fkn3,fim}.%
\footnote{This means that there is no contribution of the non-universal KP flows in the background \cite{fkn1,fkn2,fkn3,fy1,fy2,fy3}.}

This is equivalent to the standard property of the Wronskian: The rank-$p$ Wronskian of the $p$-th order differential equations should be constant in $t$ (and also in $\zeta$). In our conformal background, it is given by the unity: 
\begin{align}
W^{(p)}(t;\zeta) = 1. 
\end{align}
This property provides the concept of {\em particle/hole} for this free fermion. 
\item [2. ] 
As a result, for the indices $\{j_a\}_{a=1}^r$ and $\{l_b\}_{b=1}^{p-r}$ given by Eq.~\eq{Eqn:JLlabelingInSectionChargeConjugationTransformation}, the following two exponents differ by a sign: 
\begin{align}
\sum_{a=1}^r \hat \phi^{(j_a)}(\zeta) = - 1\times  \sum_{b=1}^{p-r} \hat \phi^{(l_b)}(\zeta),
\end{align}
and therefore, the charge conjugate pair of the Wronskian operators are related by {\em the charge conjugation of free-fermions}: 
\begin{align}
\hat W^{[j_1,\cdots,j_r]}(\zeta) = :\! e^{\hat \phi(\zeta)} \!:\qquad \overset{\mathcal C}{\longleftrightarrow} \qquad 
  \hat W^{[l_1,\cdots,l_{p-r}]}(\zeta) = :\! e^{-\hat \phi(\zeta)} \!: 
\end{align}
In other words, this is the relation {\em of brane/ghost-brane} \cite{OkudaTakayanagi}. 
\item [3. ] As an example, the charge conjugate of the standard determinant operator is given by the inverse of the determinant operator: 
\begin{align}
\hat W^{[1]}(\zeta)  \quad \leftrightarrow \quad \det(x-X)\qquad \overset{\mathcal C}{\longleftrightarrow}\qquad 
\hat W^{[2,3,\cdots,p]}(\zeta)  \quad \leftrightarrow \quad \frac{1}{\det(x-X)}. 
\end{align}
Similarly, the charge conjugate of the $(p-1,1)$-type FZZT-Cardy branes is the $\lfloor  \frac{p}{2} \rfloor$-th elemental FZZT-brane $\hat \psi^{(\lfloor  \frac{p}{2} \rfloor)}(\zeta)$, 
\begin{align}
\Theta^{(p-1,1)}_{\rm Cardy}(\zeta)\qquad \overset{\mathcal C}{\longleftrightarrow}\qquad 
\hat W^{[\lfloor  \frac{p}{2} \rfloor ]}(\zeta) \,\,  \Bigl(= \hat \psi^{(\lfloor  \frac{p}{2} \rfloor)}(\zeta) \Bigr). 
\end{align}
which is not the principal FZZT brane $\hat \psi^{(1)}(\zeta)  \, \bigl(= \Theta^{(1,1)}_{\rm Cardy}(\zeta) \bigr)$. Therefore, the charge conjugation is not the map which connects different FZZT-Cardy branes: 
\begin{align}
\mathcal C [\Theta_{\rm Cardy}^{(r,s)}(\zeta)] \, \, \neq \,\, \Theta_{\rm Cardy}^{(p-r,s)}(\zeta). 
\end{align}
\item [5. ] Under the abbreviation of indices (discussed in Section \ref{Subsubsection:AbbreviationAndNotations}), one can express the transformation as 
\begin{align}
\mathcal C [W^{(r)}_\varnothing (t;\zeta)] = W^{(p-r)}_\varnothing (t;\zeta). 
\end{align}
This should be understood as a map of differential equations. 
\end{itemize}

\subsubsection{Charge conjugation as a B\"acklund transformation \label{Subsubsection:Backlundtransformation}}

In addition, we can also consider the charge conjugation as a transformation of coefficients $\bigl\{u_n(t)\bigr\}_{n=2}^p$ and $\bigl\{v_n(t)\bigr\}_{n=2}^p$ in the Baker-Akhiezer systems: 
\begin{align}
\left\{
\begin{array}{ll}
\ds
u_n(t) \quad \to \quad \widetilde u_n(t) =\sum_{j=0}^{n-2} (-1)^{n-j}\binom{p-n+j}{j}  u_{n-j}^{(j)}(t)
& \quad \bigl(n=2,3,\cdots,p\bigr) \cr
\ds
v_n(t) \quad \to \quad \widetilde v_n(t) =\sum_{j=0}^{n-2} (-1)^{n-j}\binom{q-n+j}{j}  v_{n-j}^{(j)}(t)
&\quad  \bigl(n=2,3,\cdots,q\bigr) \cr
\,\,\, \beta_{p,q} \quad \to \quad - \beta_{q,p}= (-1)^{p+q+1} \beta_{p,q}
\end{array}
\right.. \label{Eqn:BacklundTransformationOfUV}
\end{align}
Since $\bigl\{u_n(t)\bigr\}_{n=2}^p$ and $\bigl\{\widetilde u_n(t)\bigr\}_{n=2}^p$ both satisfy the same string equation (up to the sign-change of $\beta_{p,q}$), this transformation is understood as {\em a B\"acklund transformation} of the string equations. This also comes from the concept of charge conjugation. 

\begin{itemize}
\item [1. ] The point is that the rank-$1$ Wronskians $W^{(1)}_{\varnothing}(t;\zeta) \, \bigl(\equiv \psi(t;\zeta) \bigr)$ and the rank-$(p-1)$ Wronskians $W^{(p-1)}_{\varnothing}(t;\zeta)$ are related by the charge conjugation transformation: 
\begin{align}
\mathcal C[W^{(1)}_{\varnothing}(t;\zeta)] = W^{(p-1)}_{\varnothing}(t;\zeta)
\end{align}
Therefore, the rank-$(p-1)$ Wronskians $W^{(p-1)}_{\varnothing}(t;\zeta)$ also similarly satisfy another Baker-Akhiezer system: 
\begin{align}
\zeta \, W^{(p-1)}_\varnothing (t;\zeta) &= \bP_{\mathcal C}(t;\del)\, W^{(p-1)}_\varnothing (t;\zeta), \qquad g \frac{\del W^{(p-1)}_\varnothing (t;\zeta)}{\del \zeta} =  \bQ_{\mathcal C}(t;\del) \,W^{(p-1)}_\varnothing (t;\zeta), \nn\\
& \text{with}\qquad \bigl(\bP_{\mathcal C}(t;\del), \bQ_{\mathcal C}(t;\del) \bigr) = \bigl(\bP^{\rm T}(t;\del),  - \bQ^{\rm T}(t;\del)\bigr). \label{Eq:PQpairChargeConjugation}
\end{align}
This is also another standard result of the Wronskian: The rank-$(p-1)$ Wronskians of the $p$-th order linear differential equations satisfy the linear differential equation with ``transposed-coefficients'' (i.e.~$\bP^{\rm  T}(t;\del)$). 
\item [2. ] By definition, the Douglas equation of the charge conjugate Baker-Akhiezer system is equivalent to the original Douglas equation: 
\begin{align}
\bigl[ \bP_{\mathcal C},\bQ_{\mathcal C}\bigr] = \bigl[ \bP^{\rm T}, -\bQ^{\rm T} \bigr] = \Bigl( \bigl[ \bP, \bQ \bigr] \Bigr)^{\rm T} = g 1. 
\end{align}
In this sense, this charge conjugation is understood as a spectral duality of the integrable systems \cite{BEH}.
\item [3. ] By the concept of charge conjugation, one can also consider the rank-$(p-1)$ Wronskian, 
\begin{align}
\psi_{\mathcal C}(t;\zeta) \equiv W_\varnothing^{(p-1)}(t;(-1)^p\zeta), 
\end{align}
as the fundamental degree of freedom, with regarding $\psi(t;\zeta)$ as the bound state of them (i.e.~$\psi(t;\zeta) =W^{(p-1)}_\varnothing [\psi_{\mathcal C}](t;(-1)^p\zeta)$). From this point of view, the system is described by the Baker-Akhiezer system with replacing coefficients by Eq.~\eq{Eqn:BacklundTransformationOfUV}. In this sense, Eq.~\eq{Eqn:BacklundTransformationOfUV} is understood as the charge conjugation of the system. 
\item [3. ] The dual cases are also similarly given as follows: 
\begin{align}
\eta \,  \widetilde W^{(q-1)}_\varnothing (t;\eta) &= \widetilde \bP_{\mathcal C}(t;\del)\,  \widetilde W^{(q-1)}_\varnothing (t;\eta), \qquad g \frac{\del  \widetilde W^{(q-1)}_\varnothing (t;\eta)}{\del \eta} =  \widetilde \bQ_{\mathcal C}(t;\del) \, \widetilde W^{(q-1)}_\varnothing (t;\eta), \nn\\
& \text{with}\qquad \bigl( \widetilde \bP_{\mathcal C}(t;\del),  \widetilde \bQ_{\mathcal C}(t;\del) \bigr) = \bigl( \widetilde \bP^{\rm T}(t;\del),  -  \widetilde \bQ^{\rm T}(t;\del)\bigr),  \label{Eqn:ChargeConjuOfChi1}
\end{align}
and therefore, 
\begin{align}
\chi_{\mathcal C}(t;\eta) \equiv \widetilde W_\varnothing^{(q-1)}(t;(-1)^q\eta) \, \left( = W_\varnothing^{(q-1)}[\chi](t;(-1)^q\eta)  \right). \label{Eqn:ChargeConjuOfChi2}
\end{align}
\end{itemize}

This aspect of charge conjugation provides two different prescriptions to define $(r,1)$-type FZZT-Cardy branes (i.e.~based on $\psi$ or $\psi_{\mathcal C}$; and also based on $\chi$ or $\chi_{\mathcal C}$). In fact, these two prescriptions are consistent with each other, which is discussed in Section \ref{Subsection:ConsistencyOfChargeConjugations}.

\section{Wronskians for the $(1,s)$ FZZT-Cardy branes \label{Section:1SFZZTCardyBranesAndKacTable}}

In this section, we discuss the next step: how the $(1,s)$-type FZZT-Cardy branes should be described. So far, we have discussed that FZZT-Cardy branes are multi-body states of elemental FZZT-branes, which are described by the Wronskians of Baker-Akhiezer systems. This idea can be straightforwardly applied to the cases of $(r,1)$-type. However, as is discussed in Section \ref{Subsubsection:TwistedFermionAsElementalFZZTbranes}, the general cases further require a new idea for the construction. 

As the key idea, we first propose {\em a generalization of the spectral $p-q$ duality} to overcome the issue. In particular, this allows us to construct $(1,s)$-type FZZT-Cardy branes and then, as will be discussed in the next section, the edge entries of the Kac table are obtained combining the generalized spectral $p-q$ duality with the reflection relation. In the next next section, we also discuss a generalization of the Wronskians for the further general cases (i.e.~the interior entries of the Kac table) as a possible candidate for the complete solution.

\subsection{General spectral $p-q$ duality: $\mathcal P \mathcal Q = \mathcal L \circ \mathcal C$ \label{SubSubSection:PQdualityLaplace}}

As is discussed in Section \ref{Subsubsection:TwistedFermionAsElementalFZZTbranes}, the success of our Wronskian construction for $(r,1)$-type FZZT-Cardy branes is  because the analytic continuations of these branes themselves (Eq.~\eq{Eq:SeibergShihRelZeta} and \eq{Eq:SeibergShihRelEta}) do not depend on the structure of the spectral curve. Therefore, our task is to find out a way to construct the general $(r,s)$-type FZZT-Cardy branes, which does not depend on the spectral curves. 

The first proposal which partly solves this problem is to generalize the spectral $p-q$ duality of the Baker-Akhiezer systems to the general $(r,s)$-type of FZZT-Cardy branes. A clue for the extension can be found in Section \ref{Section:PQdualityLegendre}, where the $p-q$ duality of spectral curves is regarded as the Legendre transformation: 
\begin{align}
\left\{
\begin{array}{c}
\ds \varphi^{(r,s)}_{\rm Cardy} (\zeta) =\sup_{\eta} \Bigl( \beta_{p,q}\, A_s \widetilde A_r \times  \zeta \,\eta - \widetilde \varphi^{(s,r)}_{\rm Cardy} (e^{\pi i q}\eta) \Bigr) \cr
\ds \widetilde \varphi^{(s,r)}_{\rm Cardy} (\eta) =\sup_{\zeta} \Bigl( \beta_{q,p}\, A_s \widetilde A_r \times  \zeta \,\eta -  \varphi^{(r,s)}_{\rm Cardy} (e^{\pi i p}\zeta) \Bigr)
\end{array} 
\right.. 
\label{Eqn:LegendreInGeneralizedSpectralPQduality}
\end{align}
This can be used to connect the $(1,s)$-type FZZT-Cardy brane with the previously studied $(s,1)$-type dual FZZT-Cardy brane. An important point here is that {\em the Legendre transformation itself does not depend on the spectral curve}. Therefore, we enhance this relation to the non-perturbative relation which connects the FZZT-Cardy-brane operators $\Theta_{\rm Cardy}^{(r,s)}(\zeta)$ with their dual operators $\widetilde \Theta_{\rm Cardy}^{(s,r)}(\eta)$. Our proposal is then given as follows: 
\begin{align}
\left\{
\begin{array}{l}
\ds \Theta_{\rm Cardy}^{(r,s)}(\zeta) = \int_\gamma d\eta \, e^{\frac{\beta_{p,q}A_s \widetilde A_r}{g}\,  \zeta \,\eta } \, \mathcal C \circ\mathcal A \bigl[ \widetilde \Theta_{\rm Cardy}^{(s,r)}(\eta) \bigr] \equiv \mathcal L_{\gamma}^{(\beta_{p,q}A_s \widetilde A_r)}\bigl[\mathcal C \mathcal A \bigl[ \widetilde \Theta_{\rm Cardy}^{(s,r)}\bigr]\bigr](\zeta) \cr
\ds \widetilde \Theta_{\rm Cardy}^{(s,r)}(\eta) = \int_{\tilde \gamma} d\zeta \, e^{\frac{\beta_{q,p}A_s \widetilde A_r}{g}\,  \zeta \,\eta } \, \mathcal C \circ\mathcal A \bigl[ \Theta_{\rm Cardy}^{(r,s)}(\zeta) \bigr] \equiv \mathcal L_{\tilde \gamma}^{(\beta_{q,p}A_s \widetilde A_r)}\bigl[\mathcal C \mathcal A \bigl[ \Theta_{\rm Cardy}^{(r,s)}\bigr]\bigr](\eta) 
\end{array}
\right.. \label{Eq:PQeqLC1st}
\end{align}
This is a combination of two spectral dualities, {\em the charge conjugation} and {\em the Laplace-Fourier transformation}. Therefore, this new transformation is also a spectral duality. We refer to this transformation as {\em general spectral $p-q$ duality} (or simply, spectral $p-q$ duality). Here each operation is given as follows: 
\begin{itemize}
\item [1) ] The operation $\mathcal A$ is an antipodal operator for the index of the Wronskian, which can act on $W^{(r)}(t;\zeta)$ or $\widetilde W^{(s)}(t;\zeta)$ respectively as %
\footnote{
Of course, one should note that $\mathcal A \bigl[ \widetilde W^{[j_1,j_2,\cdots,j_s]}(t;\eta)\bigr] \neq \widetilde W^{[j_1,j_2,\cdots,j_s]}(t;e^{\pi i q}\eta)$. 
}
\begin{align}
\left\{
\begin{array}{l}
\mathcal A \bigl[ W^{[j_1,j_2,\cdots,j_r]}(t;\zeta)\bigr]  \equiv
W^{[j_1 - \frac{p}{2},j_2- \frac{p}{2},\cdots,j_r- \frac{p}{2}]}(t;\zeta), \cr
\mathcal A \bigl[ \widetilde W^{[j_1,j_2,\cdots,j_s]}(t;\eta)\bigr]  \equiv
\widetilde W^{[j_1 - \frac{q}{2},j_2- \frac{q}{2},\cdots,j_s- \frac{q}{2}]}(t;\eta). 
\end{array}
\right. \label{Eqn:RWeqW_Roperation}
\end{align}
By definition, it satisfies $\mathcal A \circ \mathcal A = 1$. 
\item [2) ] The operation $\mathcal C$ is the charge conjugation operator on the Wronskian, discussed in Section \ref{Subsubsection:ChargeConjuTransOfWronskian/Fermions}. Therefore $\mathcal C\circ \mathcal C = 1$.
\item [3) ] The operation $\mathcal L_\gamma^{(A)}$ is the Laplace-Fourier transformation. For a given function $f(\eta)$ in $\eta$ (or $g(\zeta)$ in $\zeta$), it is respectively given as 
\begin{align}
\mathcal L_\gamma^{(A)}[f](\zeta) \equiv \int_\gamma d\eta\, e^{\frac{A}{g} \zeta \eta} \, f(\eta) \qquad \Biggl(\text{or}\quad \mathcal L_\gamma^{(A)}[g](\eta) \equiv \int_\gamma d\zeta\, e^{\frac{A}{g} \zeta \eta} \, g(\zeta)\Biggr). 
\end{align}
The contour $\gamma$ is chosen so that the integral $\mathcal L$ converges. Note that we use the same notation for $\mathcal L$ on $f(\zeta)$ and also for $\mathcal L$ on $g(\eta)$. We also note the following:
\begin{align}
 G \Bigl( \zeta,g \frac{\del}{\del \zeta} \Bigr) \circ f(\zeta) =0 \quad \Leftrightarrow \quad 
G \Bigl( - \frac{A}{B}\zeta,-g \frac{B}{A}\frac{\del}{\del \zeta} \Bigr) \circ \mathcal L_{\gamma_1}^{(A)}  \mathcal L_{\gamma_2}^{(B)} [f](\zeta) = 0. 
\end{align}
Therefore, as far as one focuses on ``differential equations'', $\mathcal L\circ \mathcal L$ gives just a coordinate transformation of $\zeta$: $\zeta \to \zeta' = - (A/B)\zeta$. 

It is also natural to consider the pair of contours $(\gamma_1,\gamma_2)$ which satisfies
\begin{align}
\mathcal L_{\gamma_1}^{(A)} \circ \mathcal L_{\gamma_2}^{(-A)} [f(\zeta)] = f(\zeta), \label{Eqn:LLeq1condition}
\end{align}
then one can identify $\mathcal L_{\gamma_1}^{(A)}\circ \mathcal L_{\gamma_2}^{(-A)} = 1$. These contours $(\gamma_1,\gamma_2)$ are said to be inverse to each other and denoted as $\gamma_2 = \gamma_1^{-1}$.%
\footnote{It is worth comparing it with the Fourier transformation: 
\begin{align}
\hat  f(\eta) \equiv \mathcal L_{\gamma_1}^{(-i)}[f(\zeta)] =  \int_{-\infty}^{\infty} d\zeta \, e^{-i \frac{ \zeta \eta}{g}} f(\zeta),\qquad f(\zeta) = \mathcal L_{\gamma_2}^{(i)}[\hat f(\eta)] =  \frac{1}{2\pi g}\int_{-\infty}^\infty d\eta \, e^{i \frac{\zeta \eta}{g}}\hat f(\eta). 
\end{align}
Therefore, the contours $(\gamma_1,\gamma_2)$ are given by $(\gamma_1,\gamma_2) = (\gamma, \frac{1 }{2\pi g} \gamma)$, where $\gamma$ is the contour defined by $\gamma: \mathbb R \ni t \mapsto \gamma(t) \in \mathbb C$ with $\gamma(t) = t\in \mathbb R$. In this sense, $\gamma^{-1} = \frac{1}{2\pi g} \gamma$. This notation is consistent with the following operation: $(a\gamma)^{-1} = \frac{1}{a} \gamma^{-1}$. 
}
We will also come back to this point in Section \ref{Subsubsection:InvolutorityOfPQandReflectionChargeConjugation}. 
\end{itemize}
By using the proposed formula, we can describe the $(1,s)$ FZZT-Cardy brane by using $(s,1)$ dual FZZT-Cardy brane as follows:
\begin{align}
\left\{
\begin{array}{l}
\ds \Theta_{\rm Cardy}^{(1,s)}(\zeta) = \int_\gamma d\eta \, e^{\frac{\beta_{p,q}A_s }{g}\,  \zeta \,\eta } \, \mathcal C \circ\mathcal A \bigl[ \widetilde \Theta_{\rm Cardy}^{(s,1)}(\eta) \bigr],\qquad \widetilde \Theta_{\rm Cardy}^{(s,1)}(\eta) \quad \leftrightarrow \quad \widetilde W^{(s)}_{\varnothing}(t;\eta) \cr
\ds \widetilde \Theta_{\rm Cardy}^{(1,r)}(\eta) = \int_{\tilde \gamma} d\zeta \, e^{\frac{\beta_{q,p}A_s }{g}\,  \zeta \,\eta } \, \mathcal C \circ\mathcal A \bigl[  \Theta_{\rm Cardy}^{(r,1)}(\zeta) \bigr],\qquad \Theta_{\rm Cardy}^{(r,1)}(\zeta) \quad \leftrightarrow \quad W^{(r)}_{\varnothing}(t;\zeta)
\end{array}
\right.. \label{Eq:PQeqLC2nd_S1}
\end{align}
In this section, based on Eq.~\eq{Eq:PQeqLC2nd_S1}, we evaluate differential equations of these $(1,s)$ (and also $(1,r)$ dual) FZZT-Cardy branes. 

\subsubsection{Consistency with the Seiberg-Shih relation}

There are several reasons to consider the above form of the generalized spectral $p-q$ duality. 
\begin{itemize}
\item The string equations obtained from integrable systems of the FZZT-Cardy brane Eq.~\eq{Eq:PQeqLC1st} are the same as those of principal FZZT brane. This guarantees that the pure-closed-string amplitudes are kept the same, and the solitonic object given by Eq.~\eq{Eq:PQeqLC1st} still belongs to the same closed string (i.e.~$(p,q)$ minimal string) theory. 

Simultaneously, the spectral curves obtained from integrable systems of the FZZT-Cardy brane Eq.~\eq{Eq:PQeqLC1st} are those of the FZZT-Cardy brane evaluated in Liouville theory (i.e.~in Section \ref{SubSection:SpectralCurveCardy}). This guarantees that the open-string amplitudes associated with the FZZT-Cardy brane are also correctly prescribed. 

\item These two points are guaranteed by the fact that the generalized spectral $p-q$ duality is given by a combination of spectral dualities  ($\mathcal A$, $\mathcal C$ and $\mathcal L$).%
\footnote{Precisely speaking, since $\mathcal A$ just exchanges solutions with other solutions within the same differential equation, it is considered to be a trivial spectral duality. In other words, this operation does not change the differential equation at all.} The concept of spectral duality is introduced in \cite{BEH}. Since there are a few spectral dualities associated with the spectral curves, it is natural to consider the proposed formula Eq.~\eq{Eq:PQeqLC1st} as the correct prescription for describing FZZT-Cardy branes. 

\item One of the important consistency checks is about the Kac table. If the construction of FZZT-Cardy branes itself is consistent, it should allow us to derive the Kac table associated with the Virasoro algebra of the worldsheet CFT. In addition, various manipulations based on Eq.~\eq{Eq:PQeqLC1st} should not cause any inconsistency in the construction of the Kac table. The construction of the Kac table and consistency checks are given in Section \ref{Section:ConsistencyOfDualities}. 

\item The formula Eq.~\eq{Eq:PQeqLC1st} also should pass some direct check of the equality. In particular, as is discussed in Section \ref{Subsubsection:Backlundtransformation} and also in Section \ref{Subsubsection:CornerEntriesOfKacTable}, the corner entries of the Kac table, 
\begin{align}
\begin{tabular}{|c|c|c|}
\hline
$\underset{(1,q-1)}{\psi_{\mathcal C}}$  & $\cdots$ & $\underset{(p-1,q-1)}{\psi}$ \\
\hline
$\vdots$ & & $\vdots$ \\
$\vdots$ & & $\vdots$ \\
\hline
$\underset{(1,1)}{\psi}$ & $\cdots$ & $\underset{(p-1,1)}{\psi_{\mathcal C}}$\\
\hline
\end{tabular} 
\quad \overset{\text{$p-q$ duality}}{\longleftrightarrow} \quad 
\begin{tabular}{|c|cc|c|}
\hline
$\underset{(1,p-1)}{\chi_{\mathcal C}}$  & $\cdots$ & $\cdots$ & $\underset{(q-1,p-1)}{\chi}$ \\
\hline
$\vdots$ & & & $\vdots$ \\
\hline
$\underset{(1,1)}{\chi}$ & $\cdots$ & $\cdots$ & $\underset{(q-1,1)}{\chi_{\mathcal C}}$\\
\hline
\end{tabular}\,, 
\end{align}
are described by the Baker-Akhiezer systems. As in Section \ref{Subsubsection:CornerEntriesOfKacTable}, the direct check can be performed and results in Eq.~\eq{Eq:PQeqLC1st}. The first primitive discussion on $\psi$ and $\chi$ is also found in \cite{BEH}. 

The direct check for the general FZZT-Cardy branes should be also possible but is reserved for future investigation. 
\end{itemize}

\subsubsection{Abbreviations and notations}

As is discussed in Section \ref{Subsubsection:AbbreviationAndNotations}, we often only specify the rank of the Wronskians (i.e.~$W^{(r)}(t;\zeta)$ or $\widetilde W^{(s)}(t;\eta)$) without specifying the actual indices $(j_1,j_2,\cdots,j_r)$. In the same sense, we also skip the antipodal operation $\mathcal A$ in the following discussions: 
\begin{align}
\mathcal A[W^{(r)}(t;\zeta)] = W^{(r)}(t;\zeta), \qquad \mathcal A[\widetilde W^{(s)}(t;\eta)] = \widetilde W^{(s)}(t;\eta), 
\end{align}
since it does not affect the differential equation.

For the same reason, the contour $\gamma$ in the Laplace-Fourier transformation $\mathcal L_{\gamma}^{(A)}$ is also skipped, $\mathcal L_\gamma^{(A)} \to \mathcal L^{(A)}$. Although it is also interesting to discuss the precise roles of the contour $\gamma$ in these correspondences, we leave it for future investigations.%
\footnote{For example, see \cite{cite:ComplexPath} for investigations along this direction. }

Similar to the usual Wronskian functions $W^{(r)}_\lambda [\psi](t;\zeta)$ which we defined in Section \ref{Subsubsection:GWronskiansYoungDiagrams}, we also define the following descendants of the Wronskian functions: 
\begin{align}
\left\{
\begin{array}{l}
\mathcal L W^{(r)}_{\lambda} [\chi](t;\zeta) \equiv \mathcal L_\gamma^{(\beta_{q,p})} \bigl[ W^{(r)}_\lambda[\chi](t;\eta) \bigr] (\zeta) \cr
\mathcal L \widetilde W^{(s)}_\lambda [\psi](t;\eta) \equiv \mathcal L_\gamma^{(\beta_{p,q})} \bigl[\widetilde W^{(s)}_\lambda[\psi](t;\zeta) \bigr] (\eta) 
\end{array}
\right..  \label{Eqn:AbbreviationOfLWformula}
\end{align}
In this definition, we practically dropped $A_s\widetilde A_r$ dependence in this definition (See also Eq.~\eq{Eq:PQeqLC2nd_S1}) because it is related to a change of coordinate $\zeta$ (or $\eta$), which is just a convention of $\zeta$ (or of $\eta$). Simultaneously, we have chosen the convention of $\zeta$ (or $\eta$) so that the coefficient $\beta_{p,q}$ (or $\beta_{q,p}$) is flipped in the Laplace-Fourier transformation: $\mathcal L^{(\beta_{p,q})} \leftrightarrow \mathcal L^{(\beta_{q,p})}$.%
\footnote{Therefore, although the coefficients of Eq.~\eq{Eqn:AbbreviationOfLWformula} look different from Eq.~\eq{Eqn:LegendreInGeneralizedSpectralPQduality} at the first sight, we should note that it is not a typo. }
This choice of coefficients makes Eq.~\eq{Eqn:SchurDifferentialLPLQ} and Eq.~\eq{Eqn:SchurDifferentialLPLQdual} simple. 

Accordingly, the Wronskian representation of $\Theta_{\rm Cardy}^{(1,s)}(\zeta)$ and $\widetilde \Theta_{\rm Cardy}^{(s,r)}(\eta)$ is denoted as 
\begin{align}
\left\{
\begin{array}{l}
\ds \Theta_{\rm Cardy}^{(1,s)}(\zeta) \quad \leftrightarrow \quad W^{(1,s)}_\lambda [\chi](t;\zeta) \equiv \mathcal L \widetilde W^{(q-s)}_\lambda [\chi](t;\zeta) \cr
\ds \widetilde \Theta_{\rm Cardy}^{(1,r)}(\eta) \quad \leftrightarrow \quad \widetilde W^{(1,r)}_\lambda [\psi](t;\eta) \equiv \mathcal L  W^{(p-r)}_\lambda [\psi](t;\eta) 
\end{array}
\right..
\label{Eqn:Cardy1S_Wronskian1S_definition}
\end{align}
Here it is worth emphasizing that this is only one of the representations to describe the FZZT-Cardy branes $\Theta_{\rm Cardy}^{(1,s)}(\zeta)$ and $\widetilde \Theta_{\rm Cardy}^{(1,r)}(\eta)$. Therefore, $W^{(1,s)}_\lambda [\chi](t;\zeta)$ represents a specific realization defined by Eq.~\eq{Eqn:Cardy1S_Wronskian1S_definition}. The consistency among different descriptions is discussed in Section \ref{Section:ConsistencyOfDualities}. 

Keeping this issue in mind, we again abbreviate the new Wronskian functions as follows: 
\begin{align}
\left\{
\begin{array}{l}
W^{(1,s)}_\lambda[\chi](t;\zeta) \quad \to \quad W^{(1,s)}_\lambda(t;\zeta)
\quad \to \quad W^{(1,s)}_\lambda \cr
\widetilde W^{(1,r)}_\lambda[\psi](t;\eta) \quad \to \quad \widetilde W^{(1,r)}_\lambda(t;\eta)
\quad \to \quad \widetilde W^{(1,r)}_\lambda
\end{array}
\right., 
\end{align}
and also 
\begin{align}
\left\{
\begin{array}{l}
\mathcal L \widetilde W^{(s)}_\lambda[\chi](t;\eta) \quad \to \quad \mathcal L \widetilde W^{(s)}_\lambda(t;\eta)
\quad \to \quad \mathcal L \widetilde W^{(s)}_\lambda \cr
\mathcal L W^{(r)}_\lambda[\psi](t;\zeta) \quad \to \quad \mathcal L W^{(r)}_\lambda(t;\zeta)
\quad \to \quad \mathcal L W^{(r)}_\lambda 
\end{array}
\right., 
\end{align}
as far as it is clear. Therefore, if we encounter 
\begin{align}
W^{(1,s)}_\lambda[\chi_{\mathcal C}](t;\zeta),
\end{align}
we cannot skip ``$[\chi_{\mathcal C}]$''.

\subsection{Schur-differential and isomonodromy systems \label{Subsection:SchurDifferentialIMSin1SFZZTCardyBranes}}

In this subsection, we study Schur-differential equations and isomonodromy systems of the $(1,s)$-type FZZT-Cardy branes (or $(1,r)$-type dual FZZT-Cardy branes), especially associated with the representation Eq.~\eq{Eqn:Cardy1S_Wronskian1S_definition}. 

\subsubsection{Schur-differential equations}

The Schur-differential equations for the new Wronskian functions $W^{(1,s)}(t;\zeta)$ and $\widetilde W^{(1,r)}(t;\eta)$ are obtained from Eq.~\eq{Eq:Pequation} and Eq.~\eq{Eqn:SchurDifferentialForDualBranes}, by taking Laplace-Fourier transformation, as 
\begin{align}
&\underline{W^{(1,s)}(t;\zeta) = \mathcal L \widetilde W^{(q-s)}(t;\zeta)} \nn\\
&\qquad \left\{
\begin{array}{rc}
\text{${\mathscr L \!\widetilde{\mathscr P}}_\lambda^{(q-s)}$-eqn.}:&\ds 
g\frac{\del \mathcal L \widetilde W_\lambda^{(q-s)}(t;\zeta)}{ \del \zeta} = \beta_{q,p}\, \widetilde {\mathscr P}_{\lambda,q-s}^{(q-s)} (t;\del) \, \mathcal L \widetilde W_\varnothing^{(q-s)}(t;\zeta)  \cr
\text{${\mathscr L \! \widetilde{\mathscr Q}}_\lambda^{(q-s)}$-eqn.}:&\ds 
-\zeta\, \mathcal L \widetilde W_\lambda^{(q-s)}(t;\zeta) = \sum_{a=1}^{q-s} \widetilde {\mathscr Q}_{\lambda,a}^{(q-s)} (t;\del) \, \mathcal L \widetilde W_\varnothing^{(q-s)}(t;\zeta) \cr
\end{array}
\right. \label{Eqn:SchurDifferentialLPLQ}\\
&\underline{\widetilde W^{(1,r)}(t;\eta) = \mathcal L W^{(p-r)}(t;\eta)} \nn\\
&\qquad \left\{
\begin{array}{rc}
\text{${\mathscr L \!{\mathscr P}}_\lambda^{(p-r)}$-eqn.}:&\ds 
g\frac{\del \mathcal L W_\lambda^{(p-r)}(t;\eta)}{ \del \eta} = \beta_{p,q}\, {\mathscr P}_{\lambda,p-r}^{(p-r)} (t;\del) \, \mathcal L W_\varnothing^{(p-r)}(t;\eta)  \cr
\text{${\mathscr L \! {\mathscr Q}}_\lambda^{(p-r)}$-eqn.}:&\ds 
-\eta\, \mathcal L W_\lambda^{(p-r)}(t;\eta) = \sum_{a=1}^{p-r} {\mathscr Q}_{\lambda,a}^{(p-r)} (t;\del) \, \mathcal L W_\varnothing^{(p-r)}(t;\eta) \cr
\end{array}
\right.\label{Eqn:SchurDifferentialLPLQdual}
\end{align}
Note that we replaced $\zeta$ as $\zeta \to (-1)^{p+q} \zeta$ so that the coefficients $\beta_{p,q}$ and $\beta_{q,p}$ are exchanged by using $\beta_{p,q} = (-1)^{p+q}\beta_{q,p}$. We also note 
\begin{align}
g \frac{\del}{\del \zeta} \mathcal L^{(A)} \bigl[ f \bigr](\zeta) = A \mathcal L\bigl[ \eta f \bigr](\zeta),\qquad - \zeta \mathcal L^{(A)} \bigl[ f \bigr](\zeta) = A^{-1} \mathcal L \bigl[ g \del_\eta f \bigr](\zeta). 
\end{align}

As is commented around Eq.~\eq{Eq:PeqNegativeLambda}, some of the ${\mathscr L \!\widetilde{\mathscr P}}_\lambda^{(q-s)}$-equation (and also ${\mathscr L \!{\mathscr P}}_\lambda^{(p-r)}$-equation) remains non-trivial even when $\lambda$ is a negative partition/Young diagram $\lambda$ (i.e.~when $\del_\zeta W_{\lambda}^{(1,s)}(t;\zeta)=0$ or $\del_\eta \widetilde W_{\lambda}^{(1,r)}(t;\eta)=0$). These equations contribute to the constraint equations of Young diagrams. 

\subsubsection{The space of Young diagrams and isomonodromy systems \label{Subsubsection:SpaceOfYDandIMSin1SFZZTCardy}}

In order to derive the isomonodromy systems of $(1,s)$-type FZZT-Cardy branes (or of $(1,r)$-type dual FZZT-Cardy branes), we should specify the independent basis of Young diagrams from the Schur-differential equations. 

We have derived all the isomonodromy systems for several examples. As a result, explicit forms of isomonodromy systems are shown in Appendix \ref{Section:ExampleOfIMS}. In this section, we summarize several points which we should mention. It is then instructive to see concrete equations in a simplest example. 

\paragraph{An example: $(p,q)=(2,3)$ and the $(1,1)$ FZZT-Cardy brane}

As an example, we demonstrate the case of $\vev{\Theta^{(1,1)}_{\rm Cardy}(\zeta)}_t =  W^{(1,1)}_{\varnothing}[\chi](t;\zeta) = \mathcal L \widetilde W^{(2)}[\chi](t;\zeta)$. This is the dual counterpart of the $(2,1)$ dual FZZT-Cardy brane (i.e.~$\widetilde W^{(2)}[\chi](t;\eta)$) which is concretely shown in Section \ref{Subsubsection:SchurAsConstraintsInIMS}. Several equations are shown as follows: 
\begin{align}
\underline{\text{$ {\mathscr L \widetilde {\mathscr P}^{(2)}}$-equation}}\qquad & \nn\\
0&=4 W^{(1,1)}_{\tiny \yng(2)} + \widetilde v_2(t) W^{(1,1)}_{\tiny \varnothing} \label{Eqn:LPeq32Eg1} \\
 g\beta_{3,2}^{-1}\, \frac{\del}{\del \zeta} W^{(1,1)}_{\tiny \varnothing} &= 4 W^{(1,1)}_{\tiny \yng(3)} + \widetilde v_2(t) W^{(1,1)}_{\tiny \yng(1)} + \bigl(\widetilde v_3(t)+\widetilde v_2'(t) \bigr) W^{(1,1)}_{\tiny \varnothing} \label{Eqn:LPeq32Eg2} \\
0 &= 4 \Bigl( W^{(1,1)}_{\tiny \yng(3)} + W^{(1,1)}_{\tiny \yng(2,1)} \Bigr) + \widetilde v_2(t) W^{(1,1)}_{\tiny \yng(1)} + \widetilde v_2'(t) W^{(1,1)}_{\tiny \varnothing} \label{Eqn:LPeq32Eg3} \\
&   \cdots \nn\\
\underline{\text{$ {\mathscr L  \widetilde{\mathscr Q}^{(2)}}$-equation}} \qquad &  \nn\\
-\zeta  W^{(1,1)}_{\tiny \varnothing} & = 2 \Bigl( W^{(1,1)}_{\tiny \yng(2)} - W^{(1,1)}_{\tiny \yng(1,1)}\Bigr) + 2\widetilde u_2(t)  W^{(1,1)}_{\tiny \varnothing} \\
-\zeta  W^{(1,1)}_{\tiny \yng(1)} & = 2  W^{(1,1)}_{\tiny \yng(3)} + 2\widetilde u_2(t)  W^{(1,1)}_{\tiny \yng(1)} + 2\widetilde u_2'(t) W^{(1,1)}_{\tiny \varnothing} \\
-\zeta  W^{(1,1)}_{\tiny \yng(1,1)} & = 2 \Bigl( W^{(1,1)}_{\tiny \yng(3,1)} - W^{(1,1)}_{\tiny \yng(2,2)}\Bigr) + 2\widetilde u_2(t)  W^{(1,1)}_{\tiny \yng(1,1)} + \widetilde u_2'(t) W^{(1,1)}_{\tiny \yng(1)} - \widetilde u_2''(t) W^{(1,1)}_{\tiny \varnothing} \\
&   \cdots \nn
\end{align}
where 
\begin{align}
\widetilde u_2(t) = u_2(t),\qquad \widetilde v_2(t) = v_2(t),\qquad \widetilde v_3(t) = -v_3(t) + u_2'(t). 
\end{align}
Note that we put Eq.~\eq{Eqn:LPeq32Eg3} as $\square \times \text{Eq.~\eq{Eqn:LPeq32Eg1}}$. 

From these equations, we consider the equations which do not include $\del_\zeta$, not only from ${\mathscr L  \widetilde{\mathscr Q}}$-equation but also from ${\mathscr L  \widetilde{\mathscr P}}$-equation, and then can observe that all the higher order Young diagrams can be expressed by the following two partitions/diagrams: 
\begin{align}
{\mathsf P \mathsf T}^{(1,1)} \equiv \Bigl\{ \varnothing, \square  \Bigr\}. 
\end{align}
Therefore, we specify the linear space $\mathcal Y^{(1,1)}[\chi]$ associated with the $(1,1)$-type FZZT-Cardy brane of Eq.~\eq{Eqn:Cardy1S_Wronskian1S_definition}: 
\begin{align}
\mathcal Y^{(1,1)}[\chi] = \mathbb C \varnothing \oplus \mathbb C \,\square, \qquad  \dim_{\mathbb C} \mathcal Y^{(1,1)}[\chi] = 2, \label{Eqn:DimOfIMScasestudyPQeq23RSeq12}
\end{align}
This system is then described by $2\times 2$ isomonodromy systems. We here put ``$[\chi]$'' in the definition of the space so that one understands it comes from the representative $W^{(1,1)}[\chi](t;\zeta)$. 

It is also important to compare it with the spectral curve of $F^{(1,1)}_{\rm Cardy}(t;\zeta)=0$, which is discussed in Section \ref{Subsubsection:GeneralFormulaOfSpectralCurveRS}: 
\begin{align}
(\text{Order in $Q$}) = \mathfrak d_{1,1}^{(2,3)} =  \binom{2}{1}\frac{1}{3} \left[ \binom{3}{1} - 0 \right] = 2. 
\end{align}
This coincides with the dimension of the isomonodromy system, Eq.~\eq{Eqn:DimOfIMScasestudyPQeq23RSeq12}. 

\paragraph{The linear space of independent Young diagrams}

In a similar way to the above example, one can specify the linear space of independent Young diagrams, which is denoted by 
\begin{align}
&\mathcal Y^{(1,s)}[\chi]\quad \text{for}\quad W^{(1,s)}[\chi](t;\zeta) \qquad 
\left(\text{or} \quad \widetilde {\mathcal Y}^{(1,r)}[\psi] \quad \text{for}\quad \widetilde W^{(1,r)}[\psi](t;\eta) \right). 
\end{align}
As a new feature, which was obscure for the $(r,1)$-types, we point out that there is no particular preference for the choice of independent basis. 

For example, the case of $(p,q)=(2,5)$ and $(1,3)$-type FZZT-Cardy brane is explicitly shown in Appendix \ref{Subsubsection:ExampleOfIMS:PQeq25_RSeq13IMS}. In this case, one can choose the following two different bases for the linear space $\mathcal Y^{(1,3)}[\chi]$: 
\begin{align}
{\mathsf P \mathsf T}^{(1,3)}_{\rm I}\equiv \Bigl\{ \varnothing, \, {\tiny \yng(1)}\, ,\, {\tiny \yng(2)}\, ,\, {\tiny \yng(2,1)}\,  \Bigr\} \qquad \text{v.s.}\qquad {\mathsf P \mathsf T}^{(1,3)}_{\rm II}\equiv \Bigl\{ \varnothing, \, {\tiny \yng(1)}\, ,\, {\tiny \yng(1,1)}\, ,\, {\tiny \yng(2,1)}\,  \Bigr\}. 
\end{align}
These bases are in fact related by the following identity which is also obtained as a Schur-differential equation: 
\begin{align}
W_{\tiny \yng(1,1)}^{(1,3)}(t;\zeta) = W_{\tiny \yng(2)}^{(1,3)}(t;\zeta) + \Bigl[\frac{\zeta}{2} + u_2(t)\Bigr]  W_{\tiny \varnothing}^{(1,3)}(t;\zeta). 
\end{align}
Therefore, the different choices of the basis (i.e.~${\mathsf P \mathsf T}^{(1,3)}_{\rm I}$ or ${\mathsf P \mathsf T}^{(1,3)}_{\rm II}$) result in the same linear space: 
\begin{align}
\mathcal Y^{(1,3)}[\chi] = \bigoplus_{\lambda \in {\mathsf P \mathsf T}^{(1,3)}_{\rm I}} \mathbb C \lambda = \bigoplus_{\lambda \in {\mathsf P \mathsf T}^{(1,3)}_{\rm II}} \mathbb C \lambda. 
\end{align}
By taking this into account, we first specify a choice of the basis ${\mathsf P \mathsf T}^{(1,s)} \bigl( \subset \mathcal Y^{(1,s)} \bigr)$, and define the corresponding isomonodromy system: 
\begin{align}
&g \frac{\del \vec W^{(1,s)}(t;\zeta)}{\del t} = \mathcal B^{(1,s)}(t;\zeta) \, \vec W^{(1,s)}(t;\zeta),\qquad 
g \frac{\del \vec W^{(1,s)}(t;\zeta)}{\del \zeta} = \mathcal Q^{(1,s)}(t;\zeta) \, \vec W^{(1,s)}(t;\zeta), \nn\\
&\qquad \qquad \qquad  \text{with}\qquad  \vec{W}^{(1,s)}(t;\zeta) \equiv 
\begin{pmatrix}
W^{(1,s)}_\varnothing(t;\zeta) \cr
W^{(1,s)}_{\tiny \yng(1)}(t;\zeta) \cr 
\vdots
\end{pmatrix}
\equiv 
\begin{pmatrix}
W^{(1,s)}_\lambda(t;\zeta)
\end{pmatrix}_{\lambda\in {\mathsf P \mathsf T}^{(1,s)}}. 
\end{align}
Since the form of the isomonodromy system itself depends on the choice of bases ${\mathsf P \mathsf T}^{(1,s)}$, 
we should always take into account the existence of equivalent isomonodromy systems that are connected by a gauge transformation $S(t;\zeta)$,%
\footnote{As a result of gravitational scaling invariance, the gauge transformation should also respect the scaling behavior. For example, $W_\lambda^{(1,s)}$ has dimension $|\lambda|$ and $u_n(t)$ has $n$, and $\zeta$ has $p$, and so on. }
\begin{align}
\text{Gauge transformation:}\qquad \vec{W}^{(1,s)}(t;\zeta)\,\, \to \,\, S^{-1}(t;\zeta) \, \vec{W}^{(1,s)}(t;\zeta), \label{Eqn:GaugeTransISM1}
\end{align} 
which causes the associated gauge transformation of the isomonodromy system: 
\begin{align}
\left\{
\begin{array}{c}
\mathcal B^{(1,s)}(t;\zeta)  \,\, \to \,\, S(t;\zeta)^{-1} \mathcal B^{(1,s)}(t;\zeta)  S(t;\zeta) - g\, S(t;\zeta)^{-1} \del_t  S(t;\zeta) \cr
\mathcal Q^{(1,s)}(t;\zeta)  \,\, \to \,\, S(t;\zeta)^{-1} \mathcal Q^{(1,s)}(t;\zeta)  S(t;\zeta) - g\, S(t;\zeta)^{-1} \del_\zeta  S(t;\zeta) 
\end{array}
\right.. \label{Eqn:GaugeTransISM2}
\end{align}
This is the principle of equivalence which we utilize in the consistency check of dualities in Section \ref{Section:ConsistencyOfDualities}. 

\paragraph{Matching of dimensionality and spectral curves}
As an important result of our analysis, we should mention the matching of dimensionality of the isomonodromy systems and the spectral curves: 
\begin{align}
\dim_{\mathbb C}\mathcal Y^{(1,s)} = \mathfrak d_{1,s}^{(p,q)} 
\end{align}
where $\mathfrak d_{1,s}^{(p,q)}$ is given in Eq.~\eq{Eqn:FormulaForDimensionDeltaRS} and \eq{Eqn:FormulaForDimensionDeltaRS_GCD}, and also matching of spectral curves: 
\begin{align}
\lim_{g\to 0}\det \bigl[ QI_{k\times k} - \mathcal Q^{(1,s)}(t;\zeta) \bigr] \propto F^{(1,s)}_{\rm Cardy}(\zeta,Q) = 0\qquad \Bigl({k} = \dim_{\mathbb C}\mathcal Y^{(1,s)} \Bigr), 
\end{align}
on the conformal background (with an appropriate choice of normalization).

\subsubsection{An example: $(p,q)=(3,4)$ and $(1,2)$-type FZZT-Cardy brane}

Here we show the case of $(1,2)$-type FZZT-Cardy brane in the $(3,4)$ system. The Kac table is given as
\begin{align}
    \begin{tabular}{|c|c|} \hline
$(1,3)$ & $(2,3)$  \\ \hline
$\bf (1,2)$ & $(2,2)$  \\ \hline
$(1,1)$ & $(2,1)$  \\ \hline
    \end{tabular}
\end{align}
The coefficients satisfy $\bar v_2(t) = \frac{4}{3} \bar u_2(t)$, $\bar v_3(t) = \frac{4}{3} \bar u_3(t) + \frac{2}{3}\bar u_2'(t)$ and $\bar v_4(t) = \frac{2}{9} \bar u_2''(t) + \frac{2}{3} \bar u_3'(t) + \frac{2}{9} \bigl(\bar u_2(t) \bigr)^2$. The Schur-differential equations and the procedure to solve these Schur-differential equations are shown in detail in Appendix \ref{Subappendix:PQ34-RS12:DerivationOfIMS}. 

As is shown in Appendix \ref{Subappendix:PQ34-RS12:DerivationOfIMS}, 
the linear space of independent Young diagrams is three-dimensional:
\begin{align}
\mathcal Y^{(1,2)} = \mathbb C \varnothing \oplus \mathbb C \square \oplus \mathbb C \square^2, 
\end{align}
where ${\, \scriptsize \yng(1)\,}^2 = {\, \scriptsize \yng(2)\,} + {\, \scriptsize \yng(1,1)\,}$. 
The vector wave-function $\vec W^{(1,2)}(t;\zeta)$ of the isomonodromy system is then given by
\begin{align}
\vec W^{(1,2)}(t;\zeta) = 
\begin{pmatrix}
\mathcal L \widetilde W_\varnothing^{(2)}[\chi](t;\zeta) \cr
\mathcal L \widetilde W_\square^{(2)}[\chi](t;\zeta) \cr
\mathcal L \widetilde W_{\square^2}^{(2)}[\chi](t;\zeta)
\end{pmatrix}. 
\end{align}
Accordingly, the $\mathcal B$ and $\mathcal Q$-operators are evaluated as 
\begin{align}
\mathcal B^{(1,2)}(t;\zeta)  &= 
\bordermatrix{
 & \varnothing & \square & \square^2 \cr
\varnothing &  & 1 & \cr
\square & & & 1 \cr
\square^2 & \frac{ \zeta}{2} & - \frac{u_2}{2} & 
}, \\
\mathcal Q^{(1,2)}(t;\zeta)  &=\beta_{4,3}\times    \nn\\
& \!\!\!\!\!\!\! \!\!\!\!\!\!\! \times \biggl[
\bordermatrix{
 & \varnothing & \square & \square^2 \cr
\varnothing & -\frac{u_2^2 + 2 u_2''}{9} - \frac{F_1}{\zeta} + \frac{F_2}{\zeta^2} & - \zeta + \frac{u_2'}{3} - \frac{F_3}{\zeta} + \frac{F_4}{\zeta^2} & -\frac{u_2}{3} - \frac{(u_2^2+2u_2'')'}{9 \zeta} + \frac{F_5}{\zeta^2} \cr
\square & - \frac{u_2}{6}\zeta -\frac{(u_2^2 + 2 u_2'')'}{18} -  \frac{F_6}{\zeta} & \frac{u_2^2 + 2 u_2''}{18} - \frac{((2u_3-u_2')^2)'}{18 \zeta}& - \zeta + \frac{(2 u_3- u_2' )^2}{9\zeta} \cr
\square^2 & - \frac{ \zeta^2}{2} - \frac{u_2'}{6 } \zeta +F_7 - \frac{F_8}{\zeta} & \frac{u_2}{3} \zeta - \frac{(2u_3' - u_2'')^2}{9\zeta} & \frac{u_2^2 + 2u_2''}{18} + \frac{((2u_3-u_2')^2)'}{18 \zeta}
} \biggr], 
\end{align}
where 
\begin{align}
F_1 &= \frac{1}{9} \left(-u_2' \left(18 u_3'-6 u_2''+u_2^2\right)+4 \left(3 u_3 \left(u_3'-u_2'' \right)-3 u_3^{(4)}+u_2^{(5)} \right)-6 u_2 u_3'' \right), \\
F_2 &= \frac{1}{18} \Bigl(u_2 \left(2 u_3-u_2'\right)+2\bigl(2 u_3- u_2'\bigr)''\Bigr)^2, \\
F_3 &= \frac{1}{9} \left(10 u_3 u_2'-2 (u_2')^2+6 u_2 u_3'- u_2 u_2''-4 u_3^2+12 u_3^{(3)}-4 u_2^{(4)} \right), \\
F_4 &= -\frac{1}{9} \left(2 u_3'-u_2'' \right) \left(u_2 \left(2 u_3-u_2'\right)+2\bigl(2 u_3- u_2'\bigr)'' \right), \\
F_5 &= \left(2 u_3-u_2'\right) \left(u_2 \left(2 u_3-u_2'\right)+ 2\bigl(2 u_3- u_2'\bigr)'' \right), \\
F_6 &= -\frac{1}{18} \left(2 u_3-u_2'\right) \left(u_2 \left(2 u_3-u_2'\right)+ 2\bigl(2 u_3- u_2'\bigr)'' \right), \\
F_7 &= \frac{1}{18} \left(2 u_3 u_2'-4 (u_2')^2+6 u_2 u_3'-5 u_2 u_2''+4 u_3^2+12 u_3^{(3)}-8 u_2^{(4)}\right), \\
F_8 &= -\frac{1}{18} \left(2 u_3'-u_2''\right) \left(u_2 \left(2 u_3-u_2'\right)+ 2\bigl(2 u_3- u_2'\bigr)'' \right). 
\end{align}

\paragraph{Simplification with utilizing string equations}
These coefficients can be actually simplified by using the string equations of this (bulk) system, 
\begin{align}
0 &= u_2 \left(u_2''-2 u_3'\right)+ (u_2')^2-2 u_3 u_2'-4 u_3^{(3)}+2 u_2^{(4)}, \label{StringEquation34-1st}\\
9 \frac{g}{\beta_{3,4}} &= 24 u_2' u_2'' + 2 u_2^2 u_2' - 24 u_3 u_3' - 12 u_2' u_3' - 12 u_2 \left(u_3''-u_2^{(3)}\right) - 24 u_3^{(4)} + 16 u_2^{(5)}. \label{StringEquation34-2nd}
\end{align}
and also an integration of Eq.~\eq{StringEquation34-1st} and of Eq.~\eq{StringEquation34-2nd}, which in our conformal background is respectively given by
\begin{align}
0 & =u_2 \left(2 u_3-u_2'\right)+ 2\bigl(2 u_3- u_2'\bigr)'',  \nn\\
\frac{9(t-2\beta_{3,4} \mu)}{\beta_{3,4}} &=
 -12 u_2 \left(u_3' -u_2'' \right)+6 (u_2')^2+\frac{2 u_2^3}{3} -12 u_3^2-24 u_3^{(3)} +16 u_2^{(4)}. 
\end{align}
Here we have chosen the integration constants of these string equations so that $t=0$ becomes the conformal background with the cosmological constant $\mu>0$: 
\begin{align}
u_2(t) = -3\mu^{1/3} +O(g),\qquad u_3(t) = 0 + O(g)\qquad (t \to 0, \, g\to 0).  \label{Eqn:ConformalBackgroundPQeq24}
\end{align}
With these equations, one can show 
\begin{align}
F_2=F_4=F_5=F_6=F_8=0,
\end{align}
and also 
\begin{align}
F_1 = -\frac{g}{2 \beta_{3,4}},\qquad 
F_3 = -2 F_7 = \frac{1}{18}\left(2u_3 - u_2'\right)^2 - \frac{1}{9}  u_2 u_2'' + \frac{1}{18} \left(u_2'\right)^2 - \frac{1}{27} u_2^3 + \frac{(t-2\beta_{3,4} \mu)}{2 \beta_{3,4}}. 
\end{align}
The $\mathcal Q$-operator then becomes simpler: 
\begin{align}
\mathcal Q^{(1,2)}(t;\zeta)  &= \beta_{4,3}\times \nn\\
\times \biggl[ &
\bordermatrix{
 & \varnothing & \square & \square^2 \cr
\varnothing & -\frac{u_2^2 + 2 u_2''}{9} + \frac{g \beta_{3,4}^{-1}}{2  \zeta}  & - \zeta + \frac{u_2'}{3} - \frac{F_3}{\zeta}  & -\frac{u_2}{3} - \frac{(u_2^2+2u_2'')'}{9 \zeta}  \cr
\square & - \frac{u_2}{6}\zeta -\frac{(u_2^2 + 2 u_2'')'}{18}  & \frac{u_2^2 + 2 u_2''}{18} - \frac{((2u_3-u_2')^2)'}{18 \zeta}&  -\zeta + \frac{(2 u_3- u_2' )^2}{9\zeta} \cr
\square^2 & - \frac{ \zeta^2}{2} - \frac{u_2'}{6 } \zeta -\frac{F_3}{2}  & \frac{u_2}{3} \zeta - \frac{(2u_3' - u_2'')^2}{9\zeta} & \frac{u_2^2 + 2u_2''}{18} + \frac{((2u_3-u_2')^2)'}{18 \zeta}
} \biggr]. 
\end{align}
In this form, it is clear that this isomonodromy system is symmetric under replacing the coefficients: 
\begin{align}
\bigl( u_2(t), u_3(t)\bigr) \quad  \to \quad \bigl( u_2(t), - u_3(t) + u_2'(t)\bigr), \qquad \beta_{4,3} \to \beta_{4,3}, 
\end{align}
which is associated with the charge conjugation on the dual side (i.e.~of $(4,3)$-system). This is in fact consistent with the self-duality of this case: 
\begin{align}
\mathcal C [\widetilde W^{(2)}(\eta)] =  \widetilde W^{(2)}(\eta). 
\end{align}

\paragraph{Check of spectral curves}

We then evaluate the spectral curve of this isomonodromy system in the conformal background Eq.~\eq{Eqn:ConformalBackgroundPQeq24}: 
\begin{align}
\lim_{g\to 0} \det \Bigl( Q I_3 - \mathcal Q^{(1,2)}(t;\zeta) \Bigr) &= Q^3
-\frac{3 \beta _{4,3}^2 \mu ^{4/3}}{4}   Q +\frac{\beta _{4,3}^3}{4} \left(2 \zeta^4 -4 \mu  \zeta^2 +\mu ^2 \right)
 \nn\\
&=\frac{(\beta_{3,4} \mu^{2/3})^3}{4}\Bigl[T_3\bigl(\frac{Q}{\beta_{3,4} \mu^{2/3}} \bigr) - T_4\bigl( \frac{\zeta}{\sqrt{2 \mu}}\bigr)\Bigr] = 0. 
\end{align}
Here we flipped the coefficient as $\beta_{4,3}=- \beta_{3,4}$ in the final line. 

On the other hand, the spectral curve according to Liouville theory (calculated in Section \ref{Subsubsection:GeneralFormulaOfSpectralCurveRS}) is given by 
\begin{align}
F^{(1,2)}(\zeta,Q) =\zeta^2 \times \frac{(\beta_{3,4} \mu^{2/3})^3}{4}\Bigl[T_3\bigl(\frac{-Q}{\beta_{3,4} \mu^{2/3}} \bigr) - T_4\bigl( \frac{-\zeta}{\sqrt{2 \mu}}\bigr)\Bigr]  =0. 
\end{align}
Therefore, one shows the equivalence (up to the coordinate transformation of $\zeta$, $\zeta \to -\zeta$):
\begin{align}
\lim_{g\to 0} \det \Bigl( Q I_3 - \mathcal Q^{(1,2)}(t;\zeta) \Bigr)  \propto 
F^{(1,2)}(-\zeta,-Q)  = 0. 
\end{align}
The order of the spectral curve also coincides with the size of the isomonodromy system: 
\begin{align}
(\text{Order in $Q$}) = \mathfrak d^{(3,4)}_{1,2} = \binom{3}{1} \frac{1}{4} \left[ \binom{4}{2} - 2 \right] = 3. 
\end{align}

\subsection{General charge conjugations \label{Subsection:GeneralChargeConjugation}}

In this subsection, we discuss a more general concept of charge conjugation which connects different isomonodromy systems with each other. 

As we have seen in Section \ref{SubSubSection:ChargeConjugation}, we have observed that there is a similarity transformation among isomonodromy $\mathcal B$-(or $\mathcal Q$-)operators of different isomonodromy systems as Eq.~\eq{Eq:ChargeConjugationBQ}. In particular, the charge conjugation matrix $\mathfrak C^{(r,1)}$ (defined by Eq.~\eq{Eqn:DefinitionOfCmatrix}) is given by a specific choice of basis for both $\mathcal Y_{r\times (p-r)}$ and $\mathcal Y_{(p-r)\times r}$. 

On the other hand, as is discussed in Section \ref{Subsubsection:SpaceOfYDandIMSin1SFZZTCardy}, the linear space of Young diagrams for the $(1,s)$-type FZZT-Cardy branes $\mathcal Y^{(1,s)}$ does not possess such an a priori basis which is suitable for charge conjugation. Therefore, we should recast the concept of charge conjugation from more general viewpoints. This leads us to {\em a dual-space pair of the isomonodromy systems}. 

\subsubsection{General charge conjugation and the dual-space pairing  \label{Subsubsection:ChargeConjugationDualSpace}}

We here consider two different $k\times k$ isomonodromy systems ($\Psi$ and $\Upsilon$) which are denoted by 
\begin{align}
\left\{
\begin{array}{lc}
\Psi) \quad & \ds g \frac{\del \vec \Psi(t;\zeta)}{\del t} = \mathcal B_\Psi(t;\zeta) \vec \Psi(t;\zeta),\qquad g \frac{\del \vec \Psi(t;\zeta)}{\del \zeta} = \mathcal Q_\Psi(t;\zeta) \vec \Psi(t;\zeta)  \cr
\Upsilon) \quad & \ds g \frac{\del \vec \Upsilon(t;\zeta)}{\del t} = \mathcal B_\Upsilon(t;\zeta) \vec \Upsilon(t;\zeta),\qquad g \frac{\del \vec \Upsilon(t;\zeta)}{\del \zeta} = \mathcal Q_\Upsilon(t;\zeta) \vec \Upsilon(t;\zeta) 
\end{array}
\right.
\end{align}
These two systems are called {\em charge conjugate to each other}, if there exists a dual-space pairing $\bigl<\vec \Upsilon, \vec \Psi \bigr>_{\mathfrak D,c}$, 
\begin{align}
\bigl<\vec \Upsilon, \vec \Psi \bigr>_{\mathfrak D,c} \equiv \left[\vec \Upsilon(t;c \,\zeta)\right]^{\rm T} \, \mathfrak D(t;\zeta)\,  \vec \Psi(t;\zeta)\qquad \Bigl( \det \mathfrak D(t;\zeta) \neq 0, \, c \neq 0 \Bigr), 
\end{align}
which satisfies 
\begin{align}
\frac{\del \bigl<\vec \Upsilon, \vec \Psi \bigr>_{\mathfrak D,c}}{\del t} = \frac{\del \bigl<\vec \Upsilon, \vec \Psi \bigr>_{\mathfrak D,c}}{\del \zeta} = 0.  \label{Eqn:DualSpacePairingCondition}
\end{align}
The $k\times k$ matrix $\mathfrak D(t;\zeta)$ is referred to as {\em the charge conjugation matrix} of this charge conjugation. 

In fact, by evaluating Eq.~\eq{Eqn:DualSpacePairingCondition}, one finds that 
\begin{align}
\left\{
\begin{array}{l}
\ds
\mathcal B_{\Psi}(t;\zeta)  = \mathfrak D(t;\zeta)^{-1} \Bigl[ - \mathcal B^{\rm T}_{\Upsilon}(t;c\,\zeta) \Bigr] \mathfrak D(t;\zeta) - g\, \mathfrak D(t;\zeta)^{-1} \del_t  \mathfrak D(t;\zeta) \cr
\ds 
\mathcal Q_{\Psi}(t;\zeta)  = \mathfrak D(t;\zeta)^{-1} \Bigl[ -c\, \mathcal Q^{\rm T}_{\Upsilon}(t;c\,\zeta) \Bigr] \mathfrak D(t;\zeta) - g\, \mathfrak D(t;\zeta)^{-1} \del_\zeta  \mathfrak D(t;\zeta) 
\end{array}
\right.. 
\end{align}
Therefore, the dual-space pairing of the previous charge conjugation (discussed in Section \ref{SubSubSection:ChargeConjugation}) is given by the charge conjugation matrix $\mathfrak C^{(r,1)}$ (i.e.~$\mathfrak D(t;\zeta) = \mathfrak C^{(r,1)}$) as 
\begin{align}
\bigl<\vec W^{(p-r)}, \vec W^{(r)} \bigr>_{\mathfrak C^{(r,1)},1} \equiv \Bigl[\vec W^{(p-r)}(t;\zeta)\Bigr]^{\rm T}\,\, \mathfrak C^{(r,1)}\, \, \Bigl[ \vec W^{(r)}(t;\zeta) \Bigr]\qquad \Bigl( \det \mathfrak C^{(r,1)} \neq 0 \Bigr). 
\end{align} 

\subsubsection{Dual charge conjugation}

By utilizing the general concept of charge conjugation, one can show that there also exists a similar concept of charge conjugation which vertically connects the Wronskian functions: 
\begin{align}
\mathcal D \bigl[W^{(1,s)}[\chi](t;\zeta)\bigr]= W^{(1,q-s)}[\chi](t;\zeta). \label{Eqn:DefinitionOfDualChargeConjugationTransformation}
\end{align}
We refer to this conjugation operation $\mathcal D$ as {\em dual charge conjugation transformation}. 
As a criterion of such a conjugate relation, we have shown that these two systems are related by {\em the dual charge conjugation matrix} $\mathfrak D^{(1,s)}_{[\chi]}$ which defines the dual-space pairing of these isomonodromy systems: 
\begin{align}
\bigl<\vec W^{(1,q-s)}[\chi], \vec W^{(1,s)}[\chi] \bigr>_{\mathfrak D^{(1,s)}_{[\chi]},c} \equiv \Bigl[\vec W^{(1,q-s)}[\chi](t;c\,\zeta)\Bigr]^{\rm T}\,\, \mathfrak D^{(1,s)}_{[\chi]}(t,\zeta)\, \, \Bigl[ \vec W^{(1,s)}[\chi](t;\zeta) \Bigr], 
\end{align}
which satisfies Eq.~\eq{Eqn:DualSpacePairingCondition}. 
The concrete solutions to the dual charge conjugation matrices are shown in Appendix \ref{Section:ExampleOfIMS}. 

\begin{itemize}
\item [1) ] For the later convenience, the $[\chi]$-dependence is shown in the matrix $\mathfrak D^{(1,s)}_{[\chi]}$. It is because we will also consider charge conjugation of $\chi \to \chi_{\mathcal C}$ in the next section. 
\item [2) ] A distinct point which is special for the dual charge conjugation is that these matrices generally depend on $\zeta$ as well as $t$. 
\item [3) ] By definition, the matrix $\mathfrak D^{(1,s)}_{[\chi]}$ satisfies 
\begin{align}
\mathfrak D^{(1,q-s)}_{[\chi]}(t,\zeta) = \Bigl[ \mathfrak D^{(1,s)}_{[\chi]}(t,\zeta) \Bigr]^{\rm T}. 
\end{align}
\item [4) ] As a precise definition of Eq.~\eq{Eqn:DefinitionOfDualChargeConjugationTransformation}, we now employ the following commutation relation: 
\begin{align}
\mathcal D \circ \mathcal P \mathcal Q = \mathcal P\mathcal Q \circ \mathcal C. 
 \label{Eqn:DefinitionOfDualChargeConjugationTransformationWithPQtransformation}
\end{align}
In Section \ref{Subsubsection:InvolutorityOfPQandReflectionChargeConjugation}, we also argue that these two charge conjugations are in fact the same operation in the Kac table. 
\end{itemize}

\section{Consistency of dualities and the Kac table \label{Section:ConsistencyOfDualities}}

So far we have discussed two prescriptions to obtain a non-perturbative definition of $(r,s)$ FZZT-Cardy branes: One is by Wronskians (in Section \ref{Section:FromSStoWronskians} and Section \ref{Section:WronskiansForR1FZZTCardyBranes}) and the other is by generalization of spectral $p-q$ duality (in Section \ref{Section:1SFZZTCardyBranesAndKacTable}). In particular, they allow us to construct the differential equation systems of the $(r,1)$ and $(1,s)$-type FZZT-Cardy branes. 

Among these descriptions, there appear several {\em different} representations which describe the same FZZT-Cardy branes. In this section, we will show that all of these different representations are, in fact, equivalent. What is more, such several descriptions reveal the shape of Kac table, which suggests the further leap toward the general $(r,s)$ FZZT-Cardy branes, which is a {\em further generalization of Wronskians}.

\subsection{Consistency of charge conjugations \label{Subsection:ConsistencyOfChargeConjugations}}

\subsubsection{Two prescriptions for the charge conjugations}

In section \ref{Subsection:ChargeConjugationR1Complex} (and also in Section \ref{Subsection:GeneralChargeConjugation}), we have discussed two prescriptions for the charge conjugation. The first prescription is given by a map which connects these two Wronskian functions: 
\begin{align}
\text{1) }\qquad W^{(r)}_{\varnothing}[\psi](t;\zeta) \quad \to \quad W^{(p-r)}_{\varnothing}[\psi](t;\zeta). 
\end{align}
The second prescription is given by exchanging the role of the following two FZZT-Cardy branes: 
\begin{align}
\text{2) }\qquad \psi(t;\zeta) \quad \to \quad \psi_{\mathcal C}(t;\zeta) = W^{(p-1)}_{\varnothing}[\psi](t;\zeta). 
\end{align}
That is, we replace the wave functions $\{\psi^{(j)}(t;\zeta)\}_{j=1}^p$ which constitute the Wronskian functions $W^{(r)}_{\varnothing}[\psi](t;\zeta)$ by the charge-conjugate wave functions $\{\psi^{(j)}_{\mathcal C}(t;\zeta)\}_{j=1}^p$: 
\begin{align}
\text{2') }\qquad W^{(r)}_{\varnothing}[\psi](t;\zeta) \quad \to \quad W^{(r)}_{\varnothing}[\psi_{\mathcal C}](t;\zeta). 
\end{align}
This transformation is understood as the B\"acklund transformation (of Section \ref{Subsubsection:Backlundtransformation}) which maps the coefficients $\bigl\{ u_n(t) \bigr\}_{n=1}^p$ and $\bigl\{ v_n(t) \bigr\}_{n=1}^q$ to the dual ones $\bigl\{ \widetilde u_n(t) \bigr\}_{n=1}^p$ and $\bigl\{ \widetilde v_n(t) \bigr\}_{n=1}^q$, with replacing $\beta_{p,q}$ by $-\beta_{q,p}$ (i.e.~Eq.~\eq{Eqn:BacklundTransformationOfUV}). In this sense, the Wronskian functions $W^{(r)}_{\varnothing}[\psi_{\mathcal C}](t;\zeta)$ are also obtained by 
\begin{align}
W^{(r)}_{\varnothing}[\psi_{\mathcal C}](t;\zeta) \,\, = \,\, W^{(r)}_{\varnothing}[\psi](t;\zeta) \Bigg|_{\ds 
\text{replace coefficients as Eq.~\eq{Eqn:BacklundTransformationOfUV}}
}. 
\end{align}
We first make sure that these two prescriptions are equivalent: 
\begin{align}
&W^{(p-r)}_\varnothing[\psi ](t; \zeta) \quad \Leftrightarrow \quad  W^{(r)}_\varnothing[\psi_{\mathcal C}](t;\zeta) \nn\\
\Bigl( \text{or} & \quad  W^{(p-r)}_\varnothing[\psi_{\mathcal C} ](t; \zeta) \quad \Leftrightarrow \quad  W^{(r)}_\varnothing[\psi](t;\zeta) \Bigr)
 \label{Eq:WronskiansChargeConjugationReflection}
\end{align}

\subsubsection{Consistency of charge conjugation}

The criterion of equivalence is given by the existence of {\em a gauge transformation} (e.g.~Eq.~\eq{Eqn:GaugeTransISM1} and Eq.~\eq{Eqn:GaugeTransISM2}). In fact, for every system which we have evaluated, there exists the following matrix ${\mathsf M}^{(r)}(t)$ of gauge transformation: 
\begin{align}
& \vec W^{(p-r)}[\psi_{\mathcal C}](t; c \, \zeta) = {\mathsf M}^{(r)}(t)\times  \vec W^{(r)}[\psi](t;\zeta) \label{Eqn:MmatrixForCC}
\end{align}
with $\det {\mathsf M}^{(r)} (t) \neq 0$ and $ {}^\exists c \,(\neq 0) \in \mathbb C$. That is, one can show
\begin{align}
\left\{
\begin{array}{rl}
\mathcal B^{(r)}[\psi](t;\zeta)  \,= &\!\! {\mathsf M}^{(r)}(t)^{-1}\Bigl[ \mathcal B^{(p-r)}[\psi_{\mathcal C}](t;c\,\zeta) \Bigr] {\mathsf M}^{(r)}(t) - g\,{\mathsf M}^{(r)}(t)^{-1} \,\del_t  {\mathsf M}^{(r)}(t) \cr
\mathcal Q^{(r)}[\psi](t;\zeta)  \, = &\!\! {\mathsf M}^{(r)}(t)^{-1}\Bigl[ c\,\mathcal Q^{(p-r)}[\psi_{\mathcal C}](t;c\,\zeta) \Bigr] {\mathsf M}^{(r)}(t)
\end{array}
\right. 
\end{align}
As a result, we observe 
\begin{align}
W^{(p-r)}_\varnothing[\psi_{\mathcal C} ](t;c\, \zeta) = W^{(r)}_\varnothing[\psi](t;\zeta) \qquad \Bigl( {}^\exists c \in \mathbb C\Bigr),  
\end{align}
up to an overall normalization constant. The results for the gauge transformation matrices are shown in Appendix \ref{Section:ExampleOfIMS}. 

By taking the B\"acklund transformation of Eq.~\eq{Eqn:MmatrixForCC}, one obtains the charge conjugate counterpart: 
\begin{align}
\vec W^{(p-r)}[\psi](t; c \, \zeta) = {\mathsf M}^{(r)}_{\mathcal C}(t)\times  \vec W^{(r)}[\psi_{\mathcal C}](t;\zeta). 
\label{Eqn:MmatrixForCCofCH}
\end{align}
with 
\begin{align}
{\mathsf M}^{(r)}_{\mathcal C}(t) \equiv {\mathsf M}^{(r)}(t) \Bigg|_{\ds 
\text{replace coefficients as Eq.~\eq{Eqn:BacklundTransformationOfUV}}
}. 
\end{align}
Then these matrices satisfy the following relations: 
\begin{align}
{\mathsf M}^{(p-r)}_{\mathcal C}(t) = \Bigl( {\mathsf M}^{(r)}(t) \Bigr)^{-1}, \qquad {\mathsf M}^{(r)}_{\mathcal C}(t) \propto \Bigl({\mathfrak C^{(p-r,1)}}\Bigr)^{-1}\,   \Bigl[ {\mathsf M}^{(r)}(t) \Bigr]^{\rm T} \, \Bigl( \mathfrak C^{(r,1)} \Bigr). 
\end{align}

\subsubsection{Consistency of dual charge conjugation}

Similar to the above cases of charge conjugations, one can also find a gauge transformation which connects the pair of dual charge conjugates: 
\begin{align}
W^{(1,s)}_\varnothing[\chi](t;\zeta)\qquad \Leftrightarrow \qquad W^{(1,q-s)}_\varnothing[\chi_{\mathcal C} ](t; \zeta) . 
\end{align}
In fact, for every system which we have evaluated, there exists the following matrix ${\mathsf N}^{(s)}(t)$ of gauge transformation: 
\begin{align}
\vec W^{(1,q-s)}[\chi_{\mathcal C}](t; c \, \zeta) = {\mathsf N}^{(s)}(t)\times  \vec W^{(1,s)}[\chi](t;\zeta), \label{Eqn:DefinitionOfNmatrix}
\end{align}
with $\det {\mathsf N}^{(s)} (t) \neq 0 $ and ${}^\exists c\, (\neq 0) \in \mathbb C$. That is, one can show 
\begin{align}
\left\{
\begin{array}{rl}
\mathcal B^{(1,s)}[\chi](t;\zeta)  \,= &\!\! {\mathsf N}^{(s)}(t)^{-1}\Bigl[ \mathcal B^{(1,q-s)}[\chi_{\mathcal C}](t;c\,\zeta) \Bigr] {\mathsf N}^{(s)}(t) - g\,{\mathsf N}^{(s)}(t)^{-1} \,\del_t  {\mathsf N}^{(s)}(t) \cr
\mathcal Q^{(1,s)}[\chi](t;\zeta)  \, = &\!\! {\mathsf N}^{(s)}(t)^{-1}\Bigl[ c\,\mathcal Q^{(1,q-s)}[\chi_{\mathcal C}](t;c\,\zeta) \Bigr] {\mathsf N}^{(s)}(t)
\end{array}
\right. 
\end{align}
As a result, we observe 
\begin{align}
W^{(1,q-s)}_\varnothing[\chi ](t;c\, \zeta) = W^{(1,s)}_\varnothing[\chi_{\mathcal C}](t;\zeta) \qquad \Bigl( {}^\exists c \in \mathbb C\Bigr). 
\end{align}
The results for the gauge transformation matrices are shown in Appendix \ref{Section:ExampleOfIMS}. 

Similar to the case of charge conjugations, these matrices satisfy the following relations: 
\begin{align}
{\mathsf N}^{(q-s)}_{\mathcal C}(t) = \Bigl( {\mathsf N}^{(s)}(t) \Bigr)^{-1}, \qquad {\mathsf N}^{(s)}_{\mathcal C}(t) \propto \mathfrak D^{(1,q-s)}_{[\chi]}(t,\zeta)^{-1}\,   \Bigl[ {\mathsf N}^{(s)}(t) \Bigr]^{\rm T} \, \mathfrak D^{(1,s)}_{[\chi_{\mathcal C}]}(t,\zeta), 
\end{align}
where
\begin{align}
{\mathsf N}^{(s)}_{\mathcal C}(t) \equiv {\mathsf N}^{(s)}(t) \Bigg|_{\ds 
\text{replace coefficients as Eq.~\eq{Eqn:BacklundTransformationOfUV}}
}. 
\end{align}

\subsection{Reflection relation and edge entries of the Kac table \label{Subsection:ConsistencyOfDualityWithReflectionRelations}}

In this section, we construct the edge entries of Kac table, by utilizing {\em the reflection relation}, Eq.~\eq{Eqn:LiouvilleReflectionRelation}. We then analyze consistency associated with this construction. 

\begin{figure}[htbp]
\centering
\includegraphics[scale=1.045]{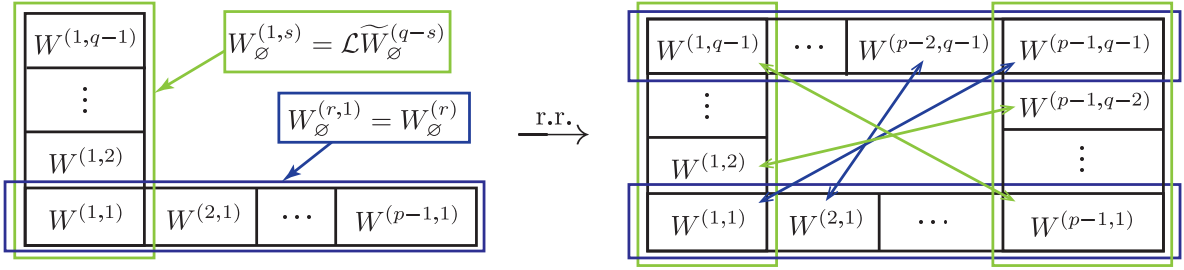}
\caption{\footnotesize Reflection relation (r.r.) and the edge entries of the Kac table}
\label{Figure:ReflectionEdgeEntries}
\end{figure}

We have proposed two prescriptions to construct $(r,1)$ and $(1,s)$ FZZT-Cardy branes: One is by the Wronskians (in Section \ref{Section:WronskiansForR1FZZTCardyBranes}); the other is by the combination of Laplace-Fourier transformation and charge conjugation (Section \ref{Section:1SFZZTCardyBranesAndKacTable}). We once again present {\em a uniform notation} $\{ W^{(r,s)}(t;\zeta) \}_{r,s}$ for the Wronskian functions associated with $(r,s)$ FZZT-Cardy branes $\Theta_{\rm Cardy}^{(r,s)}(\zeta)$: 
\begin{align}
1) \qquad &W^{(r,1)}_\lambda (t;\zeta) \equiv W^{(r)}_\lambda [\psi](t;\zeta) &(1\leq r\leq p-1)\\
2)  \qquad &W^{(1,s)}_{\lambda} (t;\zeta) \equiv \mathcal L \widetilde W^{(q-s)}_\lambda [\chi](t;\zeta) & (1\leq s \leq q-1)
\end{align}
Note that the assignment of Wronskian functions is overlapped by these two prescriptions at the Wronskian function of $W^{(1,1)}(t;\zeta)$. These two overlapped prescriptions are shown to be equivalent in Section \ref{Subsubsection:CornerEntriesOfKacTable}. These Wronskians are now aligned along the L-shape table inside the Kac table (as in the left-hand side of Fig.~\ref{Figure:ReflectionEdgeEntries}). 

As is shown in Fig.~\ref{Figure:ReflectionEdgeEntries}, we then define {\em the edge entries} of the Kac table by applying {\em the reflection relation} Eq.~\eq{Eqn:LiouvilleReflectionRelation} {\em to the Wronskians}: 
\begin{align}
W^{(r,s)}_\lambda (t;\zeta) \equiv W^{(p-r,q-s)}_\lambda (t;\zeta). 
\end{align}
This extension of the index $(r,s)$ should pass the consistency checks, which are discussed in Section \ref{Subsubsection:CornerEntriesOfKacTable} and in Section \ref{Subsubsection:InvolutorityOfPQandReflectionChargeConjugation}. 

\subsubsection{Corner entries of the Kac table \label{Subsubsection:CornerEntriesOfKacTable}}

The above two different prescriptions overlap each other, at the four corners of the Kac table. Two of the four corners are distinct points under the reflection relation. 

\paragraph{1) The $(1,1)$ FZZT-Cardy brane (i.e.~$\psi$)} The principal FZZT-brane now shares the two prescriptions:  
\begin{align}
W^{(1,1)}_\varnothing(t;\zeta): \qquad \psi(t;\zeta) = W^{(1)}_\varnothing [\psi] (t;\zeta) \qquad \Leftrightarrow \qquad \mathcal L \widetilde W^{(q-1)}_\varnothing[\chi](t;\zeta). 
\end{align}
We have also considered an example of this situation, the case of the $(2,3)$-system, in Section \ref{Subsubsection:SpaceOfYDandIMSin1SFZZTCardy}.

One way to prove this relation is to show that these two prescriptions are connected by a gauge transformation of isomonodromy systems, as in \ref{Subsubsection:SpaceOfYDandIMSin1SFZZTCardy}. In this case, however, we can see the equivalence more easily by directly showing the equivalence of the Baker-Akhiezer systems. In fact, as an equivalence of differential equations, one can see the following identification: 
\begin{align}
\psi(t;\zeta) &=  \mathcal L_{\gamma}^{(\beta_{q,p})} [\chi_{\mathcal C}](t;\zeta) = \int_\gamma d\eta \, e^{\frac{\beta_{q,p} \zeta \eta}{g}} \, \chi_{\mathcal C}(t;\eta), \nn\\
\chi_{\mathcal C}(t;\eta) &= W^{(q-1)}_\varnothing[\chi](t;(-1)^q \eta),
\label{Eqn:CornerConsistency11PsiLchiC}
\end{align}
where $\chi_{\mathcal C}$ is defined in Eq.~\eq{Eqn:ChargeConjuOfChi1} and Eq.~\eq{Eqn:ChargeConjuOfChi2}. 
This is exactly the direct check of the $p-q$ duality transformation Eq.~\eq{Eq:PQeqLC1st} at the corner entry $(r,s)=(1,1)$. 

\paragraph{2) The $(p-1,1)$ FZZT-Cardy brane (i.e.~$\psi_{\mathcal C}$)}

Another corner contains the charge conjugate of the principal FZZT-brane: 
\begin{align}
W^{(1,q-1)}_\varnothing(t;\zeta): \qquad \psi_{\mathcal C}(t;\zeta) = W^{(p-1)}_\varnothing [\psi] (t;(-1)^p\zeta) \qquad \Leftrightarrow \qquad 
 \mathcal L \widetilde W^{(1)}_\varnothing[\chi](t;\zeta). 
\end{align}
This is easily shown by taking the charge conjugation (in the form of the B\"acklund transformation Eq.~\eq{Eqn:BacklundTransformationOfUV}) of the relation Eq.~\eq{Eqn:CornerConsistency11PsiLchiC} as 
\begin{align}
\psi_{\mathcal C}(t;\zeta) &=  \mathcal L_{\gamma}^{(\beta_{p,q})} [\chi](t;\zeta) = \int_\gamma d\eta \, e^{\frac{\beta_{p,q} \zeta \eta}{g}} \, \chi(t;\eta). 
\end{align}

\subsubsection{Reflection relation and charge conjugation: $\mathcal P \mathcal Q \circ \mathcal P \mathcal Q = 1$ \label{Subsubsection:InvolutorityOfPQandReflectionChargeConjugation}}

Another consistency check is to show that the generalized spectral $p-q$ duality is an involution on the Kac table: $\mathcal P \mathcal Q \circ \mathcal P \mathcal Q  = 1$. 

Before considering this consistency check, we first comment on the charge conjugation transformation within the Kac table, especially accompanied with the reflection relation. The claim is that, with taking into account the reflection relation, there is no difference between the charge conjugation transformation $\mathcal C$ and the dual charge conjugation transformation $\mathcal D$: 
\begin{align}
\begin{array}{c}
\includegraphics{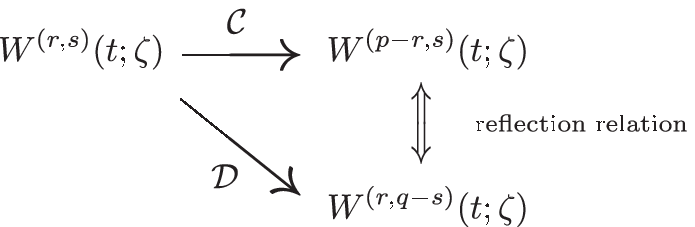}
\end{array}
\end{align}
That is, 
\begin{align}
\mathcal C [W^{(r,s)}(t;\zeta)] = \mathcal D [W^{(r,s)}(t;\zeta)]. 
\end{align}
This can be natural since the charge conjugation transformation itself should uniquely exist in the following sense: 
\begin{align}
\mathcal C[e^{\hat \phi(\zeta)}] = \mathcal D[e^{\hat \phi(\zeta)}]  = e^{- \hat \phi(\zeta)}. 
\end{align}
By taking into account Eq.~\eq{Eqn:DefinitionOfDualChargeConjugationTransformationWithPQtransformation}, one can further show that%
\footnote{$\mathcal P \mathcal Q = \mathcal P \mathcal Q \circ (\mathcal C\circ \mathcal C) = \mathcal D \circ \mathcal P \mathcal Q \circ \mathcal C =  \mathcal C \circ \mathcal P \mathcal Q \circ \mathcal C  = \mathcal C \circ (\mathcal L \circ \mathcal C) \circ \mathcal C = \mathcal C \circ \mathcal L$.  }
\begin{align}
\mathcal P\mathcal Q = \mathcal L \circ \mathcal C \circ \mathcal A= \mathcal C \circ \mathcal L \circ \mathcal A, 
\end{align}
within the Kac table, as a result of consistency with the reflection relation. In addition, according to Eq.~\eq{Eqn:RWeqW_Roperation}, the antipodal operator $\mathcal A$ also satisfies the following relation: 
\begin{align}
\mathcal A \circ \mathcal P \mathcal Q = \mathcal P \mathcal Q \circ \mathcal A, 
\end{align}
since this operation is uniformly understood by the $\tau$-coordinate, $\mathcal A: \, \tau \to \tau + \pi i$. 

With this relation, one can schematically show that the generalized spectral $p-q$ duality satisfies the following:
\begin{align}
\mathcal P \mathcal Q \circ \mathcal P \mathcal Q = \bigl( \mathcal L_{\gamma_1} \circ \mathcal C \bigr) \circ \bigl( \mathcal C \circ \mathcal L_{\gamma_2} \bigr) = \mathcal L_{\gamma_1} \circ \mathcal L_{\gamma_2}. \label{Eqn:InvolutionOfPQ}
\end{align}
Here we explicitly show the contours $\gamma_1$ and $\gamma_2$ in the Laplace-Fourier transformation. 
Therefore, up to coordinate transformation of $\zeta$, both $f(\zeta)$ and $\mathcal L_{\gamma_1} \circ \mathcal L_{\gamma_2}[f](\zeta)$ (or $g(\eta)$ and $\mathcal L_{\gamma_2} \circ \mathcal L_{\gamma_1}[g](\eta)$ on the dual side) satisfy the same differential equation. In addition, as in Eq.~\eq{Eqn:LLeq1condition}, if one chooses a proper pair of contours $(\gamma_1,\gamma_2)$ of the Laplace-Fourier transformation, then the transformation Eq.~\eq{Eqn:InvolutionOfPQ} further becomes 
\begin{align}
\mathcal P \mathcal Q \circ \mathcal P \mathcal Q = \mathcal L_{\gamma_1} \circ \mathcal L_{\gamma_2} = 1 \qquad (\gamma_2 = \gamma_1^{-1}). 
\end{align}
By this final form, we conclude that $\mathcal P \mathcal Q$ is an involution.

\section{Toward the interior entries of the Kac table \label{Section:TowardInteriorOfKacTable}}

So far, we have constructed the Wronskian descriptions of the FZZT-Cardy branes, and have proposed a consistent manipulation along the edge entries of the Kac table. These techniques are in fact enough to construct all the entries of the Kac table for various primitive models which we have elaborated in this paper. However, for the higher (or more general) $(p,q)$-systems, it becomes necessary to find another additional prescription which allows us to generate the FZZT-Cardy branes {\em aligned in the interior entries of the Kac table}. 

Note that our construction is still not completely established, but it is worth summarizing our proposal for the interior entries.

\subsection{Horizontal connections and multi-point correlators \label{Subsection:HorizontalConnectionsAndMultipointCorrelators}}

The first observation in this direction comes again from looking at the Seiberg-Shih relation Eq.~\eq{Eq:SeibergShihRelation}. The Seiberg-Shih relation can be also written in the following way:  
\begin{align}
\left|(r,s)\strut\right>_\zeta 
= 
\sum_{
\begin{subarray}{c}
k=-(r-1) \cr \text{step 2}
\end{subarray}
}^{(r-1)} 
\left|(1,s)\strut\right>_{\zeta_{k,0}(\zeta)},
\end{align}
which implies that $(r,s)$-type FZZT-Cardy branes are again constructed by the multi-body system of the {\em ``elemental''} $(1,s)$-type FZZT-Cardy branes $\ket{(1,s)}_{\zeta_{k,0}}$, where $\zeta_{0,k} = e^{\pi i k} \zeta$. In this sense, it is natural to expect that the $(r,s)$-type FZZT-Cardy branes are again described by ``some kinds of Wronskians'' associated with {\em correlators of these ``elemental'' $(1,s)$-type FZZT-Cardy branes}. 

For instance, as a non-trivial example, one can see that the $(r,q-1)$-type FZZT-Cardy branes follow this idea: That is, $W^{(r,q-1)}_\varnothing (t;\zeta)$ is given by a $r$-body state of $W^{(1,q-1)}_\varnothing (t;\zeta)$: 
\begin{align}
W^{(r,q-1)}_\varnothing (t;\zeta) = W^{(r)}_\varnothing  [\psi_{\mathcal C}](t;\zeta) = \vev{\prod_{a=1}^r \hat \psi_{\mathcal C}^{(j_a)}(\zeta)}_t,  \label{Eqn:Section7:ChargeConjuWRqMS1Relation}
\end{align}
where $\psi_{\mathcal C}(t;\zeta) = W^{(1,q-1)}_\varnothing (t;(-1)^p\zeta)$. 
Therefore, the next task is to extend this consideration to the general $(r,s)$-type FZZT-Cardy branes. 
The point is that the usual kinds of Wronskians Eq.~\eq{Eqn:GeneralizedWronskiansDefinitionInFurtherMoreWronskians} are no longer applied to the general cases. This leads us to the next generalization of the Wronskian functions. 

\subsection{Proposal: Schur-differential Wronskians \label{SDWronskianFZZTCardyProposal}}

Our proposal is then to adopt the formula for {\em the correlators of the Wronskian functions} which we have derived in Section \ref{Subsection:MultipointCorrelatorsOfWronskians}. Therefore, it is related to the Schur-differential Wronskians of Eq.~\eq{Eqn:SDWronskianSection2Correlators} and Eq.~\eq{Eqn:SDWronskianSection2lambda}. We propose that the general $(r,s)$ FZZT-Cardy branes are given by {\em normal-ordered Schur-differential Wronskians} $\star \mathscr W^{(r)}_\varnothing(t;\zeta)$ as follows: 
\begin{align}
&\Theta^{(r,s)}_{\rm Cardy}(\zeta) \qquad \leftrightarrow \qquad 
\star \mathscr W^{(r)}_{\varnothing } \bigl[W^{(1,s)}\bigr](t;\zeta), 
\label{Eqn:CardyBySDWronkians}
\end{align}
where 
\begin{align}
&\overbrace{\star \mathscr W^{(r)}_{\lambda} \bigl[W^{(1,s)}\bigr](t;\zeta)}^{\bigl[j_1^{(1)},\cdots,j_s^{(1)}\bigl|\cdots \big|\cdots \bigr|j_1^{(r)},\cdots,j_s^{(r)}\bigr]}  \equiv \sum_{
\pi \in \binom{rs}{s|s|\cdots|s} 
}
\sgn(\pi) \,  \nol \overbrace{W^{(1,s)}_{\pi^{(1)}_\lambda}(t;\zeta)}^{[j_1^{(1)},\cdots,j_s^{(1)}]}\times  \cdots \times  \overbrace{W^{(1,s)}_{\pi^{(r)}_\lambda}(t;\zeta)}^{[j_1^{(r)},\cdots,j_s^{(r)}]} \nor,\nn\\
&\quad \qquad \Bigl( \text{and } \bigl\{ \pi_\lambda^{(a)}\bigr\}_{a=1}^r \text{ is given by Eq.~\eq{DefinitionOfPiLambdaEi1} and Eq.~\eq{DefinitionOfPiLambdaEi2} }\Bigr), 
\end{align}
and $\overbrace{\qquad \qquad }^{[j_1,j_2,\cdots,j_s]}$ denotes the actual indices of the Wronskian functions: 
\begin{align}
\overbrace{W^{(1,s)}_\lambda(t;\zeta)}^{[j_1,j_2,\cdots,j_s]} \equiv \mathcal L\widetilde W^{[j_1,j_2,\cdots,j_s]}_\lambda[\chi](t;\zeta). \label{Eqn:SDWronskiansW1Sindices}
\end{align}
The new ingredient here is the normal ordering ``$\star$'', which is defined by the vanishing order $\{m_{a,b}\}_{1\leq a<b\leq r}$ in the coincide limit of $\{\zeta_a\}_{a=1}^r$ to $\zeta$: 
\begin{align}
\star \mathscr W^{(r)}_\lambda [W](t;\zeta) \equiv 
\lim_{
\left\{
\begin{subarray}{c}
\zeta_a \to \zeta \cr
1\leq a \leq r
\end{subarray}
\right\}
} \dfrac{\mathscr W^{(r)}_\lambda [W](t;\zeta_1,\zeta_2,\cdots,\zeta_r)}{\ds \prod_{1\leq a <b \leq r} (\zeta_a-\zeta_b)^{m_{a,b}}}. 
\end{align}
If $W(t;\zeta)$ is the normal Wronskian: $W(t;\zeta) = W^{(r)}(t;\zeta)$, then the vanishing order $\{m_{a,b}\}_{1\leq a<b\leq r}$ is the overlapping number of Eq.~\eq{Eq:OverlapNumber}. 
Because of this normal ordering, the normal-ordered Schur-differential Wronskians are identified with the correlators of Wronskians (as discussed in Section \ref{Subsection:MultipointCorrelatorsOfWronskians}). As a result, the following relation is automatically satisfied: 
\begin{align}
W^{(r)}_\varnothing  [\psi_{\mathcal C}](t;\zeta) &= \vev{\prod_{a=1}^r \hat \psi_{\mathcal C}^{(j_a)}(\zeta)}_t = \vev{\prod_{a=1}^r \Bigl[ \prod_{b=1}^{p-1} \hat \psi^{(j_a+b)}(\zeta) \Bigr]} \nn\\
&= \vev{\prod_{a=1}^r \Bigl[ \hat W^{[j_a+1,\cdots,j_a+p-1]}[\psi](\zeta) \Bigr]} \nn\\
&= \star \mathscr W^{(r)}_{\varnothing} \bigl[W^{(p-1)}\bigr](t;\zeta),  \label{Eqn:WandSDWinPsiCequivalence}
\end{align}
which is the relation Eq.~\eq{Eqn:Section7:ChargeConjuWRqMS1Relation} mentioned above. 
Therefore, we propose that this normal-ordered Schur-differential Wronskian is the prescription which generates all the general $(r,s)$ FZZT-Cardy branes. 

Another important point of this normal ordering is that the normal ordering creates a hierarchy among the normal-ordered Schur-differential Wronskians. It is because the normal-ordering operation introduces some derivatives with respect to $\zeta$ depending on the vanishing order of the Wronskians. This results in some sets of normal-ordered Schur-differential Wronskians. Each set of Wronskians satisfies each differential equation and is completely decoupled from the other sets of Wronskians. Therefore, these sets of Wronskians result in several decoupled isomonodromy systems. This non-perturbative decoupling phenomenon is also expected from the viewpoint of spectral curves, as is discussed in Section \ref{Subsubsection:GeneralFormulaOfSpectralCurveRS}. 

As in Section \ref{Subsubsection:AbbreviationSchurDifferentialWronskians}, we use the abbreviation of indices: 
\begin{align}
\star \mathscr W^{(r)}_{\lambda} \bigl[W^{(1,s)}\bigr](t;\zeta)  \equiv \sum_{
\pi \in \binom{rs}{s|s|\cdots|s} 
}
\sgn(\pi) \,  \nol W^{(1,s)}_{\pi^{(1)}_\lambda}(t;\zeta)\otimes  \cdots \otimes  W^{(1,s)}_{\pi^{(r)}_\lambda}(t;\zeta) \nor.
\end{align}

\subsection{Consistency of $(p,q)=(3,4)$ and $(r,s)=(2,2)$ \label{ConsistencyOfSDWronskiansProposal}}

In this section, we perform a non-trivial check of the correspondence Eq.~\eq{Eqn:CardyBySDWronkians} in the case of $(p,q)=(3,4)$.  The Kac table is given as
\begin{align}
    \begin{tabular}{|c|c|} \hline
$(1,3)$ & $(2,3)$  \\ \hline
$(1,2)$ & $\bf (2,2)$  \\ \hline
$(1,1)$ & $(2,1)$  \\ \hline
    \end{tabular}
\end{align}
and we consider the $(2,2)$-type FZZT-Cardy brane. This brane has two different prescriptions: One is given by the normal-ordered Schur-differential Wronskian $\mathscr W^{(2)}_\varnothing$; the other is given by $W^{(2,2)}_\varnothing = W^{(1,2)}_\varnothing = \mathcal L \widetilde W^{(2)}_\varnothing$: 
\begin{align}
\Theta_{\rm Cardy}^{(2,2)}(\zeta): \qquad
\star \mathscr W^{(2)}_{\varnothing} \bigl[\mathcal L \widetilde W^{(2)}\bigr](t;\zeta) \quad \Leftrightarrow \quad \mathcal L \widetilde W^{(2)}_{\varnothing}[\chi](t;\zeta). 
\end{align}
Therefore, these two prescriptions should give the same differential equation. For the coincidence of the differential equation, it is enough to consider the normal-ordered Schur-differential Wronskians with null vanishing order, that is the normal Schur-differential Wronskian: 
\begin{align}
\star \mathscr W^{(2)}_{\varnothing} \bigl[\mathcal L \widetilde W^{(2)}\bigr](t;\zeta) \quad \to \quad  \mathscr W^{(2)}_{\varnothing} \bigl[\mathcal L \widetilde W^{(2)}\bigr](t;\zeta),
\end{align}
because the null-order sector already gives the expected results. 

By the definition, the Schur-differential Wronskian is written as
\begin{align}
\mathscr W^{(2)}_\varnothing\bigl[W^{(1,2)}\bigr](t;\zeta) &= W^{(1,2)}_\varnothing (t;\zeta) \otimes W^{(1,2)}_{\tiny \yng(2,2)} (t;\zeta) - W^{(1,2)}_{\tiny \yng(1)}(t;\zeta) \otimes W^{(1,2)}_{\tiny \yng(2,1)} (t;\zeta) + \nn\\
&\quad+ W^{(1,2)}_{\tiny \yng(2)}(t;\zeta) \otimes W^{(1,2)}_{\tiny \yng(1,1)}(t;\zeta)  + W^{(1,2)}_{\tiny \yng(1,1)}(t;\zeta) \otimes W^{(1,2)}_{\tiny \yng(2)} (t;\zeta) - \nn\\
&\quad - W^{(1,2)}_{\tiny \yng(2,1)}(t;\zeta) \otimes W^{(1,2)}_{\tiny \yng(1)} (t;\zeta) + W^{(1,2)}_{\tiny \yng(2,2)} (t;\zeta) \otimes W^{(1,2)}_\varnothing (t;\zeta),
\end{align}
and also, with use of the Schur-differential equations for $W^{(1,2)}_\lambda (t;\zeta)$, one can show that 
\begin{align}
\square \times \mathscr W^{(2)}_\varnothing\bigl[W^{(1,2)}\bigr](t;\zeta) &= W^{(1,2)}_\varnothing (t;\zeta) \otimes W^{(1,2)}_{\tiny \yng(3,2)} (t;\zeta) - W^{(1,2)}_{\tiny \yng(1)}(t;\zeta) \otimes W^{(1,2)}_{\tiny \yng(3,1)} (t;\zeta) + \nn\\
&\quad + W^{(1,2)}_{\tiny \yng(1,1)}(t;\zeta) \otimes W^{(1,2)}_{\tiny \yng(3)}(t;\zeta)  + W^{(1,2)}_{\tiny \yng(3)}(t;\zeta) \otimes W^{(1,2)}_{\tiny \yng(1,1)} (t;\zeta) - \nn\\
& \quad - W^{(1,2)}_{\tiny \yng(3,1)}(t;\zeta) \otimes W^{(1)}_{\tiny \yng(1)} (t;\zeta) + W^{(1,2)}_{\tiny \yng(3,2)} (t;\zeta) \otimes W^{(1,2)}_\varnothing (t;\zeta) \nn\\
&= \frac{-1}{8 \beta_{4,3}} \Bigl(W^{(1,2)}_{\varnothing} (t;\zeta) \otimes \del_\zeta W^{(1,2)}_{\square} (t;\zeta)  - W^{(1,2)}_{\square} (t;\zeta) \otimes \del_\zeta W^{(1,2)}_{\varnothing} (t;\zeta) + \nn\\
& \quad +\del_\zeta W^{(1,2)}_{\square} (t;\zeta) \otimes  W^{(1,2)}_{\varnothing} (t;\zeta)  - \del_\zeta W^{(1,2)}_{\varnothing} (t;\zeta) \otimes  W^{(1,2)}_{\square} (t;\zeta)  \Bigr) \nn\\
&\equiv \frac{1}{8} \mathscr D_\zeta \Bigl(W^{(1,2)}_{\varnothing} (t;\zeta) \otimes W^{(1,2)}_{\square} (t;\zeta)  - W^{(1,2)}_{\square} (t;\zeta) \otimes W^{(1,2)}_{\varnothing} (t;\zeta) \Bigr) \nn\\
&= \frac{1}{8 } \mathscr D_\zeta\, W^{(2)}_\varnothing [W^{(1,2)}](t;\zeta), 
\end{align}
where 
\begin{align}
\mathscr D_\zeta \Bigl(Y_1 \otimes Y_2\Bigr) = \frac{1}{\beta_{4,3}} \Bigl( \frac{\del Y_1}{\del \zeta} \otimes Y_2 - Y_1 \otimes \frac{\del Y_2}{\del \zeta} \Bigr). 
\end{align}
Then, the higher derivatives  of the new Wronskian function are given as follows: 
\begin{align}
\square^2 \times \mathscr W^{(2)}_\varnothing\bigl[W^{(1,2)}\bigr](t;\zeta) & = \frac{1}{8} \mathscr D_\zeta\, W^{(2)}_\square [W^{(1,2)}](t;\zeta), \\ 
\square^3 \times \mathscr W^{(2)}_\varnothing\bigl[W^{(1,2)}\bigr](t;\zeta) & =\frac{1}{8} \mathscr D_\zeta\, W^{(2)}_{\tiny \yng(1,1)} [W^{(1,2)}](t;\zeta) - \frac{u_2(t)}{2} \square \times  \mathscr W^{(2)}_\varnothing\bigl[W^{(1,2)}\bigr](t;\zeta). 
\end{align}
On the other hand, 
\begin{align}
- \frac{\zeta}{2} \mathscr W^{(2)}_\varnothing\bigl[W^{(1,2)}\bigr](t;\zeta) &= \frac{1}{8}  \mathscr D_\zeta \biggl[ W^{(2)}_{\tiny \yng(1,1)}[W^{(1,2)}](t;\zeta) + \nn\\
&\qquad \quad + \Bigl( W^{(1,2)}_{{\tiny \yng(2,1)}}(t;\zeta) + \frac{u_2(t)}{12} W^{(1,2)}_{\tiny \yng(1)} (t;\zeta) \Bigr) \otimes W^{(1,2)}_{\varnothing} -\nn\\
&\qquad \quad -  W^{(1,2)}_{\varnothing}(t;\zeta) \otimes \Bigl( W^{(1,2)}_{{\tiny \yng(2,1)}} (t;\zeta)+ \frac{u_2(t)}{12} W^{(1,2)}_{\tiny \yng(1)} (t;\zeta) \Bigr)  \biggr] \nn\\
&= \frac{1}{8} \mathscr D_\zeta\, W^{(2)}_{\tiny \yng(1,1)} [W^{(1,2)}](t;\zeta). 
\end{align}
Here we use the Schur-differential equation: 
\begin{align}
W^{(1,2)}_{{\tiny \yng(2,1)}}(t;\zeta) + \frac{u_2(t)}{12} W^{(1,2)}_{\tiny \yng(1)} (t;\zeta) = \Bigl( \frac{\zeta}{4} + \frac{u_2'(t)}{4} - \frac{u_3(t)}{6}\Bigr) W^{(1,2)}_{\varnothing} (t;\zeta)
\end{align}
Therefore, one obtains 
\begin{align}
\Bigl[ 2\, \square^3 + u_2(t) \, \square + \zeta \Bigr] \mathscr W^{(2)}_\varnothing\bigl[W^{(1,2)}\bigr](t;\zeta) = 0. 
\end{align}
In fact, this is the same equation as that of $W^{(1,2)}_\varnothing(t;\zeta)$: 
\begin{align}
\Bigl[ 2\, \square^3 + u_2(t) \, \square - \zeta \Bigr] W^{(1,2)}_\varnothing(t;\zeta)= 0. 
\end{align}
This means that at least $\mathcal B$-operator is identical in these two prescriptions with a coordinate transformation of $\zeta: \, \zeta \to -\zeta$: 
\begin{align}
\mathscr W^{(2)}_\varnothing\bigl[W^{(1,2)}\bigr](t;\zeta) \propto W^{(1,2)}_\varnothing(t;\zeta). 
\end{align}

Although this is a good sign, further checks are necessary to support this proposal, which should be devoted to another paper and is out of the range of this paper. Before closing this section, necessary consistency checks are in order, which are left for future work: 
\begin{itemize}
\item [1. ] Check of equivalence for Eq.~\eq{Eqn:WandSDWinPsiCequivalence} 
\begin{align}
W^{(r)}_\varnothing  [\psi_{\mathcal C}](t;\zeta)\quad \Leftrightarrow \quad  \mathscr W^{(r)}_\varnothing\bigl[W^{(p-1)}\bigr](t;\zeta)
\end{align}
by showing existence of a gauge transformation. 
\item [2. ] Check of equivalence along the right edge of the Kac table: 
\begin{align}
\mathscr W^{(p-1)}_{\varnothing} \bigl[\mathcal L \widetilde W^{(q-s)}\bigr](t;\zeta) \quad \Leftrightarrow \quad 
\mathcal L \widetilde W^{(s)}_{\varnothing}[\chi](t;\zeta). 
\end{align}
\item [3. ] Check of charge conjugation: 
\begin{align}
\mathscr W^{(r)}_{\varnothing} \bigl[\mathcal L \widetilde W^{(q-s)}\bigr](t;\zeta) \quad \Leftrightarrow \quad 
\mathscr W^{(p-r)}_{\varnothing} \bigl[\mathcal L \widetilde W^{(s)}\bigr](t;\zeta), 
\end{align}
and also with the B\"acklund transformation, 
\begin{align}
\mathscr W^{(r)}_{\varnothing} \bigl[\mathcal L \widetilde W^{(q-s)}[\chi]\bigr](t;\zeta) \quad \Leftrightarrow \quad 
\mathscr W^{(p-r)}_{\varnothing} \bigl[\mathcal L \widetilde W^{(q-s)} [\chi_{\mathcal C}]\bigr](t;\zeta). 
\end{align}
\item [4. ] Check of triviality
\begin{align}
\frac{\del \mathscr W^{(p)}_{\varnothing} \bigl[\mathcal L \widetilde W^{(s)}\bigr](t;\zeta)}{\del \zeta} = \frac{\del \mathscr W^{(p)}_{\varnothing} \bigl[\mathcal L \widetilde W^{(s)}\bigr](t;\zeta)}{\del t} = 0. 
\end{align}
\item [5. ] Comparison of spectral curves between the isomonodromy systems of the new Wronskians and the Liouville calculus of Section \ref{Subsubsection:GeneralFormulaOfSpectralCurveRS}: 
\begin{align}
\lim_{g\to 0}\det\Bigl[Q I_{k_{r,s}\times k_{r,s}} - \mathcal Q^{(r,s)}(t;\zeta)\Bigr] = 0 \quad \Leftrightarrow \quad F^{(r,s)}_{\rm Cardy}(\zeta,Q) = 0. 
\end{align}
\end{itemize}

\section{Conclusion and discussion \label{Section:ConclusionAndDiscussion}}

In this paper, motivated by the Seiberg-Shih relation of FZZT-Cardy branes, we study three variants of Wronskian functions associated with the Baker-Akhiezer systems of $(p,q)$ minimal sting theory. The quantitative description is based on the Schur-differential equations and isomonodromy systems. In particular, we have derived the concrete forms of these differential equation systems, and discussed their mutual relationships and constructed the associated Kac table. 

As a nice feature of our Wronskian functions, as far as along the edge entries, they are consistently aligned in the Kac table, 
\begin{align}
W^{(r,s)}_\varnothing(t;\zeta),\qquad 1\leq r\leq p-1,\qquad 1\leq s\leq q-1. 
\end{align}
Moreover, they also enjoy the charge conjugation relation $\mathcal C$, the dual charge conjugation relation $\mathcal D$ and the reflection relation as their consistency condition ($ \mathcal D \circ \mathcal C = 1 $): 
\begin{align}
\mathcal C [W^{(r,s)}_\varnothing] = W^{(p-r,s)}_\varnothing, \qquad \mathcal D [W^{(r,s)}_\varnothing] = W^{(r,q-s)}_\varnothing, \qquad W^{(p-r,q-s)}_\varnothing = W^{(r,s)}_\varnothing, 
\end{align}
and also the generalized spectral $p-q$ duality transformation, 
\begin{align}
\mathcal P \mathcal Q[W^{(r,s)}_\varnothing] =  \widetilde W^{(s,r)}_\varnothing,\qquad 
\mathcal P \mathcal Q[\widetilde W^{(s,r)}_\varnothing] =  W^{(r,s)}_\varnothing,\qquad \mathcal P \mathcal Q = \mathcal L \circ \mathcal C\circ \mathcal A. \label{Conclusion:GeneralizedSpectralPQDuality}
\end{align}
These transformations are also compatible with the B\"acklund transformation of the string equation. Along this direction, we have proposed that the general Wronskian functions associated with the Kac table are given by the Schur-differential Wronskians: 
\begin{align}
W^{(r,s)}_\varnothing(t;\zeta) = \mathscr W^{(r)}_\varnothing[W^{(1,s)}](t;\zeta), 
\end{align}
and at least one of the non-trivial checks are successfully carried out. 

There are two points in our construction which deserve further clarify/improve in the future study: 
\begin{itemize}
\item The first point is about the generalized spectral $p-q$ duality Eq.~\eq{Conclusion:GeneralizedSpectralPQDuality}. It is necessary to carry out its direct check. At the level of annulus amplitudes of the FZZT-Cardy branes, there appear ``non-universal terms'' which purely depend on $\zeta$ or $\eta$. According to the free-fermion formulation \cite{fy1,fy2,fy3}, these contributions are understood as normal ordering corrections. As we have seen in Section \ref{Subsubsection:CorrelatorsOfFZZTCardyBranes}, however, such normal-ordering contributions are not quite precisely equal to those of Liouville theory calculus \cite{KOPSS, Gesser}. This part therefore affects the annulus-order contributions of the generalized spectral $p-q$ duality Eq.~\eq{Conclusion:GeneralizedSpectralPQDuality}. Comprehensive understanding of this issue would lead us to ask what is the non-perturbatively duality consistent annulus amplitude in matrix models. 
\item 
The second point is about the Schur-differential Wronskians. As is commented in Section \ref{ConsistencyOfSDWronskiansProposal}, the Schur-differential Wronskians have only passed one non-trivial check; there remain several non-trivial checks to be carried out. Not only such non-trivial checks, but also the mathematical analysis of the Schur-differential Wronskians would be interesting by itself. We hope that our results would become a new starting point toward the complete description of the matrix models and non-perturbative string theory. 
\end{itemize}
Further directions are also in order: 
\begin{itemize}
\item 
It is also interesting to extend our results to multi-cut matrix models, since the Seiberg-Shih relation of $(p,q)$ minimal superstring theory has different features \cite{Irie1}. 
\item 
It should be also important to identify the explicit forms of matrix-model determinant operators which represent the FZZT-Cardy branes. 

\item 
It should be important to analyze the Stokes phenomena of the isomonodromy systems which are obtained in this paper. This allows us to extract the full non-perturbative behavior of FZZT-Cardy branes and new implications for non-perturbative string theory. In particular, 
\begin{itemize}
\item Our new isomonodromy systems allow us to compare the non-perturbative behavior of FZZT-Cardy branes with that of principal FZZT branes with utilizing the Deift-Zhou's Riemann-Hilbert graph/method \cite{DZmethod, ItsKapaev} (i.e.~the spectral networks, see e.g.~\cite{ItsBook}), as discussed in \cite{CIY5}. As is discussed in Section \ref{Subsection:ZZCardyBranes}, focusing on the bulk physics, the total ZZ-Cardy brane contributions in the FZZT-Cardy system should coincide with the pure ZZ-brane contributions of the principal FZZT system. This consideration should allow us to extract the non-perturbative behavior of FZZT-Cardy branes. 
\item In a further elaboration, it is interesting if one can extract the multi-cut boundary condition \cite{CIY2} of the FZZT-Cardy branes. It is also nice if one can re-interpret the complementary boundary condition of \cite{CIY2} and/or find another duality constraint on string theory \cite{CIY5}.
\end{itemize}
\item As is argued in \cite{fy3}, the coordinate $\zeta$ of the spectral curve is understood as the 2-dimensional spacetime of Liouville theory, which is identified as 
\begin{align}
\zeta \sim e^{-b( \phi_{\rm Liou} + i X_{\rm FF})}, 
\end{align}
where $\phi_{\rm Liou}$ is the Liouville spacetime coordinate and $X_{\rm FF}$ is the Feign-Fuchs  spacetime coordinate (i.e.~the coordinate of minimal models). In particular, as is also discussed in \cite{MMSS}, the weak-coupling limit $\phi_{\rm Liou} \to -\infty$ is naturally identified with the asymptotic infinity $\zeta\to + \infty$ of the spectral curve. The Liouville potential is then generated by the closest ZZ-brane located around the asymptotic infinity. In this sense, the spectral curves in matrix models have been understood as the semi-classical spacetime geometry of $(p,q)$ minimal string theory. 

In particular, in our analysis, we obtained new spectral curves of FZZT-Cardy branes with their isomonodromy systems (which allow us to control ``the non-perturbative completions of the spectral curves''). 
As an analog of the spectral curves found in fractional minimal string theory/multi-cut matrix models \cite{CIY1}, these new spectral curves are constituted by several irreducible algebraic curves. 
\begin{itemize}
\item Therefore, it is interesting to discuss the meaning of these new spacetimes, which are enlarged with respect to the original spacetime generated by the principal FZZT brane. 
\item As another application of these new spectral curves (i.e.~spacetimes), it is also interesting to discuss the interpretation of the black-hole geometry \cite{MMSS}. An essential example of the black-hole interpretation of the spectral curve is demonstrated by \cite{MMSS} in the $(2,1)$ minimal string theory which is the Airy system. In particular, according to the non-perturbative connection rule of the Airy function, they argued that the quantum-gravity wave function inside the horizon of the black hole should be described by the superposition of two quantum-gravity states: two semi-classical branches associated with the imaginary spacetime coordinates, 
rather than the naive analytic continuation of the classical geometry outside the horizon. That is, the non-perturbative role of the black-hole horizon is the quantum mechanical turning-point of quantum-gravity wave functions \cite{MMSS}. 

A new entry in our new spectral curves is the {\em wormhole}, which is associated with the connecting ZZ-branes discussed in Section \ref{Subsection:ZZCardyBranes}. Enough non-perturbative analysis to discuss such considerations in general $(p,q)$ minimal string theory is already in \cite{CIY5}. Therefore, by utilizing the results of \cite{CIY5}, it is interesting to study the non-perturbative role of the wormhole in the non-perturbative completions of string theory. 
\end{itemize}
\end{itemize}

\vspace{1cm}
\noindent
{\bf \large Acknowledgment}
\vspace{0.2cm}

\noindent
The authors would like to thank Jean-Emile Bourgine for useful discussions in his visiting Taiwan which were essential in initiating this project. 
B.~Niedner would like to thank Yukawa Institute for Theoretical Physics for the kind hospitality and Japan Society for the Promotion of Science for financial support during the early stages of this project. In particular, H.~Irie and B.~Niedner would like to thank Hiroshi Kunitomo for his support which makes this collaboration possible. H.~Irie and C.-H.~Yeh would like to thank people in Taiwan String Theory Focus Group, especially Pei-Ming Ho, for kind supports and hospitality during their stay in NCTS (north) at NTU, where some of the progress has been done. H.~Irie and C.-H.~Yeh would like to thank people in Tunghai University for their kind hospitality during their stay, where some of the progress has been done. H.~Irie would like to express his gratitude to all the people who support his series of works. 
C.-T.~Chan, H.~Irie and C.-H.~Yeh are supported in part by Ministry of Science and Technology (MOST) of Taiwan under the contract 
No.~104-2112-M-029 -001 and No.~103-2112-M-029 -002 (C.-T.~Chan), No.~104-2811-M-029 -001 (H.~Irie) and No.~102-2119-M-007-003 and No.~104R8700-2 (C.-H.~Yeh). The authors would like to acknowledge the support from the thematic group (strings) at NCTS. H.~Irie is supported in part by Japan Society for the Promotion of Science (JSPS) and Grant-in-Aid for JSPS Fellows No.~24-3610.
The work of B.~Niedner is supported in part by the German National Academic Foundation and the STFC grant ST/J500641/1.


\appendix 

\section{Two-matrix models \label{Appendix:Two-matrixModels}}

In this appendix, in order to fix our conventions and notations, the basics of two-matrix models \cite{Mehta} are summarized. For more details, the reader should look through \cite{CIY5}. 
\begin{itemize}
\item Partition function: 
\begin{align}
\mathcal Z_{n} \equiv \int_{\mathscr C_X^{(n)}\times \mathscr C_Y^{(n)}} dX dY e^{-N \tr [V_1(X) + V_2(Y) -XY]}, 
\end{align}
The matrix contours $\mathscr C_X^{(n)}$ and $\mathscr C_Y^{(n)}$ are the sets of $n\times n$ normal matrices. The eigenvalues of them pass along particular paths $\mathscr C_x \subset \mathbb C$ and $\mathscr C_y \subset \mathbb C$. If $n=N$, then it is the matrix model partition function, 
\begin{align}
\mathcal Z=\mathcal Z_N. 
\end{align}
\item Expectation values: 
\begin{align}
\vev{\mathcal O(X,Y) \strut}_{n\times n} \equiv \frac{1}{\mathcal Z_n} \int_{\mathscr C_X^{(n)}\times \mathscr C_Y^{(n)}} dX dY
\, e^{-N \tr [V_1(X) + V_2(Y) -XY]} \,\mathcal O(X,Y).
\end{align}
\item Orthonormal polynomials \cite{GrossMigdal2}: 
\begin{align}
\psi_n(x) \equiv & \frac{1}{\sqrt{h_n}} \vev{\det\bigl(x-X\bigr)}_{n\times n} e^{-NV_1(x)},\\
\chi_n(y) \equiv& \frac{1}{\sqrt{h_n}} \vev{\det\bigl(y-Y\bigr)}_{n\times n} e^{-NV_2(y)},
\end{align}
with respect to the inner product of $\ds \int_{\mathscr C_x \times \mathscr C_y} dx dy \, e^{Nxy}\, \psi_n(x)\,\chi_m(y) = \delta_{n,m}$. 
The linear equations of these wave-functions are 
\begin{align}
x\, \psi_n(x) = \mathcal A(n;Z) \, \psi_n(x),&\qquad N^{-1} \frac{\del}{\del x} \psi_n(x) = \mathcal B(n;Z) \, \psi_n(x) \\
y\, \chi_n(y) = \mathcal C(n;Z) \, \chi_n(y),&\qquad N^{-1} \frac{\del}{\del y} \chi_n(y) = \mathcal D(n;Z) \, \chi_n(y). 
\end{align}
Here $Z = e^{-\del_n}$ and $Z \, \psi_n(x) = \psi_{n-1}(x)$ and the so-called ``discrete string equation'' is given as 
\begin{align}
\mathcal B(n;Z) &= V_1'\bigl(\mathcal A(n;Z)\bigr) - \mathcal C^{\rm T}(t;Z), \\
\mathcal D(n;Z) &= V_2'\bigl(\mathcal C(n;Z)\bigr) - \mathcal A^{\rm T}(t;Z). 
\end{align}
\item The double scaling limit \cite{DSL}  for $(p,q)$ critical points is given as 
\begin{align}
& x= x_c (1 +a^\frac{p}{2} \zeta),\qquad y= y_c (1 +a^\frac{q}{2} \eta), \nn\\
&N^{-1} = a^{\frac{p+q}{2}} g,\qquad \frac{n}{N} = 1-a^{\frac{p+q-1}{2}} t,\nn\\
&\del_n = - a^{\frac{1}{2}} \del,\qquad  \del\equiv g\del_t \qquad \bigl(a \to 0\bigr), 
\end{align}
and the linear system becomes
\begin{align}
\zeta \,\psi(t;\zeta) = \bP(t;\del) \, \psi(t;\zeta),&\qquad g\frac{\del}{\del \zeta} \psi(t;\zeta) = \bQ(t;\del) \, \psi(t;\zeta) \\
\eta\, \chi(t;\eta) = \widetilde \bP(t;\del) \, \chi(t;\eta),&\qquad g \frac{\del}{\del \eta} \chi(t;\eta) = \widetilde \bQ(t;\del) \, \chi(t;\eta). 
\end{align}
With properly changing the normalization of $\zeta$ and $\eta$, one obtains 
\begin{align}
\bP(t;\del) &= 2^{p-1} \del^p + \sum_{n=2}^{p} u_n(t) \del^{p-n},\qquad &\bQ(t;\del) &= \beta_{p,q}\Bigl[2^{q-1} \del^q + \sum_{n=2}^{q} v_n(t) \del^{q-n}\Bigr], \label{Eq:MatrixAppendix:PQoperators}\\
\widetilde \bP(t;\del) &= 2^{q-1} \del^q + \sum_{n=2}^{q} \widetilde v_n(t) \del^{q-n},\qquad & \widetilde \bQ(t;\del) &= \beta_{q,p}\Bigl[2^{p-1} \del^p + \sum_{n=2}^{p}\widetilde u_n(t) \del^{p-n}\Bigr]. \nn\\
&\quad  \Bigl( = (-1)^q \beta_{p,q}^{-1}\, \bQ^{\rm T}(t;\del) \Bigr) & & \quad \Bigl(  = (-1)^q  \beta_{p,q}\, \bP^{\rm T}(t;\del) \Bigr)
\label{Eq:MatrixAppendix:dualPQoperators}
\end{align}
We also obtain 
\begin{align}
\beta_{p,q} = (-1)^{p+q} \beta_{q,p}, 
\end{align}
and 
\begin{align}
\left\{
\begin{array}{cc}
\ds \widetilde u_n(t) \equiv \sum_{j=0}^{n-2} (-1)^{n-j}\binom{p-n+j}{j}  u_{n-j}^{(j)}(t)
\qquad &\bigl(n=2,3,\cdots,p\bigr) \cr
\ds \widetilde v_n(t) \equiv \sum_{j=0}^{n-2} (-1)^{n-j}\binom{q-n+j}{j}  v_{n-j}^{(j)}(t)
\qquad  &\bigl(n=2,3,\cdots,q\bigr) 
\end{array}
\right.. 
\end{align}
\end{itemize}

\section{Examples of the isomonodromy systems \label{Section:ExampleOfIMS}}
Here we show the concrete form of the isomonodromy systems for the cases of $p,q=2,3,4,5$ shown in Table \ref{Table:CaseStudyIM}.  For simplicity, we sometimes use the following rescaled convention for the coefficients: 
\begin{align}
\frac{u_n(t)}{2^{p-1}} \equiv \bar u_n(t),\qquad \frac{v_n(t)}{2^{q-1}} \equiv \bar v_n(t),\qquad 
\frac{\zeta}{2^{p-1}} \equiv \bar \zeta,\qquad \frac{\eta}{2^{q-1}} \equiv \bar \eta, 
\qquad \frac{\beta_{p,q}}{2^{2-p-q}} \equiv \bar \beta_{p,q}. 
\end{align}
As in \cite{CIY5}, we use the following notation for the KP Lax operators.
\begin{itemize}
\item [1. ] We keep the magnitude of the two integers $(p,q)$ as
\begin{align}
p<q
\end{align}
and consider the following two dual systems: 
\begin{align}
\text{$(p,q)$-system} \qquad \overset{\text{$p-q$ dual}}{\longleftrightarrow} \qquad \text{$(q,p)$-system}. 
\end{align}
\item [2. ] The KP lax operators are chosen as follows: 
\begin{itemize}
\item $(p,q)$-system:  $\bigl(\bP,\bQ\bigr)$ of  Eq.~\eq{Eq:MatrixAppendix:PQoperators} with use of $\zeta$. 
\item $(q,p)$-system:  $\bigl(\widetilde \bP, \widetilde \bQ\bigr)$ of Eq.~\eq{Eq:MatrixAppendix:dualPQoperators} with use of $\eta$. 
\end{itemize}
That is, the $(q,p)$-systems are treated as the dual $(p,q)$-systems. In this sense, the coefficients $\{v_n(t)\}_{n=2}^q$ of all the Lax operators are solved by $\{u_n(t)\}_{n=2}^p$. 
\item [3. ] In the analysis throughout this paper, we consider {\em the conformal background} as a specific choice of KP flows in the string equation. The conformal background is introduced in \cite{MSS} as a choice of KP flows which produces the results of Liouville theory, with the worldsheet cosmological constant $\mu>0$. The precise identification is found in Section 3.4 of \cite{fim}. 

As another concept, we also use the term, {\em (string) vacuum} as introduced in \cite{CIY4,CIY5}. For a fixed {\em background} (i.e.~such as conformal background), we can choose the spectral curves within a non-perturbative completion. This spectral curve is understood as a string vacuum in the string theory landscape (i.e.~the space of meta-stable string vacua) \cite{CIY4,CIY5}. Among these string vacua, there are special vacua which satisfy the dispersionless Douglas equation \cite{DKK}, which are called {\em DKK vacua} \cite{CIY6}, named after Daul-Kazakov-Kostov. In the conformal background, one of the DKK vacua is given by the Chebyshev polynomials (i.e.~Eq.~\eq{Eqn:KostovSolutionOfChebyshevPolynomialSpectralCurve}) \cite{CIY6}. 
\end{itemize}

\begin{table}[htb]
  \begin{center}
    \begin{tabular}{c|cccc} 
$p\diagdown q$ & 2 & 3 & 4 & 5 \\ \hline 
2 & $\times $ & $(2,3)$ & $\times $ & $(2,5)$ \\ 
3 & $(3,2)$ & $\times$ & $(3,4)$ & $(3,5)$ \\ 
4 & $\times$ & $(4,3)$ & $\times$ & \ttfamily N/A  \\ 
5 & $(5,2)$ & $(5,3)$ & \ttfamily N/A  & $\times$ 
    \end{tabular}
  \end{center}
\caption{The cases of study}
\label{Table:CaseStudyIM}
\end{table}

\subsection{The case of $(p,q)=(2,3)$: Pure-gravity}

\begin{itemize}
\item [I. ] Kac table: 
\begin{align}
    \begin{tabular}{|c|} \hline
$\bf (1,2)$  \\ \hline
$\bf (1,1)$   \\ \hline
    \end{tabular}
\end{align}
\item [II. ] The coefficients (as the solution of string equation): 
\begin{align}
&\bar v_2(t)= \frac{3}{2}\bar u_2(t), \qquad \bar v_3(t) = \frac{3}{4} \bar u_2'(t),\qquad \bar \beta_{2,3} = \frac{16}{3},\qquad \bar \zeta = \frac{1}{2} \zeta \nn\\
&\quad \Leftrightarrow \qquad v_2(t) = 3u_2(t),\qquad v_3(t) = \frac{3}{2} u_2'(t),\qquad \beta_{2,3} = \frac{2}{3}.  \label{Eqn:Appendix:CoefficientsPQ23}
\end{align}
\item [III. ] The Baker-Akhiezer systems: 
\begin{align}
& \left\{
\begin{array}{c}
\ds \bar \zeta  \, \psi = \Bigl[ \del^2 + \bar u_2(t) \Bigr] \psi, \qquad  g \frac{\del}{\del \bar \zeta} \, \psi = \bar \beta_{2,3} \Bigl[\del^3 + \frac{3}{2} \bar u_2(t)\, \del + \frac{3}{4} \bar u_2'(t) \Bigr]\, \psi \cr
\ds \bar \eta \, \chi = \Bigl[\del^3 + \frac{3}{2} \bar u_2(t)\, \del + \frac{3}{4} \bar u_2'(t) \Bigr]\, \chi, \qquad 
g \frac{\del}{\del \bar \eta} \, \chi = \bar \beta_{3,2}\Bigl[ \del^2 + \bar u_2(t) \Bigr] \chi
\end{array}
\right. \\
\Leftrightarrow  & \left\{
\begin{array}{c}
\ds \zeta  \, \psi = \Bigl[ 2 \del^2 +  u_2(t) \Bigr] \psi, \qquad  g \frac{\del}{\del \zeta} \, \psi =  \beta_{2,3} \Bigl[4 \del^3 + 3 u_2(t)\, \del + \frac{3}{2} u_2'(t) \Bigr]\, \psi \cr
\ds  \eta \, \chi = \Bigl[4 \del^3 + 3 u_2(t)\, \del + \frac{3}{2} u_2'(t) \Bigr]\, \chi, \qquad 
g \frac{\del}{\del \eta} \, \chi = \beta_{3,2}\Bigl[2  \del^2 + u_2(t) \Bigr] \chi
\end{array}
\right.
\label{Eqn:Appendix:BAsystemsPQ23}
\end{align}
\item [IV. ] String equation: 
\begin{align}
2 u_2'' + 3(u_2^2+t)= 0 \qquad \Leftrightarrow\qquad \bar u_2'' + 3 \bar u_2^2 + \frac{3}{4} t =0. \label{Eqn:Appendix:StringEquationPQ23}
\end{align}
\end{itemize}

\subsubsection{1) $(r,s) =(1,1)$ FZZT-Cardy brane: $W^{(1,1)} = \mathcal L \widetilde W^{(2)}[\chi]$}

\begin{align}
\vec{W}^{(1,1)} & = \Bigl( \mathcal L \widetilde W^{(2)}_\varnothing, \mathcal L \widetilde W^{(2)}_{\tiny \yng(1)} \Bigr)^{\rm T} \\
\mathcal B^{(1,1)} &= 
\bordermatrix{
 & \varnothing & {\tiny \yng(1)} \cr
\varnothing &  0 & 1  \cr
{\tiny \yng(1)} &  \frac{\zeta}{2}-\frac{u_2}{2}  &  0 
} 
\\
\mathcal Q^{(1,1)} &= \beta_{3,2} \times
\bordermatrix{
 & \varnothing & {\tiny \yng(1)}  \cr
\varnothing &   \frac{ u_2'}{2}  & - 2 \zeta - u_2  \cr
{\tiny \yng(1)} &   - \zeta^2 + \frac{u_2}{2} \zeta  +\frac{ u_2^2+ u_2''}{2}  & -\frac{u_2'}{2} 
}
\end{align}

\subsubsection{2) $(r,s) =(1,2)$ FZZT-Cardy brane: $W^{(1,2)} = \mathcal L \widetilde W^{(1)}[\chi]$}

\begin{align}
\vec{W}^{(1,2)} &= \Bigl( \mathcal L \widetilde W^{(1)}_\varnothing, \mathcal L \widetilde W^{(1)}_{\tiny \yng(1)} \Bigr)^{\rm T}   \\
\mathcal B^{(1,2)} &= 
\bordermatrix{
 & \varnothing & {\tiny \yng(1)} \cr
\varnothing & 0  & 1  \cr
{\tiny \yng(1)} & - \frac{\zeta}{2}-\frac{u_2}{2} &  0 
} 
\\
\mathcal Q^{(1,2)} &= \beta_{3,2} \times
\bordermatrix{
 & \varnothing & {\tiny \yng(1)}  \cr
\varnothing & -  \frac{ u_2'}{2}  & - 2 \zeta  + u_2 \cr
{\tiny \yng(1)} & \zeta^2 + \frac{u_2}{2} \zeta  - \frac{ u_2^2+ u_2''}{2}  & \frac{u_2'}{2} 
}
\end{align}

\subsubsection{Dualities of $(p,q)=(2,3)$}

\noindent\underline{Reflection:} \qquad $\vec W^{(1,2)}[\chi_{\mathcal C}](t; -\zeta) = {\mathsf N}^{(1)}(t)\times  \vec W^{(1,1)}[\chi](t;\zeta)$
\begin{align}
{\mathsf N}^{(1)}(t) = \bordermatrix{
 & \varnothing & {\tiny \yng(1)} \cr
\varnothing & 1  & 0  \cr
{\tiny \yng(1)} & 0  &  1
} . 
\end{align}
\noindent\underline{Dual charge conjugation:} \qquad $\Bigl[\vec W^{(1,2)}[\chi](t;-\zeta)\Bigr]^{\rm T}\,\, \mathfrak D^{(1,1)}_{[\chi]}(t;\zeta)\, \, \Bigl[ \vec W^{(1,1)}[\chi](t;\zeta) \Bigr] $
\begin{align}
\mathfrak D^{(1,1)}_{[\chi]}(t;\zeta) = 
\bordermatrix{
 & \varnothing & {\tiny \yng(1)} \cr
\varnothing & 0  & 1  \cr
{\tiny \yng(1)} & - 1  &  0
} 
\end{align}

\subsection{The case of $(p,q)=(3,2)$: Dual pure-gravity}  

\begin{itemize}
\item [I. ] Kac table: 
\begin{align}
    \begin{tabular}{|c|c|} \hline
$\bf (1,1)$ & $ \bf (2,1)$  \\ \hline
    \end{tabular}
\end{align}
\item [II. ] The coefficients (as the solution of string equation): Eq.~\eq{Eqn:Appendix:CoefficientsPQ23}. 
\item [III. ] The Baker-Akhiezer systems: Eq.~\eq{Eqn:Appendix:BAsystemsPQ23}. 
\item [IV. ] String equation: Eq.~\eq{Eqn:Appendix:StringEquationPQ23}. 
\end{itemize}

\subsubsection{1) $(r,s)=(1,1)$ FZZT-Cardy brane: $\widetilde W^{(1,1)} = \widetilde W^{(1)}[\chi]$} 

\begin{align}
\vec{W}^{(1)} &= \Bigl( \widetilde W^{(1)}_\varnothing, \widetilde W^{(1)}_{\tiny \yng(1)}, \widetilde W^{(1)}_{\tiny \yng(2)} \Bigr)^{\rm T}  \\
\mathcal B^{(1,1)} &= 
 \bordermatrix{
 & \varnothing & {\tiny \yng(1)} & {\tiny \yng(2)} \cr
\varnothing &  0 & 1 & 0 \cr
{\tiny \yng(1)} & 0 & 0 & 1 \cr
{\tiny \yng(2)} &\bar \eta-  \frac{3}{4}\bar u_2' & - \frac{3}{2}\bar u_2 & 0 
} 
\\
\mathcal Q^{(1,1)} &= \bar \beta_{3,2} \times
\bordermatrix{
 & \varnothing & {\tiny \yng(1)} & {\tiny \yng(2)} \cr
\varnothing &  \bar u_2 & 0 & 1 \cr
{\tiny \yng(1)} & \frac{\bar u'_2}{4} + \bar \eta & -\frac{\bar u_2}{2} & 0 \cr
{\tiny \yng(2)} & \frac{\bar u''_2}{4} & - \frac{\bar u_2'}{4} + \bar \eta & - \frac{\bar u_2}{2}
}
\end{align}

\subsubsection{2) $(r,s)=(2,1)$ FZZT-Cardy brane: $\widetilde W^{(1,2)} = \widetilde W^{(2)}[\chi]$} 

\begin{align}
\vec{W}^{(2)} & = \Bigl(\widetilde W^{(2)}_\varnothing,\widetilde W^{(2)}_{\tiny \yng(1)}, \widetilde W^{(2)}_{\tiny \yng(1,1)} \Bigr)^{\rm T} \\
\mathcal B^{(2,1)} &= 
 \bordermatrix{
 & \varnothing & {\tiny \yng(1)} & {\tiny \yng(1,1)}  \cr
\varnothing &  0 & 1 &0  \cr
{\tiny \yng(1)} & -\frac{3}{2}\bar{u}_2 &0  & 1 \cr
{\tiny \yng(1,1)} &-\bar \eta + \frac{3}{4}\bar{u_2'} & 0  & 0 
} 
\\
\mathcal Q^{(2,1)} &= \bar \beta_{3,2} \times
\bordermatrix{
 & \varnothing &{\tiny \yng(1)} & {\tiny \yng(1,1)} \cr
\varnothing & \frac{\bar u_2}{2} & 0 & -1 \cr
{\tiny \yng(1)} & -\frac{\bar u'_2}{4} + \bar \eta & \frac{\bar u_2}{2} & 0 \cr
{\tiny \yng(1,1)} & -\frac{\bar u''_2}{4} &  \frac{\bar u_2'}{4} + \bar \eta & - \bar u_2 
}
\end{align}

\subsubsection{Duality matrices}

\noindent\underline{Reflection:} \qquad $\vec W^{(2)}[\chi_{\mathcal C}](t; -\bar \eta) = {\mathsf M}^{(1)}(t)\times  \vec W^{(1)}[\chi](t;\bar \eta)$
\begin{align}
{\mathsf M}^{(1)}(t) = 
\bordermatrix{
 & \varnothing & {\tiny \yng(1)} & {\tiny \yng(2)} \cr
\varnothing & 1 & 0 & 0  \cr
{\tiny \yng(1)} &  0 & 1 & 0 \cr
{\tiny \yng(1,1)} &   \frac{3 \bar u_2(t)}{2} & 0 & 1
} 
\end{align}
\noindent\underline{Charge conjugation:}  \qquad $\Bigl[\vec W^{(2)}[\chi](t; \bar \eta)\Bigr]^{\rm T}\,\, \mathfrak C^{(1,1)}_{[\chi]}(t)\, \, \Bigl[ \vec W^{(1)}[\chi](t;\bar \eta) \Bigr] $
\begin{align}
\mathfrak C^{(1,1)}_{[\chi]}(t) = 
\bordermatrix{
 & \varnothing & {\tiny \yng(1)} & {\tiny \yng(2)} \cr
\varnothing & 0 & 0 & 1  \cr
{\tiny \yng(1)} &  0 & -1 & 0 \cr
{\tiny \yng(1,1)} &   1 & 0 & 0
} 
\end{align}


\subsection{The case of $(p,q)=(2,5)$: Yang-Lee edge}

\begin{itemize}
\item [I. ] Kac table: 
\begin{align}
    \begin{tabular}{|c|c|} \hline
$(1,4)$ \\ \hline
$\bf (1,3)$   \\ \hline
$\bf (1,2)$  \\ \hline
$(1,1)$   \\ \hline
    \end{tabular}
\end{align}
\item [II. ] The coefficients (with conformal background): 
\begin{align}
&v_2(t)= 20 u_2(t), \qquad v_3(t) = 30  u_2'(t), \nn\\
& v_4(t)= \frac{5 \left(10 u_2''(t) + 3 u_2^2(t) - \mu \right) }{2}, 
\qquad v_5(t) = \frac{15}{2} \left(u_2(t) u_2'(t)+u_2^{(3)}(t)\right).  \label{Eqn:Appendix:CoefficientsStringEqnPQ25}
\end{align}
\item [III. ] The Baker-Akhiezer systems: 
\begin{align}
\left\{
\begin{array}{rl}
\ds \zeta \, \psi \!\!\!&= \ds  \Bigl[
2 \del^2 + u_2  \Bigr] \psi, \cr
\ds g \frac{\del}{\del \zeta}\,  \psi \!\!\!& \ds = \beta_{2,5}\times \Bigl[16 \del^5+20 u_2 \del^3  +30  u_2' \del^2 + \cr
& \ds \qquad \qquad  + \frac{5 \left(10 u_2'' + 3 u_2^2 - \mu \right) }{2}  \del +\frac{15}{2} \left(u_2 u_2'+u_2^{(3)}\right) \Bigr] \psi, \cr
\ds \eta \, \chi \!\!\!& \ds =\Bigl[16 \del^5+20 u_2 \del^3  +30  u_2' \del^2 + \cr
& \ds \qquad \qquad  + \frac{5 \left(10 u_2'' + 3 u_2^2 - \mu \right) }{2}  \del +\frac{15}{2} \left(u_2 u_2'+u_2^{(3)}\right) \Bigr] \chi, \cr
\ds g \frac{\del}{\del \eta}\,  \chi \!\!\!&= \ds \beta_{5,2} \times   \Bigl[
2 \del^2 + u_2  \Bigr]\chi. 
\end{array}
\right. 
\end{align}
\item [IV. ] String equations (with conformal background):  
\begin{align}
&\frac{2\, g }{\beta_{2,5}}  =  5 u_2'(t) \left(\mu -4 u_2''(t)-3 u_2(t){}^2\right)-2 \left(5 u_2(t) u_2{}^{(3)}(t)+u_2{}^{(5)}(t)\right), \nn\\
\Bigl(\Leftrightarrow&\quad  \frac{2 t}{\beta_{2,5} } = 5 \mu  u_2(t) -5 u_2'(t){}^2-10 u_2(t) u_2''(t)-5 u_2(t){}^3-2 u_2{}^{(4)}(t) \Bigr).  \label{Eqn:Appendix:StringEquationPQ25}
\end{align}
\end{itemize}

\subsubsection{2) $(r,s)=(1,2)$ FZZT-Cardy brane: $W^{(1,2)} = \mathcal L \widetilde W^{(3)}[\chi]$ }

\begin{align}
\vec{W}^{(1,2)} &= \Bigl( \mathcal L \widetilde W^{(3)}_\varnothing, \mathcal L \widetilde W^{(3)}_{\tiny \yng(1)}, \mathcal L \widetilde W^{(3)}_{\tiny \yng(1,1)}, \mathcal L \widetilde W^{(3)}_{\tiny \yng(2,1)} \Bigr)^{\rm T} \\
\mathcal B^{(1,2)}(t;\zeta)  &= 
\bordermatrix{
 & \varnothing & {\tiny \yng(1)} & {\tiny \yng(1,1)} & {\tiny \yng(2,1)} \cr
\varnothing &   0 & 1 & 0 & 0 \cr
 {\tiny \yng(1)} &    - \frac{\zeta }{2}-\frac{3 u_2}{2} & 0 & 2 & 0  \cr
 {\tiny \yng(1,1)} &   -\frac{3u_2'  }{8} & \frac{\zeta }{2} + \frac{u_2}{4} & 0 & 1 \cr
{\tiny \yng(2,1)} &  -\frac{\zeta^2}{4} - \frac{7 u_2}{8}  \zeta +\frac{9 u_2^2-5 \mu +20 u_2''}{16} & \frac{7 u_2'}{8}  & \frac{3 \zeta }{2}-\frac{u_2}{2} & 0
}, \\
\mathcal Q^{(1,2)}(t;\zeta)  &=\beta_{5,2} \times    \nn\\
 &  \!\!\!\!\!\!\!\! \!\!\!\!\!\!\!\! \!\!\!\! \!\!\! 
\times \biggl[ \mbox{ \scriptsize $\bordermatrix{
 & \mbox{\normalsize $\varnothing$} & {\tiny \yng(1)} & {\tiny \yng(1,1)} & {\tiny \yng(2,1)} \cr
\mbox{\normalsize $\varnothing$}
&    - 8   u_2' \zeta  +  \frac{u_2^{(3)} -3 u_2 u_2'}{2} 
&   0  
& -2 u_2' 
&  8 \zeta  + 4 u_2   \cr
 {\tiny \yng(1)} 
&   - 2 \zeta ^3
-8 u_2(t) \zeta ^2
- \widetilde H_3 \zeta 
+\widetilde F_4 
&  - 2  u_2'  \zeta +\frac{3 u_2 u_2' +u_2^{(3)}}{2} 
&   12 \zeta ^2
+2 u_2 \zeta 
- \widetilde H_5 
&  2 u_2'  \cr
 {\tiny \yng(1,1)} 
&     -8 u_2'  \zeta ^2
- \widetilde H_2 \zeta 
+\widetilde F_3 
&   2 \zeta ^3
-2 u_2 \zeta ^2
- \widetilde H_4 \zeta 
+ \widetilde F_6 
&   \frac{-5 u_2 u_2' -u_2^{(3)}}{2} 
&  8 \zeta ^2 + 8 u_2 \zeta +2 u_2^2  \cr
{\tiny \yng(2,1)} 
&     -2 \zeta ^4
-\frac{13 u_2 }{2} \zeta ^3
+ \widetilde H_1 \zeta ^2
-
\widetilde F_1
 \zeta 
+\widetilde F_2 
&  -2 u_2' \zeta ^2
+ \frac{u_2 u_2'}{2}  \zeta 
+\widetilde F_5 
&  10 \zeta ^3
- \frac{5 \mu  }{2} \zeta
+\widetilde F_7 
&  10  u_2' \zeta 
+\frac{5 u_2 u_2'- u_2^{(3)}}{2}   \cr
}$} \biggr], 
\end{align}
where the coefficients are given as 
\begin{align}
\widetilde F_1 &\equiv 
\frac{-5 u_2 \left(7 \mu -2 u_2'' \right)+160 (u_2')^2+11 u_2^3 }{16} + \frac{t}{2 \beta_{2,5} }, \\
\widetilde F_2 &\equiv
\frac{-5 u_2^2 \left(\mu -u_2'' \right)-5 \mu  u_2''+16 u_2^{(3)} u_2'+2 (u_2'')^2+3 u_2^4 }{16}  + \frac{3 t u_2}{8 \beta_{2,5} }, \\ 
\widetilde F_3 &\equiv 
\frac{u_2' \left(-5 \mu +2 u_2''+3 u_2^2\right)+8 u_2 u_2^{(3)} }{16} -\frac{g}{2 \beta _{2,5}}, \\
\widetilde F_4 &\equiv 
-2 (u_2')^2+u_2  u_2'' +u_2^3 -\frac{t}{2 \beta _{2,5}}, \qquad
\widetilde F_5 \equiv 
\frac{ u_2'  \left(5 \mu -2 u_2'' -3 u_2^2\right)}{16} -\frac{g}{2 \beta _{2,5}}, \\
\widetilde F_6 &\equiv 
 -\frac{u_2 \left(-5 \mu +2 u_2'' +3 u_2^2\right)}{8}  -\frac{t}{2 \beta _{2,5}}, \qquad
\widetilde F_7  \equiv 
-2 (u_2')^2 -\frac{t}{2 \beta _{2,5}}, \\
\widetilde H_1 &\equiv 
\frac{11 u_2^2-5 \mu +10 u_2''}{8}, \qquad 
\widetilde H_2 \equiv 
\frac{11 u_2 u_2' - 2 u_2^{(3)}}{2}, \qquad 
\widetilde H_3 \equiv 
\frac{- 2 u_2^2+5 \mu  -4 u_2''}{2}, \\
 \widetilde H_4 &\equiv 
\frac{-u_2^2+5 \mu +2 u_2''}{4}, \qquad
 \widetilde H_5 \equiv 
 2 \left(u_2^2+u_2''\right). 
\end{align}
Note that $\widetilde F$ are simplified by the string equations. 

\subsubsection{3) $(r,s)=(1,3)$ FZZT-Cardy brane: $W^{(1,3)} = \mathcal L \widetilde W^{(2)}[\chi]$ \label{Subsubsection:ExampleOfIMS:PQeq25_RSeq13IMS} }

\begin{align}
\vec{W}^{(1,3)} &= \Bigl( \mathcal L \widetilde W^{(2)}_\varnothing, \mathcal L \widetilde W^{(2)}_{\tiny \yng(1)}, \mathcal L \widetilde W^{(2)}_{\tiny \yng(2)}, \mathcal L \widetilde W^{(2)}_{\tiny \yng(2,1)} \Bigr)^{\rm T} \\
\mathcal B^{(1,3)}(t;\zeta)  &= 
\bordermatrix{
 & \varnothing & {\tiny \yng(1)} & {\tiny \yng(2)} & {\tiny \yng(2,1)} \cr
\varnothing &   0 & 1 & 0 & 0 \cr
 {\tiny \yng(1)} &  \frac{\zeta }{2} + u_2 & 0 & 2 & 0 \cr
 {\tiny \yng(2)} & -u_2' & -\frac{\zeta }{2}-u_2 & 0 & 1 &  \cr
{\tiny \yng(2,1)} &  -\frac{\zeta ^2}{4} - u_2 \zeta + \frac{10 u_2''-u_2^2-5 \mu }{16}  & \frac{1}{4} u_2' & -\frac{3 \zeta }{2}-\frac{u_2}{2} & 0
}, \\
\mathcal Q^{(1,3)}(t;\zeta)  &=\beta_{5,2} \times    \nn\\
 &  \!\!\!\!\!\!\!\! \!\!\!\!\!\!\!\! \!\!\!\!\!\!\!\! 
\times \biggl[ \,\mbox{\scriptsize $\bordermatrix{
 & \mbox{\normalsize $\varnothing$}  & {\tiny \yng(1)} & {\tiny \yng(2)} & {\tiny \yng(2,1)} \cr
 \mbox{\normalsize $\varnothing$}
&   -3 \zeta  u_2' +\frac{3}{2} u_2  u_2' - \frac{1}{2} u_2^{(3)} 
&  0 
&  2 u_2'  
&  8 \zeta -4 u_2   \cr
 {\tiny \yng(1)} 
&   -2 \zeta ^3-7 u_2 \zeta ^2 - H_3 \zeta +F_4 
&  -2 \zeta  u_2' - \frac{3u_2 u_2' + u_2^{(3)}}{2}   
&  -12 \zeta ^2 + 2 u_2 \zeta + H_5 
&  -2 u_2'  \cr
 {\tiny \yng(2)} 
&   3 u_2' \zeta ^2 + H_2  \zeta + F_3  
&  2 \zeta ^3+2 u_2 \zeta ^2 + H_4  \zeta +F_6 
&  \frac{1}{2} u_2^{(3)} 
&   -8 \zeta ^2 - 2 u_2 \zeta +3 u_2^2  \cr
{\tiny \yng(2,1)} 
&    2 \zeta ^4 + 6 u_2 \zeta ^3 - H_1  \zeta ^2 - F_1 \zeta + F_2 
&  2 u_2' \zeta ^2 + \frac{1}{2} u_2 u_2' \zeta +F_5 
&  10 \zeta ^3 - \frac{5 \mu  \zeta }{2}+F_7 
&   5  u_2' \, \zeta  +  \frac{u_2^{(3)}}{2}  \cr
}$} \biggr], 
\end{align}
where the coefficients are given as 
\begin{align}
F_1 &= \frac{5 u_2 \left(3 \mu +2 u_2''\right)+30 (u_2')^2+11 u_2^3 }{16} + \frac{t}{2 \beta _{2,5}}, \\
F_2 &= \frac{ 5 u_2^2 \left(\mu -u_2''\right)+5 \mu  u_2''-6 u_2^{(3)} u_2' -2 (u_2'')^2-3 u_2^4 }{16}  + \frac{t u_2(t)}{4 \beta _{2,5}}, \\
F_3 &= \frac{ u_2' \left(5 \mu -2 u_2'' -3 u_2^2\right)+12 u_2 u_2^{(3)} }{16} + \frac{g}{2 \beta _{2,5}}, \\
F_4 &= \frac{3 \bigl((u_2')^2+2 u_2 u_2'' +2 u_2^3\bigr)}{4}  + \frac{t}{2 \beta _{2,5}}, \qquad
F_5 = \frac{u_2'  \left(-5 \mu +2 u_2'' +3 u_2^2\right) }{16} + \frac{g}{2 \beta _{2,5}}, \\ 
F_6 &= \frac{u_2 \left(-5 \mu +2 u_2'' +3 u_2^2\right) }{8} + \frac{t}{2 \beta _{2,5}}, \qquad
F_7 = \frac{3 (u_2')^2}{4} + \frac{t}{2 \beta _{2,5}}, \\
H_1 &= \frac{ 11 u_2^2 +10 u_2'' -5 \mu }{8}, \qquad
H_2 = \frac{ 13 u_2 u_2'+4 u_2^{(3)} }{4}, \qquad
H_3 = \frac{5 \mu -4 u_2''-7 u_2^2 }{2}, \\
H_4 &= \frac{u_2^2 - 2 u_2'' - 5 \mu }{4}, \qquad
H_5 = 2 \left(u_2^2+u_2'' \right), 
\end{align}
Note that $F$ are simplified by the string equations.

\subsubsection{Duality matrices}

\noindent\underline{Reflection:} \qquad $\vec W^{(1,3)}[\chi_{\mathcal C}](t; -\zeta) = {\mathsf N}^{(2)}(t)\times  \vec W^{(1,2)}[\chi](t;\zeta)$
\begin{align}
,\qquad {\mathsf N}^{(2)}(t) = \bordermatrix{
 & \varnothing & {\tiny \yng(1)} &   {\tiny \yng(1,1)} &  {\tiny \yng(2,1)}\cr
\varnothing &  1 & 0 & 0 & 0  \cr
{\tiny \yng(1)} &  0 & 1 & 0 & 0 \cr
{\tiny \yng(2)} &  -\frac{ 5 u_2}{4} & 0 & 1 & 0 \cr
 {\tiny \yng(2,1)} & - \frac{5 u_2'}{8}  & 0 & 0 & 1 
} . 
\end{align}
\noindent\underline{Dual charge conjugation:} \qquad $\Bigl[\vec W^{(1,3)}[\chi](t;-\zeta)\Bigr]^{\rm T}\,\, \mathfrak D^{(1,2)}_{[\chi]}(t,\zeta)\, \, \Bigl[ \vec W^{(1,2)}[\chi](t;\zeta) \Bigr] $
\begin{align}
{\mathfrak D}^{(1,2)}_{[\chi]}(t,\zeta) = 
\bordermatrix{
 & \varnothing & {\tiny \yng(1)} &   {\tiny \yng(1,1)} &  {\tiny \yng(2,1)}\cr
\varnothing &  - \frac{5 u_2'}{8}  & - \zeta +  \frac{ 3 u_2}{4}  & 0 & 1\cr
{\tiny \yng(1)} &  \zeta + \frac{u_2}{2} & 0 & -1 & 0  \cr
{\tiny \yng(2)} &   0 & 1 & 0 & 0 \cr
 {\tiny \yng(2,1)} &  -1 & 0 & 0 & 0
}. 
\end{align}

\subsection{The case of $(p,q)=(5,2)$: Dual Yang-Lee edge}

\begin{itemize}
\item [I. ] Kac table: 
\begin{align}
 \begin{tabular}{|c|c|c|c|c|} \hline
 $ \bf (1,1)$ & $ \bf (2,1)$ & $ \bf (3,1)$ & $ \bf (4,1)$   \\ \hline
    \end{tabular}
\end{align}
\item [II. ] The coefficients: Eq.~\eq{Eqn:Appendix:CoefficientsStringEqnPQ25} 
\item [III. ] The Baker-Akhiezer systems: 
\begin{align}
\left\{
\begin{array}{rl}
\ds \bar \zeta \, \psi \!\!\!&= \ds  \Bigl[
\del^2 + \bar u_2  \Bigr] \psi, \cr
\ds g \frac{\del}{\del \bar \zeta}\,  \psi \!\!\!& \ds = \bar \beta_{2,5}\times \Bigl[\del^5 +\frac{5 \bar{u}_2}{2} \del^3  +\frac{15 \bar{u}_2'}{4} \del^2 + \cr
& \ds \qquad \qquad  +  \frac{5  \left(20 \bar{u}_2'' +12 \bar{u}_2^2-\mu \right)}{32}  \del +\frac{15}{16} \left(2 \bar{u}_2 \bar{u}_2'+\bar{u}_2^{(3)}\right) \Bigr] \psi, \cr
\ds \bar \eta \, \chi \!\!\!& \ds =\Bigl[\del^5 +\frac{5 \bar{u}_2}{2} \del^3  +\frac{15 \bar{u}_2'}{4} \del^2 + \cr
& \ds \qquad \qquad  +  \frac{5  \left(20 \bar{u}_2'' +12 \bar{u}_2^2-\mu \right)}{32}  \del +\frac{15}{16} \left(2 \bar{u}_2 \bar{u}_2'+\bar{u}_2^{(3)}\right) \Bigr] \chi, \cr
\ds g \frac{\del}{\del \bar \eta}\,  \chi \!\!\!&= \ds \bar \beta_{5,2} \times   \Bigl[
\del^2 + \bar u_2  \Bigr]\chi. 
\end{array}
\right. 
\end{align}
\item [IV. ] String equations: Eq.~\eq{Eqn:Appendix:StringEquationPQ25}
\end{itemize}

\subsubsection{1) $(r,s)=(1,1)$ FZZT-Cardy brane: $\widetilde W^{(1,1)}=\widetilde W^{(1)}[\chi]$}

\begin{align}
\vec{W}^{(1,1)} &= \Bigl( \widetilde W^{(1)}_\varnothing, \widetilde W^{(1)}_{\tiny \yng(1)}, \widetilde W^{(1)}_{\tiny \yng(2)}, \widetilde W^{(1)}_{\tiny \yng(3)}, \widetilde W^{(1)}_{\tiny \yng(4)} \Bigr)^{\rm T}, \\
\mathcal B^{(1,1)} &= 
\bordermatrix{
 & \varnothing & {\tiny \yng(1)} & {\tiny \yng(2)} & {\tiny \yng(3)} & {\tiny \yng(4)}\cr
\varnothing &  & 1 & \cr
{\tiny \yng(1)} & & & 1 \cr
{\tiny \yng(2)} &  & & & 1 \cr 
{\tiny \yng(3)} &  & & & & 1 \cr 
{\tiny \yng(4)} &\bar \eta- w_5 & - w_4 &  - w_3 &  - w_2 &\cr
}, \\
\mathcal Q^{(1,1)} &= \bar \beta_{5,2}\times   \nn\\
\times \biggl[&\bordermatrix{
 & \varnothing & {\tiny \yng(1)} & {\tiny \yng(2)} & {\tiny \yng(3)} & {\tiny \yng(4)}\cr
\varnothing &    \bar{u}_2 & 0 & 1 & 0 & 0 \cr
{\tiny \yng(1)} &  \bar{u}_2' & \bar{u}_2 & 0 & 1 & 0 \cr
{\tiny \yng(2)} &  \bar{u}_2'' & 2 \bar{u}_2' & \bar{u}_2 & 0 & 1 \cr
{\tiny \yng(3)} & \bar \eta + \frac{\bar{u}_2'''-30 \bar{u}_2 \bar{u}_2'}{16}  & \frac{-60 \bar{u}_2^2+5 \mu -4 \bar{u}_2''}{32}  & - \frac{3 \bar{u}_2'}{4} & -\frac{3 \bar{u}_2}{2}  & 0 \cr
{\tiny \yng(4)} &  
 \frac{-30 (\bar{u}_2')^2-30 \bar{u}_2 \bar{u}_2''+\bar{u}_2''''}{16}  & \bar \eta -  \frac{90 \bar{u}_2 \bar{u}_2'+\bar{u}_2'''}{16}  & \frac{-60 \bar{u}_2^2+5 \mu -28 \bar{u}_2''}{32} & - \frac{9 \bar{u}_2'}{4}  & -\frac{3 \bar{u}_2}{2}  
}\biggr],
\end{align}
where the coefficients $\{w_n\}_{n=2}^5$ are given by 
\begin{align}
w_2 = \frac{5 \bar{u}_2}{2} , \quad 
w_3 = \frac{15 \bar{u}_2'}{4}, \quad 
w_4 = \frac{5 }{32} \left(20 \bar{u}_2'' +12 \bar{u}_2^2-\mu \right) , \quad
w_5 = \frac{15}{16} \left(2 \bar{u}_2  \bar{u}_2' +\bar{u}_2^{(3)} \right). 
\label{Eqn:Appendix:PQ52RS12CoefficientsWWW}
\end{align}

\subsubsection{2) $(r,s)=(2,1)$ FZZT-Cardy brane: $\widetilde W^{(2,1)} =\widetilde  W^{(2)}[\chi]$}

\begin{align}
\vec{W}^{(2,1)} &= \Bigl( \widetilde W^{(2)}_\varnothing, \widetilde W^{(2)}_{\tiny \yng(1)}, \widetilde W^{(2)}_{\tiny \yng(2)}, \widetilde W^{(2)}_{\tiny \yng(1,1)}, \widetilde W^{(2)}_{\tiny \yng(3)}, \widetilde W^{(2)}_{\tiny \yng(2,1)}, \widetilde W^{(2)}_{\tiny \yng(3,1)}, \widetilde W^{(2)}_{\tiny \yng(2,2)}, \widetilde W^{(2)}_{\tiny \yng(3,2)}, \widetilde W^{(2)}_{\tiny \yng(3,3)} \Bigr)^{\rm T}, \\
\mathcal B^{(2,1)} &= 
\mbox{\small $\bordermatrix{
 & \mbox{\normalsize $\varnothing$} & {\tiny \yng(1)} & {\tiny \yng(2)} & {\tiny \yng(1,1)} & {\tiny \yng(3)} & {\tiny \yng(2,1)} & {\tiny \yng(3,1)} & {\tiny \yng(2,2)} & {\tiny \yng(3,2)} & {\tiny \yng(3,3)} \cr
\mbox{\normalsize $\varnothing$} &  & 1 & \cr
{\tiny \yng(1)} & & & 1 &1  \cr
{\tiny \yng(2)} &  & & & & 1 & 1 \cr 
{\tiny \yng(1,1)} &  & & & & & 1 \cr 
{\tiny \yng(3)} &- w_4 & -w_3 &  -w_2 & &  & & 1\cr
{\tiny \yng(2,1)} &  & & & & & & 1 &1 \cr 
{\tiny \yng(3,1)} & w_5 - \bar \eta & & & -w_3& & -w_2& & &1 \cr 
{\tiny \yng(2,2)} &  & & & & & & & & 1 \cr 
{\tiny \yng(3,2)} & & w_5 - \bar \eta & & w_4 & & & & -w_2 & & 1 \cr 
{\tiny \yng(3,3)} &  & & w_5 - \bar \eta & & & w_4 & & w_3
}$}, \\
\mathcal Q^{(2,1)} &=  \bar \beta_{5,2} \times  \nn\\
& 
\times\Bigl[\, \mbox{\scriptsize $\bordermatrix{
 & \mbox{\normalsize $\varnothing$} & {\tiny \yng(1)} & {\tiny \yng(2)} & {\tiny \yng(1,1)} & {\tiny \yng(3)} & {\tiny \yng(2,1)} & {\tiny \yng(3,1)} & {\tiny \yng(2,2)} & {\tiny \yng(3,2)} & {\tiny \yng(3,3)} \cr
\mbox{\normalsize $\varnothing$} &   2 \bar{u}_2  &   0  &  1  &  -1  &  0  &  0  &  0  &  0  &  0  &  0  \cr
{\tiny \yng(1)} &   2 \bar{u}_2'  &  2 \bar{u}_2  &   0  &  0  &  1  &  0  &  0  &  0  &  0  &  0  \cr 
{\tiny \yng(2)} &    f_1^{(1)}  &  -\frac{3 \bar{u}_2'}{4}  &   -\frac{\bar{u}_2}{2} &  0  &  0  &  0  &  0  &   1 &  0  &  0  \cr 
{\tiny \yng(1,1)} &  -\bar{u}_2''  &   \bar{u}_2'  &   0  &   2 \bar{u}_2  &  0  &  0  &  1  &  -1 &  0  &  0  \cr 
{\tiny \yng(3)} &  \bar \eta - f_2^{(1)}   &   f_3^{(1)}  &   \frac{-9 \bar{u}_2'}{4}  &  0  &    -\frac{\bar{u}_2}{2}   &  0  &  0  &  0  & 1 &  0  \cr 
{\tiny \yng(2,1)} &    f_4^{(1)}  - \bar \eta   &  0  &   \bar{u}_2'  &   \frac{-3 \bar{u}_2'}{4}  &  0  &   -\frac{\bar{u}_2}{2}  &  0  &  0  &  0  &  0  \cr 
{\tiny \yng(3,1)} &   f_5^{(1)}  &  0   &    0  &   f_3^{(1)}  &    \bar{u}_2'  &   \frac{-9 \bar{u}_2'}{4}   &    -\frac{\bar{u}_2}{2}   &  0  &  0  &  1 \cr 
{\tiny \yng(2,2)} &   0  &   f_4^{(1)} - \bar \eta   &   \bar{u}_2''  &   -  f_1^{(1)}   &  0  &    2 \bar{u}_2'   &  0  &    -\frac{\bar{u}_2}{2}    &   0   &   -1  \cr 
	{\tiny \yng(3,2)} &   0   &  f_5^{(1)}   &  0  &   f_{2}^{(1)} - \bar \eta    &   \bar{u}_2''  &   0 &   2 \bar{u}_2'   &   \frac{-9 \bar{u}_2'}{4}  &  -\frac{\bar{u}_2}{2}   &     0   \cr 
{\tiny \yng(3,3)} &    0   &  0  &    f_{5}^{(1)}  &    0  &    \bar \eta - f_{4}^{(1)}  &       f_{2}^{(1)} - \bar \eta  &     f_{1}^{(1)}   &   -f_{3}^{(1)}   &    \frac{-3 \bar{u}_2'}{4}  &   -3 \bar{u}_2  \cr 
}$}
\Bigr], 
\end{align}
where the coefficients $\{w_n\}_{n=2}^5$ are given by Eq.~\eq{Eqn:Appendix:PQ52RS12CoefficientsWWW}, and the coefficients $\{f_n^{(1)}\}_{n=1}^5$ are given by 
\begin{align}
f_1^{(1)} &\equiv\frac{1}{32} \left(-60 \bar{u}_2^2+5 \mu -4 \bar{u}_2''\right), \qquad 
f_2^{(1)} \equiv \frac{1}{16} \left(90 \bar{u}_2 \bar{u}_2'+ \bar{u}_2'''\right), \nn\\
f_3^{(1)} &\equiv\frac{1}{32} \left(-60 \bar{u}_2^2+5 \mu -28 \bar{u}_2''\right), \qquad
f_4^{(1)} \equiv\frac{1}{16} \left(30 \bar{u}_2 \bar{u}_2'-\bar{u}_2'''\right), \nn\\
f_5^{(1)} &\equiv \frac{1}{16} \left(30 (\bar{u}_2')^2+30 \bar{u}_2 \bar{u}_2''-\bar{u}_2''''\right). \label{Eqn:AppendixB:CoefficientsffPQ52RS21}
\end{align}

\subsubsection{3) $(r,s)=(3,1)$ FZZT-Cardy brane: $\widetilde W^{(3,1)}= \widetilde W^{(3)}[\chi]$}

\begin{align}
\vec{W}^{(3,1)} &= \Bigl( \widetilde W^{(3)}_\varnothing, \widetilde W^{(3)}_{\tiny \yng(1)}, \widetilde W^{(3)}_{\tiny \yng(2)}, \widetilde W^{(3)}_{\tiny \yng(1,1)}, \widetilde W^{(3)}_{\tiny \yng(2,1)}, \widetilde W^{(3)}_{\tiny \yng(1,1,1)}, \widetilde W^{(3)}_{\tiny \yng(2,2)}, \widetilde W^{(3)}_{\tiny \yng(2,1,1)}, \widetilde W^{(3)}_{\tiny \yng(2,2,1)}, \widetilde W^{(3)}_{\tiny \yng(2,2,2)} \Bigr)^{\rm T}, \\
\mathcal B^{(3,1)} &= 
\bordermatrix{
 & \varnothing & {\tiny \yng(1)} & {\tiny \yng(2)} & {\tiny \yng(1,1)} & {\tiny \yng(2,1)} & {\tiny \yng(1,1,1)} & {\tiny \yng(2,2)} & {\tiny \yng(2,1,1)} & {\tiny \yng(2,2,1)} & {\tiny \yng(2,2,2)} \cr
\varnothing &  & 1 & \cr
{\tiny \yng(1)} & & & 1 &1  \cr
{\tiny \yng(2)} & - w_3 & -  w_2 & & & 1 &  \cr 
{\tiny \yng(1,1)} &  & & & & 1 & 1 \cr 
{\tiny \yng(2,1)} &  w_4 & &  & - w_2 &  & & 1 & 1\cr
{\tiny \yng(1,1,1)} &  & & & & & &  &1 \cr 
{\tiny \yng(2,2)} &  &  w_4 & & w_3& & & & &1 \cr 
{\tiny \yng(2,1,1)} &  \bar \eta - w_5  & & & & & - w_2 & & & 1 \cr 
{\tiny \yng(2,2,1)} & & \bar \eta - w_5  & &  & & w_3 & & & & 1 \cr 
{\tiny \yng(2,2,2)} &  & &  & \bar \eta - w_5  & & - w_4 & & 
}, \\
\mathcal Q^{(3,1)} &= \bar \beta_{5,2} \times  \nn\\
\times \Bigl[ & \mbox{ \scriptsize  $
\bordermatrix{
 & \varnothing & {\tiny \yng(1)} & {\tiny \yng(2)} & {\tiny \yng(1,1)} & {\tiny \yng(2,1)} & {\tiny \yng(1,1,1)} & {\tiny \yng(2,2)} & {\tiny \yng(2,1,1)} & {\tiny \yng(2,2,1)} & {\tiny \yng(2,2,2)} \cr
\varnothing &   3 \bar{u}_2 & 0 & 1 & -1 & 0 & 0 & 0 & 0 & 0 & 0 \cr
{\tiny \yng(1)} & \frac{-3 \bar{u}_2'}{4}  & \frac{\bar{u}_2}{2}  & 0 & 0 & 0 & -1 & 0 & 0 & 0 & 0 \cr 
{\tiny \yng(2)} & f_{3}^{(1)} & \frac{-9 \bar{u}_2'}{4} & \frac{\bar{u}_2}{2}  & 0 & 0 & 0 & 1 & -1 & 0 & 0 \cr 
{\tiny \yng(1,1)} &  - f_{1}^{(1)} & 2 \bar{u}_2' & 0 & \frac{\bar{u}_2}{2}  & 0 & 0 & -1 & 0 & 0 & 0 \cr 
{\tiny \yng(2,1)} & f_{2}^{(1)}  - \bar \eta & 0 & 2 \bar{u}_2' & \frac{-9 \bar{u}_2'}{4}  & \frac{\bar{u}_2}{2}  & 0 & 0 & 0 & 0 & 0 \cr 
{\tiny \yng(1,1,1)} &  \bar \eta - f_{4}^{(1)} & -\bar{u}_2'' & 0 & \bar{u}_2' & 0 & \frac{\bar{u}_2}{2}  & 0 & 0 & -1 & 0 \cr 
{\tiny \yng(2,2)} &   0 & f_{2}^{(1)} - \bar \eta & f_{1}^{(1)} & - f_{3}^{(1)} & \frac{-3 \bar{u}_2'}{4}  & 0 & -2 \bar{u}_2 & 0 & 0 & 1 \cr 
{\tiny \yng(2,1,1)} &  -f_{5}^{(1)} & 0 & -\bar{u}_2'' & 0 & \bar{u}_2' & \frac{-9 \bar{u}_2'}{4}  & 0 & \frac{\bar{u}_2}{2}  & 0 & -1 \cr 
{\tiny \yng(2,2,1)} &  0 & -f_{5}^{(1)}&  f_{4}^{(1)} - \bar \eta & 0 & 0 & -f_{3}^{(1)} & \bar{u}_2' & \frac{-3 \bar{u}_2'}{4} & -2 \bar{u}_2 & 0 \cr 
{\tiny \yng(2,2,2)} &   0 & 0 & 0 & - f_{5}^{(1)} & f_{4}^{(1)}- \bar \eta  & \bar \eta - f_{2}^{(1)} & \bar{u}_2'' & - f_{1}^{(1)}  & 2 \bar{u}_2' & -2 \bar{u}_2 \cr 
} $}
\Bigr], 
\end{align}
where the coefficients $\{w_n\}_{n=2}^5$ are given by Eq.~\eq{Eqn:Appendix:PQ52RS12CoefficientsWWW}, and the coefficients $\bigl\{ f_n^{(1)} \bigr\}_{n=1}^5$ are given by Eq.~\eq{Eqn:AppendixB:CoefficientsffPQ52RS21}.

\subsubsection{4) $(r,s)=(4,1)$ FZZT-Cardy brane: $\widetilde W^{(4,1)} = \widetilde W^{(4)}[\chi]$}

\begin{align}
\vec{W}^{(4,1)} &= \Bigl( \widetilde W^{(4)}_\varnothing, \widetilde W^{(4)}_{\tiny \yng(1)}, \widetilde W^{(4)}_{\tiny \yng(1,1)}, \widetilde W^{(4)}_{\tiny \yng(1,1,1)}, \widetilde W^{(4)}_{\tiny \yng(1,1,1,1)} \Bigr)^{\rm T}, \\
\mathcal B^{(4,1)} &= 
\bordermatrix{
 & \varnothing & {\tiny \yng(1)} & {\tiny \yng(1,1)} & {\tiny \yng(1,1,1)} & {\tiny \yng(1,1,1,1)}\cr
\varnothing &  & 1 & \cr
{\tiny \yng(1)} & - w_2 & & 1 \cr
{\tiny \yng(1,1)} & w_3 & & & 1 \cr 
{\tiny \yng(1,1,1)} &- w_4  & & & & 1 \cr 
{\tiny \yng(1,1,1,1)} & - \bar \eta + w_5 &  &   &  &\cr
}, \\
\mathcal Q^{(4,1)} &= \bar \beta_{5,2}\times 
\bordermatrix{
 & \varnothing & {\tiny \yng(1)} & {\tiny \yng(1,1)} & {\tiny \yng(1,1,1)} & {\tiny \yng(1,1,1,1)}\cr
\varnothing &   \frac{3\bar{u}_2}{2}  & 0 & -1 & 0 & 0 \cr 
{\tiny \yng(1)} &  \frac{-9 \bar{u}_2'}{4} & \frac{3 \bar{u}_2(t)}{2}  & 0 & -1 & 0 \cr 
{\tiny \yng(1,1)} & \frac{28 \bar{u}_2''+60 \bar{u}_2^2-5 \mu }{32} & \frac{-3 \bar{u}_2'}{4}  & -\bar{u}_2 & 0 & -1 \cr 
{\tiny \yng(1,1,1)} & \bar{\eta } + \frac{-90 \bar{u}_2 \bar{u}_2'-\bar{u}_2'''}{16}  &\frac{4 \bar{u}_2''+60 \bar{u}_2^2-5 \mu }{32}  & 2 \bar{u}_2' & -\bar{u}_2 & 0 \cr 
{\tiny \yng(1,1,1,1)} &\frac{30 (\bar{u}_2')^2+30 \bar{u}_2 \bar{u}_2''-\bar{u}_2''''}{16} & \bar{\eta } +\frac{\bar{u}_2'''-30 \bar{u}_2 \bar{u}_2'}{16} & -\bar{u}_2'' & \bar{u}_2' & -\bar{u}_2 \cr 
},
\end{align}
where the coefficients $\{w_n\}_{n=2}^5$ are given by Eq.~\eq{Eqn:Appendix:PQ52RS12CoefficientsWWW}

\subsubsection{Duality matrices}

\noindent\underline{Reflection:}  \qquad $\vec W^{(5-s)}[\chi_{\mathcal C}](t; -\eta) = {\mathsf M}^{(s)}(t)\times  \vec W^{(s)}[\chi](t;\eta)$
\begin{align}
{\mathsf M}^{(1)}(t) &= 
\bordermatrix{
 & \varnothing & {\tiny \yng(1)} & {\tiny \yng(2)} & {\tiny \yng(3)} & {\tiny \yng(4)} \cr
\varnothing &  1 & 0 & 0 & 0 & 0 \cr
{\tiny \yng(1)} & 0 & 1 & 0 & 0 & 0 \cr
{\tiny \yng(1,1)} &   \frac{5 \bar u_2(t)}{2} & 0 & 1 & 0 & 0 \cr
{\tiny \yng(1,1,1)} & - \frac{5 \bar u_2'(t)}{4}  & \frac{5 \bar u_2(t)}{2} & 0 & 1 & 0 \cr
{\tiny \yng(1,1,1,1)} &  \frac{5 \left(12 \bar u_2(t){}^2 - \mu +12\bar u_2''(t)\right)}{32}  & \frac{5 \bar u_2'(t)}{4} & \frac{5 \bar u_2(t)}{2} & 0 & 1 
}, \\
{\mathsf M}^{(2)}(t) &= 
\bordermatrix{
 & \varnothing & {\tiny \yng(1)} & {\tiny \yng(2)} & {\tiny \yng(1,1)} & {\tiny \yng(3)} & {\tiny \yng(2,1)} & {\tiny \yng(3,1)} & {\tiny \yng(2,2)} & {\tiny \yng(3,2)} & {\tiny \yng(3,3)} \cr
\varnothing &   1 & 0 & 0 & 0 & 0 & 0 & 0 & 0 & 0 & 0 \cr
{\tiny \yng(1)} &   0 & 1 & 0 & 0 & 0 & 0 & 0 & 0 & 0 & 0 \cr
{\tiny \yng(2)} & - \frac{5 \bar{u}_2}{2}  & 0 & 0 & 1 & 0 & 0 & 0 & 0 & 0 & 0 \cr
{\tiny \yng(1,1)} &   \frac{5 \bar{u}_2}{2}  & 0 & 1 & 0 & 0 & 0 & 0 & 0 & 0 & 0 \cr
{\tiny \yng(2,1)} &    \frac{5 \bar{u}_2'}{4}  & 0 & 0 & 0 & 0 & 1 & 0 & 0 & 0 & 0 \cr
{\tiny \yng(1,1,1)} &    \frac{5 \bar{u}_2'}{4}  & \frac{5}{2} \bar{u}_2 & 0 & 0 & 1 & 0 & 0 & 0 & 0 & 0 \cr
{\tiny \yng(2,2)} &   \frac{25 \bar{u}_2^2}{4}  & \frac{5}{4} \bar{u}_2' & \frac{5}{2} \bar{u}_2 & -\frac{5 \bar{u}_2}{2}  & 0 & 0 & 0 & 1 & 0 & 0 \cr
{\tiny \yng(2,1,1)} &  h^{(1)}_1   & 0 & 0 & \frac{5 \bar{u}_2}{2}  & 0 & 0 & 1 & 0 & 0 & 0 \cr
{\tiny \yng(2,2,1)} &  \frac{25 \bar{u}_2 \bar{u}_2'}{8}  & h^{(1)}_2   & 0 & -\frac{5 \bar{u}_2' }{4}  & \frac{5 \bar{u}_2}{2}  & 0 & 0 & 0 & 1 & 0 \cr
{\tiny \yng(2,2,2)} & h^{(1)}_3   & -  \frac{25 \bar{u}_2 \bar{u}_2'}{8}  & h^{(1)}_1  & \frac{25}{4} \bar{u}_2^2 & -\frac{5 \bar{u}_2' }{4} &- \frac{5 \bar{u}_2'}{4}  & \frac{5 \bar{u}_2}{2}  & -\frac{5 \bar{u}_2 }{2} & 0 & 1 
}, 
\end{align}
where the coefficients $\{h_n^{(1)}\}_{n=1}^3$ are given by 
\begin{align}
h^{(1)}_1 &= -\frac{5 \left(12 \bar{u}_2^2-\mu +12 \bar{u}_2''\right)}{32}, \quad 
h^{(1)}_2 = \frac{5 \left(28 \bar{u}_2^2+\mu -12 \bar{u}_2''\right)}{32}, \nn\\
h^{(1)}_3 &= \frac{25 \left(-12 \bar{u}_2^3+\left(\mu -12 \bar{u}_2''\right) \bar{u}_2 -4 (\bar{u}_2')^2\right)}{64}. 
\end{align}

\noindent\underline{Charge conjugation:} \qquad $\Bigl[\vec W^{(5-s)}[\chi](t;\eta)\Bigr]^{\rm T}\,\, \mathfrak C^{(s,1)}_{[\chi]}(t)\, \, \Bigl[ \vec W^{(s)}[\chi](t;\eta) \Bigr] $
\begin{align}
{\mathfrak C}^{(1,1)}_{[\chi]}(t) &= 
\bordermatrix{
 & \varnothing & {\tiny \yng(1)} & {\tiny \yng(2)} & {\tiny \yng(3)} & {\tiny \yng(4)} \cr
\varnothing &  0 & 0 & 0 & 0 & 1 \cr
{\tiny \yng(1)} & 0 & 0 & 0 & -1 & 0 \cr
{\tiny \yng(1,1)} &   0 & 0 & 1 & 0 & 0 \cr
{\tiny \yng(1,1,1)} & 0  & -1 & 0 & 0 & 0 \cr
{\tiny \yng(1,1,1,1)} &  1  & 0 &0 & 0 & 0 
}, \\
{\mathfrak C}^{(2,1)}_{[\chi]}(t) &= 
\bordermatrix{
 & \varnothing & {\tiny \yng(1)} & {\tiny \yng(2)} & {\tiny \yng(1,1)} & {\tiny \yng(3)} & {\tiny \yng(2,1)} & {\tiny \yng(3,1)} & {\tiny \yng(2,2)} & {\tiny \yng(3,2)} & {\tiny \yng(3,3)} \cr
\varnothing &   0 & 0 & 0 & 0 & 0 & 0 & 0 & 0 & 0 & 1 \cr
{\tiny \yng(1)} &   0 & 0 & 0 & 0 & 0 & 0 & 0 & 0 &- 1 & 0 \cr
{\tiny \yng(2)} & 0  & 0 & 0 & 0 & 0 & 0 & 0 & 1 & 0 & 0 \cr
{\tiny \yng(1,1)} &  0 & 0 & 0 & 0 & 0 & 0 & 1 & 0 & 0 & 0 \cr
{\tiny \yng(2,1)} &   0  & 0 & 0 & 0 & 0 & -1 & 0 & 0 & 0 & 0 \cr
{\tiny \yng(1,1,1)} &    0  & 0 & 0 & 0 & -1 & 0 & 0 & 0 & 0 & 0 \cr
{\tiny \yng(2,2)} &  0  & 0 & 0 & 1  & 0 & 0 & 0 & 0 & 0 & 0 \cr
{\tiny \yng(2,1,1)} &  0  & 0 & 1 & 0  & 0 & 0 & 0 & 0 & 0 & 0 \cr
{\tiny \yng(2,2,1)} &  0  & -1   & 0 & 0 &0  & 0 & 0 & 0 & 0 & 0 \cr
{\tiny \yng(2,2,2)} & 1   & 0  & 0  & 0 & 0 & 0  & 0 & 0 & 0 & 0
}
\end{align}



\subsection{The case of $(p,q)=(3,4)$: Ising model}  

\begin{itemize}
\item [I. ] Kac table: 
\begin{align}
    \begin{tabular}{|c|c|} \hline
$ (1,3)$ & $(2,3)$  \\ \hline
$ \bf (1,2)$ & $ (2,2)$  \\ \hline
$\bf (1,1)$ & $ \bf (2,1)$  \\ \hline
    \end{tabular}
\end{align}
\item [II. ] The coefficients (with conformal background): 
\begin{align}
v_2(t)&=\frac{8 u_2(t)}{3}, 
\qquad v_3(t)=\frac{4 u_2'(t) + 8 u_3(t)}{3}, \nn\\
\qquad v_4(t)&= \frac{4 u_2''(t) + 12 u_3'(t) + u_2(t)^2}{9}. 
\label{Eqn:Appendix:CoefficientsStringEquationPQ34}
\end{align}
\item [III. ] The Baker-Akhiezer systems: 
\begin{align}
\left\{
\begin{array}{rl}
\ds \zeta \, \psi \!\!\!&= \ds  \Bigl[
4\, \del^3 +  u_2 \del +  u_3  \Bigr] \psi, \cr
\ds g \frac{\del}{\del \zeta}\,  \psi \!\!\!& \ds = \beta_{3,4}\times \Bigl[ 8\, \del^4+ \frac{8 u_2}{3}  \, \del^2  + \frac{4 (2 u_3+ u_2')}{3} \, \del +  \frac{4 u_2'' + 12 u_3' + u_2^2}{9} \Bigr] \psi, \vspace{0.2cm} \cr
\ds \eta \, \chi \!\!\!& \ds =\Bigl[8 \,\del^4+\frac{8 u_2}{3}  \del^2 +\frac{4(3 u_2' -2 u_3) }{3} \del + \frac{16 u_2''   -12 u_3'  +u_2^2 }{9} \Bigr] \chi, \cr
\ds g \frac{\del}{\del \eta}\,  \chi \!\!\!&= \ds \beta_{4,3} \times   \Bigl[
4\,\del^3+ u_2 \del +u_2'-u_3  \Bigr]\chi. 
\end{array}
\right. \label{Eqn:Appendix:BAsystemsPQ34}
\end{align}
\item [IV. ] String equations (with conformal background): 
\begin{align}
0 &= u_2 \left(u_2''-2 u_3'\right)+ (u_2')^2-2 u_3 u_2'-4 u_3^{(3)}+2 u_2^{(4)}, \\
9 \frac{g}{\beta_{3,4}} &= 24 u_2' u_2'' + 2 u_2^2 u_2' - 24 u_3 u_3' - 12 u_2' u_3' - 12 u_2 \left(u_3''-u_2^{(3)}\right) - 24 u_3^{(4)} + 16 u_2^{(5)}. \label{Eqn:Appendix:StringEquationPQ34}
\end{align}
\end{itemize}

\subsubsection{1) $(r,s)=(1,1)$ FZZT-Cardy brane: $W^{(1,1)} = W^{(1)}[\psi]$}

\begin{align}
\vec{W}^{(1)} &= \Bigl( W^{(1)}_\varnothing, W^{(1)}_{\tiny \yng(1)}, W^{(1)}_{\tiny \yng(2)} \Bigr)^{\rm T}, \\
\mathcal B^{(1,1)} &= 
\bordermatrix{
 & \varnothing & {\tiny \yng(1)} & {\tiny \yng(2)} \cr
\varnothing &  & 1 & \cr
{\tiny \yng(1)} & & & 1 \cr
{\tiny \yng(2)} &\bar \zeta- \bar u_3 & -\bar u_2 & 
}, 
\\
\mathcal Q^{(1,1)} &= \bar \beta_{3,4} \times
\bordermatrix{
 & \varnothing & {\tiny \yng(1)} & {\tiny \yng(2)} \cr
\varnothing & \frac{2}{9} \bar u_2^2 - \frac{1}{3} \bar u_3' + \frac{2}{9} \bar u_2'' & \bar \zeta + \frac{\bar u_3}{3} - \frac{\bar u_2'}{3} & \frac{\bar u_2}{3} \cr
{\tiny \yng(1)} & \frac{\bar u_2}{3} \bar \zeta + f_{1}^{(2)} & - \frac{\bar u_2^2}{9}  - \frac{\bar u_2''}{9} & \bar \zeta + \frac{\bar u_3}{3} \cr
{\tiny \yng(2)} &\bar  \zeta^2 - \frac{2 \bar u_3 - \bar u_2'}{3} \bar \zeta + f_{3}^{(2)} & -\frac{2 \bar u_2}{3}\bar \zeta + f_{2}^{(2)} & - \frac{\bar u_2^2}{9} + \frac{\bar u_3'}{3} - \frac{\bar u_2''}{9}
}. 
\end{align}
where 
\begin{align}
f_{1}^{(2)} & = \frac{2}{9} \bar u_2''' - \frac{1}{3} \bar u_3'' + \frac{4}{9} \bar u_2 \bar u_2' - \frac{1}{3} \bar u_2 \bar u_3, \\
f_{2}^{(2)} & = \frac{1}{9} \bar u_2''' - \frac{1}{3} \bar u_3'' + \frac{2}{9} \bar u_2 \bar u_2' - \frac{2}{3} \bar u_2 \bar u_3, \\
f_{3}^{(2)} & = \frac{2}{9} \bar u_2'''' - \frac{1}{3} \bar u_3''' + \frac{4}{9} \bar u_2 \bar u_2'' - \frac{1}{3} \bar u_2 \bar u_3' + \frac{4}{9} (\bar u_2')^2 - \frac{1}{3} \bar u_3 \bar u_2'- \frac{1}{3} \bar u_3^2.  \label{Eqn:Appendix:CoefficientsffOfPQ34RS32}
\end{align}

\subsubsection{2) $(r,s)=(2,1)$ FZZT-Cardy brane: $W^{(2,1)} = W^{(2)}[\psi]$}

\begin{align}
\vec{W}^{(2)} &= \Bigl( W^{(2)}_\varnothing, W^{(2)}_{\tiny \yng(1)}, W^{(2)}_{\tiny \yng(1,1)} \Bigr)^{\rm T}, \\
\mathcal B^{(2,1)} &= 
\bordermatrix{
 & \varnothing &{\tiny \yng(1)} & {\tiny \yng(1,1)}  \cr
\varnothing &  & 1 & \cr
{\tiny \yng(1)} & -\bar{u}_2 & & 1 \cr
{\tiny \yng(1,1)} &-\bar \zeta + \bar{u}_3 &  & 
}, 
\\
\mathcal Q^{(2,1)} &= \bar \beta_{3,4} \times
\bordermatrix{
 & \varnothing &{\tiny \yng(1)} & {\tiny \yng(1,1)} \cr
\varnothing & \frac{\bar u_2^2}{9} - \frac{\bar u_3'}{3} + \frac{\bar u_2''}{9} & \bar \zeta + \frac{\bar u_3}{3} & -\frac{\bar u_2}{3} \cr
{\tiny \yng(1)} & -\frac{2\bar u_2}{3} \bar \zeta + f_{2}^{(2)} & \frac{\bar u_2^2}{9} + \frac{\bar u_2''}{9} & \bar \zeta + \frac{\bar u_3}{3} - \frac{\bar u_2'}{3} \cr
{\tiny \yng(1,1)} & -\bar \zeta^2 + \frac{2 \bar u_3 - \bar u_2'}{3} \bar \zeta - f_{3}^{(2)} &  \frac{\bar u_2}{3} \bar \zeta + f_{1}^{(2)} & - \frac{2}{9} \bar u_2^2 + \frac{\bar u_3'}{3} - \frac{2}{9} \bar u_2''
}, 
\end{align}
where the coefficients $\{f_n^{(2)}\}_{n=1}^3$ are given by Eq.~\eq{Eqn:Appendix:CoefficientsffOfPQ34RS32}. 


\subsubsection{3) $(r,s)=(1,2)$ FZZT-Cardy brane: $W^{(1,2)} = \mathcal L \widetilde W^{(2)}[\chi]$}

\begin{align}
 \vec{W}^{(1,2)} &= \Bigl( \mathcal L \widetilde W^{(2)}_\varnothing, \mathcal L  \widetilde W^{(2)}_{\square}, \mathcal L  \widetilde W^{(2)}_{\square^2} \Bigr)^{\rm T}, \\
\mathcal B^{(1,2)}(t;\zeta)  &= 
\bordermatrix{
 & \varnothing & \square & \square^2 \cr
\varnothing &  & 1 & \cr
\square & & & 1 \cr
\square^2 & \frac{ \zeta}{2} & - \frac{u_2}{2} & 
}, \\
\mathcal Q^{(1,2)}(t;\zeta)  &=   \beta_{4,3}
\bordermatrix{
 & \varnothing & \square & \square^2 \cr
\varnothing & -\frac{u_2^2 + 2 u_2''}{9} + \frac{g \beta_{3,4}^{-1}}{2  \zeta}  & -\zeta + \frac{u_2'}{3} - \frac{F_3}{\zeta}  & -\frac{u_2}{3} - \frac{(u_2^2+2u_2'')'}{9 \zeta}  \cr
\square & - \frac{u_2}{6}\zeta -\frac{(u_2^2 + 2 u_2'')'}{18}  & \frac{u_2^2 + 2 u_2''}{18} - \frac{((2u_3-u_2')^2)'}{18 \zeta}&  - \zeta + \frac{(2 u_3- u_2' )^2}{9\zeta} \cr
\square^2 & - \frac{ \zeta^2}{2} - \frac{u_2'}{6 } \zeta -\frac{F_3}{2}  & \frac{u_2}{3} \zeta - \frac{(2u_3' - u_2'')^2}{9\zeta} & \frac{u_2^2 + 2u_2''}{18}+\frac{((2u_3-u_2')^2)'}{18 \zeta}
}, 
\end{align}
with $F_3 \equiv \frac{1}{18}\left(2u_3 - u_2'\right)^2 - \frac{1}{9}  u_2 u_2'' + \frac{1}{18} \left(u_2'\right)^2 - \frac{1}{27} u_2^3 + \frac{(t-2\beta_{3,4} \mu)}{2 \beta_{3,4}}$. 

\subsubsection{Duality matrices}

\noindent\underline{Reflection 1:} \qquad $\vec W^{(2)}[\psi_{\mathcal C}](t; -\zeta) = {\mathsf M}^{(1)}(t)\times  \vec W^{(1)}[\psi](t;\zeta)$
\begin{align}
{\mathsf M}^{(1)}(t) = 
\bordermatrix{
 & \varnothing & {\tiny \yng(1)} & {\tiny \yng(2)} \cr
\varnothing & 1 & 0 & 0  \cr
{\tiny \yng(1)} &  0 & 1 & 0 \cr
{\tiny \yng(1,1)} &   \bar u_2 & 0 & 1
}. 
\end{align}
\noindent\underline{Reflection 2:} \qquad $\vec W^{(1,2)}[\chi_{\mathcal C}](t; \zeta) = {\mathsf N}^{(2)}(t)\times  \vec W^{(1,2)}[\chi](t;\zeta)$
\begin{align}
{\mathsf N}^{(2)}(t) = \bordermatrix{
 & \varnothing & {\square} &   {\square^2} \cr
\varnothing &  1 & 0 & 0  \cr
{\square} &  0 & 1 & 0  \cr
{\square^2} & 0 & 0 & 1
}. 
\end{align}
\noindent\underline{Charge conjugation:} \qquad $\Bigl[\vec W^{(2)}[\chi](t; \zeta)\Bigr]^{\rm T}\,\, \mathfrak C^{(1,1)}(t)\, \, \Bigl[ \vec W^{(1)}[\chi](t;\zeta) \Bigr] $
\begin{align}
{\mathfrak C}^{(1,1)}(t) = 
\bordermatrix{
 & \varnothing & {\tiny \yng(1)} & {\tiny \yng(2)} \cr
\varnothing & 0 & 0 & 1  \cr
{\tiny \yng(1)} &  0 & -1 & 0 \cr
{\tiny \yng(1,1)} &   1 & 0 & 0
}. 
\end{align}


\subsection{The case of $(p,q)=(4,3)$: Dual Ising model}

\begin{itemize}
\item [I. ] Kac table: 
\begin{align}
 \begin{tabular}{|c|c|c|} \hline
  $(1,2)$ & $(2,2)$ & $(3,2)$  \\ \hline
 $ \bf (1,1)$ & $ \bf (2,1)$ & $ \bf (3,1)$   \\ \hline
\end{tabular}
\end{align}
\item [II. ] The coefficients: Eq.~\eq{Eqn:Appendix:CoefficientsStringEquationPQ34}
\item [III. ] The Baker-Akhiezer systems: Eq.~\eq{Eqn:Appendix:BAsystemsPQ34} 
\item [IV. ] String equations: Eq.~\eq{Eqn:Appendix:StringEquationPQ34} 
\end{itemize}

\subsubsection{1) $(r,s)=(1,1)$ FZZT-Cardy brane: $\widetilde W^{(1,1)} =\widetilde W^{(1)}[\chi]$}

\begin{align}
\vec{W}^{(1,1)} &= \Bigl( \widetilde W^{(1)}_\varnothing, \widetilde W^{(1)}_{\tiny \yng(1)}, \widetilde W^{(1)}_{\tiny \yng(2)}, \widetilde W^{(1)}_{\tiny \yng(3)} \Bigr)^{\rm T}, \\
\mathcal B^{(1,1)} &
= \bordermatrix{
 & \varnothing & {\tiny \yng(1)} & {\tiny \yng(2)} & {\tiny \yng(3)} \cr
\varnothing &  & 1 & \cr
{\tiny \yng(1)} & & & 1 \cr
{\tiny \yng(2)} &  & & & 1 \cr 
{\tiny \yng(3)} &\bar \eta- \frac{2(\bar u_2^2-3\bar u_3' + 4 \bar u_2'')}{9} & \frac{2(2 \bar u_3- 3\bar u_2' )}{3} &  -\frac{4}{3} \bar u_2 & \cr
} \\
\mathcal Q^{(1,1)} &
= \bar \beta_{4,3} \times \bordermatrix{
 & \varnothing & {\tiny \yng(1)} & {\tiny \yng(2)} & {\tiny \yng(3)} \cr
\varnothing &- \bar u_3 + \bar u_2' & \bar u_2 & 0 & 1 \cr
{\tiny \yng(1)} & \bar \eta - \frac{2 \bar u_2^2 + 3 \bar u_3' - \bar u_2''}{9}& \frac{\bar u_3}{3} &-\frac{\bar u_2}{3} & 0 \cr
{\tiny \yng(2)} & \frac{\bar u_2''' - 3 \bar u_3'' - 4 \bar u_2 \bar u_2'}{9} & \bar \eta - \frac{2 \bar u_2^2 - \bar u_2''}{9} & \frac{\bar u_3 - \bar u_2'}{3} & -\frac{\bar u_2}{3} \cr 
{\tiny \yng(3)} & - \frac{\bar u_2}{3} \eta + f_2^{(3)} & f_1^{(3)} & \bar \eta + \frac{2 \bar u_2^2 + 3 \bar u_3' - 2 \bar u_2''}{9} & \frac{\bar u_3 - 2 \bar u_2'}{3} \cr
},
\end{align}
where 
\begin{align}
f_1^{(3)}&= \frac{2}{9} \bar u_2''' - \frac{1}{3} \bar u_3'' - \frac{2}{9} \bar u_2 \bar u_2' - \frac{4}{9} \bar u_2 \bar u_3, \nn\\
f_2^{(3)}&= \frac{1}{9} \bar u_2'''' - \frac{1}{3} \bar u_3''' - \frac{4}{27} \bar u_2 \bar u_2'' - \frac{2}{9} \bar u_2 \bar u_3' - \frac{4}{9} (\bar u_2')^2 + \frac{2}{27} \bar u_2^3. 
\label{Eqn:Appendix:CoefficientsffInPQ43RS11}
\end{align}

\subsubsection{2) $(r,s)=(2,1)$ FZZT-Cardy brane: $\widetilde W^{(2,1)} = \widetilde W^{(2)}[\chi]$}

\begin{align}
\vec{W}^{(2,1)} &= \Bigl( \widetilde W^{(2)}_\varnothing, \widetilde W^{(2)}_{\tiny \yng(1)}, \widetilde W^{(2)}_{\tiny \yng(2)}, \widetilde W^{(2)}_{\tiny \yng(1,1)}, \widetilde W^{(2)}_{\tiny \yng(2,1)}, \widetilde W^{(2)}_{\tiny \yng(2,2)} \Bigr)^{\rm T}, \\
\mathcal B^{(2,1)} &= 
\mbox{\scriptsize $ \bordermatrix{
 & \mbox{\normalsize $\varnothing$} & {\tiny \yng(1)} & {\tiny \yng(2)} & {\tiny \yng(1,1)} & {\tiny \yng(2,1)} & {\tiny \yng(2,2)}  \cr
\mbox{\normalsize$\varnothing$} &  & 1 & \cr
{\tiny \yng(1)} & & & 1 & 1 \cr
{\tiny \yng(2)} &\frac{4}{3} \bar{u}_3-2 \bar{u}_2' & - \frac{4}{3} \bar{u}_2 &  &  & 1 & \cr
{\tiny \yng(1,1)} &    &  &  &  & 1 &  \cr
{\tiny \yng(2,1)} &  - \bar{\eta } + \frac{2}{9} \bar{u}_2^2-\frac{2}{3} \bar{u}_3'+\frac{8}{9} \bar{u}_2'' &  &  & - \frac{4}{3} \bar{u}_2 &  & 1 \cr
{\tiny \yng(2,2)} &  & -\bar{\eta }+ \frac{2}{9} \bar{u}_2^2-\frac{2}{3} \bar{u}_3'+\frac{8}{9} \bar{u}_2'' &  & 2 \bar{u}_2'-\frac{4}{3} \bar{u}_3 &  &
} $}, \\
\mathcal Q^{(2,1)} &= \bar \beta_{4,3} \times \nn\\
\times \bigg[\,\,\, \,\,\,\,\,& \!\!\!\!\!\!\!
\mbox{\scriptsize $\bordermatrix{
 & \mbox{\normalsize $\varnothing$} & {\tiny \yng(1)} & {\tiny \yng(2)} & {\tiny \yng(1,1)} & {\tiny \yng(2,1)} & {\tiny \yng(2,2)}  \cr
\mbox{\normalsize $\varnothing$} &  \bar{u}_2'-\frac{2}{3} \bar{u}_3 & - \frac{\bar u_2}{3} & 0 & 0 & -1 & 0 \cr
{\tiny \yng(1)} &  \bar \eta + \frac{-2 \bar{u}_2^2+\bar{u}_2''}{9}  
& \frac{2(\bar{u}_2'- \bar{u}_3) }{3} 
& - \frac{\bar u_2}{3} & \bar{u}_2 & 0 & -1 \cr
{\tiny \yng(2)} & h_{1}^{(2)} 
& \bar \eta +\frac{2 \bar{u}_2^2+3 \bar{u}_3'-2 \bar{u}_2''}{9} 
& \frac{\bar{u}_2'-2 \bar{u}_3}{3} 
& 0 & \bar{u}_2 & 0 \cr
{\tiny \yng(1,1)} &  \frac{4 \bar{u}_2 \bar{u}_2'+3 \bar{u}_3''-\bar{u}_2'''}{9}  
&\bar \eta + \frac{-2 \bar{u}_2^2-3 \bar{u}_3'+\bar{u}_2''}{9}
& 0 & \frac{2 \bar{u}_3-\bar{u}_2'}{3}  & - \frac{\bar u_2}{3} & 0 \cr
{\tiny \yng(2,1)} &  \frac{\bar u_2}{3} \bar \eta + h_{2}^{(2)} 
& 0 
& \bar \eta + \frac{-2 \bar{u}_2^2-3 \bar{u}_3'+\bar{u}_2''}{9} 
& \bar \eta + \frac{2 \bar{u}_2^2+3 \bar{u}_3'-2 \bar{u}_2''}{9}  & \frac{2(\bar{u}_3-\bar{u}_2')}{3}  & - \frac{\bar u_2}{3} \cr
{\tiny \yng(2,2)} &  0 &  \frac{\bar u_2}{3} \bar \eta + h_{2}^{(2)} 
& \frac{-4 \bar{u}_2 \bar{u}_2'-3 \bar{u}_3''+\bar{u}_2'''}{9} 
& - h_{1}^{(2)} 
& \bar \eta + \frac{-2 \bar{u}_2^2+\bar{u}_2''}{9} 
& \frac{2}{3} \bar{u}_3-\bar{u}_2' \cr
}$} \bigg], 
\end{align}
where the coefficients $h_1^{(2)}$ and $h_2^{(2)}$ are given by 
\begin{align}
h_{1}^{(2)} & =   \frac{ -2 \bar{u}_2 \left(2 \bar{u}_3+\bar{u}_2'\right)-3 \bar{u}_3''+2 \bar{u}_2''' }{9}, \nn\\
h_{2}^{(2)} &=  \frac{-2 \bar{u}_2^3+\left(6 \bar{u}_3'+4 \bar{u}_2''\right) \bar{u}_2+12 (\bar{u}_2')^2+9 \bar{u}_3'''-3 \bar{u}_2'''' }{27}. 
\end{align}

\subsubsection{3) $(r,s)=(3,1)$ FZZT-Cardy brane: $\widetilde W^{(3,1)} = \widetilde W^{(3)}[\chi]$}

\begin{align}
\vec{W}^{(3,1)} & = \Bigl( \widetilde W^{(3)}_\varnothing, \widetilde W^{(3)}_{\tiny \yng(1)}, \widetilde W^{(3)}_{\tiny \yng(1,1)}, \widetilde W^{(3)}_{\tiny \yng(1,1,1)} \Bigr)^{\rm T}, \\
\mathcal B^{(3,1)} &= 
\bordermatrix{
 & \varnothing & {\tiny \yng(1)} & {\tiny \yng(1,1)} & {\tiny \yng(1,1,1)} \cr
\varnothing &  & 1 & \cr
{\tiny \yng(1)} & - \frac{4}{3}\bar u_2 & & 1 \cr
{\tiny \yng(1,1)} & -\frac{2(2\bar u_3 - 3 \bar u_2')}{3} & & & 1 \cr 
{\tiny \yng(1,1,1)} &\bar \eta- \frac{2(\bar u_2^2 - 3 \bar u_3' + 4 \bar u_2'')}{9} &  &   & \cr
}, \\
\mathcal Q^{(3,1)} &= \bar \beta_{4,3} \times
\bordermatrix{
 & \varnothing & {\tiny \yng(1)} & {\tiny \yng(1,1)} & {\tiny \yng(1,1,1)} \cr
\varnothing &  \frac{2 \bar{u}_2'-\bar{u}_3}{3}  & -\frac{1}{3} \bar{u}_2 & 0 & 1 \cr
{\tiny \yng(1)} & \bar{\eta } +  \frac{2 \bar{u}_2^2+3 \bar{u}_3'-2 \bar{u}_2''}{9} & \frac{\bar{u}_2'-\bar{u}_3}{3} & -\frac{1}{3} \bar{u}_2 & 0 \cr
{\tiny \yng(1,1)} & -f_{1}^{(3)}
& \bar{\eta } + \frac{-2 \bar{u}_2^2+\bar{u}_2''}{9}
& -\frac{1}{3} \bar{u}_3 & \bar{u}_2 \cr
{\tiny \yng(1,1,1)} &- \frac{\bar u_2}{3} \bar \eta + f_2^{(3)}
& \frac{4 \bar{u}_2 \bar{u}_2'+3 \bar{u}_3''-\bar{u}_2'''}{9}
& \bar{\eta } + \frac{-2 \bar{u}_2^2-3 \bar{u}_3'+\bar{u}_2''}{9} 
& \bar{u}_3-\bar{u}_2' \cr
}, 
\end{align}
where the coefficients $f_1^{(3)}$ and $f_2^{(3)}$ are given by Eq.~\eq{Eqn:Appendix:CoefficientsffInPQ43RS11}. 

\subsubsection{Duality matrices}

\noindent\underline{Reflection:} \qquad $\vec W^{(4-s)}[\chi_{\mathcal C}](t; \eta) = {\mathsf M}^{(s)}(t)\times  \vec W^{(s)}[\chi](t;\eta)$
\begin{align}
{\mathsf M}^{(1)}(t) & = 
\bordermatrix{
 & \varnothing & {\tiny \yng(1)} & {\tiny \yng(2)} & {\tiny \yng(3)} \cr
\varnothing &  1 & 0 & 0 & 0 \cr
{\tiny \yng(1)} &  0 & 1 & 0 & 0 \cr
{\tiny \yng(1,1)} &   \frac{4}{3} \bar{u}_2 & 0 & 1 & 0 \cr
{\tiny \yng(1,1,1)} &  -  \frac{4}{3} \bar{u}_3+ \frac{2}{3} \bar{u}_2' & \frac{4}{3} \bar{u}_2 & 0 & 1
} \\
{\mathsf M}^{(2)}(t) & = 
\bordermatrix{
 & \varnothing & {\tiny \yng(1)} & {\tiny \yng(2)} & {\tiny \yng(1,1)} & {\tiny \yng(2,1)} & {\tiny \yng(2,2)} \cr
\varnothing &  1 & 0 & 0 & 0 & 0 & 0  \cr
{\tiny \yng(1)} &   0 & 1 & 0 & 0 & 0 & 0 \cr
{\tiny \yng(2)} &    -\frac{4 \bar{u}_2}{3}  & 0 & 0 & 1 & 0 & 0 \cr
{\tiny \yng(1,1)} &  \frac{4 \bar{u}_2}{3}  & 0 & 1 & 0 & 0 & 0 \cr
{\tiny \yng(2,1)} &  \frac{2 (2 \bar{u}_3-\bar{u}_2' ) }{3} & 0 & 0 & 0 & 1 & 0 \cr
{\tiny \yng(2,2)} &  \frac{16 \bar{u}_2^2}{9}  & \frac{2(2 \bar{u}_3-\bar{u}_2')}{3}  & \frac{4 \bar{u}_2}{3}  & -\frac{4 \bar{u}_2}{3}   & 0 & 1
}. 
\end{align}

\noindent\underline{Charge conjugation:} \qquad $\Bigl[\vec W^{(4-s)}[\chi](t;\eta)\Bigr]^{\rm T}\,\, \mathfrak C^{(s,1)}_{[\chi]}(t)\, \, \Bigl[ \vec W^{(s)}[\chi](t;\eta) \Bigr] $
\begin{align}
{\mathfrak C}^{(1,1)}_{[\chi]}(t) & = 
\bordermatrix{
 & \varnothing & {\tiny \yng(1)} & {\tiny \yng(2)} & {\tiny \yng(3)} \cr
\varnothing & 0 & 0 & 0 & 1  \cr
{\tiny \yng(1)} & 0 &  0 & -1 & 0 \cr
{\tiny \yng(1,1)} & 0&    1 & 0 & 0 \cr
{\tiny \yng(1,1,1)} &   -1 & 0 & 0 & 0
}, \\
{\mathfrak C}^{(2,1)}_{[\chi]}(t) & = 
\bordermatrix{
 & \varnothing & {\tiny \yng(1)} & {\tiny \yng(2)} & {\tiny \yng(1,1)} & {\tiny \yng(2,1)} & {\tiny \yng(2,2)} \cr
\varnothing & 0& 0& 0 & 0 & 0 & 1  \cr
{\tiny \yng(1)} &  0& 0& 0 &  0 & -1 & 0 \cr
{\tiny \yng(2)} &  0& 0& 0&    1 & 0 & 0 \cr
{\tiny \yng(1,1)} &  0& 0&   1 & 0 & 0 & 0 \cr
 {\tiny \yng(2,1)} &  0&   -1 & 0 & 0 & 0 & 0 \cr
{\tiny \yng(2,2)} & 1 & 0 & 0 & 0 & 0  & 0 
}. 
\end{align}



\subsection{The case of $(p,q)=(3,5)$  \label{Section:Appendix:PQ35} }

\begin{itemize}
\item [I. ] Kac table: 
\begin{align}
    \begin{tabular}{|c|c|} \hline
$(1,4)$ & $ (2,4)$  \\ \hline
$ \bf (1,3)$ & $ (2,3)$  \\ \hline
$ \bf (1,2)$ & $ (2,2)$  \\ \hline
$\bf (1,1)$ & $ \bf (2,1)$  \\ \hline
    \end{tabular}
\end{align}
\item [II. ] The coefficients (with conformal background): 
\begin{align}
&v_2(t)=\frac{20 u_2(t)}{3},\quad v_3(t)=\frac{20}{3} u_2'(t)+\frac{20 u_3(t)}{3}, \nn\\
\quad &v_4(t)=\frac{20}{3} u_3'(t)+\frac{40}{9} u_2''(t)+\frac{5}{9} u_2(t){}^2, 
\quad v_5(t)=\frac{40}{9} u_3''(t)+\frac{10}{9} u_2(t) u_3(t). \label{Eqn:Appendix:CoefficientsStringEquationCBPQ35}
\end{align}
\item [III. ] The Baker-Akhiezer systems: 
\begin{align}
\left\{
\begin{array}{rl}
\ds \zeta \, \psi \!\!\!&= \ds  \Bigl[
4\, \del^3 +  u_2 \del +  u_3  \Bigr] \psi, \cr
\ds g \frac{\del}{\del \zeta}\,  \psi \!\!\!& \ds = \beta_{3,5}\times \Bigl[ 16 \del^5+\frac{20  u_2}{3} \del^3  + \frac{20  (u_2' + u_3) }{3} \del^2 + \nn\\
&\qquad \qquad \ds + \frac{5 (12 u_3' +8 u_2'' +u_2^2 )}{9} \del    +   \frac{10 (4 u_3'' +u_2  u_3)}{9} \Bigr] \psi, \vspace{0.2cm} \cr
\ds \eta \, \chi \!\!\!& \ds =\Bigl[   16 \del^5+\frac{20 u_2 }{3} \del ^3  -\frac{20 (u_3 -2 u_2' )}{3} \del ^2 + \nn\\
&\quad \ds + \frac{5 \left(-12 u_3' +20 u_2'' +u_2^2\right)}{9} \del     +\frac{10 \left(u_2 \left(u_2' -u_3 \right)-4 u_3'' +4 u_2^{(3)} \right) }{9}   \Bigr] \chi, \cr
\ds g \frac{\del}{\del \eta}\,  \chi \!\!\!&= \ds \beta_{4,3} \times   \Bigl[
4\,\del^3+ u_2 \del +u_2'-u_3  \Bigr]\chi. 
\end{array}
\right.  \label{Eqn:Appendix:BAsystemCBPQ35equation}
\end{align}
\item [IV. ] String equations (with conformal background): 
\begin{align}
&
\mbox{ $\left\{
\begin{array}{rl}
0 \!\! &=\, 60 u_3 \left(u_2'' -2 u_3' \right)+5 u_2'  \left(4 \left(3 u_3' +u_2'' \right)+u_2^2\right)+20 u_2 u_2^{(3)} +16 u_2^{(5)}, \cr
\ds \frac{9 g}{\beta_{3,5} } \!\! &=\, 80 u_3' u_2''+60 \left(u_2'-u_3\right) u_3'' +5 u_2^2 u_3'+  \cr 
&\qquad \quad \quad   + 10 u_2 \left(u_3 u_2'+2 u_3^{(3)}\right) - 60 (u_3')^2+40 u_3 u_2^{(3)}+16 u_3^{(5)}
\end{array}
\right. $}  \\
\Leftrightarrow & \qquad 
\mbox{ $\left\{
\begin{array}{rl}
0 \!\! &=\, 45 \mu +60 u_3  \left(u_2' -u_3 \right)+20 u_2  u_2'' +\frac{5}{3} u_2^3+16 u_2^{(4)} \cr
\ds \frac{9 t}{\beta_{3,5} } \!\! &=\, 20 \left(2 u_2' -3 u_3 \right) u_3' +5 u_3  \left(8 u_2'' +u_2^2\right)  +20 u_2 u_3'' +16 u_3^{(4)} 
\end{array}
\right. $}. \label{Eqn:Appendix:StringEquationCBPQ35equation}
\end{align}
\end{itemize}

\subsubsection{1) $(r,s)=(1,1)$ FZZT-Cardy brane:  $W^{(1,1)} = W^{(1)}[\psi]$}

\begin{align}
\vec{W}^{(1)} &= \Bigl( W^{(1)}_\varnothing, W^{(1)}_{\tiny \yng(1)}, W^{(1)}_{\tiny \yng(2)} \Bigr)^{\rm T}, \\
\mathcal B^{(1,1)} &= 
\bordermatrix{
 & \varnothing & {\tiny \yng(1)} & {\tiny \yng(2)} \cr
\varnothing &  & 1 & \cr
{\tiny \yng(1)} & & & 1 \cr
{\tiny \yng(2)} &\bar \zeta- \bar u_3 & -\bar u_2 & 
}, 
\\
\mathcal Q^{(1,1)} &= \bar \beta_{3,5} \times
\bordermatrix{
 & \varnothing & {\tiny \yng(1)} & {\tiny \yng(2)} \cr
\varnothing & \frac{2\bar u_2}{3} \bar \zeta + \frac{4}{9} \bar u_2 \bar u_3 + \frac{1}{9} \bar u_3'' & -\frac{\bar u_2^2}{9} - \frac{\bar u_3'}{3} + \frac{\bar u_2''}{9} & \bar \zeta + \frac{2 \bar u_3-\bar u_2'}{3} \cr
{\tiny \yng(1)} & \bar \zeta^2 - \frac{\bar u_3-\bar u_2'}{3} \bar \zeta + f_3^{(3)} &  - \frac{\bar u_2}{3} \bar \zeta + f_1^{(3)} & -\frac{\bar u_2^2}{9} + \frac{\bar u_3'}{3} - \frac{2 \bar u_2''}{9} \cr
{\tiny \yng(2)} & \frac{\bar u_2''-\bar u_2^2}{9} \bar \zeta + f_5^{(3)} & \bar \zeta^2 - \frac{\bar u_3}{3} \bar \zeta + f_{4}^{(3)} & -\frac{\bar u_2}{3} \bar \zeta + f_2^{(3)}
}. 
\end{align}

\subsubsection{2) $(r,s)=(2,1)$ FZZT-Cardy brane:  $W^{(2,1)} = W^{(2)}[\psi]$}

 $\vec{W}^{(2)} = \Bigl( W^{(2)}_\varnothing, W^{(2)}_{\tiny \yng(1)}, W^{(2)}_{\tiny \yng(1,1)} \Bigr)^{\rm T}$.
\begin{align}
\mathcal B^{(2,1)} &= 
\bordermatrix{
 & \varnothing &{\tiny \yng(1)} & {\tiny \yng(1,1)}  \cr
\varnothing &  & 1 & \cr
{\tiny \yng(1)} & -\bar{u}_2 & & 1 \cr
{\tiny \yng(1,1)} &-\bar \zeta + \bar{u}_3 &  & 
}, 
\\
\mathcal Q^{(2,1)} &= \bar \beta_{3,5} \times
\bordermatrix{
 & \varnothing &{\tiny \yng(1)} & {\tiny \yng(1,1)} \cr
\varnothing & \frac{\bar u_2}{3} \bar \zeta - f_2^{(3)} & -\frac{\bar u_2^2}{9} + \frac{\bar u_3'}{3} - \frac{2 \bar u_2''}{9} &  - \bar \zeta - \frac{2 \bar u_3- \bar u_2'}{3} \cr
{\tiny \yng(1)} & \bar \zeta^2 - \frac{\bar u_3}{3} \bar \zeta + f_4^{(3)} &\frac{\bar u_2}{3} \bar \zeta - f_1^{(3)} & -\frac{\bar u_2^2}{9} - \frac{\bar u_3'}{3} +\frac{\bar u_2''}{9} \cr
{\tiny \yng(1,1)} & - \frac{\bar u_2'' - \bar u_2^2}{9} \bar \zeta -f_5^{(3)} &\bar \zeta^2 - \frac{\bar u_3 - \bar u_2'}{3} \bar \zeta +  f_3^{(3)} & -\frac{2 \bar u_2}{3}\bar  \zeta - \frac{4}{9} \bar u_2 \bar u_3 - \frac{1}{9} \bar u_3''
}, 
\end{align}
where $\{f_n^{(3)}\}_{n=1}^{5}$ are given by 
\begin{align}
f_{1}^{(3)} & \equiv \frac{1}{9} \bar u_2''' - \frac{2}{9} \bar u_3'' +\frac{1}{9} \bar u_2 \bar u_2' - \frac{2}{9} \bar u_2 \bar u_3, \quad
f_{2}^{(3)}  \equiv -\frac{1}{9} \bar u_2''' + \frac{1}{9} \bar u_3'' - \frac{1}{9} \bar u_2 \bar u_2' - \frac{2}{9} \bar u_2 \bar u_3, \\
f_{3}^{(3)} & \equiv \frac{1}{9} \bar u_3''' + \frac{4}{9} \bar u_2 \bar u_3' + \frac{7}{9} \bar u_3 \bar u_2' -\frac{2}{3} \bar u_3^2, \\
f_{4}^{(3)} & \equiv \frac{1}{9} \bar u_2'''' - \frac{1}{9} \bar u_3''' + \frac{1}{3} \bar u_2 \bar u_2'' - \frac{1}{9} \bar u_2 \bar u_3' + \frac{1}{9} (\bar u_2')^2 + \frac{5}{9} \bar u_3 \bar u_2'- \frac{2}{3} \bar u_3^2+ \frac{1}{9} \bar u_2^3, \\
f_{5}^{(3)} & \equiv \frac{1}{9} \bar u_3'''' + \frac{4}{9} \bar u_2 \bar u_3'' + \bar u_3 \bar u_2'' + \frac{11}{9} \bar u_2' \bar u_3' - \frac{5}{3} \bar u_3 \bar u_3'+ \frac{1}{9} \bar u_2^2 \bar u_3. 
\end{align}


\subsubsection{4) $(r,s)=(1,3)$ FZZT-Cardy brane:  $W^{(1,3)} = \mathcal L \widetilde W^{(2)}[\chi]$}


\begin{align}
& \quad\,\, \,\, \vec{W}^{(1,3)} = \Bigl( \mathcal L \widetilde W^{(2)}_\varnothing, \mathcal L \widetilde W^{(2)}_{\tiny \yng(1)}, \mathcal L \widetilde W^{(2)}_{\tiny \yng(2)}, \mathcal L \widetilde W^{(2)}_{\tiny \yng(1,1)}, \mathcal L \widetilde W^{(2)}_{\tiny \yng(2,1)}, \mathcal L \widetilde W^{(2)}_{\tiny \yng(3,1)} \Bigr)^{\rm T}, \\
&\mathcal B^{(1,3)}(t;\zeta)  =  \nn\\
&  = \mbox{ \scriptsize $\bordermatrix{
 & \mbox{\normalsize $\varnothing$} & {\tiny \yng(1)} & {\tiny \yng(2)} & {\tiny \yng(1,1)} & {\tiny \yng(2,1)} & {\tiny \yng(3,1)}\cr
\mbox{\normalsize $\varnothing$} &    0 & 1 & 0 & 0 & 0 & 0 \cr
 {\tiny \yng(1)} &  0 & 0 & 1 & 1 & 0 & 0 \cr
 {\tiny \yng(2)} &  - \frac{\zeta }{4}+\frac{2 u_3-3 u_2'}{4}  & -\frac{1}{4} u_2(t) & 0 & 0 & 2 & 0 \cr
{\tiny \yng(1,1)} &   0 & 0 & 0 & 0 & 1 & 0 \cr
{\tiny \yng(2,1)} &  \frac{-12 u_3'+8 u_2''-5 u_2^2}{144} & \frac{\zeta }{4} + \frac{2 u_2'-u_3}{12}  & -\frac{u_2}{6}  & \frac{u_2}{4} & 0 & 1 \cr
{\tiny \yng(3,1)} &  \frac{u_2}{12} \zeta + \frac{4 u_2 u_2'-15 u_3''-6 u_2 u_3+13 u_2^{(3)}}{36}   & \frac{-6 u_3'+14 u_2''+u_2^2}{72} &  \frac{\zeta }{2}-\frac{u_3}{6} &  \frac{\zeta }{4}+\frac{4 u_3-5 u_2'}{12} & - \frac{5 u_2}{12}  & 0 
}$} , \\
& \mathcal Q^{(1,3)}(t;\zeta)  =\beta_{5,3} \times \biggl[   \nn\\
 &  
\!\!\!\!\! \!\!\!\!
\mbox{\scriptsize $ \bordermatrix{
 & \mbox{\normalsize $\varnothing $} & {\tiny \yng(1)} & {\tiny \yng(2)} & {\tiny \yng(1,1)} & {\tiny \yng(2,1)} & {\tiny \yng(3,1)}\cr
\mbox{\normalsize $\varnothing $}
&  -  \frac{2 u_2}{3} \zeta   +  f_{15} 
& \frac{4 u_2''-12 u_3' -u_2^2}{9} 
& -4 \zeta +\frac{4 u_3}{3} 
& 0 
& -\frac{4 u_2 }{3} 
& 0 \cr
 {\tiny \yng(1)} 
& \zeta ^2 - \frac{7 (u_3 -u_2' )}{3}  \zeta + f_{9}
& - \frac{2 (u_2 u_2' +2 u_3'' ) }{9} 
& \frac{u_2^2+4 u_2''}{9}
&-  \frac{4 (u_2^2+3 u_3'-u_2'' ) }{9} 
& - 8 \zeta +\frac{8 u_3- 4 u_2'}{3}
& - \frac{4 u_2}{3}  \cr
 {\tiny \yng(2)} 
& - f_{22} \zeta  + f_4
& -\zeta ^2 -  f_{27} \zeta + f_{10}
&  u_2 \zeta  + f_{16}
& - \frac{4 u_2}{3} \zeta   + f_{17}
& \frac{3 u_2^2+12 u_3' -4 u_2''}{9}
& - 4 \zeta +\frac{4(u_3 -u_2' )}{3}  \cr
{\tiny \yng(1,1)} 
& - f_{23} \zeta  + f_5
& -\frac{u_2 (u_2^2+12 u_3'-4 u_2'') }{108} 
& - \frac{u_2}{3} \zeta  +\frac{u_3 u_2 }{9} 
& - u_2 \zeta   + f_{18}
& \frac{4 u_2'' }{9} 
& - 4 \zeta +\frac{4(u_3 -u_2' )}{3}  \cr
{\tiny \yng(2,1)} 
&   -\frac{u_2}{4} \zeta ^2 - f_{20} \zeta + f_2 
& -  f_{24} \zeta  + f_{6}
& -2 \zeta ^2 - f_{28} \zeta + f_{11}
& -\zeta ^2 -  f_{29} \zeta + f_{12}
& - \frac{4 (u_3'' -u_2^{(3)})}{9} 
& \frac{4(3 u_3'-2 u_2'' )}{9}  \cr
{\tiny \yng(3,1)} 
&  \frac{\zeta ^3}{4}+ f_{30} \zeta ^2 -  f_{14} \zeta
+ f_1
& \frac{u_2}{12} \zeta ^2 
-  f_{21} \zeta 
+ f_{3}
& - f_{25} \zeta  + f_{7}
& -  f_{26} \zeta  + f_{8}
& -3 \zeta ^2 + \frac{u_3}{3}  \zeta + f_{13}
& \frac{2 u_2}{3} \zeta  + f_{19}
}$} \biggr], 
\end{align}
where 
\begin{align}
f_{1} &= -\frac{t u_2 }{48 \beta_{3,5} } + \frac{1}{1296} \Bigl\{  -u_3 \left(405 \mu +80 u_2 \left(3 u_3' -u_2'' \right)  -708 (u_2')^2+11 u_2^3+192 u_3^{(3)}\right)+  \nn\\
&\qquad \qquad \qquad  +u_2' \left(405 \mu +20 u_2 \left(6 u_3' +5 u_2'' \right)-6 u_2^3+240 u_3^{(3)} \right) - \nn\\
&\quad -8 \left(6 u_2^{(3)} u_3' +u_2^2 \bigl(3 u_3'' +u_2^{(3)} \bigr)-6 u_2'' u_3'' \right) -1092 u_3^2 u_2' -72 (u_2')^3+468 u_3^3 \Bigr\}, \\
f_{2} &= \frac{1}{4 \beta _{3,5}}+\frac{1}{1296} \Bigl\{ -3 u_2  \Bigl(45 \mu -40 u_3 u_2' +8 (u_2')^2+12 u_3^2\Bigr)+4 u_2^2 \bigl(u_2'' -9 u_3' \bigr) + \nn\\
&\quad \quad   +16 \left(-48 u_3' u_2''-3 \bigl(2 u_3 -3 u_2' \bigr) \bigl(3 u_3'' -2 u_2^{(3)} \bigr)  +27 (u_3')^2+20 (u_2'')^2\right)-5 u_2^4 \Bigr\}, \\
f_{3} &= \frac{1}{4 \beta _{3,5}}+\frac{1}{1296}\Bigl\{ -3 u_2 \left(45 \mu +52 u_3 u_2' -8 (u_2')^2-44 u_3^2+16 u_3^{(3)} \right) + \nn\\
&\qquad \qquad \quad  +8 u_2^2 \left(3 u_3' -8 u_2'' \right)+16 \left(-39 u_3'  u_2'' +27 (u_3')^2+10 (u_2'')^2\right)-2 u_2^4 \Bigr\}, \\
f_{4} &= \frac{t}{4 \beta _{3,5}}+\frac{ 8 u_2'  \bigl(7 u_2'' -9 u_3' \bigr)+u_3  \bigl(84 u_3' -68 u_2'' \bigr)  +u_2^2 \bigl(3 u_2' -7 u_3 \bigr)-4 u_2  \bigl(6 u_3'' -5 u_2^{(3)} \bigr)}{108}  , \\
f_{5} &= \frac{t}{4 \beta _{3,5}}+\frac{ 12 \left(3 u_3 -2 u_2' \right) \left(u_3' -u_2'' \right)+u_2^2 \left(7 u_2' -3 u_3 \right)+4 u_2^{(3)} u_2 }{108} , \\
f_{6} &= \frac{t}{4 \beta _{3,5}}+\frac{ 2 \Bigl(2 u_2'  \bigl(u_2'' -3 u_3' \bigr)+u_3  \bigl(6 u_3' -3 u_2'' \bigr)\Bigr)+u_2^2 \bigl(u_2' -2 u_3 \bigr)+u_2  \bigl(4 u_2^{(3)} -6 u_3'' \bigr) }{54} , \\
f_{7} &= \frac{t}{4 \beta _{3,5}}+\frac{-4 \Bigl(9 u_3 \bigl(u_3' -u_2'' \bigr)+u_2'  \bigl(u_2'' -3 u_3' \bigr)\Bigr)+u_2^2 \bigl(u_2' -7 u_3 \bigr)+4 u_2  \bigl(u_2^{(3)} -3 u_3'' \bigr)}{108} , \\
f_{8} &= \frac{1}{27} \left(4 u_2' u_2^2+12 u_3'' u_2 -4 u_2^{(3)} u_2-u_2^2 u_3-6 u_3 u_3' +3 u_2' u_3' +2 u_3 u_2'' -u_2' u_2''\right), \\
f_{9} &= \frac{ 3 \left(45 \mu +32 u_3 u_2' +8 (u_2')^2-36 u_3^2+32 u_3^{(3)}\right)+4 u_2 \left(9 u_3'+19 u_2'' \right)+10 u_2^3 }{108} , \\
f_{10} &= \frac{3 \left(45 \mu +52 u_3 u_2' -8 (u_2')^2-44 u_3^2+16 u_3^{(3)} \right)+u_2 \left(60 u_3' +8 u_2'' \right)+6 u_2^3 }{108} , \\
f_{11} &= \frac{1}{108} \left(-u_2^3-4 u_2'' u_2-24 u_3^2+36 u_3 u_2'\right), \\
f_{12} &= \frac{1}{108} \left(-u_2^3+24 u_3' u_2 -44 u_2'' u_2+48 u_3^2-72 u_3 u_2' -48 u_3^{(3)} \right), \\
f_{13} &= \frac{-3 \left(45 \mu +80 u_3  u_2' -8 (u_2')^2-68 u_3^2+32 u_3^{(3)} \right)-12 u_2 \left(2 u_3' +7 u_2'' \right)-7 u_2^3}{108} , \\
f_{14} &= \frac{ -405 \mu +u_2  \left(80 u_2'' -300 u_3' \right)+48 (u_2')^2-444 u_3 u_2' -21 u_2^3+384 u_3^2-192 u_3^{(3)} }{432} , \\
f_{15} &= \frac{2}{9} \left(u_2 u_3 +u_2 u_2' +4 u_3'' -2 u_2^{(3)} \right), \quad
f_{16} = \frac{-3 u_2 u_3 +2 u_2 u_2' -4 u_3'' +4 u_2^{(3)}}{9}, \\
f_{17} &= \frac{1}{9} \left(-u_2 u_3 -3 u_2 u_2'-12 u_3'' +4 u_2^{(3)}\right), \qquad
f_{18} = \frac{3 u_2 u_3 -5 u_2 u_2' -4 u_3''}{9}, \\
f_{19} &= \frac{1}{9} \left(-2 u_2 u_3+3 u_2 u_2'+8 u_3''-4 u_2^{(3)}\right), \qquad 
f_{20} = \frac{-21 u_2 u_3 +15 u_2 u_2' -24 u_3'' +16 u_2^{(3)} }{36}, \\
f_{21} &= \frac{u_2 u_2'}{18}  -\frac{u_2 u_3}{12}, \qquad
f_{22} = \frac{-7 u_2^2+24 u_3' -28 u_2''}{36}, \qquad
f_{23} = \frac{1}{18} u_2^2+u_3' -u_2'', \\
f_{24} &= \frac{u_2^2+24 u_3' -12 u_2''}{36}, \quad
f_{25} = -\frac{1}{18} u_2^2-u_3'+u_2'', \quad
f_{26} = -\frac{1}{9} u_2^2+u_3'-\frac{1}{3} u_2'', \\
f_{27} &= u_3-\frac{2}{3} u_2', \qquad
f_{28} = -\frac{4}{3} u_3 + u_2', \qquad
f_{29} = u_3-\frac{u_2'}{3}, \qquad
f_{30} = \frac{ 9 u_2'-8 u_3 }{12}. 
\end{align}


\subsubsection{5) $(r,s)=(1,2)$ FZZT-Cardy brane:  $W^{(1,2)} = \mathcal L \widetilde W^{(3)}[\chi]$}


\begin{align}
& \quad\,\, \,\, \vec{W}^{(1,2)} = \Bigl( \mathcal L \widetilde W^{(3)}_\varnothing, \mathcal L \widetilde W^{(3)}_{\tiny \yng(1)}, \mathcal L \widetilde W^{(3)}_{\tiny \yng(2)}, \mathcal L \widetilde W^{(3)}_{\tiny \yng(1,1)}, \mathcal L \widetilde W^{(3)}_{\tiny \yng(2,1)}, \mathcal L \widetilde W^{(3)}_{\tiny \yng(2,1,1)} \Bigr)^{\rm T}, \\
&\mathcal B^{(1,2)}(t;\zeta) = \mbox{ \scriptsize $\bordermatrix{
 & \mbox{\normalsize $\varnothing$} & {\tiny \yng(1)} & {\tiny \yng(2)} & {\tiny \yng(1,1)} & {\tiny \yng(2,1)} & {\tiny \yng(2,1,1)}\cr
\mbox{\normalsize $\varnothing$} &    0 & 1 & 0 & 0 & 0 & 0 \cr
 {\tiny \yng(1)} &  0 & 0 & 1 & 1 & 0 & 0 \cr
 {\tiny \yng(2)} &  \frac{5 (u_3 -2 u_2')}{12}  & - \frac{5 u_2}{12}   & 0 & 0 & 1 & 0 \cr
{\tiny \yng(1,1)} &  -  \frac{\zeta }{4}+\frac{ u_3 -2 u_2' }{3}  & \frac{u_2}{6} & 0 & 0 & 2 & 0 \cr
{\tiny \yng(2,1)} &  \frac{2 u_2'' - u_3' }{3}    &  \frac{\zeta }{4}+\frac{ 3 u_2'  - 2 u_3}{6} & -\frac{u_2}{6}  & -\frac{u_2}{6}  & 0 & 1 \cr
{\tiny \yng(2,1,1)} & - \frac{ u_2}{48} \zeta +\frac{-u_2 u_3 +2 u_2 u_2' +10 u_3'' -8 u_2^{(3)} }{24}  & \frac{ 4 u_3' -8 u_2'' -u_2^2 }{48}  &  \frac{\zeta }{4} + \frac{u_2' -u_3 }{3} &  \frac{\zeta }{2} + \frac{u_3 -u_2' }{6} & 0 & 0
}$} , \\
& \mathcal Q^{(1,2)}(t;\zeta)  =\beta_{5,3} \times \biggl[   \nn\\
 &  
\!\!\!\!\! \!\!\!\!
\mbox{\scriptsize $ \bordermatrix{
 & \mbox{\normalsize $\varnothing $} & {\tiny \yng(1)} & {\tiny \yng(2)} & {\tiny \yng(1,1)} & {\tiny \yng(2,1)} & {\tiny \yng(2,1,1)}\cr
\mbox{\normalsize $\varnothing $}
&   - u_2 \zeta      + \tilde f_{15}
& \frac{u_2^2-12 u_3'+8 u_2''}{9}
& 0 
& 4 \zeta   +  \frac{4(u_3-u_2')}{3} 
& \frac{4 u_2}{3} 
& 0 \cr
 {\tiny \yng(1)} 
& -\zeta ^2    -  \frac{10 u_2' - 3 u_3}{3} \zeta     +\tilde f_9 
& \frac{2 (u_2 u_2' -2 u_3''+2 u_2^{(3)})}{9} 
& \frac{-u_2^2-12 u_3' +8 u_2'' }{9} 
& \frac{-u_2^2-4 u_2'' }{9}  
& 8 \zeta     +\frac{8 u_3 -4  u_2' }{3} 
& \frac{4 u_2(t)}{3}  \cr
 {\tiny \yng(2)} 
&  - \tilde f_{22} \zeta     +\tilde f_{4} 
& -\frac{ u_2 \left(u_2^2-12 u_3' +8 u_2''\right) }{27}
& - \frac{2  u_2}{3} \zeta     +\tilde f_{16}
& - \frac{4 u_2}{3} \zeta      -\frac{4u_2 (u_3- u_2')}{9} 
& \frac{-5 u_2^2-4 u_2''}{9} 
& 4 \zeta  +  \frac{4 u_3}{3}  \cr
{\tiny \yng(1,1)} 
& - \tilde f_{23} \zeta      +\tilde  f_5 
& \zeta ^2    - \tilde f_{27} \zeta     +\tilde  f_{10} 
& - \frac{u_2}{3} \zeta      + \tilde f_{17} 
&  \frac{2  u_2 }{3} \zeta      + \tilde f_{18}
& \frac{2(u_2^2+6 u_3'-4 u_2'')}{9} 
& 4 \zeta +  \frac{4 u_3}{3} \cr
{\tiny \yng(2,1)} 
&    -\frac{u_2}{6}  \zeta ^2    -  \tilde f_{20} \zeta     + \tilde f_2
& - \tilde  f_{24} \zeta     +\tilde  f_6 
& \zeta ^2    -  \tilde f_{28} \zeta     +\tilde  f_{11}
& 2 \zeta ^2    - \tilde  f_{29} \zeta     + \tilde f_{12} 
& \frac{-5 u_2 \left(u_3-u_2'\right)-4 u_3'' }{9} 
& \frac{4 (3 u_3'-u_2'')}{9}  \cr
{\tiny \yng(2,1,1)} 
&  - \frac{\zeta ^3}{4}      + \tilde f_{30} \zeta ^2   -  \tilde f_{14} \zeta    + \tilde  f_1 
& -\frac{ u_2}{12} \zeta ^2   -  \tilde f_{21}\zeta   +  \tilde f_3 
& - \tilde f_{25}\zeta    +  \tilde f_7 
&  - \tilde f_{26}\zeta   + \tilde  f_8 
& 3 \zeta ^2  - \frac{u_2' -u_3 }{3} \zeta  +  \tilde f_{13}
& u_2 \zeta  +  \tilde f_{19}
}$} \biggr], 
\end{align}
where 
\begin{align}
\tilde f_{1} &=-\frac{t u_2 }{48 \beta _{3,5}} + \frac{1}{648} \Bigl\{ -u_3 \left(6 \left(45 \mu -94 (u_2')^2+24 u_3^{(3)} \right)+60 u_2  \left(u_3' +u_2'' \right)+13 u_2^3\right) + \nn\\
&\qquad \qquad \qquad \qquad  +3 u_2'  \left(135 \mu +20 u_2  \left(u_3' +3 u_2'' \right)+7 u_2^3+88 u_3^{(3)}\right) + \nn\\
&\quad + 3 \left(-12 (u_2')^3-8 u_2^{(3)} u_3' +8 u_2''  u_3'' +u_2^2 \left(2 u_2^{(3)} -4 u_3'' \right)\right)-756 u_3^2 u_2' +264 u_3^3\Bigr\}, \\
\tilde f_{2} &= \frac{1}{4 \beta _{3,5}}+\frac{1}{324} \Bigl\{ -3 u_2 \left(5 \bigl(9 \mu +2 (u_2')^2+4 u_3^{(3)}\bigr)+26 u_3 u_2' -38 u_3^2\right)  -3 u_2^2 \left(13 u_3' +9 u_2'' \right) - \nn\\
&\qquad \qquad -36 \left(-4 u_3'  u_2'' +\left(u_3 -2 u_2' \right) \left(3 u_3'' -u_2^{(3)} \right)+3 (u_3')^2+(u_2'')^2\right)-5 u_2^4\Bigr\}, \\
\tilde f_{3} &= \frac{1}{4 \beta _{3,5}}+\frac{1}{432} \Bigl\{ 4 u_2^2 \left(2 u_3'-5 u_2''\right)-48 \left(-5 u_3' u_2''+3 (u_3')^2+2 (u_2'')^2\right)+ \nn\\
&\qquad \qquad +8 u_2 \left(-3 u_3 u_2'+2 u_3^2-2 u_3^{(3)} \right)-u_2^4 \Bigr\}, \\
\tilde f_{4} &= \frac{t}{4 \beta _{3,5}}+\frac{4 \Bigl(u_3  \big(5 u_2'' -9 u_3' \bigr)+u_2'  \bigl(3 u_3' +5 u_2'' \bigr)\Bigr)+u_2^2 \bigl(31 u_2' -23 u_3 \bigr)+u_2  \bigl(36 u_2^{(3)} -60 u_3'' \bigr) }{108}  , \\
\tilde f_{5} &= \frac{t}{4 \beta _{3,5}}+\frac{4 \Bigl(u_3  \left(2 u_2'' -3 u_3' \right)+u_2'  \left(5 u_2'' -6 u_3' \right)\Bigr)+u_2^2 \left(7 u_2' -11 u_3 \right)-12 u_2  \bigl(u_3'' -u_2^{(3)} \bigr) }{54} , \\
\tilde f_{6} &= \frac{t}{4 \beta _{3,5}}+\frac{4 \Bigl(u_2'  \bigl(11 u_2'' -15 u_3' \bigr)+u_3 \bigl(9 u_3' -7 u_2'' \bigr)\Bigr)+u_2^2 \bigl(7 u_2' -9 u_3 \bigr)+4 u_2  \bigl(u_2^{(3)} -3 u_3'' \bigr)}{108}, \\
\tilde f_{7} &= \frac{4 \bigl(2 u_3 -u_2' \bigr) \bigl(3 u_3' -2 u_2'' \bigr)+u_2^2 \bigl(3 u_2' -4 u_3 \bigr)+u_2  \bigl(8 u_2^{(3)} -12 u_3'' \bigr) }{108}, \\
\tilde f_{8} &= \frac{t}{4 \beta _{3,5}}+\frac{ 4 \Bigl(2 u_2' \bigl(4 u_2'' -3 u_3' \bigr)+u_3  \bigl(9 u_3' -10 u_2'' \bigr)\Bigr)+u_2^2 \bigl(6 u_2' -7 u_3 \bigr)+u_2  \bigl(8 u_2^{(3)} -12 u_3'' \bigr) }{108}, \\
\tilde f_{9} &= \frac{3 \left(45 \mu +20 u_3  u_2' +16 (u_2')^2-44 u_3^2+32 u_3^{(3)} \right)-12 u_2 \left(2 u_3' -9 u_2'' \right)+5 u_2^3}{108} , \\
\tilde f_{10} &= \frac{6 \left(3 u_3  u_2' -2 u_3^2+2 u_3^{(3)} \right)+8 u_2 u_2'' +u_2^3 }{27}, \\
\tilde f_{11} &= \frac{-3 \left(45 \mu +52 u_3 u_2' -8 (u_2')^2-44 u_3^2+16 u_3^{(3)} \right)-4 u_2 \left(9 u_3' +5 u_2'' \right)-4 u_2^3 }{108} , \\
\tilde f_{12} &= \frac{-36 \Bigl(-3 u_3  u_2' +2 (u_2')^2+u_3^2\Bigr)+4 u_2 u_2'' +u_2^3}{108} , \\
\tilde f_{13} &= \frac{ -3 \left(45 \mu +64 u_3 u_2' -4 (u_2')^2-52 u_3^2+32 u_3^{(3)}\right)-24 u_2 \left(u_3' +3 u_2'' \right)-8 u_2^3}{108}, \\
\tilde f_{14} &= \frac{ -3 \left(45 \mu +92 u_3  u_2' +16 (u_2')^2-92 u_3^2+64 u_3^{(3)} \right)+20 u_2 \left(12 u_3' -19 u_2'' \right)-4 u_2^3 }{432}, \\
\tilde f_{15} &= \frac{2}{9} \left(u_2 \left(u_3 -2 u_2' \right)+4 u_3'' -2 u_2^{(3)} \right), \qquad
\tilde f_{16} = - \frac{2}{9}  \left(u_2  \left(u_3 -u_2' \right)+2 \bigl(u_3'' -u_2^{(3)}\bigr)\right), \\
\tilde f_{17} &= - \frac{2}{9}  \left(u_2  \left(3 u_3 -2 u_2' \right)+6 u_3'' -4 u_2^{(3)} \right), \qquad
\tilde f_{18} = \frac{2}{9} \Bigl(u_2  \left(u_3 -2 u_2' \right)-2 u_3'' \Bigr), \\
\tilde f_{19} &= \frac{u_2  \left(3 u_3 -u_2' \right)+8 u_3'' -4 u_2^{(3)} }{9}, \qquad 
\tilde f_{20} = \frac{ 8 \bigl(u_2^{(3)} -3 u_3'' \bigr)-u_2  \left(u_3 -6 u_2' \right) }{36} , \\
\tilde f_{21} &= \frac{u_2 \left(u_2' -3 u_3 \right)}{36} , \qquad
\tilde f_{22} = u_3' -\frac{10 u_2''}{9}  -\frac{7 u_2^2}{36}, \qquad
\tilde f_{23} = \frac{2  u_3'}{3} -u_2''+\frac{7 u_2^2}{36} , \\
\tilde f_{24} &= \frac{ 24 u_3' -12 u_2'' -u_2^2 }{36}, \quad
\tilde f_{25} = \frac{ 9 u_3' -6 u_2'' +u_2^2 }{9}, \quad
\tilde f_{26} =\frac{ -18 u_3' +20 u_2'' +u_2^2 }{18}, \\
\tilde f_{27} &= u_3 -\frac{u_2'}{3}, \qquad
\tilde f_{28} = u_3 -\frac{2  u_2'}{3}, \qquad
\tilde f_{29} = \frac{41 u_2'}{3}  -3 u_3, \qquad
\tilde f_{30} =\frac{7 u_3}{12}-\frac{4 u_2'}{3}. 
\end{align}


\subsubsection{Duality matrices}

\noindent\underline{Reflection:} \qquad $\vec W^{(2)}[\psi_{\mathcal C}](t;- \zeta) = {\mathsf M}^{(1)}(t)\times  \vec W^{(1)}[\psi](t;\zeta)$
\begin{align}
{\mathsf M}^{(1)}(t) = 
\bordermatrix{
 & \varnothing & {\tiny \yng(1)} & {\tiny \yng(2)} \cr
\varnothing & 1 & 0 & 0  \cr
{\tiny \yng(1)} &  0 & 1 & 0 \cr
{\tiny \yng(1,1)} &    \bar u_2  & 0 & 1
}. 
\end{align}

\noindent\underline{Reflection:} \qquad $\vec W^{(1,3)}[\chi_{\mathcal C}](t; \zeta) = {\mathsf N}^{(2)}(t)\times  \vec W^{(1,2)}[\chi](t;\zeta)$
\begin{align}
{\mathsf N}^{(2)}(t) = \bordermatrix{
 & \varnothing & {\tiny \yng(1)} &   {\tiny \yng(2)} &  {\tiny \yng(1,1)} &   {\tiny \yng(2,1)} &  {\tiny \yng(2,1,1)} \cr
\varnothing &  1 & 0 & 0 & 0 & 0 & 0  \cr
{\tiny \yng(1)} &  0 & 1 & 0 & 0 & 0 & 0  \cr
{\tiny \yng(2)} &    -\frac{5 u_2}{12}  & 0 & 0 & 1 & 0 & 0   \cr
 {\tiny \yng(1,1)} &   \frac{5 u_2}{12} & 0 & 1 & 0 & 0 & 0   \cr
 {\tiny \yng(2,1)} &   \frac{5 ( u_3- u_2' )}{12} & 0 & 0 & 0 & 1 & 0 \cr
 {\tiny \yng(3,1)} &   \frac{10  u_2'' - 5 u_2^2}{36} & 0 & -\frac{5 u_2}{12}  & 0 & 0 & 1 \cr
}. 
\end{align}

\noindent\underline{Charge conjugation:} \qquad $\Bigl[\vec W^{(2)}[\psi](t;\zeta)\Bigr]^{\rm T}\,\, \mathfrak C^{(1,1)}_{[\psi]}(t)\, \, \Bigl[ \vec W^{(1)}[\psi](t;\zeta) \Bigr] $
\begin{align}
\mathfrak C^{(1,1)}_{[\psi]}(t) = 
\bordermatrix{
 & \varnothing & {\tiny \yng(1)} & {\tiny \yng(2)} \cr
\varnothing & 0 & 0 & 1  \cr
{\tiny \yng(1)} &  0 & -1 & 0 \cr
{\tiny \yng(1,1)} &   1 & 0 & 0
}. 
\end{align}

\noindent\underline{Dual charge conjugation:} \qquad $\Bigl[\vec W^{(1,3)}[\chi](t;-\zeta)\Bigr]^{\rm T}\,\, \mathfrak D^{(1,2)}_{[\chi]}(t,\zeta)\, \, \Bigl[ \vec W^{(1,2)}[\chi](t;\zeta) \Bigr] $
\begin{align}
\mathfrak D^{(1,2)}_{[\chi]}(t,\zeta) = 
 \bordermatrix{
 & \varnothing & {\tiny \yng(1)} &   {\tiny \yng(2)} &  {\tiny \yng(1,1)} &   {\tiny \yng(2,1)} &  {\tiny \yng(2,1,1)} \cr
\varnothing &   \frac{5 u_2^2-2 u_2''}{72}  & - \frac{\zeta}{2}+  \frac{ -2 u_3+3 u_2'}{12}  & \frac{u_2}{6} & -\frac{u_2}{12}  & 0 & 1 \cr
{\tiny \yng(1)} &   \frac{\zeta}{2}+ \frac{ -u_3 +2 u_2' }{4}  & -\frac{u_2}{12}  & 0 & 0 & -1 & 0  \cr
{\tiny \yng(2)} &  \frac{u_2}{3} & 0 & 1 & 0 & 0 & 0  \cr
 {\tiny \yng(1,1)} &   -\frac{u_2}{4}  & 0 & -1 & 1 & 0 & 0   \cr
 {\tiny \yng(2,1)} &   0 & -1 & 0 & 0 & 0 & 0 \cr
 {\tiny \yng(3,1)} &    1 & 0 & 0 & 0 & 0 & 0 \cr
}. 
\end{align}


\subsection{The case of $(p,q)=(5,3)$  }

\begin{itemize}
\item [I. ] Kac table: 
\begin{align}
 \begin{tabular}{|c|c|c|c|c|} \hline
  $(1,2)$ & $(2,2)$ & $(3,2)$ & $(4,2)$  \\ \hline
 $ \bf (1,1)$ & $ \bf (2,1)$ & $ \bf (3,1)$ & $ \bf (4,1)$   \\ \hline
    \end{tabular}
\end{align}
\item [II. ] The coefficients: Eq.~\eq{Eqn:Appendix:CoefficientsStringEquationCBPQ35}
\item [III. ] The Baker-Akhiezer systems: Eq.~\eq{Eqn:Appendix:BAsystemCBPQ35equation}
\item [IV. ] String equations: Eq.~\eq{Eqn:Appendix:StringEquationCBPQ35equation}
\end{itemize}

\subsubsection{1) $(r,s)=(1,1)$ FZZT-Cardy brane:  $\widetilde W^{(1,1)} = \widetilde W^{(1)}[\chi]$}

\begin{align}
\vec{W}^{(1)} &= \Bigl( \widetilde W^{(1)}_\varnothing, \widetilde W^{(1)}_{\tiny \yng(1)}, \widetilde W^{(1)}_{\tiny \yng(2)}, \widetilde W^{(1)}_{\tiny \yng(3)}, \widetilde W^{(1)}_{\tiny \yng(4)} \Bigr)^{\rm T}, \\
\mathcal B^{(1,1)} &= 
\bordermatrix{
 & \varnothing & {\tiny \yng(1)} & {\tiny \yng(2)} & {\tiny \yng(3)} & {\tiny \yng(4)}\cr
\varnothing &  & 1 & \cr
{\tiny \yng(1)} & & & 1 \cr
{\tiny \yng(2)} &  & & & 1 \cr 
{\tiny \yng(3)} &  & & & & 1 \cr 
{\tiny \yng(4)} &\bar \eta- w_5 & - w_4 &  - w_3 &  - w_2 &\cr
}, \\
\mathcal Q^{(1,1)} &= \bar \beta_{5,3}\times     \nn\\
\times \biggl[ &\bordermatrix{
 & \varnothing & {\tiny \yng(1)} & {\tiny \yng(2)} & {\tiny \yng(3)} & {\tiny \yng(4)}\cr
\varnothing &   \bar{u}_2'-\bar{u}_3 &  \bar{u}_2 & 0 & 1 & 0 \cr 
{\tiny \yng(1)} & \bar{u}_2''-\bar{u}_3' & 2 \bar{u}_2'-\bar{u}_3 &  \bar{u}_2 & 0 & 1 \cr 
{\tiny \yng(2)} &  \bar \eta +  f_1^{(4)}   & \frac{-5 \bar{u}_2^2-3 \bar{u}_3'+2 \bar{u}_2''}{9} & \frac{2 \bar{u}_3-\bar{u}_2'}{3} & - \frac{2 \bar u_2}{3} & 0 \cr 
{\tiny \yng(3)} & f_4^{(4)} 
& \bar \eta + f_2^{(4)} 
& \frac{-5 \bar{u}_2^2+3 \bar{u}_3'-\bar{u}_2''}{9} 
& \frac{2}{3} \bar{u}_3-\bar{u}_2' 
& - \frac{2 \bar u_2}{3} \cr 
{\tiny \yng(4)} &  
- \frac{2 \bar u_2}{3} \bar \eta + f_6^{(4)}
& f_5^{(4)}
& \bar \eta +  f_3^{(4)} 
&  \frac{5  \bar{u}_2^2 + 9 \bar{u}_3' -10  \bar{u}_2'' }{9}
& \frac{2 \bar{u}_3-5 \bar{u}_2'}{3}  \cr
}\biggr],
\end{align}
where 
\begin{align}
w_2  &= \frac{5 \bar{u}_2 }{3},\qquad w_3  = - \frac{5(\bar{u}_3 - 2\bar{u}_2')}{3}, \qquad
w_4  = \frac{5(\bar{u}_2^2-3 \bar{u}_3'+5 \bar{u}_2'')}{9}, \nn\\
w_5  &= - \frac{10(\bar u_2 \bar u_3 - \bar u_2 \bar u_2' + \bar u_3'' - \bar u_2^{(3)})}{9}, 
\label{Eqn:Appendix:PQ53CoefficientsW}
\end{align}
and 
\begin{align}
f_1^{(4)} &= \frac{10 \bar{u}_2 \left(\bar{u}_3-\bar{u}_2'\right)+\bar{u}_3''-\bar{u}_2^{(3)}}{9},  \qquad
f_2^{(4)} =\frac{10 \bar{u}_2 \left(\bar{u}_3-2 \bar{u}_2'\right)-2 \bar{u}_3''+\bar{u}_2^{(3)}}{9},   \nn\\
f_3^{(4)} &= \frac{\bar{u}_3''-10 \bar{u}_2 \bar{u}_2'}{9},   \qquad 
f_4^{(4)} = \frac{-10 (\bar{u}_2')^2+10 \bar{u}_3 \bar{u}_2'+10 \bar{u}_2 \bar{u}_3'-10 \bar{u}_2 \bar{u}_2''+\bar{u}_3^{(3)} -\bar{u}_2^{(4)}}{9},  \nn\\
f_5^{(4)} &= \frac{10 \bar{u}_2^3+10 \left(3 \bar{u}_3'-4 \bar{u}_2''\right) \bar{u}_2+60 \bar{u}_3 \bar{u}_2'-3 \left(30 (\bar{u}_2')^2+\bar{u}_3^{(3)}\right)}{27}, \nn\\
f_6^{(4)} &= \frac{-20 \left(\bar{u}_3-\bar{u}_2'\right) \bar{u}_2^2+10 \left(\bar{u}_3''-\bar{u}_2^{(3)}(t)\right) \bar{u}_2+3 \left(10 \bar{u}_3 \bar{u}_2''+10 \bar{u}_2' \left(2 \bar{u}_3'-3 \bar{u}_2''\right)+\bar{u}_3^{(4)}-\bar{u}_2^{(5)}\right)}{27}.  \label{Eqn:Appendix:PQ53RS11CoefficientsFFFFF}
\end{align}

\subsubsection{2) $(r,s)=(2,1)$ FZZT-Cardy brane:  $\widetilde W^{(2,1)} =\widetilde W^{(2)}[\chi]$}

\begin{align}
\vec{W}^{(2)} &= \Bigl( \widetilde W^{(2)}_\varnothing, \widetilde W^{(2)}_{\tiny \yng(1)}, \widetilde W^{(2)}_{\tiny \yng(2)}, \widetilde W^{(2)}_{\tiny \yng(1,1)}, \widetilde W^{(2)}_{\tiny \yng(3)}, \widetilde W^{(2)}_{\tiny \yng(2,1)}, \widetilde W^{(2)}_{\tiny \yng(3,1)}, \widetilde W^{(2)}_{\tiny \yng(2,2)}, \widetilde W^{(2)}_{\tiny \yng(3,2)}, \widetilde W^{(2)}_{\tiny \yng(3,3)} \Bigr)^{\rm T}, \\
\mathcal B^{(2,1)} &= 
\mbox{\scriptsize $ \bordermatrix{
 & \mbox{\normalsize $\varnothing$} & {\tiny \yng(1)} & {\tiny \yng(2)} & {\tiny \yng(1,1)} & {\tiny \yng(3)} & {\tiny \yng(2,1)} & {\tiny \yng(3,1)} & {\tiny \yng(2,2)} & {\tiny \yng(3,2)} & {\tiny \yng(3,3)} \cr
\mbox{\normalsize $\varnothing$} &  & 1 & \cr
{\tiny \yng(1)} & & & 1 &1  \cr
{\tiny \yng(2)} &  & & & & 1 & 1 \cr 
{\tiny \yng(1,1)} &  & & & & & 1 \cr 
{\tiny \yng(3)} &- w_4 & -w_3 &  -w_2 & &  & & 1\cr
{\tiny \yng(2,1)} &  & & & & & & 1 &1 \cr 
{\tiny \yng(3,1)} & w_5 - \bar \eta & & & -w_3& & -w_2& & &1 \cr 
{\tiny \yng(2,2)} &  & & & & & & & & 1 \cr 
{\tiny \yng(3,2)} & & w_5 - \bar \eta & & w_4 & & & & -w_2 & & 1 \cr 
{\tiny \yng(3,3)} &  & & w_5 - \bar \eta & & & w_4 & & w_3
}$}, \\
\mathcal Q^{(2,1)} &=  \bar \beta_{5,3} \times \Bigl[  \nn\\
 & \!\!\!\!\!\! \!\!\!\!\!\! \!\!\!\!\!\! 
\mbox{ \tiny $ \bordermatrix{
 & \mbox{\normalsize $\varnothing$} & {\tiny \yng(1)} & {\tiny \yng(2)} & {\tiny \yng(1,1)} & {\tiny \yng(3)} & {\tiny \yng(2,1)} & {\tiny \yng(3,1)} & {\tiny \yng(2,2)} & {\tiny \yng(3,2)} & {\tiny \yng(3,3)} \cr
\mbox{\normalsize $\varnothing$} & 3 \bar{u}_2'-2 \bar{u}_3 & \bar{u}_2  & 0 & 0 & 1 & -1 & 0 & 0 & 0 & 0 \cr 
{\tiny \yng(1)} & h_{7}^{(3)}  &  \frac{2 \bar{u}_2'-\bar{u}_3}{3}  & - \frac{2\bar u_2}{3} &  \bar{u}_2   & 0 & 0 & 0 & -1 & 0 & 0 \cr 
{\tiny \yng(2)} & \bar \eta + h_{4}^{(3)} &  h_{8}^{(3)}  & -\frac{\bar{u}_3}{3}  & 0 & - \frac{2\bar u_2}{3} &  \bar{u}_2  & 0 & 0 & 0 & 0 \cr 
{\tiny \yng(1,1)} &  -\bar \eta - h_{5}^{(3)}  & h_{9}^{(3)} & 0 & \frac{5 \bar{u}_2'-\bar{u}_3}{3}  & 0 & - \frac{2\bar u_2}{3} & 0 & 0 & -1 & 0 \cr 
{\tiny \yng(3)} & h_2^{(3)}  &  \bar \eta + h_{6}^{(3)} & h_{10}^{(3)}  & 0 &  \frac{-\bar{u}_3-2 \bar{u}_2'}{3}  & 0 &  \bar{u}_2   & 0 & 0 & 1 \cr 
{\tiny \yng(2,1)} &  h_{3}^{(3)} & 0 & h_{9}^{(3)} & h_{8}^{(3)}  & 0 &  \frac{3\bar{u}_2'-\bar{u}_3}{3}   & - \frac{2\bar u_2}{3} & \bar{u}_2   & 0 & -1 \cr 
{\tiny \yng(3,1)} & \frac{2\bar u_2}{3} \bar \eta + h_{1}^{(3)}  & 0 & 0 &  \bar \eta + h_{6}^{(3)}  & h_{9}^{(3)} & h_{10}^{(3)}&\frac{\bar{u}_2'-\bar{u}_3}{3}  & 0 &  \bar{u}_2   & 0 \cr 
{\tiny \yng(2,2)} &   0 &  h_{3}^{(3)}  & \bar \eta + h_{5}^{(3)}  & - \bar \eta - h_{4}^{(3)}  & 0 & h_{7}^{(3)}  & 0 & \frac{4(\bar{u}_3-\bar{u}_2')}{3} & - \frac{2\bar u_2}{3} & 0 \cr 
{\tiny \yng(3,2)} &  0 & \frac{2\bar u_2}{3} \bar \eta +  h_{1}^{(3)}  & 0 & - h_{2}^{(3)}  & \bar \eta + h_{5}^{(3)} & 0 & h_{7}^{(3)} & h_{10}^{(3)}  & \frac{4 \bar{u}_3- 6 \bar{u}_2'}{3} & - \frac{2\bar u_2}{3} \cr 
{\tiny \yng(3,3)} &  0 & 0 & \frac{2\bar u_2}{3} \bar \eta +  h_{1}^{(3)} & 0 & - h_{3}^{(3)} & - h_{2}^{(3)}  & \bar \eta + h_{4}^{(3)}  & - \bar \eta -  h_{6}^{(3)}  & h_{8}^{(3)}  &\frac{4(\bar{u}_3-2 \bar{u}_2')}{3} \cr 
} $}
\Bigr]
\end{align}
where the coefficients $\{w_n\}_{n=2}^5$ are given by Eq.~\eq{Eqn:Appendix:PQ53CoefficientsW} and the coefficients $\{h_n^{(3)}\}_{n=1}^{10}$ are given as follows: 
\begin{align}
h_{1}^{(3)} &\equiv \frac{1}{27} \Bigl(20 \left(\bar{u}_3-\bar{u}_2'\right) \bar{u}_2^2-10 \left(\bar{u}_3''-\bar{u}_2'''\right) \bar{u}_2  -3 \left(10 \bar{u}_3 \bar{u}_2''+10 \bar{u}_2' \left(2 \bar{u}_3'-3 \bar{u}_2''\right)+\bar{u}_3^{(4)}-\bar{u}_2^{(5)}\right)\Bigr), \nn\\
h_{2}^{(3)} &\equiv \frac{1}{27} \left(10 \bar{u}_2^3+10 \left(3 \bar{u}_3'-4 \bar{u}_2''\right) \bar{u}_2+60 \bar{u}_3 \bar{u}_2'-3 \left(30 (\bar{u}_2')^2+\bar{u}_3'''\right)\right), \nn\\
h_{3}^{(3)} &\equiv \frac{1}{9} \left(10 (\bar{u}_2')^2-10 \bar{u}_3 \bar{u}_2'-10 \bar{u}_2 \bar{u}_3'+10 \bar{u}_2 \bar{u}_2''-\bar{u}_3'''+\bar{u}_2''''\right), \nn\\
h_{4}^{(3)} &\equiv  \frac{1}{9} \left(10 \bar{u}_2 \left(\bar{u}_3-2 \bar{u}_2'\right)-2 \bar{u}_3''+\bar{u}_2'''\right), \quad 
h_{5}^{(3)} \equiv \frac{1}{9} \left(10 \bar{u}_2 \left(\bar{u}_3-\bar{u}_2'\right)+\bar{u}_3''-\bar{u}_2'''\right), \nn\\
h_{6}^{(3)} &\equiv \frac{1}{9} \left(\bar{u}_3''-10 \bar{u}_2 \bar{u}_2'\right), \quad
h_{7}^{(3)} \equiv \frac{1}{9} \left(-5 \bar{u}_2^2-3 \bar{u}_3'+2 \bar{u}_2''\right), \nn\\
h_{8}^{(3)} &\equiv \frac{1}{9} \left(-5 \bar{u}_2^2+3 \bar{u}_3'-\bar{u}_2''\right), \quad
h_{9}^{(3)} \equiv \bar{u}_2''-\bar{u}_3', \quad
h_{10}^{(3)} \equiv \frac{5}{9} \bar{u}_2^2+\bar{u}_3'-\frac{10}{9} \bar{u}_2''. 
\label{Eqn:Appendix:PQ53RS21CoefficientsHH}
\end{align}

\subsubsection{3) $(r,s)=(3,1)$ FZZT-Cardy brane:  $\widetilde W^{(3,1)} = \widetilde W^{(3)}[\chi]$}

\begin{align}
\vec{W}^{(3)} &= \Bigl( \widetilde W^{(3)}_\varnothing, \widetilde W^{(3)}_{\tiny \yng(1)}, \widetilde W^{(3)}_{\tiny \yng(2)}, \widetilde W^{(3)}_{\tiny \yng(1,1)}, \widetilde W^{(3)}_{\tiny \yng(2,1)}, \widetilde W^{(3)}_{\tiny \yng(1,1,1)}, \widetilde W^{(3)}_{\tiny \yng(2,2)}, \widetilde W^{(3)}_{\tiny \yng(2,1,1)}, \widetilde W^{(3)}_{\tiny \yng(2,2,1)}, \widetilde W^{(3)}_{\tiny \yng(2,2,2)} \Bigr)^{\rm T}, \\
\mathcal B^{(3,1)} &= 
\bordermatrix{
 & \varnothing & {\tiny \yng(1)} & {\tiny \yng(2)} & {\tiny \yng(1,1)} & {\tiny \yng(2,1)} & {\tiny \yng(1,1,1)} & {\tiny \yng(2,2)} & {\tiny \yng(2,1,1)} & {\tiny \yng(2,2,1)} & {\tiny \yng(2,2,2)} \cr
\varnothing &  & 1 & \cr
{\tiny \yng(1)} & & & 1 &1  \cr
{\tiny \yng(2)} & - w_3 & -  w_2 & & & 1 &  \cr 
{\tiny \yng(1,1)} &  & & & & 1 & 1 \cr 
{\tiny \yng(2,1)} &  w_4 & &  & - w_2 &  & & 1 & 1\cr
{\tiny \yng(1,1,1)} &  & & & & & &  &1 \cr 
{\tiny \yng(2,2)} &  &  w_4 & & w_3& & & & &1 \cr 
{\tiny \yng(2,1,1)} &  \bar \eta - w_5  & & & & & - w_2 & & & 1 \cr 
{\tiny \yng(2,2,1)} & & \bar \eta - w_5  & &  & & w_3 & & & & 1 \cr 
{\tiny \yng(2,2,2)} &  & &  & \bar \eta - w_5  & & - w_4 & & 
}, \\
\mathcal Q^{(3,1)} &= \bar \beta_{5,3}\times \Bigl[  \nn\\
& \!\!\!\!\!\!  \!\!\!\!\!\! \!\!  \mbox{\tiny $\bordermatrix{
 & \mbox{\normalsize $\varnothing$} & {\tiny \yng(1)} & {\tiny \yng(2)} & {\tiny \yng(1,1)} & {\tiny \yng(2,1)} & {\tiny \yng(1,1,1)} & {\tiny \yng(2,2)} & {\tiny \yng(2,1,1)} & {\tiny \yng(2,2,1)} & {\tiny \yng(2,2,2)} \cr
\mbox{\normalsize $\varnothing$} &  \frac{-4 (\bar{u}_3-2 \bar{u}_2')}{3}  & - \frac{2 \bar u_2}{3} & 0 & 0 & -1 & 1 & 0 & 0 & 0 & 0 \cr
{\tiny \yng(1)} & h_{8}^{(3)}  & \frac{ 6 \bar{u}_2'-4 \bar{u}_3}{3} &- \frac{2 \bar u_2}{3} & \bar{u}_2   & 0 & 0 & -1 & 0 & 0 & 0 \cr
{\tiny \yng(2)} & \bar \eta + h_{6}^{(3)} & h_{10}^{(3)}  & \frac{4 (\bar{u}_2' - \bar{u}_3)}{3}  & 0 & \bar{u}_2   & 0 & 0 & 0 & -1 & 0 \cr
{\tiny \yng(1,1)} & - \bar \eta - h_{4}^{(3)}  & h_{7}^{(3)} & 0 &  \frac{\bar{u}_3-\bar{u}_2'}{3}  & - \frac{2 \bar u_2}{3} & \bar{u}_2  & 0 & 0 & 0 & 0 \cr
{\tiny \yng(2,1)} &  - h_{2}^{(3)}  & 0 & h_{7}^{(3)} & h_{10}^{(3)}  &  \frac{\bar{u}_3-3 \bar{u}_2'}{3}  & 0 & - \frac{2 \bar u_2}{3} & \bar{u}_2   & 0 & -1 \cr
{\tiny \yng(1,1,1)} & - h_{3}^{(3)}  & - \bar \eta - h_{5}^{(3)} & 0 & h_{9}^{(3)}  & 0 & \frac{\bar{u}_3+2 \bar{u}_2'}{3} & 0 & - \frac{2 \bar u_2}{3} & 0 & 1 \cr
{\tiny \yng(2,2)} &   0 &- h_{2}^{(3)} & \bar \eta+ h_{4}^{(3)} & - \bar \eta - h_{6}^{(3)} & h_{8}^{(3)} & 0 &  \frac{\bar{u}_3-5 \bar{u}_2'}{3} & 0 &  \bar{u}_2   & 0 \cr
{\tiny \yng(2,1,1)} & - \frac{2 \bar u_2}{3}\bar \eta - h_{1}^{(3)}  & 0 & - \bar \eta - h_{5}^{(3)} & 0 & h_{9}^{(3)} & h_{10}^{(3)} & 0 & \frac{\bar{u}_3}{3}  &  - \frac{2 \bar u_2}{3} & 0 \cr
{\tiny \yng(2,2,1)} &  0 & - \frac{2 \bar u_2}{3} \bar \eta- h_{1}^{(3)} & h_{3}^{(3)} & 0 & 0 & - \bar \eta - h_{6}^{(3)} & h_{9}^{(3)} & h_{8}^{(3)}  & \frac{\bar{u}_3-2 \bar{u}_2'}{3} & \bar{u}_2   \cr
{\tiny \yng(2,2,2)} &   0 & 0 & 0 &  - \frac{2 \bar u_2}{3} \bar \eta - h_{1}^{(3)} & h_{3}^{(3)} & h_{2}^{(3)}  & \bar \eta + h_{5}^{(5)} & - \bar \eta - h_{4}^{(3)} & h_{7}^{(3)} &  2 \bar{u}_3-3 \bar{u}_2'\cr
} $} \Bigr], 
\end{align}
where the coefficients $\{w_n\}_{n=2}^5$ are given by Eq.~\eq{Eqn:Appendix:PQ53CoefficientsW} and the coefficients $\{h_n^{(3)}\}_{n=1}^{10}$ are given by Eq.~\eq{Eqn:Appendix:PQ53RS21CoefficientsHH}.

\subsubsection{4) $(r,s)=(4,1)$ FZZT-Cardy brane:  $\widetilde W^{(4,1)} = \widetilde W^{(4)}[\chi]$}

\begin{align}
\vec{W}^{(1)} &= \Bigl(\widetilde  W^{(1)}_\varnothing, \widetilde W^{(1)}_{\tiny \yng(1)}, \widetilde W^{(1)}_{\tiny \yng(1,1)}, \widetilde W^{(1)}_{\tiny \yng(1,1,1)}, \widetilde W^{(1)}_{\tiny \yng(1,1,1,1)} \Bigr)^{\rm T}, \\
\mathcal B^{(4,1)} &= 
\bordermatrix{
 & \varnothing & {\tiny \yng(1)} & {\tiny \yng(1,1)} & {\tiny \yng(1,1,1)} & {\tiny \yng(1,1,1,1)}\cr
\varnothing &  & 1 & \cr
{\tiny \yng(1)} & - w_2 & & 1 \cr
{\tiny \yng(1,1)} & w_3 & & & 1 \cr 
{\tiny \yng(1,1,1)} &- w_4  & & & & 1 \cr 
{\tiny \yng(1,1,1,1)} & - \bar \eta + w_5 &  &   &  &\cr
}, \\
\mathcal Q^{(4,1)} &= \bar \beta_{5,3}  \Bigl[
\bordermatrix{
 & \varnothing & {\tiny \yng(1)} & {\tiny \yng(1,1)} & {\tiny \yng(1,1,1)} & {\tiny \yng(1,1,1,1)}\cr
\varnothing &  \frac{5 \bar{u}_2'-2 \bar{u}_3}{3}  &  - \frac{2 \bar u_2}{3}  & 0 & 1 & 0 \cr
{\tiny \yng(1)} &  \frac{5 \bar{u}_2^2 +9 \bar{u}_3' - 10 \bar{u}_2''}{9} & \frac{3 \bar{u}_2' -2 \bar{u}_3}{3}  & - \frac{2 \bar u_2}{3}  & 0 & 1 \cr
{\tiny \yng(1,1)} & - \bar \eta - f_3^{(4)}  & \frac{-5 \bar{u}_2^2+3 \bar{u}_3'-\bar{u}_2''}{9}  & \frac{\bar{u}_2'-2 \bar{u}_3}{3}  & \bar{u}_2 & 0 \cr
{\tiny \yng(1,1,1)} & f_5^{(4)}  &  -\bar \eta - f_2^{(4)} & \frac{-5 \bar{u}_2^2-3 \bar{u}_3'+2 \bar{u}_2''}{9} & \bar{u}_3-2 \bar{u}_2' &  \bar{u}_2 \cr
{\tiny \yng(1,1,1,1)} &   \frac{2 \bar u_2}{3} \bar \eta -  f_6^{(4)}  & f_4^{(4)}  & - \bar \eta - f_1^{(4)}  & \bar{u}_2''-\bar{u}_3' & \bar{u}_3-\bar{u}_2' \cr
}\Bigr]
\end{align}
where the coefficients $\{w_n\}_{n=2}^5$ are given by Eq.~\eq{Eqn:Appendix:PQ53CoefficientsW} and the coefficients $\{ f_n^{(4)}\}_{n=1}^{6}$ are given by Eq.~\eq{Eqn:Appendix:PQ53RS11CoefficientsFFFFF}.


\subsubsection{Duality matrices}

\noindent\underline{Reflection:} \qquad $\vec W^{(5-s)}[\chi_{\mathcal C}](t; - \eta) = {\mathsf M}^{(s)}(t)\times  \vec W^{(s)}[\chi](t;\eta)$

\begin{align}
{\mathsf M}^{(1)}(t) = \bordermatrix{
 & \mbox{\normalsize $\varnothing$} & {\tiny \yng(1)} & {\tiny \yng(2)}& {\tiny \yng(3)} & {\tiny \yng(4)}  \cr
\mbox{\normalsize $\varnothing$} &  1 & 0 & 0 & 0 & 0  \cr
{\tiny \yng(1)} & 0 & 1 & 0 & 0 & 0  \cr 
{\tiny \yng(1,1)} &  \frac{5 \bar{u}_2 }{3} & 0 & 1 & 0 & 0\cr
{\tiny \yng(1,1,1)} &  - \frac{5 \bar{u}_3}{3}  & \frac{5  \bar{u}_2}{3} & 0 & 1 & 0 \cr
{\tiny \yng(1,1,1,1)} &    \frac{5  (\bar{u}_2^2+2 \bar{u}_2'' ) }{9}& \frac{5(\bar{u}_2' - \bar{u}_3)}{3}  & \frac{5 \bar{u}_2}{3}  & 0 & 1
}, \quad \qquad \qquad \qquad \qquad \\
\!
{\mathsf M}^{(2)}(t) = 
\mbox{ \tiny $\bordermatrix{
 & \mbox{\normalsize $\varnothing$} & {\tiny \yng(1)} & {\tiny \yng(2)}& {\tiny \yng(1,1)} & {\tiny \yng(3)}  & {\tiny \yng(2,1)} & {\tiny \yng(3,1)}& {\tiny \yng(2,2)} & {\tiny \yng(3,2)} & {\tiny \yng(3,3)} \cr
\mbox{\normalsize $\varnothing$} &  1 & 0 & 0 & 0 & 0 & 0 & 0 & 0 & 0 & 0  \cr
{\tiny \yng(1)} &  0 & 1 & 0 & 0 & 0 & 0 & 0 & 0 & 0 & 0 \cr 
{\tiny \yng(2)} & - \frac{5 \bar{u}_2}{3}   & 0 & 0 & 1 & 0 & 0 & 0 & 0 & 0 & 0 \cr
{\tiny \yng(1,1)} &  \frac{5 \bar{u}_2}{3}   & 0 & 1 & 0 & 0 & 0 & 0 & 0 & 0 & 0 \cr
{\tiny \yng(2,1)} &  \frac{5 \bar{u}_3}{3}   & 0 & 0 & 0 & 0 & 1 & 0 & 0 & 0 & 0 \cr
{\tiny \yng(1,1,1)} &  - \frac{5 (\bar{u}_3 -\bar{u}_2' )}{3}  
& \frac{5 \bar{u}_2}{3}  & 0 & 0 & 1 & 0 & 0 & 0 & 0 & 0 \cr 
{\tiny \yng(2,2)} &   \frac{25 \bar{u}_2^2}{9}  
& \frac{5 \bar{u}_3}{3}  
& \frac{5 \bar{u}_2}{3}  
& - \frac{5 \bar{u}_2}{3}  & 0 & 0 & 0 & 1 & 0 & 0 \cr 
{\tiny \yng(2,1,1)} &  - \frac{5 (\bar{u}_2^2+2 \bar{u}_2'' )}{9}  
& 0 & 0 
& \frac{5 \bar{u}_2}{3}  & 0 & 0 & 1 & 0 & 0 & 0  \cr
{\tiny \yng(2,2,1)} &  -  \frac{25 \bar{u}_2 (\bar{u}_3-\bar{u}_2' )}{9}  
& \frac{10(2 \bar{u}_2^2-\bar{u}_2'' )}{9}  
& 0 
& \frac{5 (\bar{u}_3 -\bar{u}_2' )}{3}  
& \frac{5 \bar{u}_2}{3}  & 0 & 0 & 0 & 1 & 0 \cr
{\tiny \yng(2,2,2)} & h_1^{(4)}
& - \frac{25 \bar{u}_2 \bar{u}_3}{9}  
& -\frac{5 (\bar{u}_2^2+2 \bar{u}_2'' ) }{9} 
& \frac{25 \bar{u}_2^2}{9}  
& -\frac{5 \bar{u}_3}{3}  
& \frac{5 (\bar{u}_3 -\bar{u}_2' )}{3}  
& \frac{5 \bar{u}_2}{3}  
& - \frac{ 5 \bar{u}_2}{3} & 0 & 1 
} $} \nn\\
\text{with}\qquad h_1^{(4)} \equiv - \frac{ 25 \left(\bar{u}_2^3+2 \bar{u}_2'' \bar{u}_2+3 \bar{u}_3 (\bar{u}_2'-\bar{u}_3)\right)}{27}. \qquad \qquad \qquad 
\end{align}

\noindent\underline{Charge conjugation:} \qquad $\Bigl[\vec W^{(5-s)}[\chi](t;\eta)\Bigr]^{\rm T}\,\, \mathfrak C^{(1,s)}_{[\chi]}(t)\, \, \Bigl[ \vec W^{(s)}[\chi](t;\eta) \Bigr] $
\begin{align}
\mathfrak C^{(1,1)}_{[\chi]}(t) & = 
\bordermatrix{
 & \varnothing & {\tiny \yng(1)} & {\tiny \yng(2)}& {\tiny \yng(3)} & {\tiny \yng(4)}  \cr
\varnothing &0&0& 0 & 0 & 1  \cr
{\tiny \yng(1)} & 0&0&  0 & -1 & 0 \cr 
{\tiny \yng(1,1)} &0&0&   1 & 0 & 0 \cr
{\tiny \yng(1,1,1)} &0& -1 &  0 & 0 & 0 \cr
{\tiny \yng(1,1,1,1)} &   1 & 0 & 0&0&0
}, \\
\mathfrak C^{(1,2)}_{[\chi]}(t) & = 
\bordermatrix{
 & \varnothing & {\tiny \yng(1)} & {\tiny \yng(2)}& {\tiny \yng(1,1)} & {\tiny \yng(3)}  & {\tiny \yng(2,1)} & {\tiny \yng(3,1)}& {\tiny \yng(2,2)} & {\tiny \yng(3,2)} & {\tiny \yng(3,3)} \cr
\varnothing &0&0& 0 & 0 & 0 & 0&0& 0 & 0 & 1  \cr
{\tiny \yng(1)} &0&0& 0 & 0 & 0 & 0&0&  0 & -1 & 0 \cr 
{\tiny \yng(2)} & 0&0& 0 & 0 & 0 & 0&0&   1 & 0 & 0 \cr
{\tiny \yng(1,1)} &0&0& 0 & 0 & 0 & 0& 1 &  0 & 0 & 0 \cr
{\tiny \yng(2,1)} & 0&0& 0 & 0 & 0 &   -1 & 0 & 0&0&0 \cr
{\tiny \yng(1,1,1)} &0&0& 0 & 0 & -1 & 0&0&  0 & 0 & 0 \cr 
{\tiny \yng(2,2)} &0&0& 0 & 1 & 0 & 0&0&  0 & 0 & 0 \cr 
{\tiny \yng(2,1,1)} & 0&0& 1 & 0 & 0 & 0&0&  0 & 0 & 0 \cr
{\tiny \yng(2,2,1)} &0&- 1& 0 & 0 & 0 & 0& 0 &  0 & 0 & 0 \cr
{\tiny \yng(2,2,2)} & 1&0& 0 & 0 & 0 &  0 & 0 & 0&0&0
}. 
\end{align}


\section{Derivation of isomonodromy systems \label{Appendix:DerivationOfIMS}}

In this appendix, we show the explicit solutions of ${\mathscr L \! \widetilde{\mathscr Q}}_\lambda^{(p-r)}$- and ${\mathscr L \!\widetilde{\mathscr P}}_\lambda^{(p-r)}$-equations which we obtained in the evaluation of isomonodromy systems. We show the detailed steps of the $(1,2)$-type FZZT-Cardy brane of the $(3,4)$-system. For the other cases, we only show the solutions of the ${\mathscr L \! \widetilde{\mathscr Q}}_\lambda^{(p-r)}$- and ${\mathscr L \!\widetilde{\mathscr P}}_\lambda^{(p-r)}$-equations. Since the evaluation itself is quite complicated, the results in this appendix should be useful in future developments of the isomonodromy systems of the Wronskians. 

\subsection{$(p,q)=(3,4)$ and $(r,s)=(1,2)$ \label{Subappendix:PQ34-RS12:DerivationOfIMS}}

In this system, we show that all the Young diagrams are solved by the following three diagrams: 
\begin{align}
\Bigl\{ \varnothing, \square,\square^2 \Bigr\}. \label{AppendixC1BasicBaseForPQ34RS12}
\end{align}
We first write down ${\mathscr L \! \widetilde{\mathscr Q}}_\lambda^{(2)}$- and ${\mathscr L \!\widetilde{\mathscr P}}_\lambda^{(2)}$-equations which do not include $\del_\zeta$. Here the size of the largest Young diagram in the differential equation is called the {\em order} of schur-differential equations. 

\paragraph{1) $\bigl|\lambda\bigr|=3$-rd order equations}
\begin{align}
\underline{\text{${\mathscr L \!\widetilde{\mathscr P}}_{(-1,0)}^{(2)}$-eqn.: }} && 0 &= 8 {\, \scriptsize \yng(3)\,} + \frac{8 u_2}{3} {\, \scriptsize\yng(1)\,} - \frac{4}{3} \bigl(2u_3 - 3 u_2'\bigr) \varnothing \\
\underline{\text{${\mathscr L \!\widetilde{\mathscr Q}}_{(0,0)}^{(2)}$-eqn.: }} && - \zeta \varnothing &= 4 {\, \scriptsize \yng(3)\,} - 4 {\, \scriptsize \yng(2,1)\,} + u_2 {\, \scriptsize \yng(1)\,} -(2 u_3 - 3 u_2' ) \varnothing 
\end{align}
With these equations, one can solve ${\scriptsize \yng(3)}$ and ${\scriptsize \yng(2,1)}$ in terms of  Eq.~\eq{AppendixC1BasicBaseForPQ34RS12}. 

\paragraph{2) $\bigl|\lambda\bigr|=4$-th order equations}
\begin{align}
&\underline{\text{$ \square\times {\mathscr L \!\widetilde{\mathscr P}}_{(-1,0)}^{(2)}$-eqn.: }} \nn\\
 &\quad 0 = 8 \bigl({\, \scriptsize \yng(4)\,} +{\, \scriptsize \yng(3,1) \,} \bigr) + \frac{8 u_2}{3} \bigl({\, \scriptsize\yng(2)\,} + {\, \scriptsize \yng(1,1) \,} \bigr) - \frac{4}{3} \bigl(2u_3 -  3u_2'\bigr) {\, \scriptsize \yng(1) \,} - \frac{4}{3} \bigl(2u_3' - 3 u_2''\bigr) \varnothing \\
&\underline{\text{${\mathscr L \!\widetilde{\mathscr Q}}_{(1,0)}^{(2)}$-eqn.: }} \nn\\
 &\quad - \zeta {\, \scriptsize \yng(1)\,} = 4 \bigl({\, \scriptsize \yng(4)\,} -  {\, \scriptsize \yng(2,2)\,} \bigr) + u_2 \bigl({\, \scriptsize \yng(2)\,} + {\, \scriptsize \yng(1,1)\,} \bigr) -(2 u_3 - 4 u_2' ) {\, \scriptsize \yng(1)\,} - (2u_3' - 3 u_2'') \varnothing 
\end{align}
With these equations, one can solve two size-4 Young diagrams ${\,\scriptsize \yng(3,1) \,}$ and ${\, \scriptsize \yng(2,2)\, }$; but ${\,\scriptsize \yng(4) \,}$ remains as an unsolved Young diagram (this diagram can be solved by the 7-th order equations). 

\paragraph{3) $\bigl|\lambda\bigr|=5$-th order equations}
\begin{align}
&\underline{\text{$ \square^2\times {\mathscr L \!\widetilde{\mathscr P}}_{(-1,0)}^{(2)}$-eqn.: }} \nn\\
&\quad  0= 8 \bigl({\, \scriptsize \yng(5)\,}+  2 {\, \scriptsize \yng(4,1) \,} +{\, \scriptsize \yng(3,2) \,}  \bigr) + \frac{8 u_2}{3} \bigl({\, \scriptsize\yng(3)\,} + 2 {\, \scriptsize \yng(2,1) \,} \bigr) - \nn\\
&\quad \quad  - \frac{4}{3} \bigl(2u_3 - 7 u_2'\bigr) \bigl( {\, \scriptsize \yng(2) \,} + {\, \scriptsize \yng(1,1) \,} \bigr) - \frac{16}{3} \bigl(u_3' - 2 u_2''\bigr) {\, \scriptsize \yng(1) \,} - \frac{4}{3} (2 u_3'' - 3 u_2''') \varnothing \\
&\underline{\text{${\mathscr L \!\widetilde{\mathscr P}}_{(0,1)}^{(2)}$-eqn.: }} \nn\\
& \quad  0 = 8 {\, \scriptsize \yng(4,1)\,} +  \frac{8 u_2 }{3} {\, \scriptsize \yng(2,1)\,} - \frac{4}{3} (2 u_3 - 5 u_2' ) {\, \scriptsize \yng(1,1)\,} -\frac{2}{9} (u_2 u_2' - 6 u_3'' + 8 u_2''') \varnothing \\
&\underline{\text{${\mathscr L \!\widetilde{\mathscr Q}}_{(2,0)}^{(2)}$-eqn.: }} \nn\\
&\quad  - \zeta {\, \scriptsize \yng(2)\,} = 4 {\, \scriptsize \yng(5)\,}  + u_2 \bigl( {\, \scriptsize \yng(3)\,} + {\, \scriptsize \yng(2,1)\,} \bigr) -(2 u_3 - 5 u_2' ) {\, \scriptsize \yng(2)\,} - \nn\\
&\qquad\qquad \qquad \qquad \qquad \qquad \qquad \qquad \qquad  - 3(u_3' - 2 u_2'')  {\, \scriptsize \yng(1)\,} - (3u_3'' - 4 u_2''')\varnothing \\
&\underline{\text{${\mathscr L \!\widetilde{\mathscr Q}}_{(1,1)}^{(2)}$-eqn.: }} \nn\\
&\quad  - \zeta {\, \scriptsize \yng(1,1)\,} = 4  \bigl( {\, \scriptsize \yng(4,1)\,} - {\, \scriptsize \yng(3,2)\,} \bigr) + u_2 {\, \scriptsize \yng(2,1)\,}  -(2 u_3 - 5 u_2' ) {\, \scriptsize \yng(1,1)\,} - \nn\\
&\qquad\qquad \qquad \qquad \qquad \qquad \qquad \qquad \qquad- (u_3' -  u_2'') {\, \scriptsize \yng(1)\,} + (u_3'' -  u_2''') \varnothing 
\end{align}
With these equations, one can solve size-5 Young diagrams, ${\, \scriptsize \yng(5)\, }$,  ${\, \scriptsize \yng(4,1)\, }$ and ${\, \scriptsize \yng(3,2) \, }$, and the remaining size-2 diagram ${\scriptsize \yng(2)}$ as well. 

\paragraph{4) $\bigl|\lambda\bigr|=6$-th order equations}
\begin{align}
&\underline{\text{$ \square^3\times {\mathscr L \!\widetilde{\mathscr P}}_{(-1,0)}^{(2)}$-eqn.: }} \nn\\
&\quad  0= 8 \bigl({\, \scriptsize \yng(6)\,}+  3 {\, \scriptsize \yng(5,1) \,}+ 3 {\, \scriptsize \yng(4,2) \,} +{\, \scriptsize \yng(3,3) \,}  \bigr) + \frac{8 u_2}{3} \bigl({\, \scriptsize\yng(4)\,} +  3 {\, \scriptsize \yng(3,1) \,} + 2 {\, \scriptsize \yng(2,2) \,} \bigr) - \nn\\
&\quad \quad  - \frac{4}{3} \bigl(2u_3 - 9 u_2'\bigr) \bigl( {\, \scriptsize \yng(3) \,} + 2{\, \scriptsize \yng(2,1) \,} \bigr) - 4 \bigl(2 u_3' - 5 u_2''\bigr) \bigl( {\, \scriptsize \yng(2) \,} + {\, \scriptsize \yng(1,1) \,} \bigr) - \frac{4}{3} (6 u_3'' - 11 u_2''') {\, \scriptsize \yng(1) \,} \nn\\
&\quad \quad - \frac{4}{3} (2 u_3''' - 3 u_2'''') \varnothing \\
&\underline{\text{$\square\times {\mathscr L \!\widetilde{\mathscr P}}_{(0,1)}^{(2)}$-eqn.: }} \nn\\
& \quad  0 = 8 \bigl( {\, \scriptsize \yng(5,1)\,} + {\, \scriptsize \yng(4,2)\,}  \bigr) +  \frac{8 u_2 }{3} \bigl( {\, \scriptsize \yng(3,1)\,} + {\, \scriptsize \yng(2,2)\,}\bigr)  - \frac{4}{3} (2 u_3 - 7 u_2' ) {\, \scriptsize \yng(2,1)\,} - \nn\\
&\qquad - \frac{4}{3} (2 u_3' - 5 u_2'' ) {\, \scriptsize \yng(1,1)\,} -\frac{2(u_2 u_2' - 6 u_3'' + 8 u_2''')}{9}  {\, \scriptsize \yng(1)\,} -\frac{2( (u_2')^2 + u_2 u_2'' - 6 u_3''' + 8 u_2'''')}{9}  \varnothing \\
&\underline{\text{${\mathscr L \!\widetilde{\mathscr Q}}_{(3,0)}^{(2)}$-eqn.: }} \nn\\
&\quad  - \zeta {\, \scriptsize \yng(3)\,} = 4 \bigl({\, \scriptsize \yng(6)\,} + {\, \scriptsize \yng(3,3)\,} \bigr)  + u_2 \bigl( {\, \scriptsize \yng(4)\,} + {\, \scriptsize \yng(3,1)\,} \bigr) - 2( u_3 - 3 u_2' ) {\, \scriptsize \yng(3)\,} - \nn\\ 
&\qquad \qquad \qquad    -  2(2u_3' - 5 u_2'')  {\, \scriptsize \yng(2)\,} - 2 (3u_3'' - 5 u_2''') {\, \scriptsize \yng(1)\,} - (4 u_3''' - 5 u_2'''') \varnothing \\
&\underline{\text{${\mathscr L \!\widetilde{\mathscr Q}}_{(2,1)}^{(2)}$-eqn.: }} \nn\\
&\quad  - \zeta {\, \scriptsize \yng(2,1)\,} = 4  \bigl( {\, \scriptsize \yng(5,1)\,} - {\, \scriptsize \yng(3,3)\,} \bigr) + u_2  \bigl({\, \scriptsize \yng(2,2)\,} + {\, \scriptsize \yng(3,1)\,} \bigr) -2 (u_3 - 3 u_2' ) {\, \scriptsize \yng(2,1)\,} - \nn\\
&\qquad  \qquad \qquad  - (u_3' -  u_2'') {\, \scriptsize \yng(2)\,} - 3 (u_3' - 2 u_2'') {\, \scriptsize \yng(1,1)\,} + (u_3''' -  u_2'''') \varnothing. 
\end{align}
With these equations, one can solve size-6 Young diagrams, ${\, \scriptsize \yng(5,1)\, }$,  ${\, \scriptsize \yng(4,2)\, }$ and ${\, \scriptsize \yng(3,3) \, }$; but ${\, \scriptsize \yng(6)\, }$ remains as an unsolved Young diagram (this diagram can be solved by the 9-th order equations). Also note that $ \square^3\times {\mathscr L \!\widetilde{\mathscr P}}_{(-1,0)}^{(2)}$-equation is redundant and equivalent to the string equation Eq.~\eq{StringEquation34-1st}.

\paragraph{5) General cases and solutions} In general, one can see the following: 
\begin{itemize}
\item With use of $2n$-th order equations, one can solve size-$2n$ Young diagrams except for $\lambda= (2n,0,0,\cdots)$. 
\item With use of $(2n+1)$-th order equations, one can solve size-$(2n+1)$ Young diagrams and the Young diagram $\lambda= (2n-2,0,0,\cdots)$. 
\end{itemize}
The solutions to these equations are also shown up to $|\lambda|=5$-th order as follows: 
\begin{align}
{\, \scriptsize \yng(2)\, } &=  
\varnothing \left[ -\frac{u_2 }{6} - \frac{\frac{u_2 u_2'}{12}  -\frac{u_3''}{3}  -\frac{ u_2  u_3}{6} +\frac{u_2^{(3)}}{6} }{\zeta } \right]   
+  \square \left[ - \frac{ \frac{u_3'}{3} -  \frac{ u_2''}{6} }{\zeta }\right] 
+\square^2 \left[ \frac{1}{2} - \frac{ \frac{u_2'}{6} - \frac{ u_3}{3} }{\zeta }\right], \\
{\, \scriptsize \yng(1,1)\, } &=  
\varnothing \left[ \frac{u_2 }{6} + \frac{\frac{u_2 u_2'}{12}  -\frac{u_3''}{3}  -\frac{ u_2  u_3}{6} +\frac{u_2^{(3)}}{6} }{\zeta } \right]  
+  \square \left[  \frac{ \frac{u_3'}{3} -  \frac{ u_2''}{6} }{\zeta }\right] 
+\square^2 \left[ \frac{1}{2} + \frac{ \frac{u_2'}{6} - \frac{ u_3}{3} }{\zeta }\right], \\
{\, \scriptsize \yng(3)\, } &=  
\varnothing \left[-\frac{u_2'}{2} +\frac{u_3}{3} \right]  
+ \square \left[ -\frac{1}{3}  u_2\right], \qquad
{\, \scriptsize \yng(2,1)\, } =  
\varnothing \left[ \frac{\zeta }{4} +\frac{u_2'}{4} -\frac{u_3}{6}\right] 
+ \square \left[-\frac{u_2}{12}  \right], \\
{\, \scriptsize \yng(4)\, } &= 
\varnothing \Biggl[\frac{1}{36} \left(18 u_3'-27 u_2''+u_2^2\right) + 
\frac{\left(u_2 \left(2 u_3-u_2'\right)+4 u_3''-2 u_2^{(3)}\right)^2}{144 \zeta ^2}- \nn\\
- &\frac{6 \left(u_2'-2 u_3\right) u_2''+\left(u_2^2 +6 u_3'\right) \left(2 u_3-3 u_2'\right)  +2 u_2^2u_3+2 u_2 \left(u_3''-2 u_2^{(3)}\right)+4 \left(u_2^{(5)} - 3 u_3^{(4)}\right)}{72 \zeta } \Biggr] + \nn\\
+ &\square \Biggl[- \frac{\zeta }{8}+\frac{1}{24} \left(8 u_3-19 u_2'\right)   - \frac{u_2 \left(3 u_2''-2 u_3'\right)+10 u_3 u_2'-2 (u_2')^2-4 u_3^2+12 u_3^{(3)}-4 u_2^{(4)}}{72 \zeta }- \nn\\
& \qquad\quad  -\frac{\left(2 u_3'-u_2''\right) \left(u_2 \left(2 u_3 -u_2' \right)+4 u_3''-2 u_2^{(3)}\right)}{72 \zeta ^2}\Biggr]  + \nn\\
+ &\square^2 \Biggl[-\frac{5 u_2}{24} 
- \frac{u_2 \left(4 u_3 - 3 u_2'\right)-u_2^{(3)}}{36 \zeta } +  \frac{\left(2 u_3 - u_2'\right) \left(u_2 \left(2 u_3-u_2'\right)+4 u_3''-2 u_2^{(3)}\right)}{72 \zeta ^2} \Biggr], \\
{\, \scriptsize \yng(3,1)\, } &= \varnothing \Biggl[\frac{1}{36} \left(-6 u_3'+9 u_2''-u_2^2\right)-\frac{\left(u_2 \left(2 u_3-u_2'\right)+4 u_3''-2 u_2^{(3)}\right)^2}{144 \zeta ^2} - \nn\\
- & \frac{-2 \left(6 u_3 \left(u_3'-u_2''\right)+u_2' \left(3 u_2''-9 u_3'\right)-6 u_3^{(4)}+2 u_2^{(5)}\right)+u_2^2 \left(3 u_2'-4 u_3\right)-2 u_2 \left(u_3''-2 u_2^{(3)}\right)}{72 \zeta } \Biggr] + \nn\\
+ & \square \Biggl[\frac{\zeta }{8}-\frac{u_2'}{24}-\frac{-10 u_3 u_2'+2 (u_2')^2+2 u_2 u_3' -3 u_2 u_2''+4 u_3^2-12 u_3^{(3)}+4 u_2^{(4)}}{72 \zeta }+ \nn\\
&\qquad \quad +\frac{\left(2 u_3'-u_2''\right) \left(u_2 \left(2 u_3-u_2'\right)+4 u_3''-2 u_2^{(3)}\right)}{72 \zeta ^2} \Biggr] + \nn\\
+ &\square^2 \Biggl[ -\frac{u_2}{8} -\frac{u_2 \left(3 u_2'-4 u_3\right)+u_2^{(3)}}{36 \zeta }  -\frac{\left(2 u_3-u_2'\right) \left(u_2 \left(2 u_3-u_2'\right)+4 u_3''-2 u_2^{(3)}\right)}{72 \zeta ^2} \Biggr], \\
{\, \scriptsize \yng(2,2)\, } &= \varnothing \Biggl[ \frac{u_2^2}{36}  + \frac{\left(u_2 \left(2 u_3-u_2'\right)+4 u_3''-2 u_2^{(3)}\right)^2}{144 \zeta ^2} - \nn\\
-& \frac{6 \left(u_2'-2 u_3\right) u_2'' +\left(u_2^2 +6 u_3' \right) \left(2 u_3-3 u_2'\right)  + 2 u_2^2 u_3+ 2 u_2 \left(u_3''-2 u_2^{(3)}\right)+4 \left(u_2^{(5)}-3 u_3^{(4)}\right)}{72 \zeta } \Biggr] + \nn\\
+& \square \Biggl[ \frac{\zeta }{8} +\frac{ 5 u_2'-4 u_3 }{24} -\frac{u_2 \left(3 u_2''-2 u_3'\right)+10 u_3 u_2' -2 (u_2')^2-4 u_3^2+12 u_3^{(3)}-4 u_2^{(4)}}{72 \zeta } - \nn\\
&\qquad \quad -\frac{\left(2 u_3'-u_2''\right) \left(u_2 \left(2 u_3 -u_2' \right)+4 u_3'' -2 u_2^{(3)}\right)}{72 \zeta ^2} \Biggr] + \nn\\
+ & \square^2 \Biggl[\frac{u_2}{24} - \frac{u_2 \left(4 u_3-3 u_2'\right)-u_2^{(3)}}{36 \zeta } + \frac{\left(2 u_3-u_2'\right) \left(u_2 \left(2 u_3-u_2'\right)+4 u_3''-2 u_2^{(3)}\right)}{72 \zeta ^2} \Biggr], \\
{\, \scriptsize \yng(5)\, } &= \varnothing \Biggl[ - \frac{u_2}{48} \zeta  +\frac{1}{24} \left(u_2 \left(7 u_2'-4 u_3\right)+16 u_3''-23 u_2^{(3)}\right)-\nn\\
&\qquad \quad + \frac{\left(2 u_3-5 u_2'\right) \left(u_2 \left(2 u_3-u_2'\right)+4 u_3''-2 u_2^{(3)}\right)}{48 \zeta } \Biggr] + \nn\\
&+ \square \Biggl[ -  \frac{\left(2 u_3-5 u_2'\right) \left(2 u_3'-u_2''\right)}{24 \zeta }+\frac{1}{48} \left(40 u_3'-74 u_2''+5 u_2^2\right)\Biggr] + \nn\\
&+ \square^2 \Biggl[- \frac{\zeta }{8}+\frac{\left(2 u_3-7 u_2'\right)}{12} - \frac{\frac{1}{2} u_3 u_2'-\frac{5}{24} (u_2')^2-\frac{1}{6} u_3^2}{\zeta }\Biggr],  \\
{\, \scriptsize \yng(4,1)\, } &= \varnothing \Biggl[ - \frac{u_2}{12} \zeta +\frac{1}{36} \left(u_2 \left(4 u_3-7 u_2'\right)-6 u_3''+8 u_2^{(3)}\right) - \nn\\
&\qquad \quad - \frac{\left(2 u_3-5 u_2'\right) \left(u_2 \left(2 u_3-u_2'\right)+4 u_3''-2 u_2^{(3)}\right)}{72 \zeta } \Biggr]  + \nn\\
+\,  & \square \Biggl[ \frac{u_2^2}{36} + \frac{\left(2 u_3 - 5 u_2'\right) \left(2 u_3'-u_2''\right)}{36 \zeta }\Biggr]+ \square^2 \Biggl[ - \frac{-12 u_3 u_2'+5 (u_2')^2+4 u_3^2}{36 \zeta }+\frac{\left(2 u_3 - 5 u_2'\right)}{12} \Biggr],  \\
{\, \scriptsize \yng(3,2)\, } &= \varnothing \Biggl[ \frac{u_2}{48} \zeta +\frac{1}{72} \left(u_2 \left(7 u_2'-4 u_3\right)+u_2^{(3)}\right) + \frac{\left(2 u_3-5 u_2'\right) \left(u_2 \left(2 u_3-u_2'\right)+4 u_3'' - 2 u_2^{(3)}\right)}{144 \zeta } \Biggr] + \nn\\
& + \square \Biggl[ \frac{\left(-24 u_3'+30 u_2''+u_2^2\right)}{144}  -  \frac{\left(2 u_3-5 u_2'\right) \left(2 u_3'-u_2''\right)}{72 \zeta }\Biggr] +  \nn\\
&+\square^2 \Biggl[ \frac{\zeta }{8}+\frac{3u_2'-2 u_3}{12} + \frac{-12 u_3 u_2'+5 (u_2')^2+4 u_3^2}{72 \zeta } \Biggr]. 
\end{align}

\subsection{$(p,q)=(2,5)$ and $(r,s)=(1,3)$}

Every Young diagram is expressed by a superposition of the following four basis vectors: 
\begin{align}
\Bigl\{ \varnothing, {\, \scriptsize \yng(1)\, }, {\, \scriptsize \yng(2)\, }, {\, \scriptsize \yng(2,1)\, } \Bigr\}. 
\end{align}
Here the procedure to solve the Schur-differential equations is briefly summarized: 
\begin{itemize}
\item From the $|\lambda| = 3$-rd order equations to the $5$-th order equations, one can determine: 
\begin{align}
\Bigl\{ {\, \scriptsize \yng(3)\, }, {\, \scriptsize \yng(4)\, }, {\, \scriptsize \yng(3,1)\, }, {\, \scriptsize \yng(2,2)\, }, {\, \scriptsize \yng(5)\, }, {\, \scriptsize \yng(4,1)\, }, {\, \scriptsize \yng(3,2)\, }\Bigr\}. 
\end{align}
\item With use of the $6$-th order equations, one can determine ${\, \scriptsize \yng(1,1)\, }$ as well as the size-$6$ Young diagrams. 
With use of the $7$-th order equations, one can determine the size-$7$ Young diagrams. There appears a redundant equation in the $7$-th order equations. 
\item In the $8$- and $9$-th order equations, there is an equation which is equivalent to the bulk string equation. In the $10$- and $11$-th order equations, there are two equations which are equivalent to the bulk string equation. With use of these equations of each order can determine Young diagrams of each size, and so forth. 
\end{itemize}
The solutions to the Schur-differential equations are also shown up to $|\lambda|=6$-th order as follows: 
\begin{align}
{\, \scriptsize \yng(1,1)\, } &= \varnothing \Bigl[ \frac{\zeta}{2} + u_2  \Bigr]+{\, \scriptsize \yng(2)\, }, \\
{\, \scriptsize \yng(3)\, } &= \varnothing \Bigl[-u_2'(t)\Bigr] + {\, \scriptsize \yng(1)\, } \Bigl[ - \frac{\zeta  }{2} - u_2 \Bigr], \\
{\, \scriptsize \yng(4)\, } &= \varnothing \Bigl[ \frac{5 \left(\mu -10 u_2''-3 u_2^2\right)}{32}  \Bigr] + {\, \scriptsize \yng(1)\, } \Bigl[ -\frac{15u_2' }{8}   \Bigr] + {\, \scriptsize \yng(2)\, } \Bigl[ -\frac{5 u_2}{4}  \Bigr], \\
{\, \scriptsize \yng(3,1)\, } &= \varnothing \Bigl[ - \frac{\zeta ^2}{4} -  u_2 \zeta  - \frac{5 \mu  - 18 u_2'' + 17 u_2^2 }{32}  \Bigr] + {\, \scriptsize \yng(1)\, } \Bigl[ -\frac{u_2'}{8}   \Bigr]+ {\, \scriptsize \yng(2)\, }  \Bigl[ - \zeta -\frac{3 u_2}{4}  \Bigr], \\
{\, \scriptsize \yng(2,2)\, } &= \varnothing \Bigl[ \frac{ -5 \mu +2 u_2''+15 u_2^2 }{32}  \Bigr] + {\, \scriptsize \yng(1)\, } \Bigl[ \frac{3 u_2'}{8}  \Bigr]  + {\, \scriptsize \yng(2)\, } \Bigl[ \frac{\left(-2 \zeta +u_2\right)}{4} \Bigr], \\
{\, \scriptsize \yng(5)\, } &= \varnothing \Bigl[ - \frac{3  u_2'}{16} \zeta - \frac{u_2 u_2' + 33 u_2^{(3)} }{16}  \Bigr]  + {\, \scriptsize \yng(1)\, } \Bigl[  \frac{5 u_2}{8} \zeta + \frac{5 \left(\mu -22 u_2''+5 u_2^2\right)}{32}  \Bigr]+ \nn\\
&\quad  + {\, \scriptsize \yng(2)\, } \Bigl[ -3 u_2'(t) \Bigr]  + {\, \scriptsize \yng(2,1)\, } \Bigl[ \frac{\zeta }{2}-\frac{u_2(t)}{4} \Bigr], \\
{\, \scriptsize \yng(4,1)\, } &= \varnothing \Bigl[  - \frac{3 \zeta  u_2'}{4} - \frac{ 6 u_2 u_2' - 2 u_2^{(3)} }{4}  \Bigr] + {\, \scriptsize \yng(2)\, }
\Bigl[ -2 u_2' \Bigr] 
+ {\, \scriptsize \yng(2,1)\, } \Bigl[- \frac{\zeta }{2}-u_2 \Bigr], \\
{\, \scriptsize \yng(3,2)\, } &=  \varnothing \Bigl[ \frac{11 u_2'}{16} \zeta+ \frac{ 17 u_2 u_2' +u_2^{(3)} }{16} \Bigr] 
+{\, \scriptsize \yng(1)\, }  \Bigl[ \frac{\zeta ^2}{4} + \frac{3 u_2}{8} \zeta +\frac{-5 \mu +14 u_2''+7 u_2^2}{32} \Bigr] + \nn\\
&\quad  +  {\, \scriptsize \yng(2)\, }  \Bigl[ u_2' \Bigr] + 
{\, \scriptsize \yng(2,1)\, }  \Bigl[ - \frac{\zeta}{2} +  \frac{u_2}{4} \Bigr], \\
{\, \scriptsize \yng(6)\, } &= \varnothing \Bigl[ - \frac{\zeta ^3}{8}-\frac{7 u_2}{16} \zeta^2 - \frac{5 \mu -4 u_2''-7 u_2^2}{32} \zeta  \nn\\
&\qquad\qquad  +\frac{-15 \mu  u_2+316 (u_2')^2+2 u_2 u_2''+77 u_2^3-324 u_2^{(4)}}{128} \Bigr] + \nn\\
&\quad +{\, \scriptsize \yng(1)\, }  \Bigl[ \frac{33 u_2'}{16} \zeta  + \Bigl( \frac{137}{32} u_2 u_2' - \frac{11}{2} u_2^{(3)} \Bigr) \Bigr] + \nn\\
&\quad + {\, \scriptsize \yng(2)\, }  \Bigl[ -\frac{3 \zeta ^2}{4} + \frac{u_2}{8} \zeta + \frac{5 \mu -206 u_2'' + 39 u_2^2}{32} \Bigr] 
+ {\, \scriptsize \yng(2,1)\, }  \Bigl[-  \frac{u_2'}{8}  \Bigr], \\
{\, \scriptsize \yng(5,1)\, } &= \varnothing \Bigl[ \frac{5 u_2}{16} \zeta ^2   + \frac{5\left(\mu -22 u_2''+13 u_2^2\right)}{64} \zeta + \nn\\
&\qquad \qquad +\frac{5 \left(9 \mu  u_2+12 (u_2')^2-94 u_2 u_2''+5 u_2^3+12 u_2^{(4)}\right)}{128}  \Bigr] + \nn\\
&\quad + {\, \scriptsize \yng(1)\, }  \Bigl[ \frac{5 u_2 u_2'}{32}  \Bigr]+ 
{\, \scriptsize \yng(2)\, }  \Bigl[ \frac{5 u_2}{4} \zeta+  \frac{5\left(-22 u_2''+3  u_2^2+\mu \right)}{32}   \Bigr] + {\, \scriptsize \yng(2,1)\, }  \Bigl[- \frac{25 u_2'}{8}  \Bigr], \\
{\, \scriptsize \yng(4,2)\, } &= \varnothing \Bigl[ \frac{\zeta ^3}{8}+\frac{7 u_2}{16} \zeta ^2  + \frac{5 \mu + 42 u_2'' +u_2^2}{64} \zeta  +\frac{-5 \mu  u_2 +4 (u_2')^2+198 u_2 u_2''-17 u_2^3+4 u_2^{(4)}}{128} \Bigr] + \nn\\
&\quad + {\, \scriptsize \yng(1)\, }  \Bigl[ \frac{u_2'}{8} \zeta + \Bigl( \frac{3 u_2 u_2'}{32} +\frac{u_2^{(3)}}{2} \Bigr) \Bigr] + {\, \scriptsize \yng(2)\, }  \Bigl[ \frac{3 \zeta^2}{4} + \frac{u_2}{2} \zeta  +\frac{-5 \mu +46 u_2''+u_2^2}{32} \Bigr]  + {\, \scriptsize \yng(2,1)\, }  \Bigl[ \frac{u_2'}{8}  \Bigr], \\
{\, \scriptsize \yng(3,3)\, } &= \varnothing \Bigl[ \frac{\zeta ^3}{8}+\frac{7 u_2}{16} \zeta ^2 - \frac{u_2''-4 u_2^2}{16} \zeta +\frac{-25 \mu  u_2 +4 (u_2')^2+14 u_2 u_2''+43 u_2^3+4 u_2^{(4)}}{128} \Bigr]+ \nn\\
&\quad + {\, \scriptsize \yng(1)\, }  \Bigl[\frac{5 u_2'}{16} \zeta  + \frac{15 u_2 u_2'}{32} \Bigr] + {\, \scriptsize \yng(2)\, }  \Bigl[ \frac{\zeta^2}{2} + \frac{u_2}{8} \zeta  +\frac{-5 \mu +14 u_2''+9 u_2^2}{32} \Bigr] + {\, \scriptsize \yng(2,1)\, }  \Bigl[
\frac{9 u_2'}{8}  \Bigr]. 
\end{align}


\subsection{$(p,q)=(2,5)$ and $(r,s)=(1,2)$}

Every Young diagram is expressed by a superposition of the following four basis vectors: 
\begin{align}
\Bigl\{ \varnothing, {\, \scriptsize \yng(1)\, }, {\, \scriptsize \yng(1,1)\, }, {\, \scriptsize \yng(2,1)\, } \Bigr\}. 
\end{align}
The solutions to the Schur-differential equations are also shown up to $|\lambda|=6$-th order as follows: 
\begin{align}
 {\, \scriptsize \yng(2) \,} & = \varnothing \Bigl[ -\frac{\zeta }{2}-\frac{3 u_2}{2} \Bigr]+ {\, \scriptsize \yng(1,1)\, }, \\
 {\, \scriptsize \yng(3) \,} & =  \varnothing \Bigl[ -\frac{15 u_2'}{8}  \Bigr]+ {\, \scriptsize \yng(1)\, } \Bigl[-\frac{5 u_2}{4}  \Bigr] , \\
 {\, \scriptsize \yng(1,1,1) \,} & =  \varnothing \Bigl[-\frac{3 u_2'}{8}  \Bigr] + {\, \scriptsize \yng(1) \, }\Bigl[\frac{\zeta }{2} + \frac{u_2}{4}\Bigr], \\
 {\, \scriptsize \yng(4) \,} & = \varnothing  \Bigl[ \frac{5 u_2 }{8} \zeta   +\frac{5\left(\mu -22 u_2'' +9 u_2^2\right)}{32}   \Bigr] + {\, \scriptsize \yng(1) \, } \Bigl[ -\frac{25 u_2' }{8}   \Bigr] + {\, \scriptsize \yng(1,1)} \Bigl[-\frac{5 u_2}{4}  \Bigr], \\
 {\, \scriptsize \yng(3,1) \,} & =  \varnothing \Bigl[ -\frac{5\left(\mu -10 u_2'' -3 u_2^2\right)}{32}  \Bigr] + {\, \scriptsize \yng(1,1) \, } \Bigl[ - \frac{5 u_2}{4}   \Bigr] , \\
 {\, \scriptsize \yng(2,2) \,} & =  \varnothing \Bigl[ \frac{1}{32} \left(-5 \mu +2 u_2'' +15 u_2^2\right) \Bigr]+ {\, \scriptsize \yng(1) \, } \Bigl[ u_2'  \Bigr] + {\, \scriptsize \yng(1,1) \, } \Big[  \frac{\zeta }{2} + \frac{u_2 }{4} \Bigr], \\
 {\, \scriptsize \yng(2,1,1) \,} & = \varnothing \Bigl[ -\frac{\zeta^2}{4} - \frac{7 u_2}{8} \zeta  -\frac{3 \left(u_2'' +u_2^2\right)}{8}    \Bigr] + {\, \scriptsize \yng(1) \, } \Bigl[-\frac{u_2'  }{8}  \Bigr] + {\, \scriptsize \yng(1,1)\, }  \Bigl[ \zeta + \frac{u_2 }{2} \Bigr],  \\
 {\, \scriptsize \yng(5) \,} & =\varnothing \Bigl[ \frac{27 u_2'}{16} \zeta   +\frac{3  \bigl(69 u_2 u_2' -58 u_2^{(3)}\bigr)}{32} \Bigr] + {\, \scriptsize \yng(1)\, } \Bigl[ \frac{5\left(\mu -42 u_2'' +7 u_2^2\right)}{32}  \Bigr]  + \nn\\
&\quad + {\, \scriptsize \yng(1,1) \,} \Bigl[ -\frac{9 u_2' }{2}  \Bigr] + {\, \scriptsize \yng(2,1) \, } \Bigl[ \frac{\zeta }{2} + \frac{u_2}{4} \Bigr], \\
 {\, \scriptsize \yng(4,1) \,} & =\varnothing \Bigl[  \frac{u_2' }{2} \zeta  +\Bigl( \frac{3 u_2 u_2'}{2} +2 u_2^{(3)}\Bigr) \Bigr] + {\, \scriptsize \yng(1,1)\,} \Bigl[ -3 u_2'  \Bigr]+ {\, \scriptsize \yng(2,1) \,} \Bigl[ - \frac{\zeta }{2}-\frac{3 u_2 }{2}  \Bigr], \\
 {\, \scriptsize \yng(3,2) \,} & = {\, \scriptsize \yng(1)} \Bigl[ -\frac{5 \left(\mu -10 u_2'' -3 u_2 ^2\right)}{32}  \Bigr]+  {\, \scriptsize \yng(1,1) \,} \Bigl[ \frac{15 u_2' }{8}  \Bigr], \\
 {\, \scriptsize \yng(3,1,1) \,} & = \varnothing \Bigl[ - \frac{u_2'}{2} \zeta   - \frac{ 3 u_2 u_2' +14 u_2^{(3)} }{32}  \Bigr] + {\, \scriptsize \yng(1) \, } \Bigl[ -  \frac{5 u_2}{8} \zeta    -\frac{5 u_2^2}{16}  \Bigr]  +   {\, \scriptsize \yng(1,1) \, } \Bigl[-\frac{u_2'}{8}  \Bigr] + \nn\\
&\quad + {\, \scriptsize \yng(2,1) \, } \Bigl[ \frac{\zeta }{2} + \frac{u_2}{4}  \Bigr], \\
 {\, \scriptsize \yng(2,2,1) \,} & = \varnothing \Bigl[  - \frac{11  u_2'}{16}   \zeta  +  \frac{ 2 u_2^{(3)} -21 u_2 u_2' }{32}  \Bigr] + {\, \scriptsize \yng(1) \,}  \Bigl[ \frac{\zeta ^2}{4}      + \frac{u_2}{4} \zeta     +   \frac{ u_2^2-8 u_2'' }{16}   \Bigr]  + {\, \scriptsize \yng(1,1)\, } \Bigl[ \frac{3  u_2'}{8}  \Bigr] + \nn\\
&\quad + {\, \scriptsize \yng(2,1) \,} \Bigl[ \frac{\zeta }{2} + \frac{u_2 }{4}  \Bigr], \\
 {\, \scriptsize \yng(6) \,} & = \varnothing \Bigl[-  \frac{\zeta^3}{8}-\frac{u_2}{2} \zeta^2  - \frac{\left(15 \mu -358 u_2''+31 u_2^2\right)}{64} \zeta   + \nn\\
& \quad\quad +\frac{  \bigl(2138 u_2'' -65 \mu \bigr) u_2 +804 (u_2')^2-117 u_2^3-1016 u_2^{(4)} }{128}   \Bigr] + \nn\\
&\quad + {\, \scriptsize \yng(1) \,} \Bigl[  - \frac{u_2'}{8} \zeta   + \Bigl( \frac{31  u_2 u_2'}{4} -12 u_2^{(3)} \Bigr)  \Bigr]  + \nn\\
&\quad + {\, \scriptsize \yng(1,1) \,} \Bigl[ \frac{3 \zeta^2}{4} + \frac{u_2}{8} \zeta   + \frac{5 \mu -354 u_2'' +31 u_2^2 }{32}  \Bigr]+ {\, \scriptsize \yng(2,1) \,}  \Bigl[ \frac{u_2' }{8}  \Bigr], \\
 {\, \scriptsize \yng(5,1) \,} & = \varnothing \Bigl[ \frac{5 \Bigl(\bigl(5 \mu -2 u_2'' \bigr) u_2 +48 (u_2')^2-15 u_2^3+64 u_2^{(4)}\Bigr)}{128}  \Bigr]+ \nn\\
&\quad + {\, \scriptsize \yng(1,1) \,} \Bigl[ \frac{5\left(\mu -42 u_2'' +7 u_2^2\right)}{32}  \Bigr] + {\, \scriptsize \yng(2,1) \,} \Bigl[ - \frac{35 u_2' }{8} \Bigr], \\
 {\, \scriptsize \yng(4,2) \,} & = \varnothing \Bigl[ \frac{\zeta ^3}{8}+\frac{ u_2}{2} \zeta ^2 - \frac{\left(-5 \mu +4 u_2'' +2 u_2^2\right)}{32} \zeta  + \nn\\
&\quad \quad +\frac{ \bigl(35 \mu -38 u_2''\bigr) u_2+6 (u_2')^2-93 u_2^3-4 u_2^{(4)} }{128}   \Bigr] + \nn\\
&\quad + {\, \scriptsize \yng(1) \, } \Bigl[ \frac{u_2'}{8} \zeta   + \Bigl(\frac{ u_2 u_2'}{16} +2 u_2^{(3)}\Bigr)  \Bigr] 
 + \nn\\
&\quad + {\, \scriptsize \yng(1,1) \,} \Bigl[ -\frac{3 \zeta^2}{4} - \frac{3 u_2}{4} \zeta  + \frac{-5 \mu +114 u_2'' +9 u_2^2 }{32}  \Bigr]+ {\, \scriptsize \yng(2,1) \,} \Bigl[-\frac{u_2' }{8}  \Bigr], \\
 {\, \scriptsize \yng(4,1,1) \,} & = \varnothing \Bigl[\frac{5 u_2}{16} \zeta ^2  + \frac{35  u_2^2}{32} \zeta   +   \frac{15 \bigl(3 (u_2')^2+2 u_2^3-2 u_2^{(4)} \bigr)}{64}  \Bigr] + \nn\\
&\quad + {\, \scriptsize \yng(1) \,} \Bigl[ - \frac{25 u_2'}{16} \zeta   -\frac{5 u_2 u_2' }{8} \Bigr] + {\, \scriptsize \yng(1,1) \,} \Bigl[ - \frac{5 u_2}{4} \zeta  -\frac{5 u_2^2}{8}  \Bigr], \\
 {\, \scriptsize \yng(3,3) \,} & = \varnothing \Bigl[  \frac{5 \left(\mu -10 u_2'' -3 u_2^2\right)}{64} \zeta  +\frac{5 u_2 \left(\mu -58 u_2'' -3 u_2^2\right) }{128}   \Bigr] + {\, \scriptsize \yng(1) \,} \Bigl[ \frac{5 u_2 u_2' }{4}  \Bigr] + \nn\\
&\quad   + {\, \scriptsize \yng(1,1) \, } \Bigl[ \frac{5 u_2}{8} \zeta  -\frac{ 5 \left(\mu -10 u_2'' -5 u_2^2\right) }{32}  \Bigr]+ {\, \scriptsize \yng(2,1) \,} \Bigl[ \frac{15 u_2' }{8}  \Bigr], \\
 {\, \scriptsize \yng(3,2,1) \,} & = \varnothing \Bigl[-  \frac{\zeta^3}{8}-\frac{u_2}{2} \zeta^2  -   \frac{5 \mu -4 u_2'' -2 u_2^2 }{32}  \zeta + \nn\\
&\quad\quad  +\frac{u_2  \left(14 u_2'' -5 \mu \right)-48 (u_2')^2+9 u_2^3+2 u_2^{(4)} }{64}   \Bigr] + \nn\\
&\quad + {\, \scriptsize \yng(1) \, } \Bigl[ \frac{13 u_2' }{16}  \zeta  + \Bigl( \frac{3 u_2 u_2'}{32}  -\frac{7 u_2^{(3)} }{16} \Bigr)  \Bigr]+ \nn\\
&\quad  + {\, \scriptsize \yng(1,1) \,} \Bigl[  \frac{3 \zeta ^2}{4} + \frac{ u_2}{8} \zeta  -\frac{u_2'' + u_2^2 }{8}  \Bigr]  + {\, \scriptsize \yng(2,1) \,} \Bigl[  \frac{u_2' }{8}  \Bigr], \\
 {\, \scriptsize \yng(2,2,2) \,} & = \varnothing \Bigl[ - \frac{\zeta ^3}{8}-\frac{u_2}{2} \zeta^2  -  \frac{ 13 u_2^2-2 u_2'' }{32}  \zeta +  \frac{ -3 (u_2')^2+12 u_2 u_2'' -6 u_2^3+2 u_2^{(4)}  }{64} \Bigr]  + \nn\\
&\quad + {\, \scriptsize \yng(1) \,} \Bigl[ - \frac{5u_2' }{8} \zeta   -\frac{5u_2 u_2'}{16}  \Bigr] + {\, \scriptsize \yng(1,1) \,} \Bigl[ \frac{\zeta ^2}{2} + \frac{u_2}{2} \zeta  + \frac{u_2^2-4 u_2'' }{8}  \Bigr]  + {\, \scriptsize \yng(2,1) \,} \Bigl[  \frac{u_2' }{2}  \Bigr]. 
\end{align}


\subsection{$(p,q)=(3,5)$ and $(r,s)=(1,3)$}

Every Young diagram is expressed by a superposition of the following six basis vectors: 
\begin{align}
\Bigl\{ \varnothing, {\, \scriptsize \yng(1)\, }, {\, \scriptsize \yng(2)\, }, {\, \scriptsize \yng(1,1)\, }, {\, \scriptsize \yng(2,1)\, }, {\, \scriptsize \yng(3,1)\, } \Bigr\}. 
\end{align}
Here the procedure to solve the Schur-differential equations is briefly summarized: 
\begin{itemize}
\item From the $|\lambda| = 3$-rd order equations, one can solve the Young diagrams of each order by the above basis vectors. 
\item In the $8$- and $9$-th order equations, there is an equation which is equivalent to the bulk string equation. In the $10$- and $11$-th order equations, there are two equations which are equivalent to the bulk string equation, and so forth. 
\end{itemize}
The solutions to the Schur-differential equations are also shown up to $|\lambda|=6$-th order as follows: 
\begin{align}
{\, \scriptsize \yng(3)\, } &= \varnothing \Bigl[ - \frac{\zeta}{4}  + \frac{-3 u_2'+2 u_3}{4} \Bigr] + {\, \scriptsize \yng(1)\, } \Bigl[ -\frac{u_2}{4}  \Bigr]  + {\, \scriptsize \yng(2,1)\, }, \\
{\, \scriptsize \yng(4)\, } &= \varnothing \Bigl[-\frac{5 (-12 u_3'+20 u_2'' +u_2^2)}{144}  \Bigr] + {\, \scriptsize \yng(1)\, } \Bigl[\frac{5 (u_3-2 u_2' ) }{12} \Bigr] + {\, \scriptsize \yng(2)\, } \Bigl[ - \frac{5 u_2}{12}  \Bigr], \\
{\, \scriptsize \yng(2,2)\, } &= \varnothing \Bigl[ \frac{-12 u_3' +8 u_2'' -5 u_2^2}{144} \Bigr] 
+{\, \scriptsize \yng(1)\, } \Bigl[ \frac{\zeta}{4}  +  \frac{2 u_2'-u_3}{12}  \Bigr]
+ {\, \scriptsize \yng(2)\, } \Bigl[ -\frac{u_2}{6}   \Bigr] 
+ {\, \scriptsize \yng(1,1)\, } \Bigl[ \frac{u_2}{4} \Bigr], \\
{\, \scriptsize \yng(5)\, } &= \varnothing \Bigl[ \frac{1}{16} \left(\zeta  u_2+3 u_2 u_2' +12 u_3'' -2 u_3 u_2-16 u_2^{(3)} \right) \Bigr] +{\, \scriptsize \yng(1)\, } \Bigl[  \frac{12 u_3'-24 u_2''+u_2^2}{16} \Bigr] + \nn\\
&\quad 
+ {\, \scriptsize \yng(2)\, } \Bigl[ - \frac{\zeta}{4}- \frac{1}{4} \left(5 u_2' - 2 u_3\right) \Bigr] 
+ {\, \scriptsize \yng(2,1)\, } \Bigl[ -\frac{u_2}{2}  \Bigr], \\
{\, \scriptsize \yng(4,1)\, } &= \varnothing \Bigl[ \frac{1}{72} \left( 3 \zeta  u_2+4 u_2 u_2' -24 u_3'' -6 u_3 u_2+22 u_2^{(3)} \right) \Bigr] +{\, \scriptsize \yng(1)\, } \Bigl[ \frac{12 u_3' -4 u_2'' +u_2^2}{144} \Bigr] + \nn\\
&\quad 
+ {\, \scriptsize \yng(2)\, } \Bigl[  \frac{\zeta }{4}-\frac{u_3}{12} \Bigr]
+ {\, \scriptsize \yng(1,1)\, } \Bigl[ \frac{5 (u_3 -2 u_2' )}{12}  \Bigr]
+ {\, \scriptsize \yng(2,1)\, } \Bigl[ -\frac{u_2}{3}  \Bigr], \\
{\, \scriptsize \yng(3,2)\, } &= \varnothing \Bigl[ \frac{1}{72} \left( 3 \zeta  u_2 +4 u_2 u_2' -6 u_3'' -6 u_3 u_2 +4 u_2^{(3)}\right) \Bigr]
+{\, \scriptsize \yng(1)\, } \Bigl[ \frac{-24 u_3' +32 u_2'' +u_2^2}{144} \Bigr]+ \nn\\
&\quad 
+ {\, \scriptsize \yng(2)\, } \Bigl[ \frac{\zeta }{4}-\frac{u_3}{12} \Bigr] 
+{\, \scriptsize \yng(1,1)\, } \Bigl[  \frac{\zeta}{4}+ \frac{5 u_2'-u_3}{12} \Bigr] 
+ {\, \scriptsize \yng(2,1)\, } \Bigl[ -\frac{u_2}{12}  \Bigr], \\
{\, \scriptsize \yng(6)\, } &= \varnothing \Bigl[ \frac{1}{288} \Bigl(18 \zeta ^2  + 18(9  u_2' - 4  u_3) \zeta -4 u_2 u_3' +304 (u_2')^2-304 u_3 u_2'+26 u_2 u_2'' + \nn\\
&\qquad \quad +5 u_2^3+72 u_3^2+296 u_3^{(3)} -368 u_2^{(4)} \Bigr) \Bigr] + {\, \scriptsize \yng(1)\, } \Bigl[ \frac{13 u_2 u_2' +36 u_3'' -5 u_2 u_3 -60 u_2^{(3)}}{24} \Bigr] + \nn\\
&\quad + {\, \scriptsize \yng(2)\, } \Bigl[ \frac{60 u_3' -132 u_2'' +7 u_2^2}{48} \Bigr] + {\, \scriptsize \yng(1,1)\, } \Bigl[\frac{-3 u_3'+u_2''-u_2^2}{36} \Bigr]
+ {\, \scriptsize \yng(2,1)\, } \Bigl[ \frac{- 6 \zeta -21 u_2'+7 u_3}{12} \Bigr] + \nn\\
&\quad + {\, \scriptsize \yng(3,1)\, } \Bigl[ -\frac{u_2}{12}  \Bigr], \\
{\, \scriptsize \yng(5,1)\, } &= \varnothing \Bigl[ -\frac{5}{72} \Bigl(-(u_2')^2+u_3 u_2' +u_2 u_3' -u_2 u_2'' +4 u_3^{(3)} -4 u_2^{(4)} \Bigr) \Bigr] + \nn\\
&+ {\, \scriptsize \yng(1,1)\, } \Bigl[ \frac{5(24 u_3'  - 44 u_2''  - u_2^2)}{144}  \Bigr]
+{\, \scriptsize \yng(2,1)\, } \Bigl[  \frac{5 (u_3-3 u_2' )}{12} \Bigr]
+ {\, \scriptsize \yng(3,1)\, } \Bigl[ -  \frac{5 u_2}{12}  \Bigr], \\
{\, \scriptsize \yng(4,2)\, } &= \varnothing \Bigl[ -\frac{\zeta ^2}{16} + \frac{7(u_3-u_2')}{48} \zeta +\frac{1}{432} \Bigl(6 u_2 u_3' -6 (u_2')^2+21 u_3 u_2' -14 u_2 u_2'' +5 u_2^3 - \nn\\
&\qquad \qquad\qquad \qquad\qquad \qquad \qquad \qquad    -18 u_3^2-24 u_3^{(3)}+12 u_2^{(4)}\Bigr) \Bigr] + \nn\\
&\quad +{\, \scriptsize \yng(1)\, } \Bigl[ - \frac{5 u_2}{48} \zeta  +\frac{1}{144} \left(2 u_2 u_2' -36 u_3'' -5 u_2 u_3 +40 u_2^{(3)} \right) \Bigr] +{\, \scriptsize \yng(2)\, } \Bigl[\frac{9 u_2^2-4 u_2''}{144} \Bigr] + \nn\\
&\quad  + {\, \scriptsize \yng(1,1)\, } \Bigl[ \frac{-8 u_3' +16 u_2'' -u_2^2}{24} \Bigr]
+ {\, \scriptsize \yng(2,1)\, } \Bigl[  \frac{\zeta}{2} + \frac{ u_2' -2 u_3 }{12}  \Bigr]
+{\, \scriptsize \yng(3,1)\, } \Bigl[ \frac{u_2}{12}  \Bigr], \\
{\, \scriptsize \yng(3,3)\, } &= 
\varnothing \Bigl[ \frac{1}{576} \left(-52 u_2 u_3'+40 (u_2')^2-40 u_3 u_2'+48 u_2 u_2'' -5 u_2^3-16 u_3^{(3)} +16 u_2^{(4)} \right) \Bigr]+ \nn\\
&\quad +{\, \scriptsize \yng(1)\, } \Bigl[  \frac{u_2}{16} \zeta  +\frac{u_2 (2 u_2'+u_3)}{48}  \Bigr]
+{\, \scriptsize \yng(2)\, } \Bigl[ \frac{-6 u_3'+6 u_2'' -u_2^2}{24} \Bigr]
+{\, \scriptsize \yng(1,1)\, } \Bigl[ \frac{3 u_3' -u_2'' +u_2^2}{36} \Bigr] + \nn\\
&\quad +{\, \scriptsize \yng(2,1)\, } \Bigl[ \frac{\zeta }{4} + \frac{3 u_2'-u_3}{12} \Bigr]
+{\, \scriptsize \yng(3,1)\, } \Bigl[-\frac{u_2}{6}  \Bigr]. 
\end{align}

\subsection{$(p,q)=(3,5)$ and $(r,s)=(1,2)$}

Every Young diagram is expressed by a superposition of the following six basis vectors: 
\begin{align}
\Bigl\{ \varnothing, {\, \scriptsize \yng(1)\, }, {\, \scriptsize \yng(2)\, }, {\, \scriptsize \yng(1,1)\, }, {\, \scriptsize \yng(2,1)\, }, {\, \scriptsize \yng(2,1,1)\, } \Bigr\}. 
\end{align}
The solutions to the Schur-differential equations are also shown up to $|\lambda|=6$-th order as follows: 
\begin{align}
 {\, \scriptsize \yng(3)\, }& = \varnothing \Bigl[ \frac{5\left(u_3-2 u_2'\right)}{12}  \Bigr] + {\, \scriptsize \yng(1) \, } \Bigl[ -\frac{5 u_2}{12}  \Bigr], \\
 {\, \scriptsize \yng(1,1,1)\, } & = \varnothing \Bigl[ - \frac{\zeta }{4} -\frac{2 u_2' - u_3}{3}  \Bigr] + {\, \scriptsize \yng(1) \,} \Bigl[ \frac{ u_2}{6}  \Bigr] + {\, \scriptsize \yng(2,1) \, }, \\
  {\, \scriptsize \yng(4)\, } & =  {\varnothing} \Bigl[-\frac{5 \left(-24 u_3' +44 u_2'' +u_2^2\right)}{144} \Bigr]+ {\, \scriptsize \yng(1) \,} \Bigl[ \frac{5\left(u_3 -3 u_2' \right)}{12}  \Bigr]+ {\, \scriptsize \yng(2) \,} \Bigl[- \frac{5  u_2 }{12} \Bigr], \\
 {\, \scriptsize \yng(3,1)\, } & = \varnothing \Bigl[\frac{5 \left(-12 u_3' +20 u_2'' +u_2^2\right)}{144}  \Bigr] + {\, \scriptsize \yng(1,1)\,} \Bigl[-\frac{5 u_2  }{12} \Bigr], \\
 {\, \scriptsize \yng(2,2)\, } & = {\varnothing} \Bigl[ \frac{\left(12 u_3' -4 u_2'' -5 u_2^2\right)}{144}  \Bigr] + {\, \scriptsize \yng(1) \, } \Bigl[  \frac{\zeta }{4}+ \Bigl( \frac{u_2'}{2}  -\frac{u_3}{3} \Bigr) \Bigr] +{\, \scriptsize \yng(2) \,} \Bigl[ -\frac{ u_2 }{6}\Bigr] + {\, \scriptsize \yng(1,1) \,} \Bigl[ \frac{ u_2 }{4} \Bigr], \\ 
  {\, \scriptsize \yng(5)\, } & = \varnothing \Bigl[ - \frac{u_2}{16} \zeta    +   \frac{u_2  \bigl(16 u_2' -13 u_3 \bigr)+4 \bigl(57 u_3'' -91 u_2^{(3)} \bigr) }{144}   \Bigr] + \nn\\
&\quad
+ {\, \scriptsize \yng(1)\, } \Bigl[ \frac{7\left(24 u_3' -56 u_2'' +3 u_2^2\right) }{144} \Bigr]+{\, \scriptsize \yng(2) \,} \Bigl[ \frac{5 (u_3 - 4 u_2')}{12}   \Bigr]
 +\nn\\
&\quad 
+ {\, \scriptsize \yng(1,1) \,} \Bigl[  \frac{\zeta }{4}+\frac{- u_2' + u_3}{12} \Bigr] + {\, \scriptsize \yng(2,1) \, } \Bigl[ \frac{u_2}{12}  \Bigr], \\ 
 {\, \scriptsize \yng(4,1)\, } & =\varnothing \Bigl[ \frac{ u_2}{16} \zeta +   \Bigl( -\frac{ u_2 \left(u_3 -2 u_2' \right)}{12}-\frac{3 u_3''}{4}  +u_2^{(3)}  \Bigr) \Bigr]  + {\, \scriptsize \yng(1) \,} \Bigl[  \frac{ 12 u_3' -8 u_2'' -u_2^2 }{144}\Bigr] +\nn\\
&\quad 
+ {\, \scriptsize \yng(1,1) \, } \Bigl[ - \frac{\zeta }{4} +  \frac{- 7 u_2' + 2 u_3}{6}  \Bigr] + {\, \scriptsize \yng(2,1) \,} \Bigl[ -\frac{ u_2  }{2} \Bigr], \\
 {\, \scriptsize \yng(3,2)\, } & = {\, \scriptsize \yng(1) \,} \Bigl[ \frac{5 \left(-12 u_3' +20 u_2'' +u_2^2\right)}{144}  \Bigr] 
+ {\, \scriptsize \yng(1,1) \,} \Bigl[ -\frac{5 \left(u_3 -2 u_2' \right)}{12}   \Bigr], \\
 {\, \scriptsize \yng(3,1,1)\, } & = \varnothing \Bigl[  \frac{u_2}{24} \zeta    +  \frac{ u_2 \left(13 u_2'-4 u_3\right)+24 u_3'' -22 u_2^{(3)} }{72}  \Bigr] +{\, \scriptsize \yng(1) \,} \Bigl[  \frac{ -12 u_3' +8 u_2'' -9 u_2^2 }{144}  \Bigr] + \nn\\
&\quad 
+{\, \scriptsize \yng(1,1) \,} \Bigl[   \frac{\zeta }{4} + \frac{- u_2' + u_3 }{12}  \Bigr] +{\, \scriptsize \yng(2,1)\, } \Bigl[ -\frac{ u_2 }{3} \Bigr], \\
 {\, \scriptsize \yng(2,2,1)\, } & = \varnothing \Bigl[ - \frac{u_2 }{16} \zeta + \frac{ u_2 \left(u_3 -7 u_2' \right)+6 u_3'' -2 u_2^{(3)} }{72}  \Bigr] +{\, \scriptsize \yng(1) \,} \Bigl[ \frac{ 12 u_3' -16 u_2'' +3 u_2^2 }{72} \Bigr] + \nn\\
&\quad 
+{\, \scriptsize \yng(2) \,} \Bigl[  \frac{\zeta }{4}+\frac{u_2'  - u_3  }{3}   \Bigr]  
+{\, \scriptsize \yng(1,1) \,} \Bigl[   \frac{\zeta }{4} + \frac{- u_2' + u_3 }{12}  \Bigr] 
+{\, \scriptsize \yng(2,1) \,} \Bigl[ \frac{u_2}{3} \Bigr], \\
 {\, \scriptsize \yng(6)\, } & = \varnothing\Bigl[ -\frac{\zeta ^2}{16} -  \frac{ 10 u_2' -3 u_3 }{48}   \zeta   +  \frac{1}{1728} \Bigl\{ 25 u_2^3  +  u_2  \bigl(428 u_2''-264 u_3' \bigr) + \nn\\
&\qquad \qquad \qquad + 12 \Bigl(-155 u_3 u_2'+194 (u_2')^2+29 u_3^2+368 u_3^{(3)} -544 u_2^{(4)} \Bigr)\Bigr\} \Bigr] + \nn\\
&\quad 
  +{\, \scriptsize \yng(1) \, } \Bigl[  \frac{ u_2 \bigl(81 u_2' -20 u_3 \bigr)+198 u_3'' -378 u_2^{(3)} }{72} \Bigr] + {\, \scriptsize \yng(2) \,} \Bigl[  \frac{ 228 u_3' -632 u_2'' +19 u_2^2 }{144}  \Bigr] + \nn\\
&\quad 
+{\, \scriptsize \yng(1,1) \,} \Bigl[   \frac{ -4 u_2'' -u_2^2 }{144} \Bigr] +{\, \scriptsize \yng(2,1) \,} \Bigl[  \frac{\zeta }{2} + \frac{ - u_2' + 2 u_3(t)}{12} \Bigr] +{\, \scriptsize \yng(2,1,1) \,} \Bigl[ \frac{  u_2 }{12} \Bigr], \\
 {\, \scriptsize \yng(5,1)\, } & = \varnothing \Bigl[-\frac{5 \Bigl(5 u_2^3 -4 u_2 \left(3 u_3' -7 u_2'' \right)+ 24 \bigl(-3 (u_2')^2+2 u_3 u_2' +14 u_3^{(3)} -18 u_2^{(4)} \bigr)\Bigr)}{1728}    \Bigr] + \nn\\
&\quad 
+{\, \scriptsize \yng(1,1) \,} \Bigl[ \frac{5 \left(9 u_3' -20 u_2'' +u_2^2\right)}{36}   \Bigr] +{\, \scriptsize \yng(2,1) \,} \Bigl[ \frac{5 \left(u_3 -4 u_2' \right)}{12}   \Bigr], \\
 {\, \scriptsize \yng(4,2)\, } & = \varnothing \Bigl[ \frac{\zeta^2}{16} - \frac{ 3 u_3 -10 u_2' }{48}  \zeta   + \nn\\
&\quad \quad 
+\frac{25 u_2^3 -4 u_2 \left(9 u_3' +7 u_2'' \right)-24 \left(-5 u_3  u_2' +2 (u_2')^2+2 u_3^2+4 u_3^{(3)}-2 u_2^{(4)} \right)}{1728} \Bigr] + \nn\\
&\quad 
+{\, \scriptsize \yng(1) \,} \Bigl[ - \frac{5  u_2}{48} \zeta  + \frac{u_2  \left(5 u_3 -6 u_2' \right)-48 u_3'' +68 u_2^{(3)} }{72}  \Bigr] +{\, \scriptsize \yng(2)\, } \Bigl[ \frac{ 12 u_3' -8 u_2'' +11 u_2^2  }{144}  \Bigr] + \nn\\
&
+{\, \scriptsize \yng(1,1) \, } \Bigl[ -\frac{5 u_3' }{6} +\frac{14  u_2'' }{9}-\frac{ u_2^2}{16} \Bigr]  
+{\, \scriptsize \yng(2,1) \,} \Bigl[ - \frac{\zeta }{2}+\frac{ u_2' -2 u_3 }{12}   \Bigr]
+ {\, \scriptsize \yng(2,1,1)\,} \Bigl[ -\frac{ u_2 }{12}   \Bigr], \\
 {\, \scriptsize \yng(4,1,1)\, } & = \varnothing \Bigl[ - \frac{5 \left(u_3 -3 u_2' \right)}{48} \zeta   +\frac{5 \left(-9 u_3 u_2' +11 (u_2')^2+u_2 u_3' -u_2 u_2'' +2 u_3^2+4 u_3^{(3)} -4 u_2^{(4)}\right) }{72} \Bigr] +\nn\\
&\quad 
+{\, \scriptsize \yng(1) \,} \Bigl[ \frac{5 u_2 \left(u_3 -3 u_2' \right) }{72} \Bigr] +{\, \scriptsize \yng(2,1) \,} \Bigl[ \frac{5 \left(u_3 -3 u_2' \right) }{12}  \Bigr] +{\, \scriptsize \yng(2,1,1) \,} \Bigl[ - \frac{5   u_2 }{12} \Bigr], \\
  {\, \scriptsize \yng(3,3)\, } & = \varnothing \Bigl[ -\frac{5 u_2 \bigl(4 \left(u_2'' -3 u_3'\right)+5 u_2^2\bigr)}{1728}\Bigr]  +{\, \scriptsize \yng(1)\,} \Bigl[    \frac{5  u_2 }{48} \zeta +\frac{5 u_2 \left(3 u_2' -2 u_3 \right)}{72}  \Bigr]+ \nn\\
&\quad 
+{\, \scriptsize \yng(2) \,} \Bigl[ - \frac{5 \left(12 u_3' -20 u_2'' +u_2^2\right)}{144}  \Bigr] 
+{\, \scriptsize \yng(1,1) \,} \Bigl[ \frac{5 u_2^2 }{48}  \Bigr] 
+{\,\scriptsize \yng(2,1) \,} \Bigl[ - \frac{5  \left(u_3 -2 u_2' \right)}{12} \Bigr], \\
 {\, \scriptsize \yng(3,2,1)\, } & = \varnothing \Bigl[ -\frac{\zeta^2}{16}-  \frac{5 u_2' -2 u_3 }{12}  \zeta  + \nn\\
&\quad \quad +\frac{ 35 u_3 u_2' -38 (u_2')^2-u_2 u_3' +2 u_2  u_2'' -8 u_3^2+4 u_3^{(3)} -2 u_2^{(4)} }{72}  \Bigr] + \nn\\
&\quad 
+{\, \scriptsize \yng(1) \,} \Bigl[  \frac{ u_2 u_2' +3 u_3'' -3 u_2^{(3)} }{12} \Bigr] +{\, \scriptsize \yng(2) \,} \Bigl[  \frac{-12 u_3' +8 u_2'' -u_2^2 }{144}  \Bigr] + \nn\\
&\quad 
+{\, \scriptsize \yng(1,1)\,} \Bigl[  \frac{-4 u_2'' -u_2^2 }{144}  \Bigr] 
+{\, \scriptsize \yng(2,1)\,} \Bigl[ \frac{\zeta }{2}+\frac{3  u_2' -u_3  }{4}  \Bigr] 
+{\, \scriptsize \yng(2,1,1)\,} \Bigl[ \frac{ u_2 }{12} \Bigr], \\
 {\, \scriptsize \yng(2,2,2)\, } & = \varnothing \Bigl[ \frac{15 (u_2')^2-10 u_3 u_2' -6 u_2 u_3' +7 u_2  u_2'' +2 u_3^{(3)} }{72}  \Bigr]+ \nn\\
&\quad  
+{\, \scriptsize \yng(1) \,} \Bigl[ - \frac{u_2}{24} \zeta  +  \frac{u_2  \left(2 u_3 -3 u_2' \right)}{36} \Bigr] +{\, \scriptsize \yng(2) \, } \Bigl[  \frac{ -12 u_3' +8 u_2'' -u_2^2 }{144}\Bigr]
 + \nn\\
&\quad 
 +{\,\scriptsize \yng(1,1) \,} \Bigl[  \frac{ 36 u_3' -40 u_2'' -u_2^2 }{144}  \Bigr]
+{\, \scriptsize \yng(2,1)\,} \Bigl[ \frac{\zeta }{4} + \frac{ u_3 -3 u_2' }{12} \Bigr]
 +{\, \scriptsize \yng(2,1,1)\, } \Bigl[ \frac{ u_2 }{4}  \Bigr]. 
\end{align}


\section{Notes on boundary entropies}

In this appendix, we would like to evaluate

\begin{equation}
A_s=(-1)^{s-1}\sum_{
\begin{subarray}{c}
l=-(s-1) \cr
\text{step 2}
\end{subarray}
}^{s-1}\frac{\zeta_{0,l}}{\zeta}\ , \quad \widetilde{A}_r= (-1)^{r-1}\sum_{
\begin{subarray}{c}
k=-(r-1) \cr
\text{step 2}
\end{subarray}
}^{r-1}\frac{\eta_{0,k}}{\eta}\ .
\end{equation}

\noindent Since we sum over values of $s$ centered symmetrically around  $s=0$, we observe that the $\tau$-dependence in each term cancels:

\begin{equation}
\begin{aligned}
A_s&=\frac{(-1)^{s-1}}{2\zeta}
\sum_{
\begin{subarray}{c}
l=-(s-1) \cr
\text{step 2}
\end{subarray}
}^{s-1}
\left(e^{p\tau+i\pi lp/q}+e^{-p\tau-i\pi lp/q}\right)\\
&=\frac{(-1)^{s-1}}{2\zeta}
\sum_{
\begin{subarray}{c}
l=-(s-1) \cr
\text{step 2}
\end{subarray}
}^{s-1}
e^{i\pi lp/q}\left(e^{p\tau}+e^{-p\tau}\right)\\
&=(-1)^{s-1}
\sum_{
\begin{subarray}{c}
l=-(s-1) \cr
\text{step 2}
\end{subarray}
}^{s-1}
e^{i\pi lp/q}=(-1)^{s-1}\frac{\sin(\pi s p/q)}{\sin(\pi p/q)}\ .
\end{aligned}
\end{equation}

\noindent In an entirely analogous manner one obtains $\widetilde{A}_r=(-1)^{r-1}\sin(\pi rq/p)/\sin(\pi q/p)$. Other typical ways of expressing these real numbers is via traces of $SU(2)$ characters in the spin $j=(s-1)/2$-representation: 
\begin{align}
\frac{\sin(\pi s p/q)}{\sin(\pi p/q)}=\mathrm{tr}_{j=\frac{s-1}{2}}e^{2i \pi\frac{p}{q}J_3}=U_{s-1}(\cos\pi p/q), 
\end{align}
where $U_{n-1}(\cos\theta) = \sin(n\theta)/\sin\theta$ are the Chebyshev polynomials of the second kind. 
Note also that the explicit form of the modular $S$-matrix of the $(p,q)$ minimal model (see e.g.~\cite{FMS-CFT}),

\begin{equation}
S_{(r,s)(m,n)}=2\sqrt{\frac{2}{pq}}(-1)^{sm+rn+1}\sin(\pi r mq/p)\sin(\pi s np/q)\ ,
\end{equation}

\noindent implies the relation 

\begin{equation}
d_{r,s}=A_s\widetilde{A}_r=\frac{
S_{(r,s)(1,1)}}{
S_{(1,1)(1,1)}}\ .
\end{equation}

\noindent For this reason, $d_{r,s}$ is sometimes called the quantum dimension of the state $(r,s)$; The definition of Cardy states also gave rise to the interpretation of these numbers as ground state degeneracies, and their logarithm as ``boundary entropies'' \cite{AffleckLudwig,FriedanKonechny}.


\end{document}